\documentclass{article}

\usepackage{amsmath}
\usepackage{amssymb}
\usepackage{physics}
\usepackage{graphicx}
\usepackage{tikz-cd}
\usepackage{authblk}
\usepackage{hyperref}
\usepackage{setspace}
\usepackage{fullpage}

\newtheorem{definition}{Definition}

\title{Algorithmic Causal Sets and the Wolfram Model}
\author{Jonathan Gorard\footnote{\url{jg865@cam.ac.uk}}}

\begin{document}
\maketitle

\begin{abstract}
The formal relationship between two differing approaches to the description of spacetime as an intrinsically discrete mathematical structure, namely causal set theory and the Wolfram model, is studied, and it is demonstrated that the hypergraph rewriting approach of the Wolfram model can effectively be interpreted as providing an underlying algorithmic dynamics for causal set evolution. We show how causal invariance of the hypergraph rewriting system can be used to infer conformal invariance of the induced causal partial order, in a manner that is provably compatible with the measure-theoretic arguments of Bombelli, Henson and Sorkin. We then illustrate how many of the local dimension estimation algorithms developed in the context of the Wolfram model may be reformulated as generalizations of the midpoint scaling estimator on causal sets, and are compatible with the generalized Myrheim-Meyer estimators, as well as exploring how the presence of the underlying hypergraph structure yields a significantly more robust technique for estimating spacelike distances when compared against several standard distance and predistance estimator functions in causal set theory. We finally demonstrate how the Benincasa-Dowker action on causal sets can be recovered as a special case of the discrete Einstein-Hilbert action over Wolfram model systems (with ergodicity assumptions in the hypergraph replaced by Poisson distribution assumptions in the causal set), and also how both classical and quantum sequential growth dynamics can be recovered as special cases of Wolfram model multiway evolution with an appropriate choice of discrete measure.
\end{abstract}

\newpage

\tableofcontents

\section{Introduction}

\textit{Causal set theory} is a fundamentally discrete approach to quantum gravity, initially proposed and championed by Bombelli, Lee, Meyer and Sorkin\cite{bombelli}\cite{bombelli2}, based upon the realization that the causal structure of a Lorentzian manifold is invariant under conformal transformations\cite{sorkin}\cite{sorkin2}\cite{hawking}\cite{malament}, and therefore a continuous spacetime may be arbitrarily well-approximated by a finite set of events - a \textit{causal set} - with a causal partial order relation defined upon them; the volume of a given spacetime region (which is otherwise left undetermined by the causal structure, due to the presence of this arbitrary conformal rescaling factor) can thus be recovered simply by counting the number of events that lie within a given discrete interval in the causal set. The majority of early work within the causal set program focused on causal set geometry and kinematics, such as approaches to determining discrete geodesics\cite{brightwell} and computing dimension estimates\cite{myrheim} for causal sets (and, more generally, attempting to ``port'' important geometrical constructions from Lorentzian manifolds over to their discrete counterparts in causal sets). In such cases, the causal set is assumed a priori to be \textit{faithfully embeddable}\cite{bombelli3}\cite{noldus}\cite{noldus2} into some Lorentzian manifold, which can be guaranteed using processes such as \textit{sprinkling} (in which discrete points are sprinkled into a Lorentzian manifold via a Poisson process, with the sprinkling probability being proportional to the continuum spacetime volume), and the key question then becomes how and whether one can reconstruct various geometrical features of the manifold using only combinatorial properties of the sprinkled points.

In a similar vein, the \textit{Wolfram model}\cite{wolfram}\cite{wolfram2} (formally introduced in Section \ref{sec:Section1} of the present article) is a discrete spacetime formalism based on \textit{hypergraph transformation dynamics}\cite{gorard}\cite{gorard2}, in which a causal set is effectively generated algorithmically via an abstract rewriting system defined over hypergraphs. In addition to defining a purely algorithmic dynamics for causal sets, the presence of the underlying hypergraph/rewriting structure (in addition to the usual causal structure) leads to \textit{multiway systems} produced by Wolfram model evolution that exhibit an apparently far richer variety of mathematical features than one observes in conventional causal set dynamics. As we shall show in Section \ref{sec:Section2}, invariance of the causal structure under parametrized changes in the rewriting order for the hypergraph (otherwise known as \textit{causal invariance}) yields an invariance of the induced causal partial order relation under conformal transformations, in a manner that is provably compatible with the measure-theoretic arguments of Bombelli, Henson and Sorkin\cite{bombelli4} in relation to the invariance of causal sets (produced by Poisson sprinkling processes) under the action of the restricted Lorentz group ${SO^{+} \left( 1, 3 \right)}$. We also describe how a combination of causal invariance and a simple coarse-graining procedure allows one to approximate the non-compact hyperboloid of future-pointing timelike null vectors ${\mathcal{H}}$ by an underlying causal network of bounded valence. In Section \ref{sec:Section3}, we proceed to demonstrate (via a combination of explicit numerical simulations and exact analytical arguments) that the geodesic ball and geodesic cone approaches to local dimension estimation in causal networks generated by Wolfram model evolution are effective generalizations of the midpoint scaling estimator proposed by Bombelli, which is in turn compatible with the generalized Myrheim-Meyer global dimension estimators based on abundances of $k$-chains in causal sets. The presence of the hypergraph structure also resolves a key conceptual difficulty in defining a compatible spatial metric tensor over causal antichains, consequently (and unsurprisingly) yielding more stable numerical approximations to spacelike distances than those generated by the predistance functions of Brightwell, Gregory, Rideout and Wallden, and we provide some preliminary comparison with the more sophisticated distance function recently proposed by Eichhorn, Surya and Versteegen. In Section \ref{sec:Section4}, by introducing a modification of the discrete d'Alembertian operator proposed by Sorkin and Henson for causal sets, we show how the Ollivier-Ricci constructions of both scalar and sectional curvature in Wolfram model hypergraphs and causal networks can be recovered by simply setting a certain scalar field ${\phi}$ to be constant over all of spacetime, formally proving that the Benincasa-Dowker action follows as a special case of the discrete Einstein-Hilbert action of the Wolfram model: namely, the case in which the weak ergodicity assumption of the hypergraph rewriting dynamics is replaced with the much stronger assumption of uniform Poission distribution of the causal set elements. Finally, in Section \ref{sec:Section5}, we show how both classical and quantum sequential growth dynamics of causal sets can be recovered as special cases of Wolfram model multiway evolution (by either following a single branch of a multiway evolution graph stochastically in the classical case, or by following a superposition of all possible branches in the quantum case). More precisely, we make use of previous category-theoretic arguments to prove that the multiway evolution graph is endowed with a natural measure given by state path weights, satisfying a certain quantum summation rule, and whose associated decoherence functional satisfies the appropriate algebraic properties. We also show how this measure immediately implies Markovian dynamics in the classical case, with merging of state vertices based on causal network isomorphism implying discrete general covariance, and the locality of causal influence in \textit{branchial space} implying a generalized form of Bell causality.

We henceforth adopt the terminology of \textit{algorithmic causal sets}, as coined originally by Bolognesi \cite{bolognesi}\cite{bolognesi2}, as a means of referring to the general procedure of generating a causal set algorithmically via Wolfram model evolution. Throughout the remainder of this article, the exposition of basic concepts in causal set theory will be based in large part upon the recent excellent review article of Surya\cite{surya}. Note also that all of the code required to reproduce the simulations and visualizations presented throughout this article is freely available through the \textit{Wolfram Function Repository} (e.g. \url{https://resources.wolframcloud.com/FunctionRepository/resources/FlatSpacetimeSprinkling},\\ \url{https://resources.wolframcloud.com/FunctionRepository/resources/CurvedSpacetimeSprinkling},\\ \url{https://resources.wolframcloud.com/FunctionRepository/resources/CurvedSpacetimeRegionSprinkling}, etc.).

\section{Background: Causal Networks and Causal Sets}
\label{sec:Section1}

\begin{figure}[ht]
\centering
\includegraphics[width=0.295\textwidth]{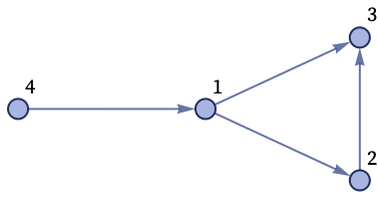}\hspace{0.1\textwidth}
\includegraphics[width=0.295\textwidth]{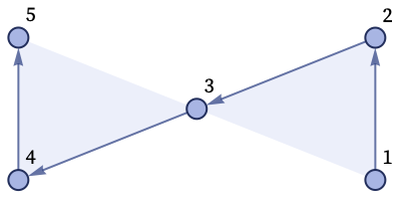}
\caption{Spatial hypergraphs corresponding to finite collections of ordered relations between elements, namely ${\left\lbrace \left\lbrace 1, 2 \right\rbrace, \left\lbrace 1, 3 \right\rbrace, \left\lbrace 2, 3 \right\rbrace, \left\lbrace 4, 1 \right\rbrace \right\rbrace}$ and ${\left\lbrace \left\lbrace 1, 2, 3 \right\rbrace, \left\lbrace 3, 4, 5 \right\rbrace \right\rbrace}$, respectively. Example taken from \cite{wolfram2}.}
\label{fig:Figure1}
\end{figure}

At its core, the Wolfram model is a discrete spacetime formalism in which the continuum structure of a Lorentzian manifold emerges as the large-scale limit of a discrete \textit{causal network}. The causal network is generated in a very natural and algorithmic way from diagrammatic rewriting rules acting on \textit{hypergraphs}, which correspond, in turn, to discrete approximations to spacelike hypersurfaces, in the following way\cite{gorard}:

\begin{definition}
A ``spatial hypergraph'', denoted ${H = \left( V , E \right)}$, is a finite, undirected hypergraph:

\begin{equation}
E \subset \mathcal{P} \left( V \right) \setminus \left\lbrace \emptyset \right\rbrace,
\end{equation}
where ${\mathcal{P} \left( V \right)}$ denotes the power set of $V$.
\end{definition}
In other words, each (directed) hyperedge corresponds to an (ordered) relation between abstract elements, allowing us to represent the hypergraph itself as an finite collection of such relations, as shown in Figure \ref{fig:Figure1}.

\begin{figure}[ht]
\centering
\includegraphics[width=0.495\textwidth]{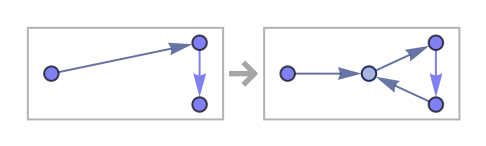}
\caption{A hypergraph transformation rule corresponding to the set substitution system ${\left\lbrace \left\lbrace x, y \right\rbrace, \left\lbrace y, z \right\rbrace \right\rbrace \to \left\lbrace \left\lbrace w, y \right\rbrace, \left\lbrace y, z \right\rbrace, \left\lbrace z, w \right\rbrace, \left\lbrace x, w \right\rbrace \right\rbrace}$. Example taken from \cite{wolfram2}.}
\label{fig:Figure2}
\end{figure}

The dynamics of a Wolfram model system are then defined in terms of hypergraph rewriting rules:

\begin{definition}
An ``update rule'', denoted $R$, for a spatial hypergraph ${H = \left( V, E \right)}$, is an abstract rewrite rule of the form ${H_1 \to H_2}$, in which a subhypergraph matching pattern ${H_1}$ is replaced by a distinct subhypergraph matching pattern ${H_2}$.
\end{definition}
Equivalently, we can formulate the rewriting rules as set substitution systems, in which a subset of ordered relations matching one pattern is replaced with a distinct subset of ordered relations matching a different pattern, as shown in Figure \ref{fig:Figure2}.

\begin{figure}[ht]
\centering
\includegraphics[width=0.695\textwidth]{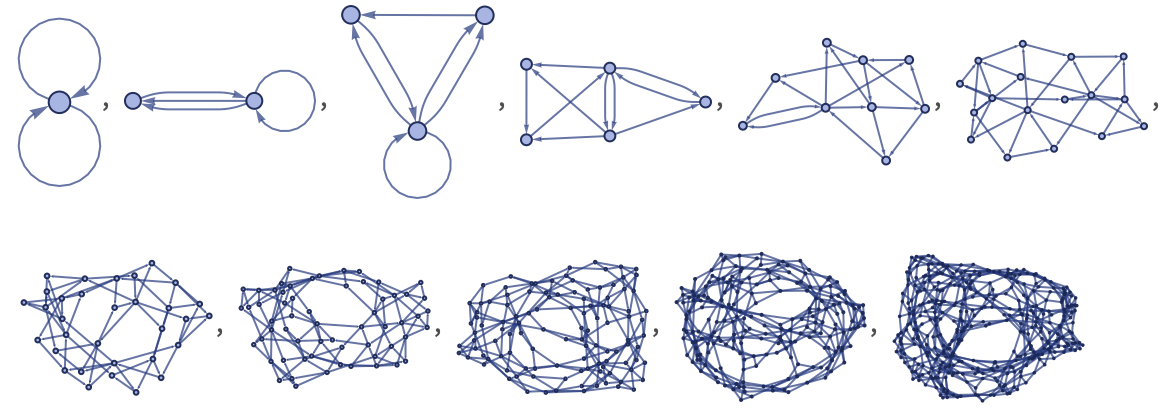}
\caption{The results of the first 10 steps in the evolution history of the set substitution system ${\left\lbrace \left\lbrace x, y \right\rbrace, \left\lbrace y, z \right\rbrace \right\rbrace \to \left\lbrace \left\lbrace w, y \right\rbrace, \left\lbrace y, z \right\rbrace, \left\lbrace z, w \right\rbrace, \left\lbrace x, w \right\rbrace \right\rbrace}$, starting from a double self-loop initial condition. Example taken from \cite{wolfram2}.}
\label{fig:Figure3}
\end{figure}

\begin{figure}[ht]
\centering
\includegraphics[width=0.495\textwidth]{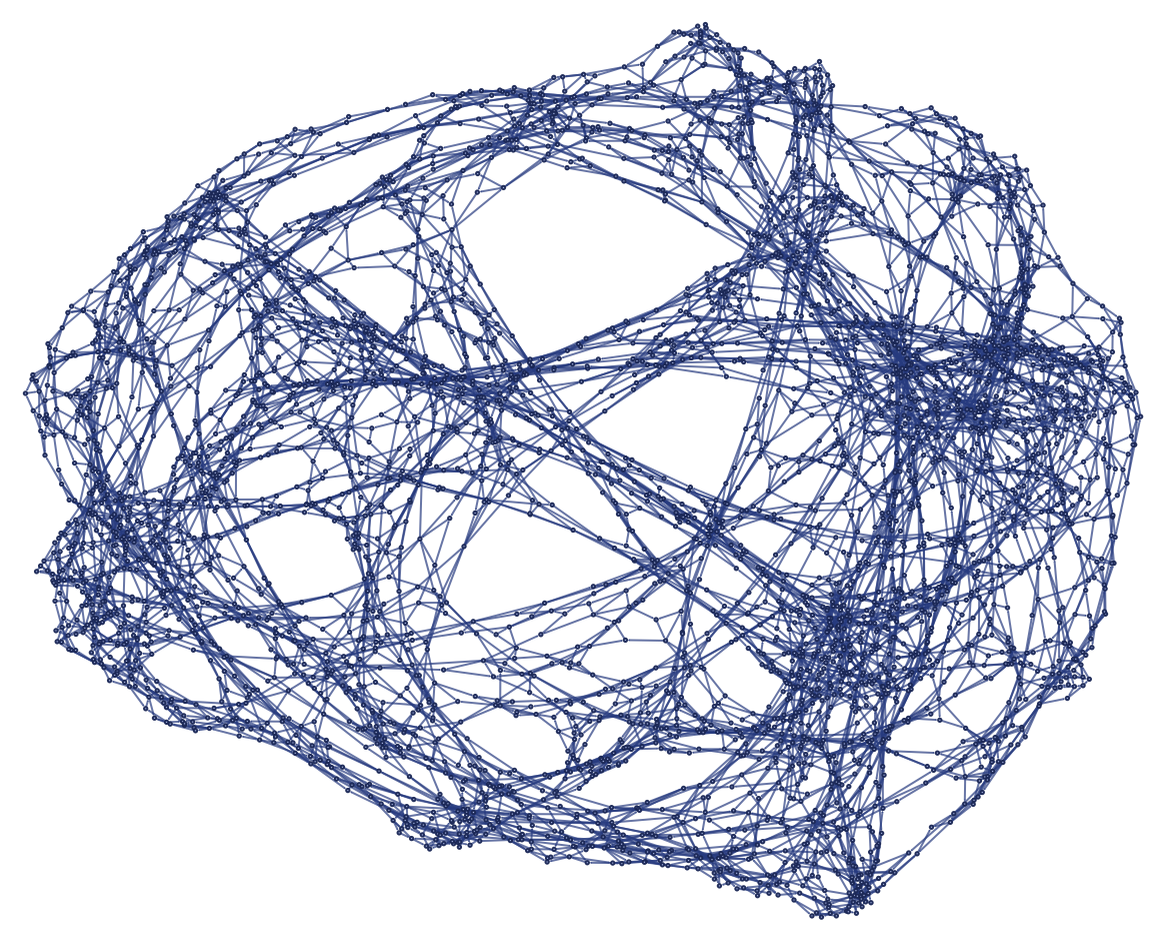}
\caption{The result after 14 steps of evolution of the set substitution system ${\left\lbrace \left\lbrace x, y \right\rbrace, \left\lbrace y, z \right\rbrace \right\rbrace \to \left\lbrace \left\lbrace w, y \right\rbrace, \left\lbrace y, z \right\rbrace, \left\lbrace z, w \right\rbrace, \left\lbrace x, w \right\rbrace \right\rbrace}$, starting from a double self-loop initial condition. Example taken from \cite{wolfram2}.}
\label{fig:Figure4}
\end{figure}

Although the order in which to apply the rewriting rules to a given hypergraph is not (in general) well-defined, we can nevertheless consider a simplified case in which we apply the rule to every possible matching (and non-overlapping) subhypergraph, yielding (for instance) the evolution shown in Figures \ref{fig:Figure3} and \ref{fig:Figure4}.

\begin{figure}[ht]
\centering
\includegraphics[width=0.395\textwidth]{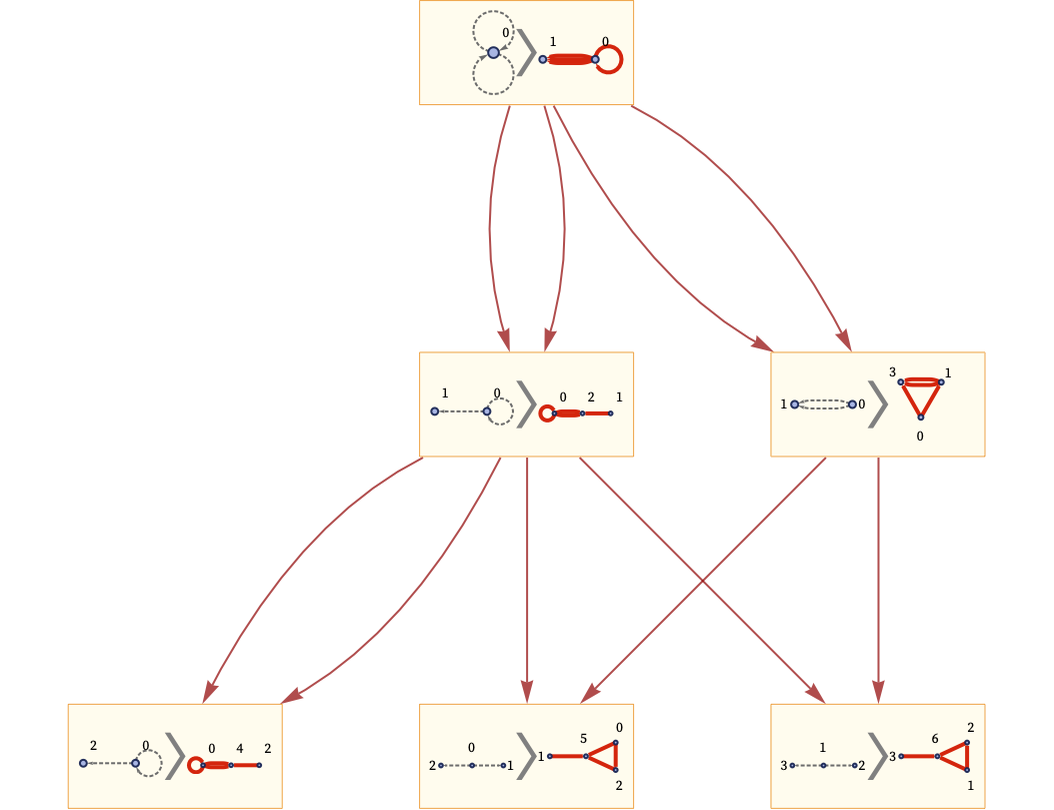}
\caption{The causal network after the first three evolution steps of the set substitution system ${\left\lbrace \left\lbrace x, y \right\rbrace, \left\lbrace x, z \right\rbrace \right\rbrace \to \left\lbrace \left\lbrace x, y \right\rbrace, \left\lbrace x, w \right\rbrace, \left\lbrace y, w \right\rbrace, \left\lbrace z, w \right\rbrace \right\rbrace}$. Example taken from \cite{wolfram2}.}
\label{fig:Figure5}
\end{figure}

\begin{figure}[ht]
\centering
\includegraphics[width=0.495\textwidth]{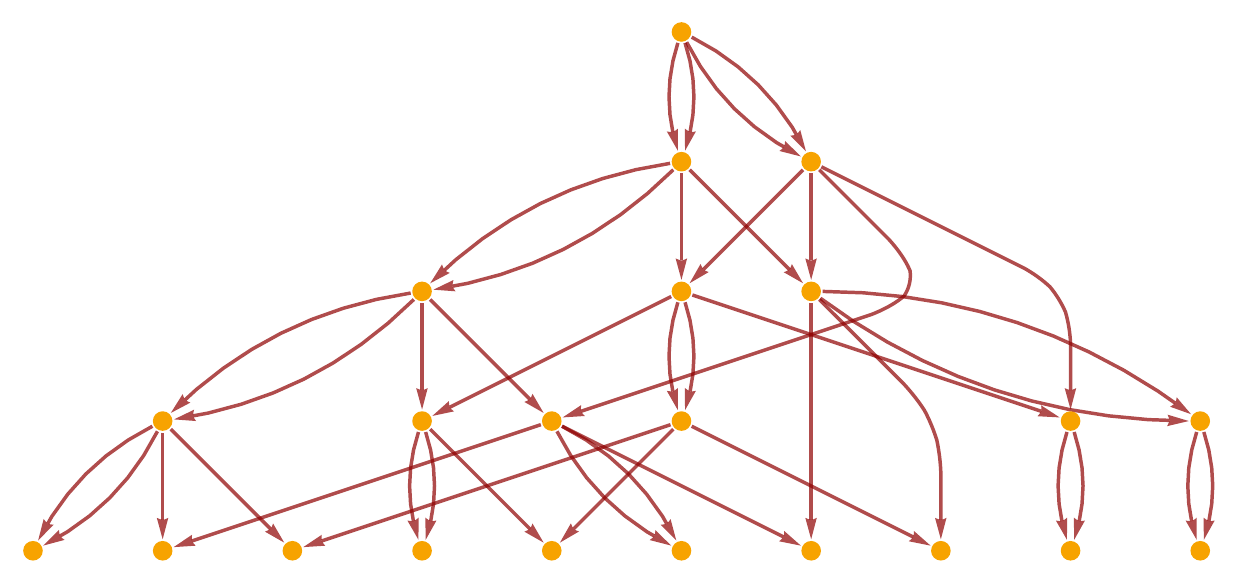}
\caption{The causal network after the first five evolution steps of the set substitution system ${\left\lbrace \left\lbrace x, y \right\rbrace, \left\lbrace x, z \right\rbrace \right\rbrace \to \left\lbrace \left\lbrace x, y \right\rbrace, \left\lbrace x, w \right\rbrace, \left\lbrace y, w \right\rbrace, \left\lbrace z, w \right\rbrace \right\rbrace}$.}
\label{fig:Figure6}
\end{figure}

\begin{figure}[ht]
\centering
\includegraphics[width=0.495\textwidth]{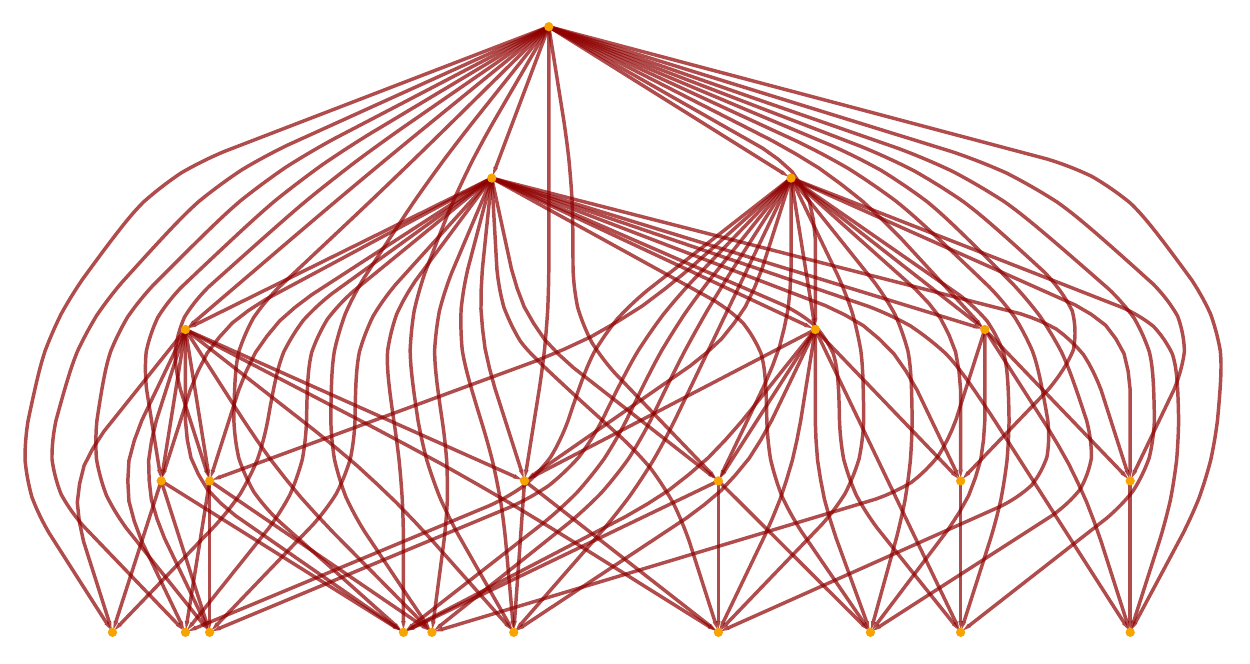}
\includegraphics[width=0.495\textwidth]{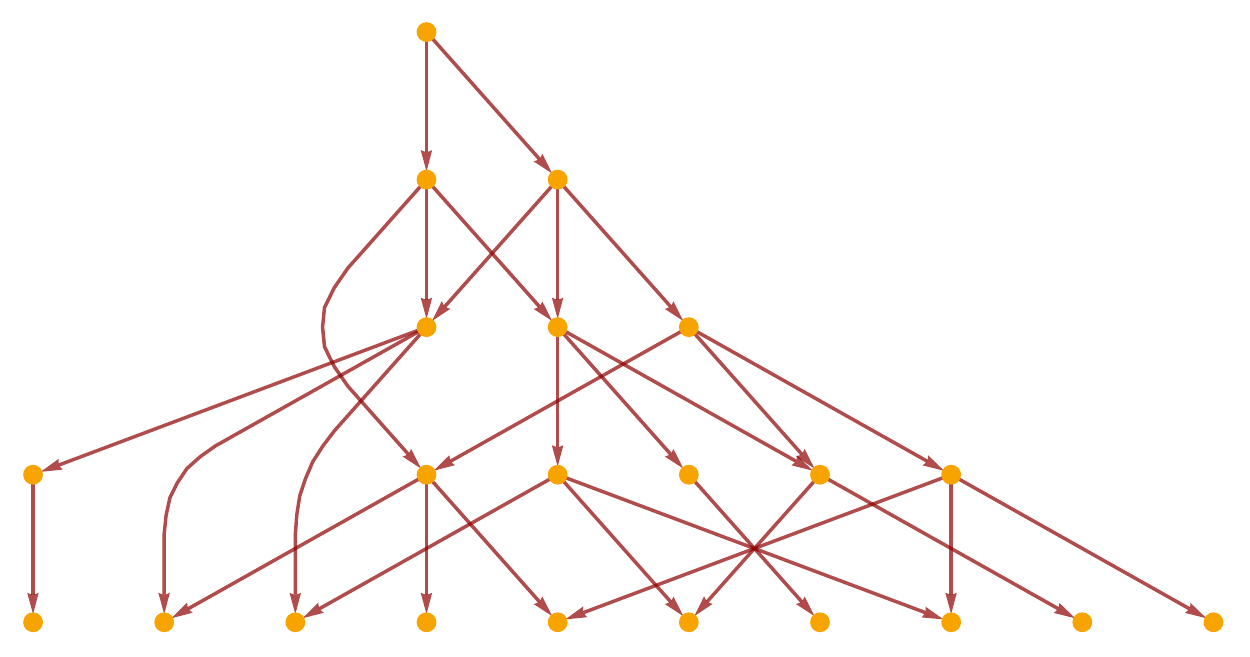}
\caption{The transitive closure (left) and the transitive reduction (right) of the causal network after the first five evolution steps of the set substitution system ${\left\lbrace \left\lbrace x, y \right\rbrace, \left\lbrace x, z \right\rbrace \right\rbrace \to \left\lbrace \left\lbrace x, y \right\rbrace, \left\lbrace x, w \right\rbrace, \left\lbrace y, w \right\rbrace, \left\lbrace z, w \right\rbrace \right\rbrace}$.}
\label{fig:Figure7}
\end{figure}

Despite the order in which the update rules are applied being, in some sense, arbitrary (as we shall discuss at length later), it is nevertheless the case that certain applications of rules may have dependencies upon prior applications, such that update event $B$ could only have occurred if update event $A$ had previously occurred. The structure of such dependencies can be captured through a purely combinatorial object known as a \textit{causal network}:

\begin{definition}
A ``causal network'', denoted ${G_{causal}}$, is a directed, acyclic graph in which every vertex corresponds to an application of an update rule (i.e. an update ``event''), and in which the directed edge ${A \to B}$ exists if and only if:

\begin{equation}
\mathrm{In} \left( B \right) \cap \mathrm{Out} \left( A \right) \neq \emptyset,
\end{equation}
i.e. if the input for event $B$ makes use of hyperedges that were produced by the output of event $A$.
\end{definition}
An example of a causal network for the first three evolution steps of a simple Wolfram model system is shown in Figure \ref{fig:Figure5}. The transitive reduction of a causal network therefore forms a Hasse diagram for the causal partial order relation on the hypergraph, which is presumed in turn to approximate the causal partial order relation for some Lorentzian manifold, as shown in Figures \ref{fig:Figure6} and \ref{fig:Figure7}.

Thus, the discrete past and future \textit{light cones} of a given updating event simply correspond to the in- and out-components of the associated vertices in the causal network, respectively. Two updating events may consequently be said to be \textit{causally related} (i.e. connected by a directed path in the causal network) if and only if they are either \textit{lightlike-separated} or \textit{timelike-separated} (with lightlike separation corresponding to directed paths which lie on the boundary of discrete light cones, and all other such paths yielding timelike separations), as shown in Figure \ref{fig:Figure9}. All other pairs of updating events that are not connected by directed edges in the causal network (and which therefore form antichains in the associated causal partial order) may be considered to be \textit{spacelike-separated}. We may therefore extend the notions of chronological and causal precedence of spacetime events in Lorentzian manifolds\cite{penrose}\cite{levichev} over to the case of updating events in causal networks in a very straightforward way (denoting, in what follows, the set of all updating events in the causal network by ${\mathcal{M}}$):

\begin{figure}[ht]
\centering
\includegraphics[width=0.495\textwidth]{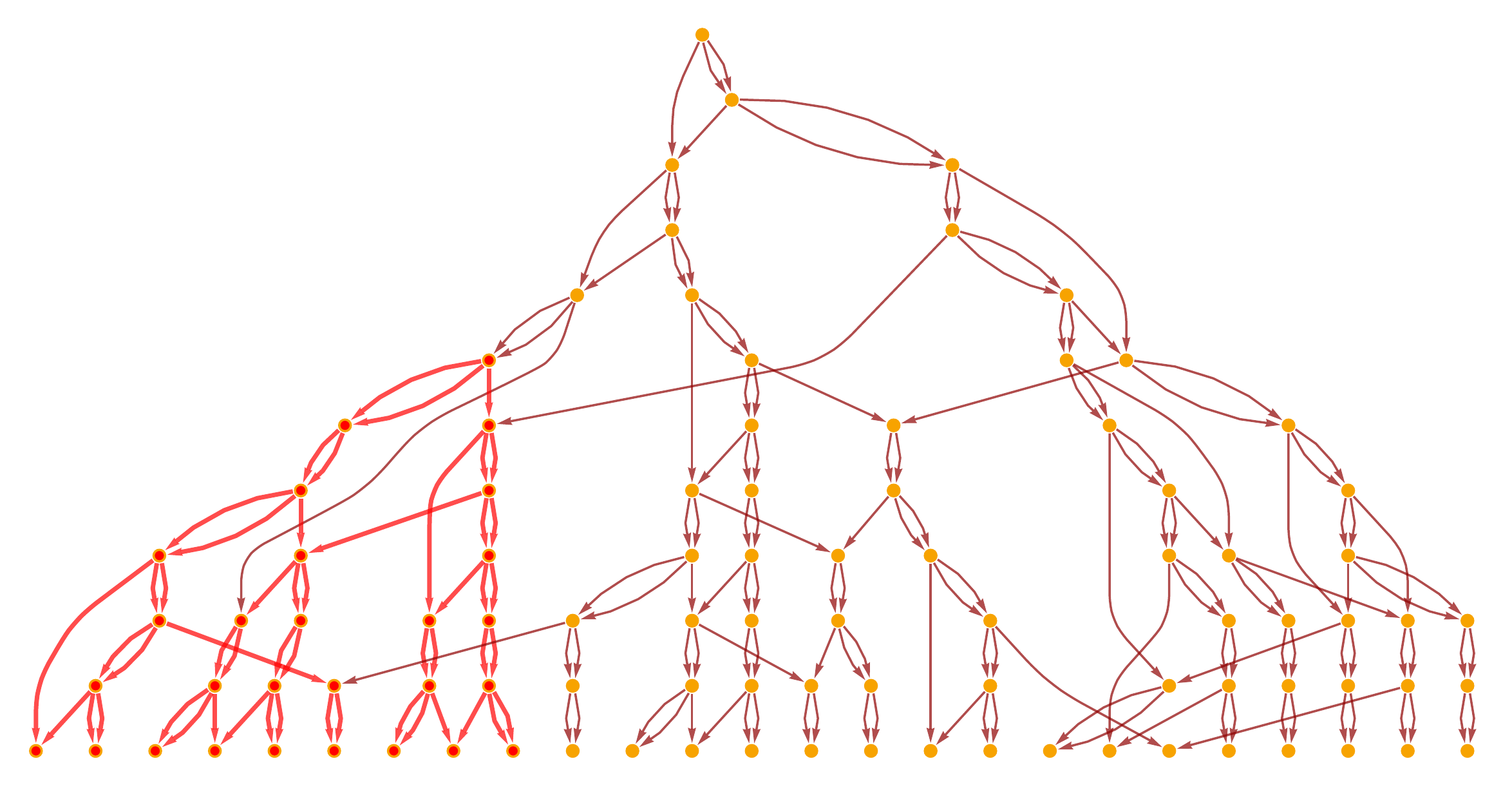}
\includegraphics[width=0.495\textwidth]{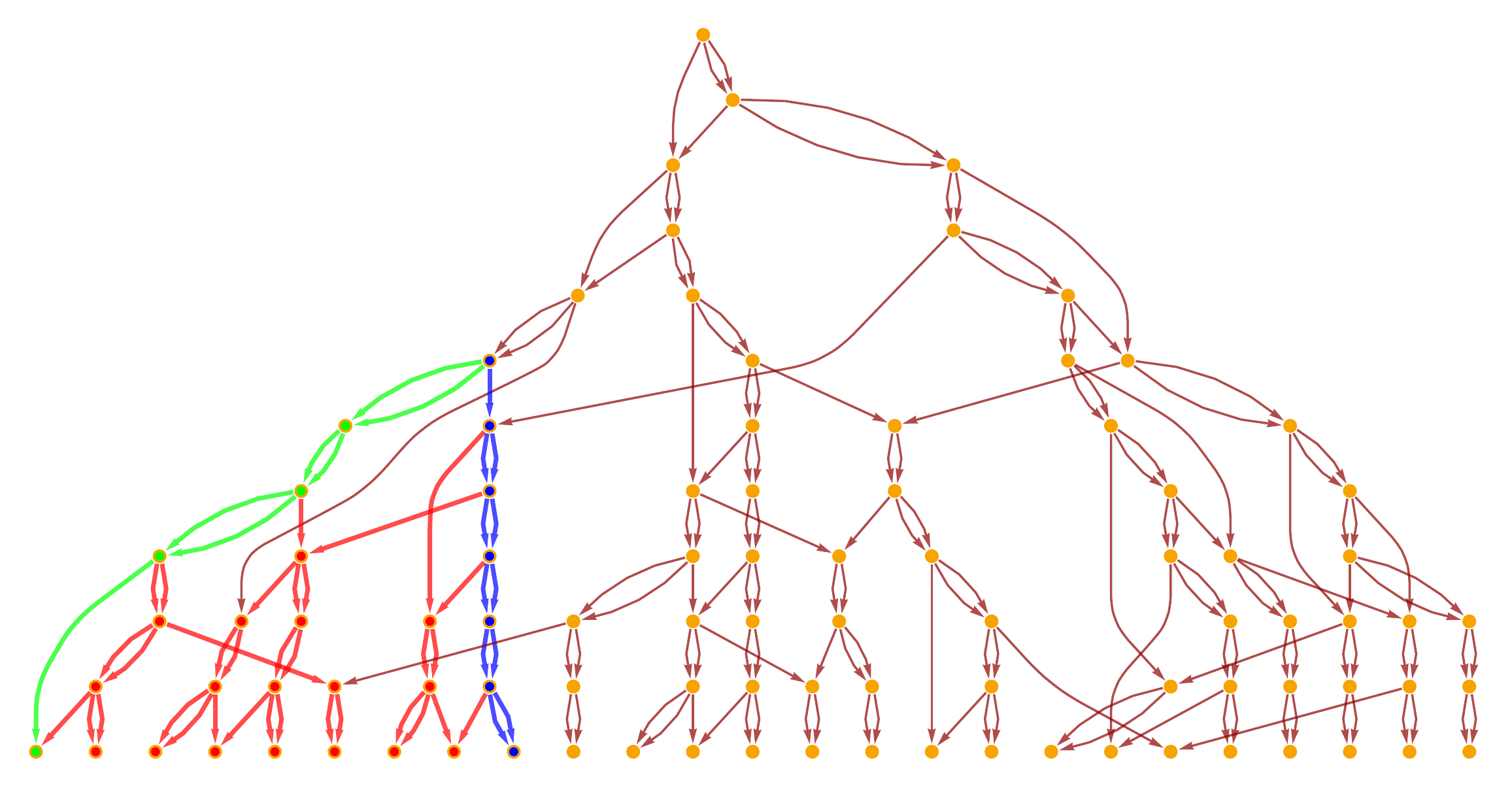}
\caption{On the left, the discrete future light cone (highlighted in red) of a given updating event in the causal network after the first five evolution steps of the set substitution system ${\left\lbrace \left\lbrace x, y \right\rbrace, \left\lbrace z, y \right\rbrace \right\rbrace \to \left\lbrace \left\lbrace x, z \right\rbrace, \left\lbrace y, z \right\rbrace, \left\lbrace w, z \right\rbrace \right\rbrace}$. On the right, the two lightlike paths are highlighted in green and blue, with all other (timelike) paths shown in red.}
\label{fig:Figure9}
\end{figure}

\begin{definition}
An updating event $x$ ``chronologically precedes'' an updating event $y$, denoted ${x \ll y}$, if there exists a future-directed chronological (i.e. timelike) path through the causal network connecting vertices $x$ and $y$.
\end{definition}

\begin{definition}
An updating event $x$ ``strictly chronologically precedes'' an updating event $y$, denoted ${x < y}$, if there exists a future-directed causal (i.e. non-spacelike) path through the causal network connecting vertices $x$ and $y$.
\end{definition}

\begin{definition}
An updating event $x$ ``causally precedes'' an updating event $y$, denoted ${x \prec y}$, if either $x$ strictly causally precedes $y$, or ${x = y}$.
\end{definition}
Here, the standard algebraic properties of chronological and causal precedence, such as transitivity:

\begin{equation}
\forall x, y, z \in \mathcal{M}, \qquad x \ll y \text{ and } y \ll z \implies x \ll z, \qquad x \prec y, y \prec z \implies x \prec z, \qquad \text{etc.},
\end{equation}
follow trivially from the definition of the causal partial order as the transitively-reduced version of the causal network. From here, the discrete past and future light cones (and their boundaries) of updating events can be specified purely in terms of the chronological and causal precedence relations:

\begin{definition}
The ``chronological future'' and ``chronological past'' of an updating event $x$, denoted ${I^{+} \left( x \right)}$ and ${I^{-} \left( x \right)}$, are defined as the sets of updating events which $x$ chronologically precedes, and which chronologically precede $x$, respectively:

\begin{equation}
I^{+} \left( x \right) = \left\lbrace y \in \mathcal{M} : x \ll y \right\rbrace, \qquad I^{-} \left( x \right) = \left\lbrace y \in \mathcal{M} : y \ll x \right\rbrace.
\end{equation}
\end{definition}

\begin{definition}
The ``causal future'' and ``causal past'' of an updating event $x$, denoted ${J^{+} \left( x \right)}$ and ${J^{-} \left( x \right)}$, are defined as the sets of updating events which $x$ causally precedes, and which causally precede $x$, respectively:

\begin{equation}
J^{+} \left( x \right) = \left\lbrace y \in \mathcal{M} : x \prec y \right\rbrace, \qquad J^{-} \left( x \right) = \left\lbrace y \in \mathcal{M} : y \prec x \right\rbrace.
\end{equation}
\end{definition}
The notions of chronological and causal future and past may also be extended to discrete sets of updating events ${S \subset \mathcal{M}}$ in the obvious way:

\begin{equation}
I^{\pm} \left( S \right) = \bigcup_{x \in S} I^{\pm} \left( x \right), \qquad J^{\pm} \left( S \right) = \bigcup_{x \in S} J^{\pm} \left( x \right).
\end{equation}
Thus, the chronological future and past sets designate the interiors of discrete light cones, whilst the causal future and past sets designate the light cones themselves (including the lightlike boundaries). As before, the standard algebraic properties of chronological and causal future and past hold trivially by our definition of event separations in the causal network, e.g. one can easily see that the interiors of discrete past and future light cones are strict supersets of the light cones themselves:

\begin{equation}
I^{+} \left( S \right) = I^{+} \left( I^{+} \left( S \right) \right) \subset J^{+} \left( S \right) = J^{+} \left( J^{+} \left( S \right) \right), \qquad I^{-} \left( S \right) = I^{-} \left( I^{-} \left( S \right) \right) \subset J^{-} \left( S \right) = J^{-} \left( J^{-} \left( S \right) \right),
\end{equation}
For the case of a Lorentzian manifold with metric $g$, the causal structure (as defined through the chronological and causal precedence relations on events) is precisely the structure that is kept invariant under conformal transformations of the form:

\begin{equation}
\hat{g} = \Omega^2 g,
\end{equation}
where ${\Omega}$ denotes the conformal factor, since the timelike, null and spacelike qualities of tangent vectors $X$ are kept invariant under this map:

\begin{equation}
g \left( X, X \right) < 0, \qquad \iff \hat{g} \left( X, X \right) = \Omega^2 g \left( X, X \right) < 0,
\end{equation}

\begin{equation}
g \left( X, X \right) = 0, \qquad \iff \hat{g} \left( X, X \right) = \Omega^2 g \left( X, X \right) = 0,
\end{equation}
and:

\begin{equation}
g \left( X, X \right) > 0, \qquad \iff \hat{g} \left( X, X \right) = \Omega^2 g \left( X, X \right) > 0,
\end{equation}
and so, if the causal network is to be faithfully embedded into a Lorentzian manifold, it must be the case that the combinatorial structure of the causal network is invariant under the discrete analog of conformal transformations (as we shall discuss later).

\begin{figure}[ht]
\centering
\includegraphics[width=0.495\textwidth]{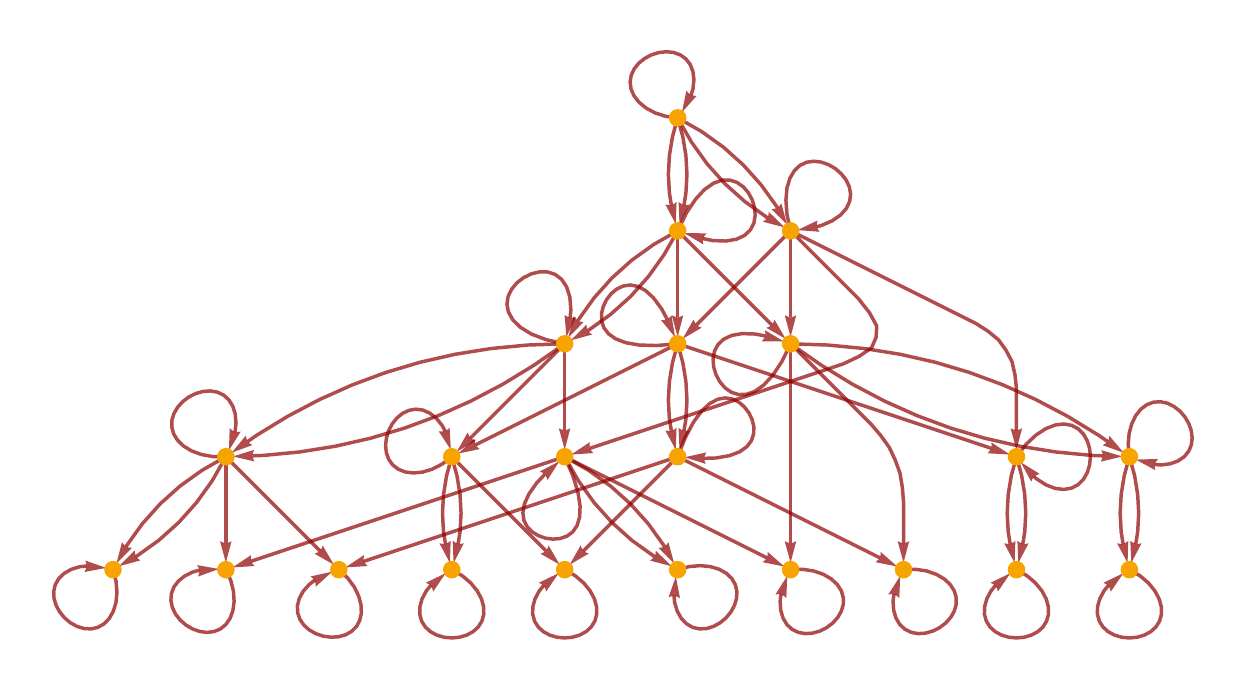}
\caption{The causal network after the first five evolution steps of the set substitution system ${\left\lbrace \left\lbrace x, y \right\rbrace, \left\lbrace x, z \right\rbrace \right\rbrace \to \left\lbrace \left\lbrace x, y \right\rbrace, \left\lbrace x, w \right\rbrace, \left\lbrace y, w \right\rbrace, \left\lbrace z, w \right\rbrace \right\rbrace}$, with self-loops added at each updating event, corresponding to a causal partial order relation obeying reflexivity.}
\label{fig:Figure8}
\end{figure}

In other words, we see explicitly that the object that one obtains by performing a transitive reduction on a causal network produced by a Wolfram model evolution is exactly a \textit{causal set} in the conventional sense:

\begin{definition}
A ``causal set'', denoted ${\mathcal{C}}$, is a set equipped with a binary relation, denoted ${\prec}$, which satisfies (most of) the axioms of a partial order relation, namely acyclicity/antisymmetry:

\begin{equation}
\forall x, y \in \mathcal{C}, \qquad x \prec y \text{ and } y \prec x \implies x = y,
\end{equation}
and transitivity:

\begin{equation}
\forall x, y, z \in \mathcal{C}, \qquad x \prec y \text{ and } y \prec z \implies x \prec z,
\end{equation}
in addition to exhibiting the property of ``local finiteness'':

\begin{equation}
\forall x, y \in \mathcal{C}, \qquad \left\lvert \mathbf{I} \left[ x, y \right] \right\rvert < \infty,
\end{equation}
where ${\mathbf{I} \left[ x, y \right]}$ denotes the discrete analog of the Alexandrov/interval topology on spacetime, namely:

\begin{equation}
\mathbf{I} \left[ x, y \right] = \mathrm{Fut} \left( x \right) \cap \mathrm{Past} \left( y \right),
\end{equation}
where ${\mathrm{Fut}}$ and ${\mathrm{Past}}$ designate the (exclusive) future and past sets for a given element, i.e:

\begin{equation}
\mathrm{Fut} \left( x \right) = \left\lbrace w \in \mathcal{C} : x \prec w \text{ and } x \neq w \right\rbrace, \qquad \text{ and } \qquad \mathrm{Past} \left( x \right) = \left\lbrace w \in \mathcal{C} : w \prec x \text{ and } x \neq w \right\rbrace.
\end{equation}
\end{definition}
In the above, the order interval ${\mathbf{I} \left[ x, y \right]}$ is taken to be the discrete analog of the Alexandrov interval ${\mathbf{A} \left[ x, y \right]}$ on continuous spacetime:

\begin{equation}
\mathbf{A} \left[ x, y \right] = I^{+} \left( x \right) \cap I^{-} \left( y \right).
\end{equation}
There are a couple of immediate mathematical issues with this definition of a causal set. The first is that, strictly speaking, the binary relation described above is not precisely a partial order relation, since (by convention) it is assumed not to satisfy the reflexivity condition:

\begin{equation}
\forall x \in \mathcal{C}, \qquad x \prec x,
\end{equation}
in which case the acyclicity/antisymmetry condition must be replaced by:

\begin{equation}
\nexists x, y \in \mathcal{C}, \qquad \text{ such that } x \prec y \text{ and } y \prec x,
\end{equation}
although in our particular case this can easily be corrected by simply introducing self-loops at each event vertex, as shown in Figure \ref{fig:Figure8}. A more significant subtlety relates to the condition of \textit{local finiteness}, which is enforced within causal set theory as means of encoding the intrinsic discreteness of spacetime, but which can equally well be replaced by a \textit{local countability} condition without sacrificing discreteness (e.g. the rational numbers ${\mathbb{Q}}$ are discrete and locally countable, without being locally finite). As we shall see, the arguments that we present in the case of the Wolfram model formalism rely only upon local countability, rather than strict local finiteness.

A causal set ${\mathcal{C}}$ may be said to approximate the Lorentzian manifold ${\left( \mathcal{M}, g \right)}$ (where ${\mathcal{C}}$ is a discrete proper subset of the uncountable set of events ${\mathcal{M}}$, i.e. ${\mathcal{C} \subset \mathcal{M}}$) if there exists a \textit{faithful embedding} of ${\mathcal{C}}$ within ${\mathcal{M}}$\cite{bombelli5}:

\begin{definition}
An ``embedding'' of a causal set ${\mathcal{C}}$ into a Lorentzian manifold ${\left( \mathcal{M}, g \right)}$ is an injective map:

\begin{equation}
\Phi : \mathcal{C} \to \left( \mathcal{M}, g \right), \qquad \text{ such that } \forall x, y \in \mathcal{C}, \qquad x \prec_{\mathcal{C}} y \iff \Phi \left( x \right) \prec_{\mathcal{M}} \Phi \left( y \right),
\end{equation}
where ${\prec_{\mathcal{C}}}$ and ${\prec_{\mathcal{M}}}$ denote the causal partial order relations on ${\mathcal{C}}$ and ${\mathcal{M}}$, respectively.
\end{definition}

\begin{definition}
An embedding ${\Phi : \mathcal{C} \to \mathcal{M}}$ is said to be ``faithful'' if ${\Phi \left( \mathcal{C} \right)}$ forms a uniform distribution (with respect to the spacetime volume measure on ${\left( \mathcal{M}, g \right)}$) with density:

\begin{equation}
\rho_c = V_{c}^{-1},
\end{equation}
for some characteristic spacetime discreteness scale (i.e. a spacetime volume cut-off) ${V_c}$.
\end{definition}
From here, the central conjecture (or \textit{Hauptvermutung}) of causal set theory can be expressed as the statement that the causal set ${\mathcal{C}}$ can be faithfully embedded (with density ${\rho_c}$) into two distinct Lorentzian manifolds ${\left( \mathcal{M}, g \right)}$ and ${\left( \mathcal{M}^{\prime}, g^{\prime} \right)}$, if and only if manifolds ${\left( \mathcal{M}, g \right)}$ and ${\left( \mathcal{M}^{\prime}, g^{\prime} \right)}$ are \textit{approximately isometric}. The intuitive notion of an approximate isometry captures the idea that the geometry of manifolds ${\left( \mathcal{M}, g \right)}$ and ${\left( \mathcal{M}^{\prime}, g^{\prime} \right)}$ should only differ at spacetime volume scales smaller than that of the characteristic discreteness scale ${V_c}$. One possible formal definition of an approximate isometry, as provided by Bombelli and Noldus\cite{bombelli6}, is as follows:

\begin{definition}
For a Lorentzian manifold ${\left( \mathcal{M}, g \right)}$ and some ${\epsilon > 0}$, a map ${f : \mathcal{M} \to \mathcal{M}}$ is said to be an ``approximate isometry'' if and only if:

\begin{equation}
\forall x, y \in \mathcal{M}, \qquad \left\lvert d \left( f \left( x \right), f \left( y \right) \right) - d \left( x, y \right) \right\rvert \leq \epsilon,
\end{equation}
i.e. if the map $f$ distorts the distances between points $x$ and $y$ by no more than ${\epsilon}$.
\end{definition}

\begin{figure}[ht]
\centering
\includegraphics[width=0.295\textwidth]{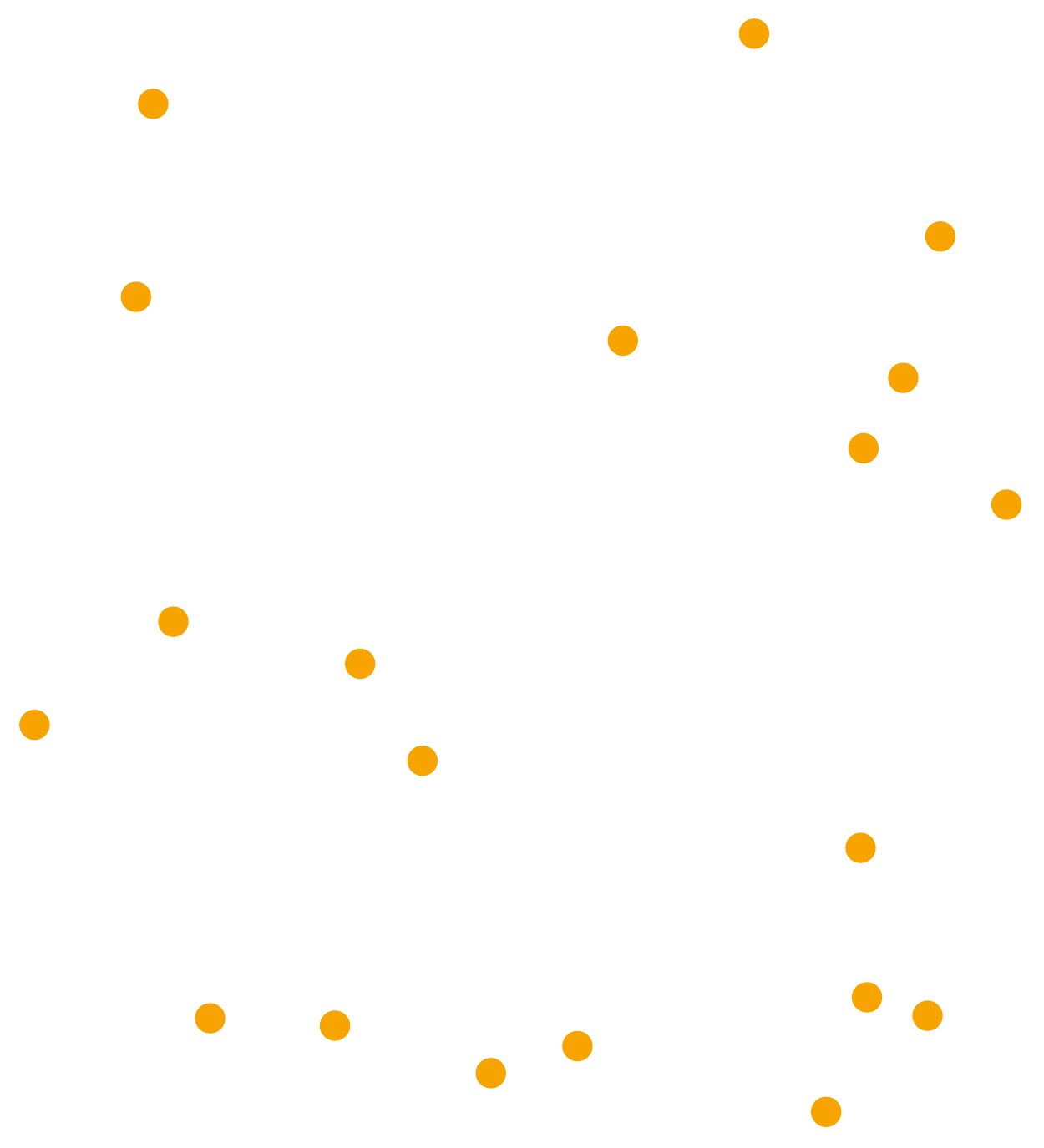}
\caption{A Poission sprinkling of 20 uniformly-selected points in a rectangular region of 1+1-dimensional flat (Minkowski) spacetime.}
\label{fig:Figure10}
\end{figure}

\begin{figure}[ht]
\centering
\includegraphics[width=0.295\textwidth]{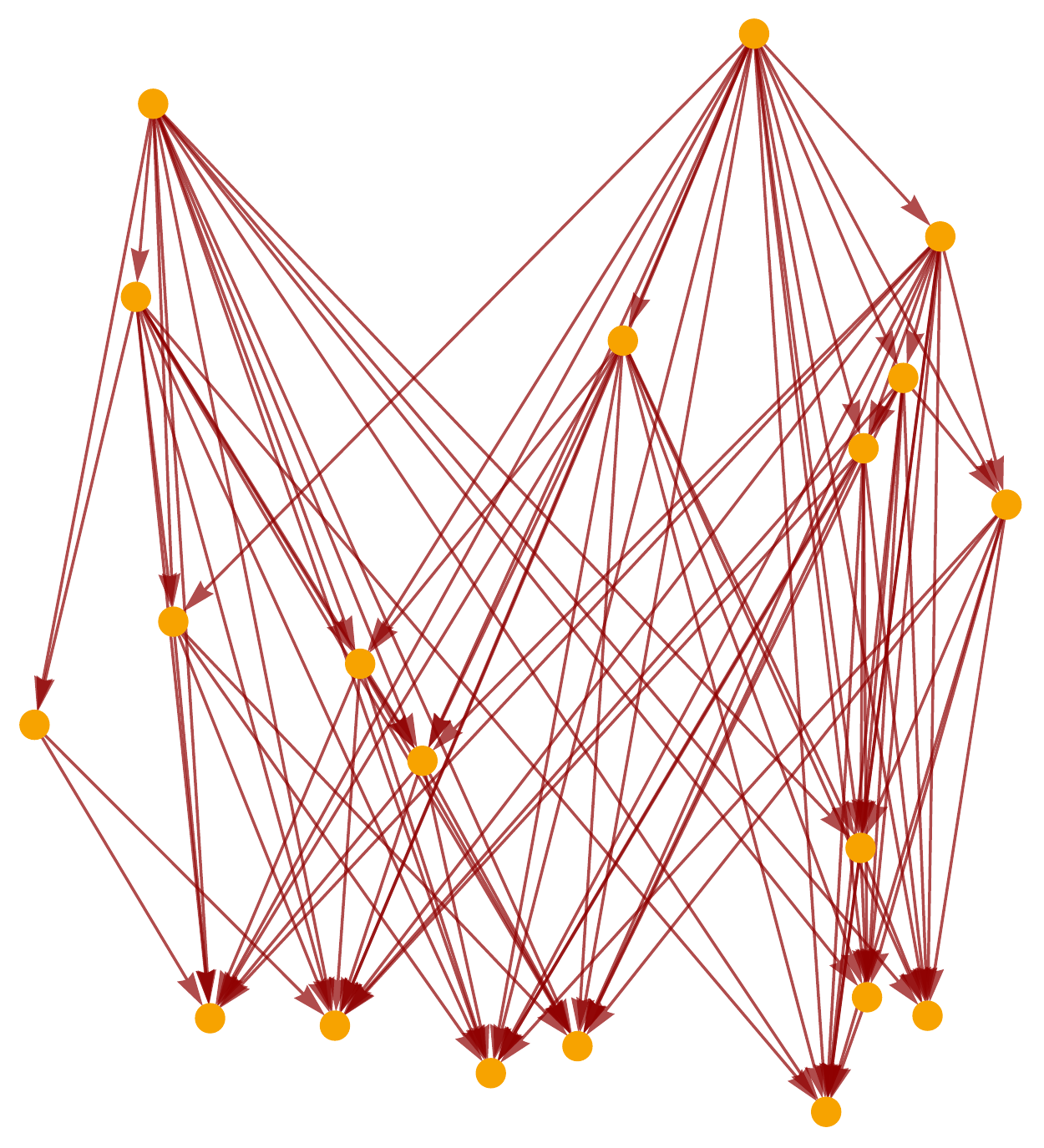}\hspace{0.1\textwidth}
\includegraphics[width=0.495\textwidth]{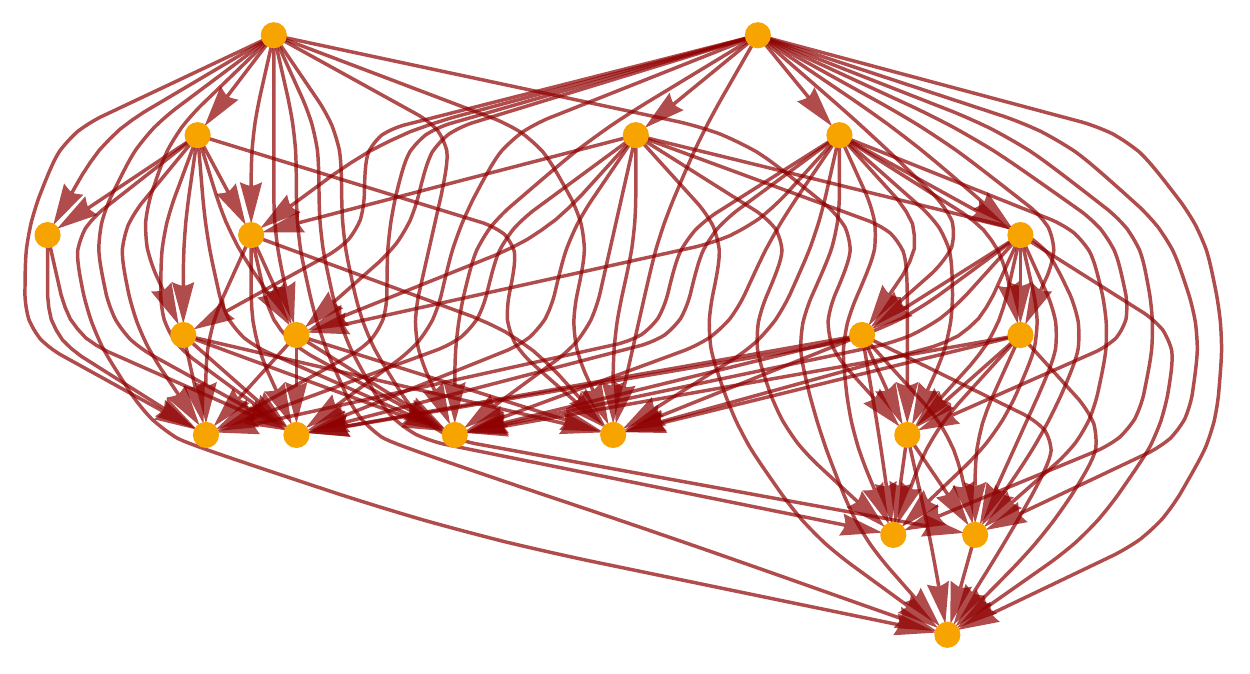}
\caption{On the left, a directed graph indicating which pairs of uniformly sprinkled points are related by the causal partial order in a rectangular region of 1+1-dimensional flat (Minkowski) spacetime. On the right, the same directed graph but with the vertex coordinate information removed.}
\label{fig:Figure11}
\end{figure}

\begin{figure}[ht]
\centering
\includegraphics[width=0.295\textwidth]{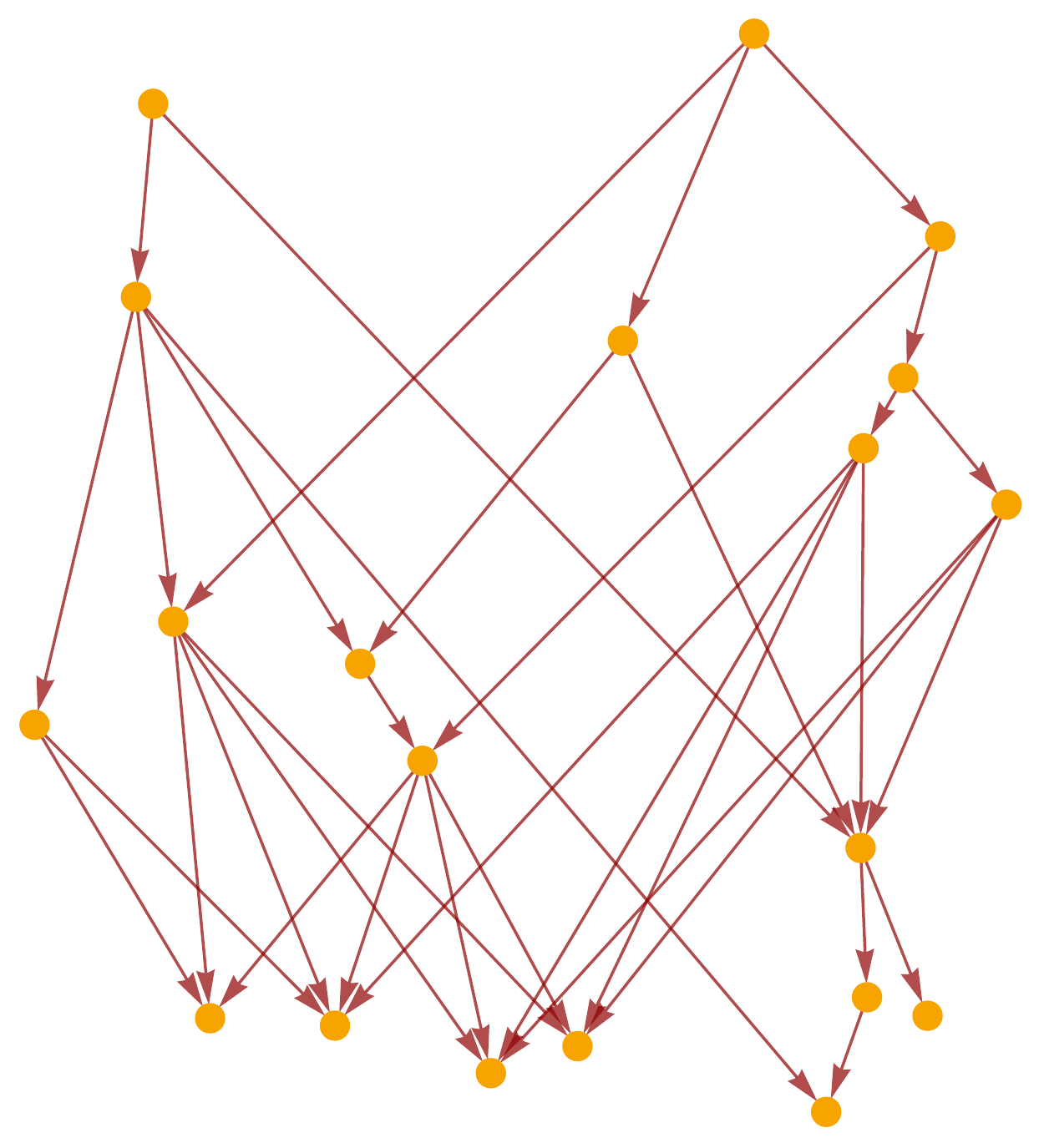}\hspace{0.1\textwidth}
\includegraphics[width=0.495\textwidth]{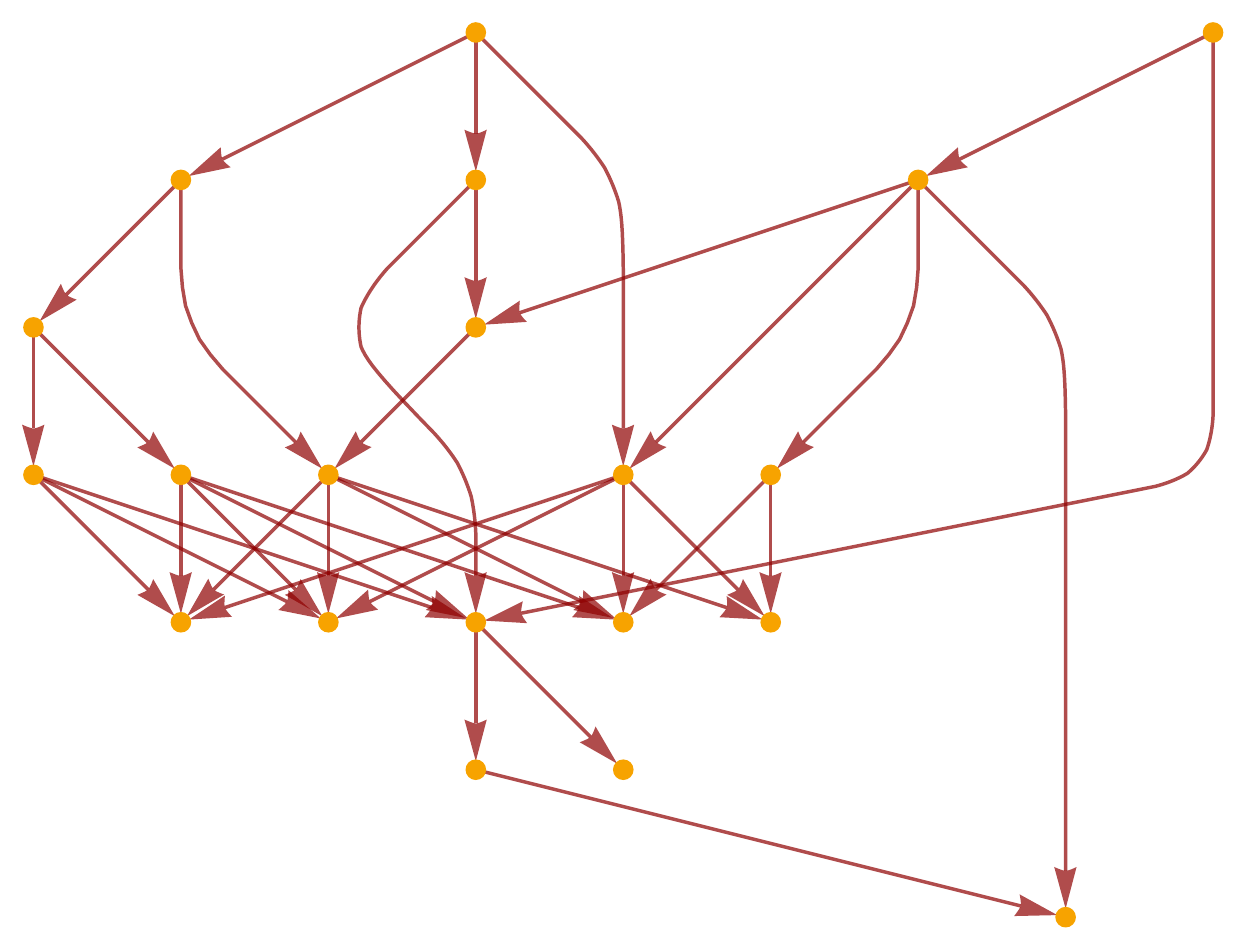}
\caption{On the left, the transitive reduction (i.e. the Hasse diagram) of the causal partial order graph for the uniformly sprinkled points in a rectangular region of 1+1-dimensional flat (Minkowski) spacetime. On the right, the same transitive reduction graph but with the vertex coordinate information removed.}
\label{fig:Figure12}
\end{figure}

\textit{Sprinkling} offers one possible method of constructing causal sets that are mathematically guaranteed to be faithfully embeddable into Lorentzian manifolds, by first starting from a manifold ${\left( \mathcal{M}, g \right)}$ and constructing ${\Phi \left( \mathcal{C} \right)}$ by means of a Poisson distribution, in which the probability of $n$ elements of ${\mathcal{C}}$ lying within a spacetime region of volume $v$ is given by:

\begin{equation}
P_v \left( n \right) = \frac{\left( \rho_c v \right)^n}{n!} \exp \left( - \rho_c v \right),
\end{equation}
such that the expectation value is simply:

\begin{equation}
\left\langle \hat{n} \right\rangle = \rho_c v,
\end{equation}
where ${\hat{n}}$ is the random variable designating the number of elements lying within a region of volume $v$ in the random causal set ${\mathcal{C}}$ in ${\Phi \left( \mathcal{C} \right)}$. Pairs of elements in ${\mathcal{C}}$ are then related by the discrete partial order relation ${\prec_{\mathcal{C}}}$ if and only if the corresponding events are related by continuous partial order relation ${\prec_{\mathcal{M}}}$ in ${\Phi \left( \mathcal{C} \right)}$, as shown in Figures \ref{fig:Figure10}, \ref{fig:Figure11} and \ref{fig:Figure12}.

\begin{figure}[ht]
\centering
\includegraphics[width=0.295\textwidth]{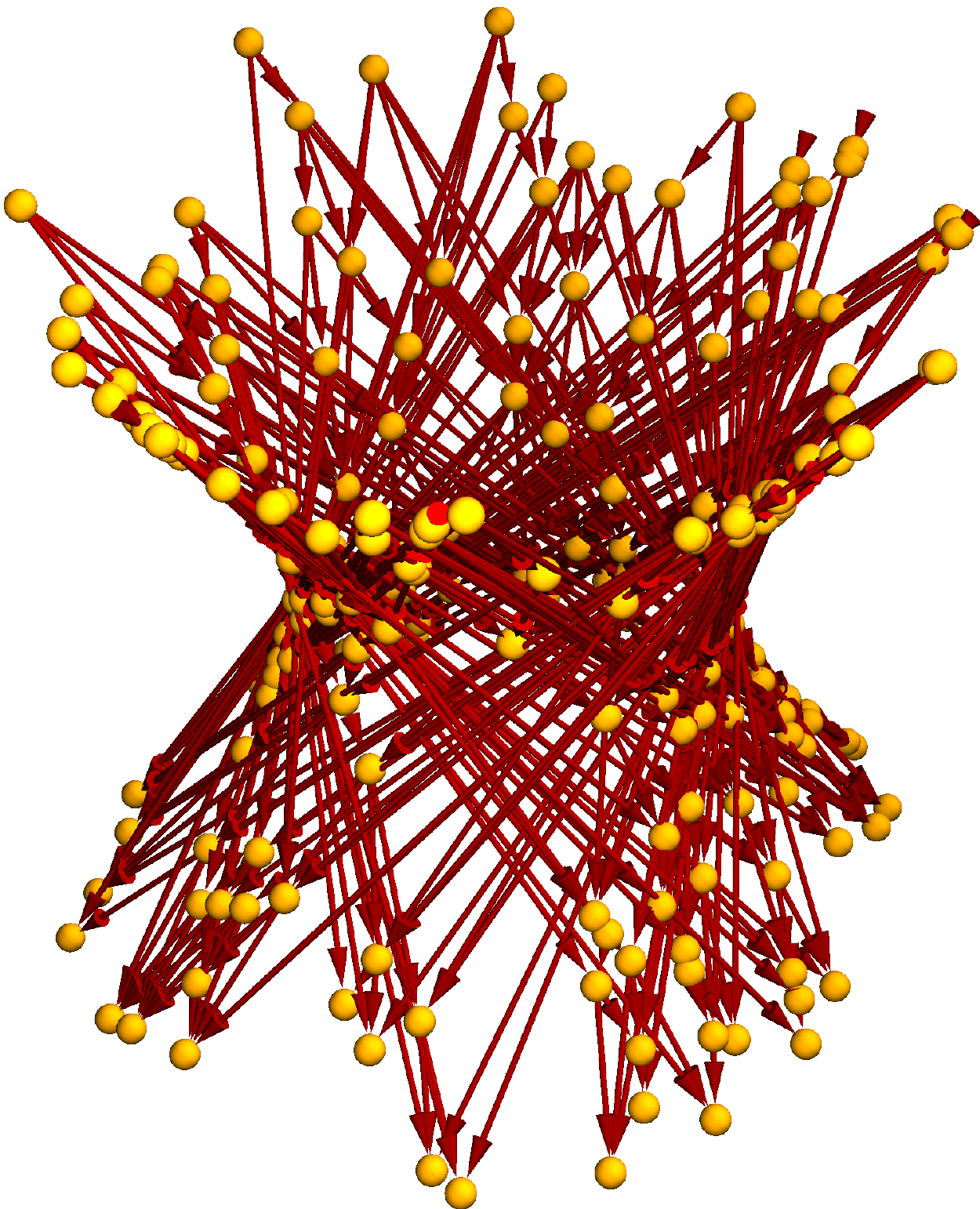}\hspace{0.1\textwidth}
\includegraphics[width=0.595\textwidth]{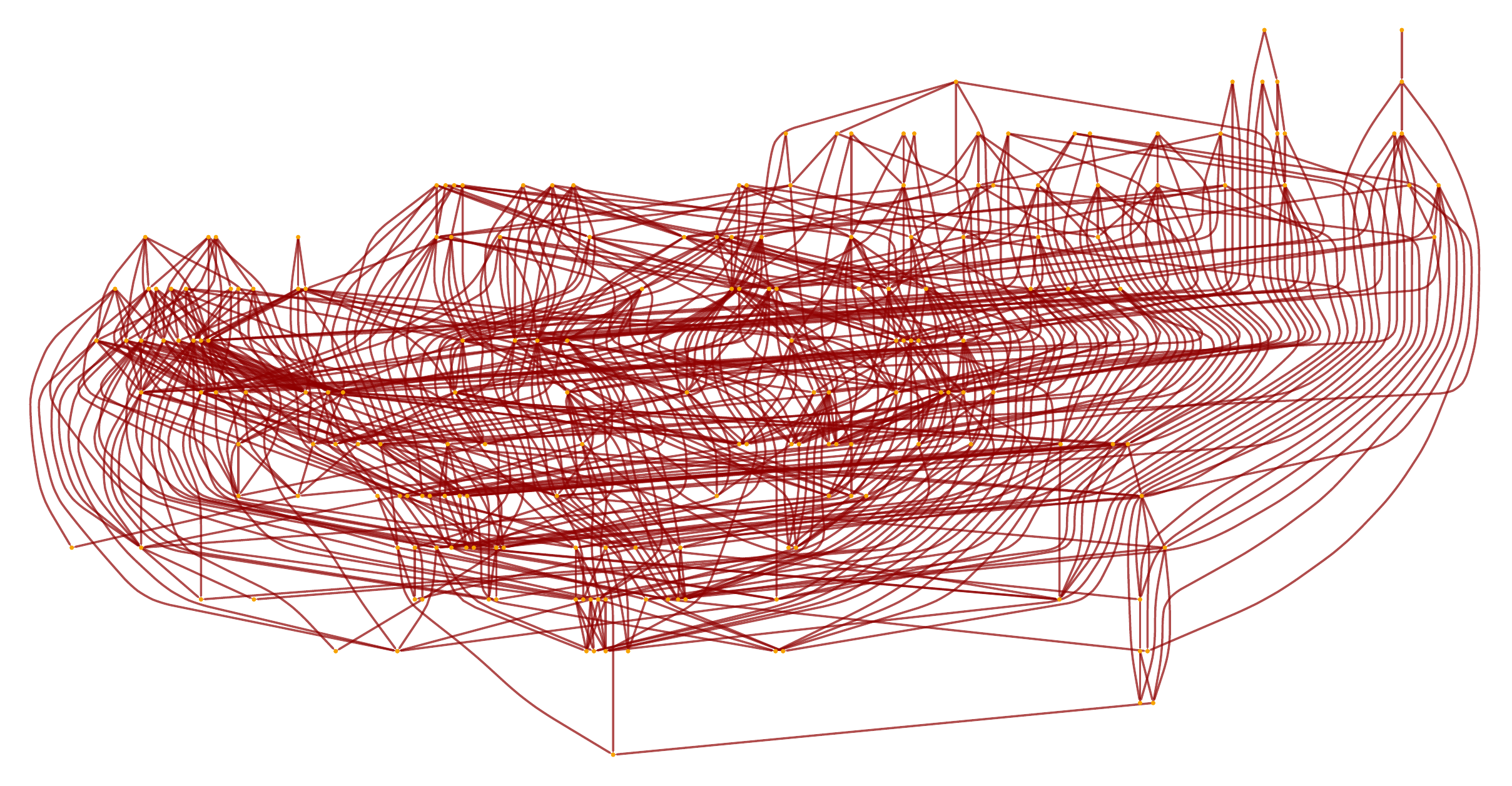}
\caption{On the left, the transitive reduction (i.e. the Hasse diagram) of the causal partial order graph for 200 uniformly sprinkled points into a hyperboloidal region of a 1+1-dimensional de Sitter spacetime embedded within a 2+1-dimensional background spacetime. On the right, the same transitive reduction graph but with the vertex coordinate information removed.}
\label{fig:Figure72}
\end{figure}

It is in this sense that the Wolfram model may be thought of as constituting a deterministic algorithmic procedure for generating causal sets, based upon rewriting rules defined over hypergraphs, that does not depend upon Poisson sprinkling into a pre-existing Lorentzian manifold (or any similar stochastic process). However, the existence of the underlying hypergraphs and the associated rewriting rules endows the resulting causal sets with a natural dynamics, as well as with additional mathematical structure that simplifies many of the derivations of properties such as Lorentz and Poincar\'e symmetry, in addition to the constructions of appropriate discretizations of spacetime geodesics, Ricci curvature and the Einstein-Hilbert action, as we shall proceed to see over the next several sections. For the sake of expository simplicity, the remainder of this article will primarily use examples of causal sets that have been produced by sprinklings into flat regions of spacetime; however, note that the numerical simulations and comparisons presented here have also been made against sprinklings into a variety of curved spacetimes also, as illustrated for the case of a de Sitter spacetime in Figure \ref{fig:Figure72}.

\section{Lorentz Symmetry and Causal Invariance}
\label{sec:Section2}

One somewhat counterintuitive aspect of the causal set approach lies in understanding how a fundamentally discrete object such as a causal set can possibly be compatible with a continuous symmetry group, such as the Lorentz and Poincar\'e groups that describe the isometries of Minkowski space. As first indicated by Bombelli, Henson and Sorkin\cite{bombelli4}, the key to this compatibility lies in the measure-theoretic properties of the Poisson distribution, as illustrated by the following argument\cite{strichartz}:

\begin{definition}
If ${f : X \to Y}$ is a function between sets $X$ and $Y$, and ${\Sigma}$ is a ${\sigma}$-algebra of subsets of $Y$, then the ``${\sigma}$-algebra generated by function $f$'', denoted ${\sigma \left( f \right)}$, is given by:

\begin{equation}
\sigma \left( f \right) = \left\lbrace f^{-1} \left( S \right) : S \in \Sigma \right\rbrace.
\end{equation}
\end{definition}

\begin{definition}
A function ${f : X \to Y}$ between measurable spaces $X$ and $Y$, with associated ${\sigma}$-algebras ${\Sigma}$ and $T$, is said to be a ``measurable map'' if:

\begin{equation}
\forall E \in T, \qquad f^{-1} \left( E \right) = \left\lbrace x \in X : f \left( x \right) \in E \right\rbrace \in \Sigma,
\end{equation}
i.e. if ${\sigma \left( f \right) \subseteq \Sigma}$, where ${\sigma \left( f \right)}$ is the ${\sigma}$-algebra generated by $f$.
\end{definition}

\begin{definition}
A ``$G$-set'' for a group $G$\cite{auslander} is a set $S$ equipped with the group action (on the left) of $G$ on $S$.
\end{definition}

\begin{definition}
A function ${f : X \to Y}$ between $G$-sets $X$ and $Y$ (for a common group $G$) is said to be an ``equivariant map'' if:

\begin{equation}
\forall g \in G, x \in X \qquad f \left( g \circ x \right) = g \circ \left( f \circ x \right).
\end{equation}
\end{definition}
We can extend the definition of equivariance to the case in which both $X$ and $Y$ are right group actions as follows:

\begin{equation}
\forall g \in G, x \in X, \qquad f \left( x \circ g \right) = f \left( x \right) \circ g,
\end{equation}
or alternatively to the case in which $X$ is a right action and $Y$ is a left action, or $X$ is a left action and $Y$ is a right action, as:

\begin{equation}
\forall g \in G, x \in X, \qquad f \left( x \circ g \right) = g^{-1} \circ f \left( x \right),
\end{equation}
or:

\begin{equation}
\forall g \in G, x \in X, \qquad f \left( g \circ x \right) = f \left( x \right) \circ g^{-1},
\end{equation}
respectively. Equivariant maps can be described quite generally as morphisms in the category of $G$-sets, and thus the condition of equivariance becomes equivalent to the statement that the following diagram:

\begin{equation}
\begin{tikzcd}
X \arrow[r, "g \circ"] \arrow[d, "f"] & X \arrow[d, "f"]\\
Y \arrow[r, "g \circ"] & Y
\end{tikzcd},
\end{equation}
commutes, where ${g \circ}$ denotes the map from elements $z$ to compositions ${g \circ z}$.

If we denote the space of all possible sprinklings into $n$-dimensional Minkowski space ${\mathbb{R}^{1, n - 1}}$ by ${\Omega}$, then the Poisson sprinkling process may be described formally as a stochastic process of the form ${\left( \Omega, \Sigma, \mu \right)}$, where ${\Sigma}$ is the ${\sigma}$-algebra of all measurable subsets of ${\Omega}$, and ${\mu}$ is the probability measure:

\begin{equation}
\mu : \Sigma \to \mathbb{R},
\end{equation}
such that the measure is trivially invariant under Lorentz boosts, i.e:

\begin{equation}
\forall \Lambda \in SO^{+} \left( 1, n -1 \right), \qquad \mu = \mu \circ \Lambda,
\end{equation}
since the definition of the Poisson process:

\begin{equation}
P_v \left( n \right) = \frac{\left( \rho_c v \right)^n}{n!} \exp \left( - \rho_c v \right),
\end{equation}
depends only upon the spacetime volume $v$\cite{chiu}, and so the measure must be preserved by any volume-preserving map. In the above description, ${SO^{+} \left( 1, n - 1 \right)}$ denotes the restricted Lorentz group, i.e. the subgroup of the full Lorentz group ${O \left( 1, n - 1 \right)}$ that preserves both orientations and the direction of time (that is, the connected component of the identity element in ${O \left( 1, n - 1 \right)}$).

The proof of compatibility of the sprinkled causal set ${\mathcal{C}}$ with Lorentz symmetry now proceeds by showing that no preferred direction map ${\mathbf{D}}$ can be defined in ${\mathcal{C}}$ for any ${n > 1}$; in other words, if ${\mathcal{H}^{n - 1}}$ denotes the hyperboloid of future-directed timelike unit vectors, then there cannot exist any equivariant measurable map:

\begin{equation}
\mathbf{D} : \Omega \to \mathcal{H}^{n - 1},
\end{equation}
i.e. there does not exist a measurable map ${\mathbf{D}}$ with the property that:

\begin{equation}
\forall \Lambda \in SO^{+} \left( 1, n - 1 \right), \qquad \mathbf{D} \circ \Lambda = \Lambda \circ \mathbf{D},
\end{equation}
which we can equivalently express as the statement that the following diagram:

\begin{equation}
\begin{tikzcd}
\Omega \arrow[r, "\Lambda"] \arrow[d, "\mathbf{D}"] & \Omega \arrow[d, "\mathbf{D}"]\\
\mathcal{H}^{n - 1} \arrow[r, "\Lambda"] & \mathcal{H}^{n - 1}
\end{tikzcd},
\end{equation}
cannot be made to commute for any ${\mathbf{D}}$. If such a map were to exist, then the inverse map ${\mathbf{D}^{-1}}$ from subsets of ${\mathcal{H}^{n - 1}}$ to subsets of ${\Omega}$ would allow us to define the following probability distribution on ${\mathcal{H}^{n -1}}$:

\begin{equation}
\mu_{\mathbf{D}} = \mu \circ \mathbf{D}^{-1},
\end{equation}
with the property that ${\mu_{\mathbf{D}}}$ would also be invariant under Lorentz boosts, since:

\begin{equation}
\forall \Lambda \in SO^{+} \left( 1, n - 1 \right), \qquad \mathbf{D} \circ \Lambda = \Lambda \circ \mathbf{D} \implies \Lambda \circ \mathbf{D}^{-1} = \mathbf{D}^{-1} \circ \Lambda,
\end{equation}
and therefore:

\begin{equation}
\forall \Lambda \in SO^{+} \left( 1, n - 1 \right), \qquad \mu_{\mathbf{D}} = \mu \circ \mathbf{D}^{-1} = \mu \circ \Lambda \circ \mathbf{D}^{-1} = \mu \circ \mathbf{D}^{-1} \circ \Lambda = \mu_{\mathbf{D}} \circ \Lambda.
\end{equation}
as required. However, since the hyperboloid ${\mathcal{H}^{n - 1}}$ is not compact, we can now force ${\mu_{\mathbf{D}} \left( U \right)}$ (for some open set ${U \subset \mathcal{H}^{n - 1}}$ whose closure is compact, and which therefore has a finite measure) to be arbitrarily large, by assuming without loss of generality that:

\begin{equation}
\mu_{\mathbf{D}} \left( U \right) > 0,
\end{equation}
and then applying a boost ${\Lambda}$ which forces $U$ and all of its images:

\begin{equation}
U_m = \Lambda^{m} U,
\end{equation}
to be disjoint sets. By Lorentz symmetry, we have:

\begin{equation}
\mu_{\mathbf{D}} \left( U_m \right) = \mu_{\mathbf{D}} \left( U \right),
\end{equation}
and moreover, by the additivity of measures:

\begin{equation}
\forall m \in \mathbb{N}, \qquad \mu_{\mathbf{D}} \left( \bigcup_{i =  1}^{m} U_i \right) = m \mu_{\mathbf{D}} \left( U \right),
\end{equation}
which can be made arbitrarily large by making $m$ arbitrarily large, and in particular can be made to exceed 1, hence violating the axioms of a probability measure. Thus, by contradiction, no such equivariant measurable map ${\mathbf{D}}$ can exist, which completes the proof.

We can see that this compatibility depends not only upon the measure-theoretic properties of the Poisson distribution, but moreover on the geometrical structure of Lorentzian manifolds in dimension ${n > 1}$. If, on the other hand, we had considered ${\Omega}$ to be the space of all possible sprinklings into the Euclidean plane ${\mathbb{R}^2}$, then we can show that there \textit{may} exist an equivariant measurable map:

\begin{equation}
\mathbf{D} : \Omega \to S^1,
\end{equation}
where ${S^1}$ is the circle of unit vectors in ${\mathbb{R}^2}$ centered at some point ${e \in \Phi \left( \omega \right)}$, and where ${\omega \in \Omega}$ is a particular sprinkling. Once again, the invariance of the measure ${\mu}$ and the assumed equivariance of the measurable map ${\mathbf{D}}$ under 2-dimensional rotations would allow us to define a probability distribution on ${S^1}$:

\begin{equation}
\mu_{\mathbf{D}} = \mu \circ \mathbf{D}^{-1},
\end{equation}
where ${\mathbf{D}^{-1}}$ is the inverse map from subsets of ${S^1}$ to subsets of ${\Omega}$, such that ${\mu_{\mathbf{D}}}$ is also invariant under 2-dimensional rotations:

\begin{equation}
\forall \Lambda \in SO \left( 2 \right), \qquad \mu_{\mathbf{D}} = \mu \circ \mathbf{D}^{-1} = \mu \circ \Lambda \circ \mathbf{D}^{-1} = \mu \circ \mathbf{D}^{-1} \circ \Lambda = \mu_{\mathbf{D}} \circ \Lambda,
\end{equation}
where the rotation group ${SO \left( 2 \right)}$ is trivially a subgroup of the 2-dimensional Euclidean group ${SO \left( 2 \right) \subset ISO \left( 2 \right)}$, although now, since ${S^1}$ is compact, no such contradiction follows. In particular, we can choose the measurable set:

\begin{equation}
U = \left( 0, \frac{2 \pi}{m} \right) \in S^1,
\end{equation}
such that applying the rotation ${\Lambda \left( \frac{2 \pi}{m} \right)}$ to $U$ once yields a disjoint set ${U^{\prime}}$, but applying it $m$ times now yields:

\begin{equation}
\Lambda^m \left( \frac{2 \pi}{m} \right) \circ U = U,
\end{equation}
such that:

\begin{equation}
\mu_{\mathbf{D}} \left( U \right) = \frac{1}{m},
\end{equation}
since ${\mu_{\mathbf{D}}}$ is invariant under each rotation, and so:

\begin{equation}
\mu_{\mathbf{D}} \left( S^1 \right) = 1.
\end{equation}
Therefore, a preferred direction map ${\mathbf{D}}$ may be constructed consistently within the sprinkling ${\omega \in \Omega}$, allowing us to break Euclidean symmetry as required.

\begin{figure}[ht]
\centering
\includegraphics[width=0.495\textwidth]{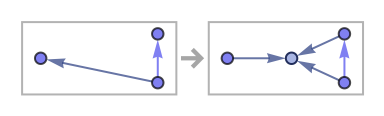}
\caption{A hypergraph transformation rule corresponding to the set substitution system ${\left\lbrace \left\lbrace x, y \right\rbrace, \left\lbrace x, z \right\rbrace \right\rbrace \to \left\lbrace \left\lbrace x, y \right\rbrace, \left\lbrace x, w \right\rbrace, \left\lbrace y, w \right\rbrace, \left\lbrace z, w \right\rbrace \right\rbrace}$. Example taken from \cite{wolfram2}.}
\label{fig:Figure13}
\end{figure}

\begin{figure}[ht]
\centering
\includegraphics[width=0.695\textwidth]{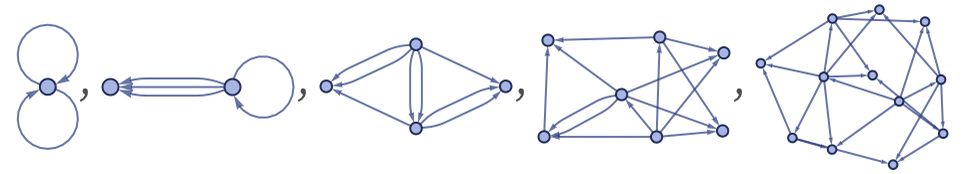}
\caption{The results of the first four steps in the evolution history of the set substitution system ${\left\lbrace \left\lbrace x, y \right\rbrace, \left\lbrace x, z \right\rbrace \right\rbrace \to \left\lbrace \left\lbrace x, y \right\rbrace, \left\lbrace x, w \right\rbrace, \left\lbrace y, w \right\rbrace, \left\lbrace z, w \right\rbrace \right\rbrace}$, starting from a double self-loop initial condition. Example taken from \cite{wolfram2}.}
\label{fig:Figure14}
\end{figure}

\begin{figure}[ht]
\centering
\includegraphics[width=0.695\textwidth]{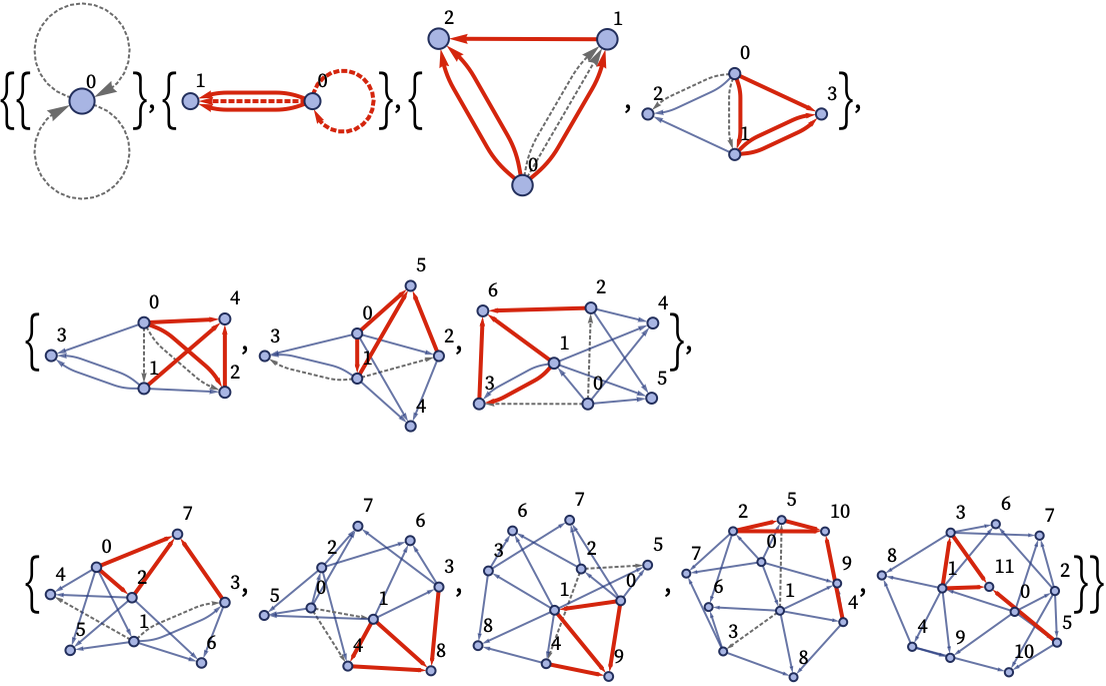}
\caption{The non-overlapping (and hence, spacelike-separated) subhypergraphs to which the transformation rules are being applied in the first four steps in the evolution history of the set substitution system ${\left\lbrace \left\lbrace x, y \right\rbrace, \left\lbrace x, z \right\rbrace \right\rbrace \to \left\lbrace \left\lbrace x, y \right\rbrace, \left\lbrace x, w \right\rbrace, \left\lbrace y, w \right\rbrace, \left\lbrace z, w \right\rbrace \right\rbrace}$, starting from a double self-loop initial condition. Example taken from \cite{wolfram2}.}
\label{fig:Figure15}
\end{figure}

In the case of causal sets generated by the evolution of Wolfram model systems, however, there exists a much more direct method of guaranteeing invariance of the causal network under the action of the Lorentz group, by exploiting the bijective correspondence between hypergraph updating orders and spacetime reference frames\cite{gorard}. As mentioned previously, our default choice of updating order simply applies a given transformation rule to every possible matching and non-overlapping subhypergraph, as shown (for example) in Figures \ref{fig:Figure13} and \ref{fig:Figure14}, with the non-overlapping (and hence, spacelike-separated) subhypergraphs corresponding to each event in the evolution shown in Figure \ref{fig:Figure15}.

However, the choice of which particular set of non-overlapping subhypergraphs to update, and hence of which particular set of spacelike-separated updating events to be applied simultaneously, is usually not pre-determined. In other words, different choices of spacelike-separated updating events to apply simultaneously will, in general, yield non-isomorphic sequences of hypergraphs in the evolution, and so the evolution will be generically non-deterministic. We can parametrize this non-determinism by using the formalism of an \textit{abstract rewriting system} from mathematical logic\cite{baader}\cite{bezem}:

\begin{definition}
An ``abstract rewriting system'' or (``ARS'') is a set, denoted $A$ (in which each element is known as an ``object''), equipped with a binary relation, denoted ${\to}$, known as the ``rewrite relation''.
\end{definition}

\begin{figure}[ht]
\centering
\includegraphics[width=0.695\textwidth]{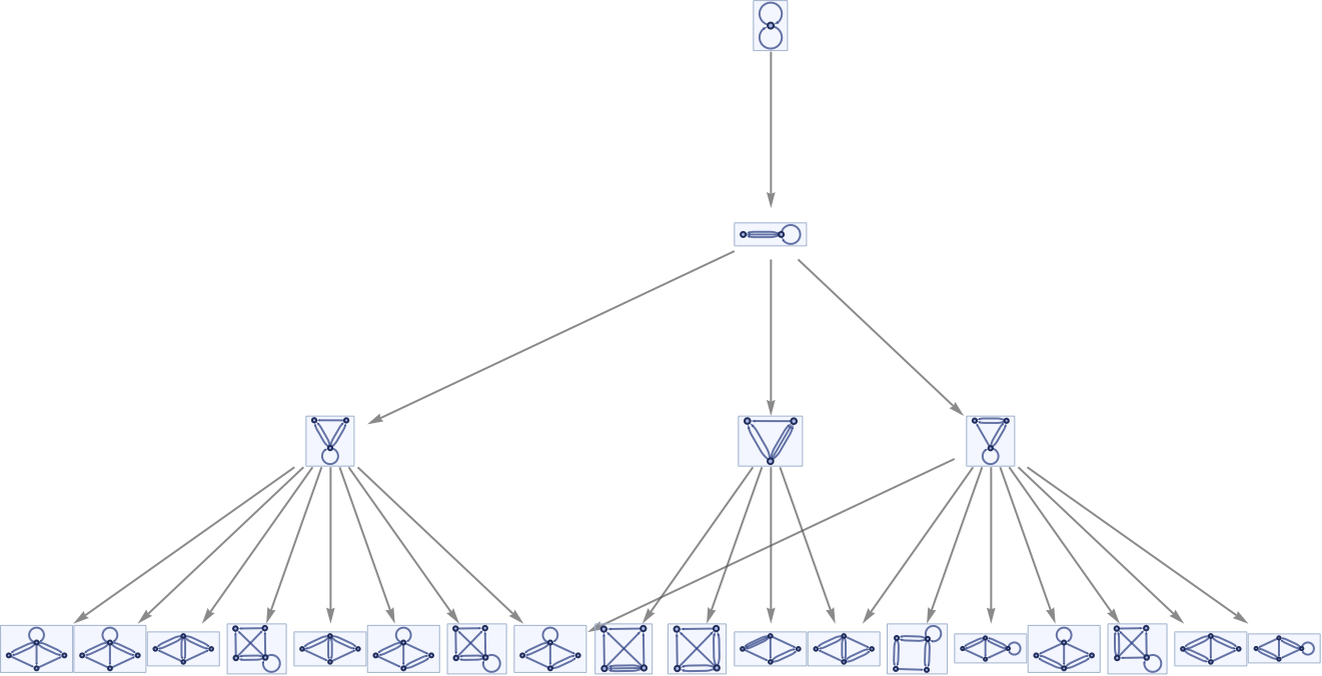}
\caption{The multiway evolution graph corresponding to the non-deterministic evolution of the set substitution system ${\left\lbrace \left\lbrace x, y \right\rbrace, \left\lbrace x, z \right\rbrace \right\rbrace \to \left\lbrace \left\lbrace x, y \right\rbrace, \left\lbrace x, w \right\rbrace, \left\lbrace y, w \right\rbrace, \left\lbrace z, w \right\rbrace \right\rbrace}$. Example taken from \cite{wolfram2}.}
\label{fig:Figure16}
\end{figure}

\begin{figure}[ht]
\centering
\includegraphics[width=0.495\textwidth]{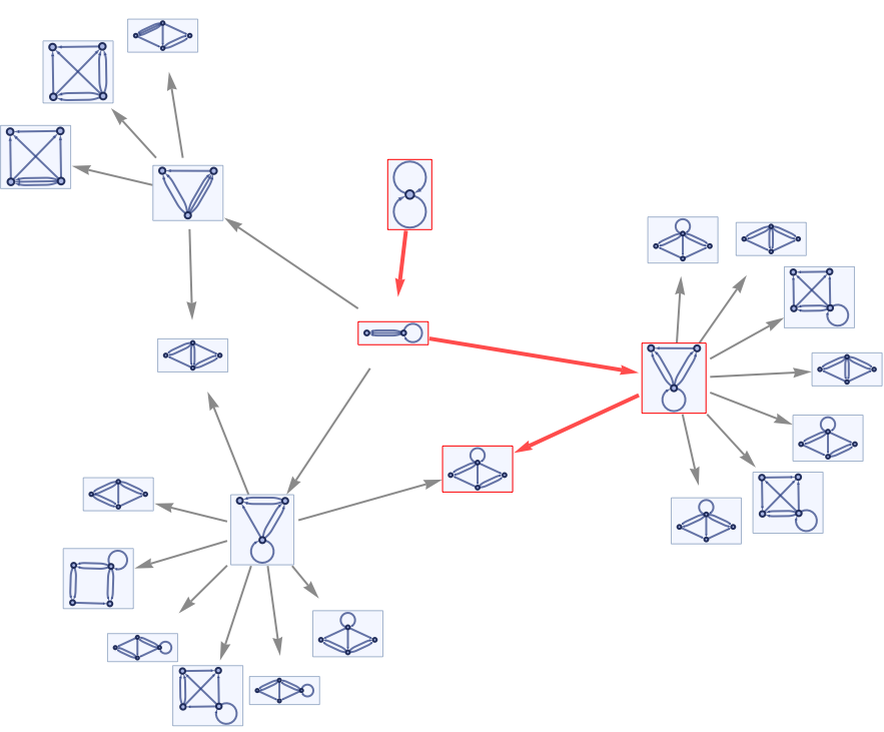}
\caption{The standard updating order for the evolution of the set substitution system ${\left\lbrace \left\lbrace x, y \right\rbrace, \left\lbrace x, z \right\rbrace \right\rbrace \to \left\lbrace \left\lbrace x, y \right\rbrace, \left\lbrace x, w \right\rbrace, \left\lbrace y, w \right\rbrace, \left\lbrace z, w \right\rbrace \right\rbrace}$, highlighted as a single path in the associated multiway evolution graph. Example taken from \cite{wolfram2}.}
\label{fig:Figure17}
\end{figure}

\begin{definition}
${\to^{*}}$ is the reflexive transitive closure of ${\to}$, i.e. it is the transitive closure of ${\to \cup =}$, where $=$ denotes the standard identity relation.
\end{definition}
In other words, ${\to^{*}}$ is the smallest preorder that contains ${\to}$, and hence the smallest binary relation that contains ${\to}$ and also satisfies the axioms of reflexivity and transitivity:

\begin{equation}
\forall a, b, c \in A, \qquad a \to^{*} a, \qquad \text{ and } \qquad a \to^{*} b, b \to^{*} c \implies a \to^{*} c.
\end{equation}
\textit{Multiway systems} are a class of combinatorial structures that allow one to represent the abstract rewriting structure of Wolfram model systems in a more concrete way:

\begin{definition}
A ``multiway system'' (or, strictly speaking, a ``multiway evolution graph''), denoted ${G_{multiway}}$, is a directed, acyclic graph in which every vertex corresponds to an object, and in which the directed edge ${a \to b}$ exists if and only if there exists a rewrite rule application transforming object $a$ to object $b$.
\end{definition}
In other words, a directed edge connects vertices $a$ and $b$ in a multiway evolution graph if and only if ${a \to b}$ in the associated abstract rewriting system, and a directed path connects vertices $a$ and $b$ if and only if ${a \to^{*} b}$, i.e. if there exists a finite rewriting sequence of the form:

\begin{equation}
a \to a^{\prime} \to a^{\prime \prime} \to \cdots \to b^{\prime} \to b.
\end{equation}
Therefore, we can express the non-deterministic evolution of a generic Wolfram model system as a multiway evolution graph, with the ``standard'' updating order shown above corresponding to a particular path in the associated graph, as shown in Figures \ref{fig:Figure16} and \ref{fig:Figure17}. Note that, in all such multiway evolution graphs, state vertices are merged based on hypergraph isomorphism (using a slightly generalized version of the algorithm presented in \cite{gorard3}).

The statement that different choices of updating order will generically yield non-isomorphic sequences of hypergraphs corresponds to the statement that the associated abstract rewriting system is generically not (globally) \textit{confluent}\cite{dershowitz}\cite{huet}:

\begin{definition}
An object ${a \in A}$ (for an abstract rewriting system $A$) is ``confluent'' if and only if:

\begin{equation}
\forall b, c \in A, \text{ such that } a \to^{*} b \text{ and } a \to^{*} c, \qquad \exists d \in A \text{ such that } b \to^{*} d \text{ and } c \to^{*} d.
\end{equation}
\end{definition}

\begin{figure}[ht]
\centering
\includegraphics[width=0.395\textwidth]{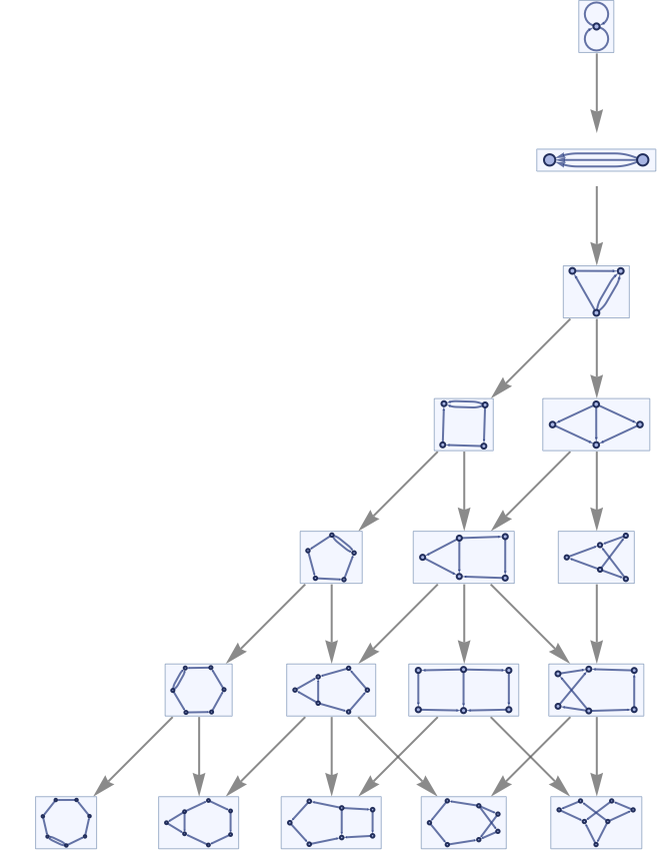}
\caption{An example of a multiway evolution graph for a (globally) confluent set substitution system ${\left\lbrace \left\lbrace x, y \right\rbrace, \left\lbrace z, y \right\rbrace \right\rbrace \to \left\lbrace \left\lbrace x, w \right\rbrace, \left\lbrace y, w \right\rbrace, \left\lbrace z, w \right\rbrace \right\rbrace}$, in which all bifurcations converge after a single step. Example taken from \cite{wolfram2}.}
\label{fig:Figure18}
\end{figure}

\begin{definition}
An abstract rewriting system $A$ is (globally) ``confluent'' (or is said to exhibit the ``Church-Rosser property'') if and only if every object ${a \in A}$ is confluent.
\end{definition}
In other words, the notion of confluence formalizes a property of certain multiway systems in which all bifurcations in the multiway evolution graph can (eventually) be made to converge, as shown in Figure \ref{fig:Figure18} for the case of a multiway evolution graph in which all bifurcations converge after a single step.

\begin{figure}[ht]
\centering
\includegraphics[width=0.395\textwidth]{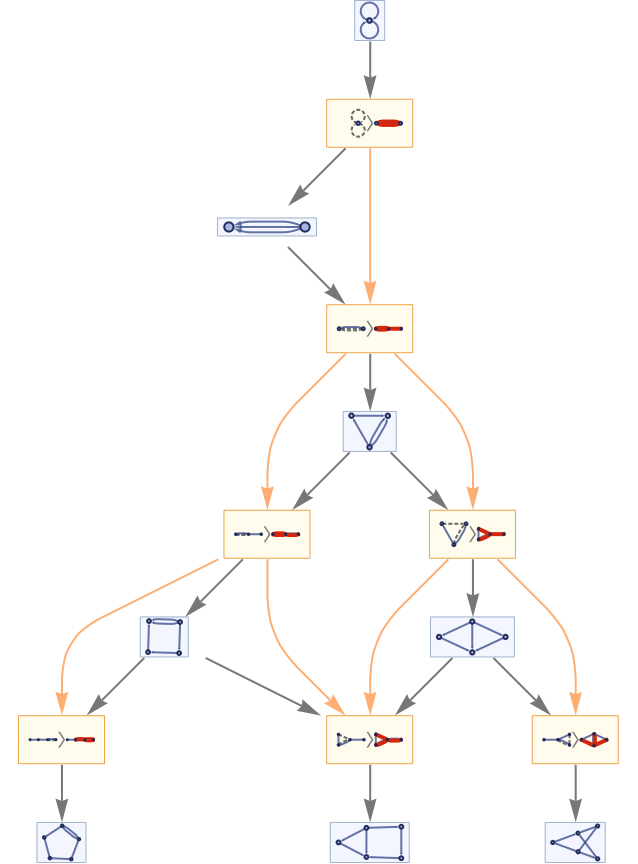}
\caption{The multiway evolution causal graph (with evolution edges shown in gray and causal edges shown in orange) for the (globally) confluent set substitution system ${\left\lbrace \left\lbrace x, y \right\rbrace, \left\lbrace z, y \right\rbrace \right\rbrace \to \left\lbrace \left\lbrace x, w \right\rbrace, \left\lbrace y, w \right\rbrace, \left\lbrace z, w \right\rbrace \right\rbrace}$, illustrating trivial causal invariance. Example taken from \cite{wolfram2}.}
\label{fig:Figure19}
\end{figure}

The condition of confluence in abstract rewriting systems is directly related to the criterion of \textit{causal invariance} in multiway evolution:

\begin{definition}
A multiway system is ``causal invariant'' if and only if the causal networks associated with all paths through the multiway system are (eventually) isomorphic as directed, acyclic graphs.
\end{definition}
Indeed, global confluence is a necessary (though not sufficient) condition for causal invariance, which can easily be proved by strong induction on the set of updating events. An example of a \textit{multiway evolution causal graph} (in which updating events are shown in yellow, state vertices are shown in blue, evolution edges are shown in gray and causal edges are shown in orange), for the confluent set substitution system shown in Figure \ref{fig:Figure18}, demonstrating that it also exhibits trivial causal invariance, is illustrated in Figure \ref{fig:Figure19}.

In order to specify the sets of spacelike-separated updating events to be applied simultaneously at each evolution step, and hence to specify which particular hypergraph evolution order to use, it suffices to define a \textit{universal time function} $t$ that maps updating events to integers, i.e:

\begin{equation}
t : \mathcal{M} \to \mathbb{Z}, \qquad \text{ such that } \Delta t \neq 0 \text{ everywhere},
\end{equation}
where ${\mathcal{M}}$ denotes, as usual, the set of all updating events in the causal network, such that the level sets of $t$ \textit{foliate} the causal network into a family of non-intersecting \textit{spacelike hypersurfaces}:

\begin{equation}
\forall t_1, t_2 \in \mathbb{Z}, \qquad \Sigma_{t_1} = \left\lbrace p \in \mathcal{M} : t \left( p \right) = t_1 \right\rbrace, \qquad \text{ and } \qquad \Sigma_{t_1} \cap \Sigma_{t_2} = \emptyset \iff t_1 \neq t_2.
\end{equation}
In the above, spacelike hypersurfaces are the discrete analog of Cauchy surfaces in spacetime, i.e. they are sets of updating events ${\Sigma \subset \mathcal{M}}$ with the property that every timelike or lightlike (i.e. null) path through the causal network that cannot be extended in either direction (taken here to be the discrete analog of a smooth lightlike or timelike curve) intersects exactly one updating event in ${\Sigma}$. We can give a slightly more formal definition as follows:

\begin{definition}
The ``discrete future Cauchy development'' of a set of updating events ${S \subset \mathcal{M}}$, denoted ${D^{+} \left( S \right)}$, is the set of all updating events $x$ for which every past-directed, inextendible, causal (i.e. non-spacelike) path in the causal network intersects at least one updating event in $S$.
\end{definition}

\begin{definition}
The ``discrete past Cauchy development'' of a set of updating events ${S \subset \mathcal{M}}$, denoted ${D^{-} \left( S \right)}$, is the set of all updating events $x$ for which every future-directed, inextendible, causal (i.e. non-spacelike) path in the causal network intersects at least one updating event in $S$.
\end{definition}

\begin{definition}
The ``discrete Cauchy development'' of a set of updating events ${S \subset \mathcal{M}}$, denoted ${D \left( S \right)}$, is the union of the discrete future and past Cauchy developments:

\begin{equation}
D \left( S \right) = D^{+} \left( S \right) \cup D^{-} \left( S \right).
\end{equation}
\end{definition}

\begin{definition}
A set of updating events ${S \subset \mathcal{M}}$ is ``achronal'' if and only if the set $S$ is disjoint from its own chronological future:

\begin{equation}
\nexists q, r \in S, \qquad \text{ such that } r \in I^{+} \left( q \right).
\end{equation}
\end{definition}

\begin{figure}[ht]
\centering
\includegraphics[width=0.495\textwidth]{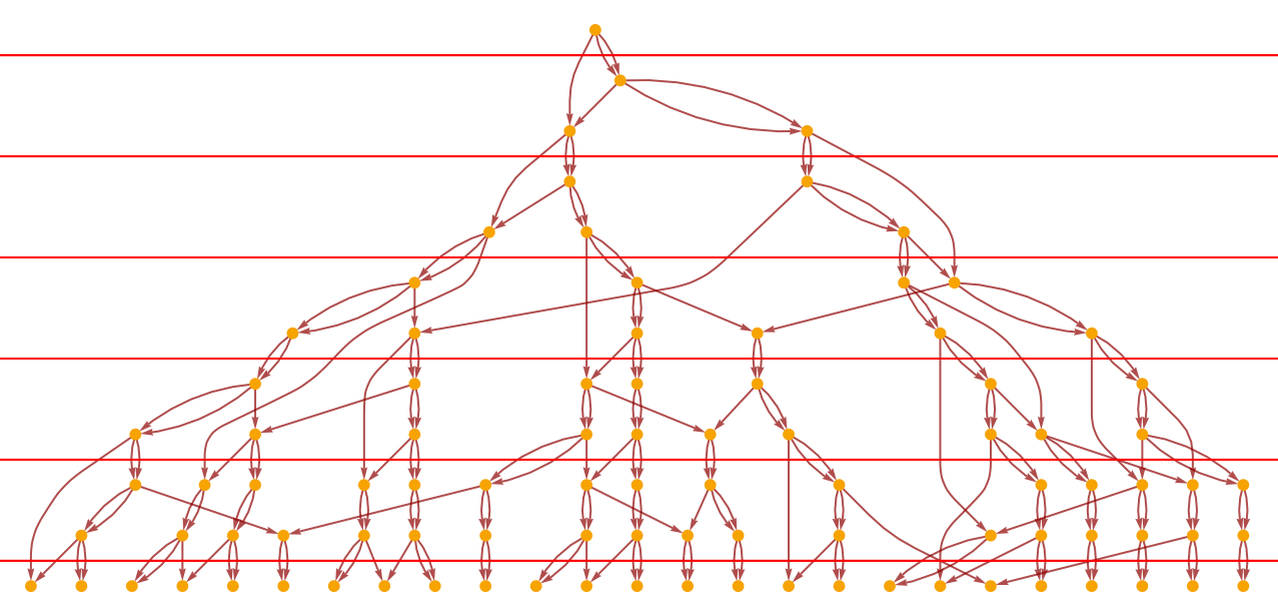}
\includegraphics[width=0.395\textwidth]{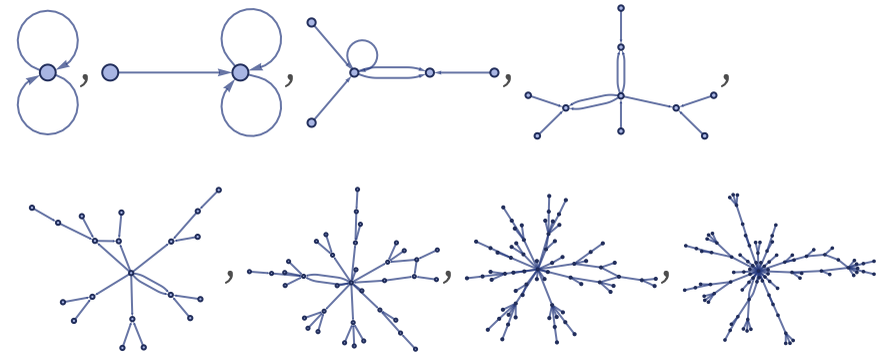}
\caption{The default choice of foliation of the causal network for the set substitution system ${\left\lbrace \left\lbrace x, y \right\rbrace, \left\lbrace z, y \right\rbrace \right\rbrace \to \left\lbrace \left\lbrace x, z \right\rbrace, \left\lbrace y, w \right\rbrace, \left\lbrace w, z \right\rbrace \right\rbrace}$, corresponding to the canonical choice of updating order. Example taken from \cite{wolfram2}.}
\label{fig:Figure20}
\end{figure}

\begin{figure}[ht]
\centering
\includegraphics[width=0.495\textwidth]{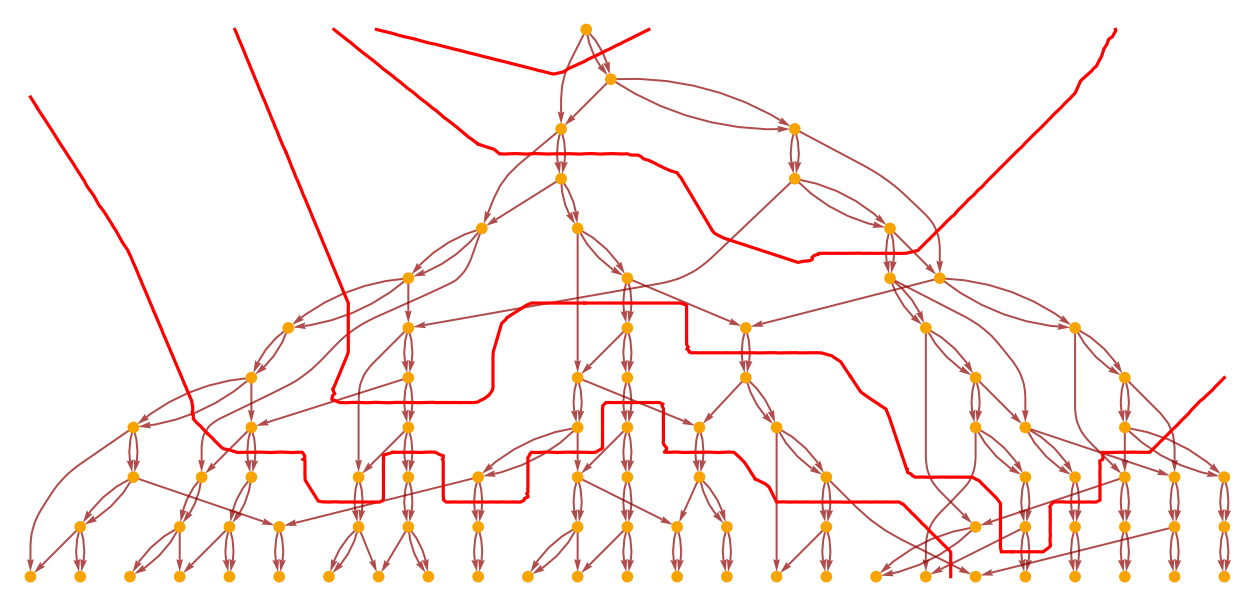}
\includegraphics[width=0.495\textwidth]{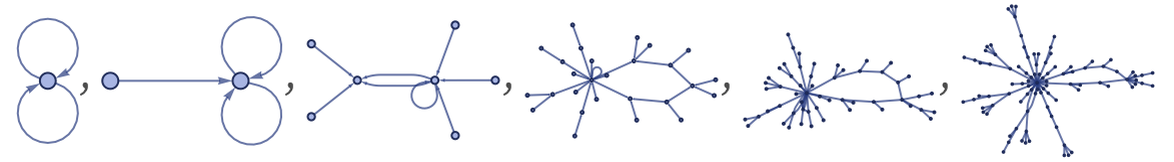}
\caption{An alternative possible choice of foliation of the causal network for the set substitution system ${\left\lbrace \left\lbrace x, y \right\rbrace, \left\lbrace z, y \right\rbrace \right\rbrace \to \left\lbrace \left\lbrace x, z \right\rbrace, \left\lbrace y, w \right\rbrace, \left\lbrace w, z \right\rbrace \right\rbrace}$, corresponding to a different possible choice of updating order. Example taken from \cite{wolfram2}.}
\label{fig:Figure21}
\end{figure}

\begin{definition}
A ``discrete Cauchy surface'' in ${\mathcal{M}}$ is an achronal set of updating events $S$ whose discrete Cauchy development ${D \left( S \right)}$ is equal to ${\mathcal{M}}$.
\end{definition}
The existence of such a universal time function $t$ is the discrete analog of the condition of global hyperbolicity in the associated Lorentzian manifold ${\left( \mathcal{M}, g \right)}$, and therefore in particular necessitates that no closed timelike curves may exist (i.e. the causal network must indeed be acyclic). An example of two possible foliations of the causal network for a simple set substitution system, corresponding to two possible choices of updating order, are shown in Figures \ref{fig:Figure20} and \ref{fig:Figure21}; it is possible to note by eye that the sequences of hypergraphs associated with the two evolutions are non-isomorphic.

\begin{figure}[ht]
\centering
\includegraphics[width=0.595\textwidth]{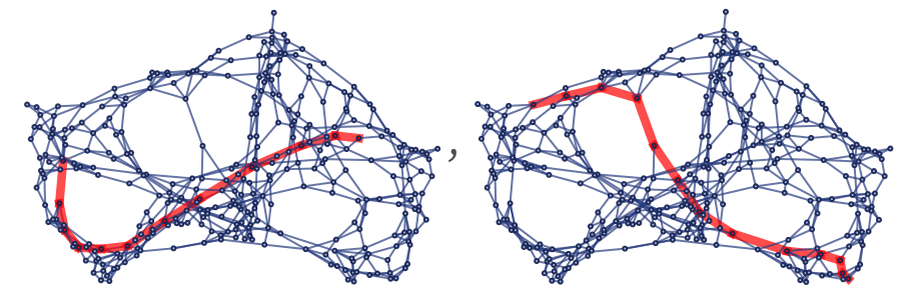}
\caption{Two examples of geodesics on a hypergraph generated by the set substitution system ${\left\lbrace \left\lbrace x, y \right\rbrace, \left\lbrace x, z \right\rbrace \right\rbrace \to \left\lbrace \left\lbrace x, z \right\rbrace, \left\lbrace x, w \right\rbrace, \left\lbrace y, w \right\rbrace, \left\lbrace z, w \right\rbrace \right\rbrace}$, illustrating the natural combinatorial distance metric. Example taken from \cite{wolfram2}.}
\label{fig:Figure22}
\end{figure}

\begin{figure}[ht]
\centering
\includegraphics[width=0.595\textwidth]{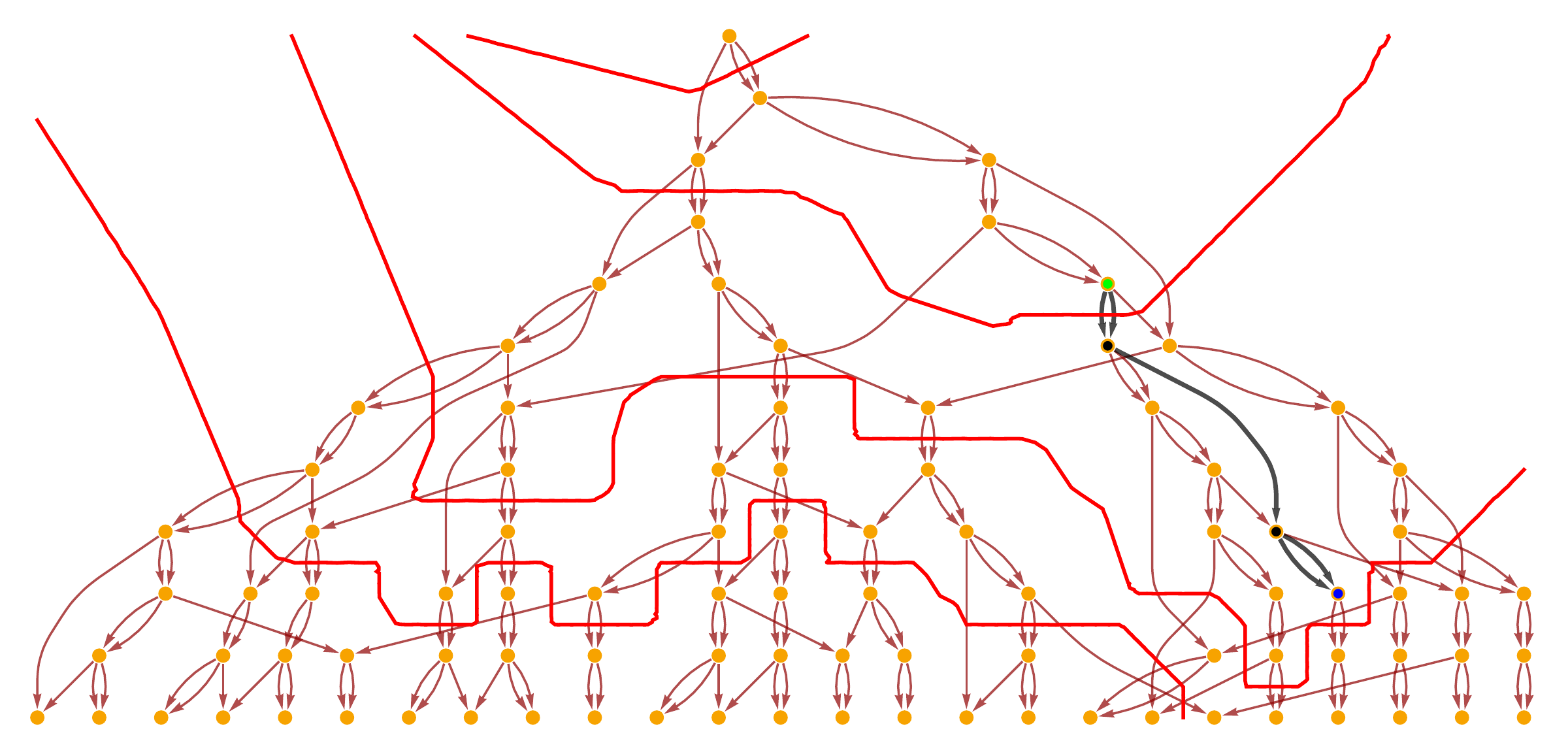}
\caption{A foliation of the causal network for the set substitution system ${\left\lbrace \left\lbrace x, y \right\rbrace, \left\lbrace z, y \right\rbrace \right\rbrace \to \left\lbrace \left\lbrace x, z \right\rbrace, \left\lbrace y, w \right\rbrace, \left\lbrace w, z \right\rbrace \right\rbrace}$, with two updating events on two neighboring hypersurfaces highlighted in green and blue, respectively, and the shortest path between them highlighted in black. The length of the black path (three causal edges) corresponds to the value of ${\Delta \tau}$, whilst the average geodesic distance between the green and blue events in the associated hypergraphs corresponds to the value of (the projection of) ${\beta^{i}}$.}
\label{fig:Figure23}
\end{figure}

Since each hypergraph corresponds to a particular spacelike hypersurface, we can parametrize the set of possible foliations of a given causal network by using the following discrete analog of the ADM formalism for canonical decomposition of Lorentzian manifolds\cite{arnowitt}\cite{misner}. The standard combinatorial (geodesic) distance, denoted ${\Delta l}$, between points in a hypergraph (i.e. the number of hyperedges that one must traverse in order to travel from one point to the other), as shown in Figure \ref{fig:Figure22}, induces a natural spatial metric on each spacelike hypersurface, which we denote ${\gamma_{i j}}$, such that:

\begin{equation}
\Delta l^2 = \gamma_{i j} \Delta x^{i} \Delta x^{j},
\end{equation}
for some choice of spatial coordinates ${x^{i} \left( t \right)}$ corresponding to a particular choice of hypergraph embedding. For each updating event in the causal network, the direction of outgoing causal edges defines a local vector, which we denote ${\mathbf{n} \in \mathbb{Z}^{1, n}}$, representing the relativistic ${\left( 1 + n \right)}$-velocity of a normal observer (i.e. an observer in geodesic normal coordinates, in which all geodesics correspond to straight lines through the origin). The number of causal edges that one must traverse between two updating events on neighboring hypersurfaces at times $t$ and ${t + \Delta t}$, denoted ${\Delta \tau}$, is therefore given by:

\begin{equation}
\Delta \tau = \alpha \Delta t,
\end{equation}
where ${\alpha}$ is a \textit{lapse function} that determines the foliation in the timelike direction. Moreover, we can relabel the spatial coordinates ${x^{i}}$ in accordance with the scheme:

\begin{equation}
x^{i} \left( t + \Delta t \right) = x^{i} \left( t \right) - \beta^{i} \left( t \right),
\end{equation}
where ${\beta^{i}}$ is a \textit{shift vector} that determines the foliation in the spacelike direction. Therefore, the overall discrete spacetime line element may be written in the following general form:

\begin{equation}
\Delta s^2 = \left( - \alpha^2 + \beta^{i} \beta_{i} \right) \Delta t^2 + 2 \beta_{i} \Delta x^{i} \Delta t + \gamma_{i j} \Delta x^{i} \Delta x^{j}.
\end{equation}
We can see this decomposition illustrated for the case of a particular causal network foliation in Figure \ref{fig:Figure23}; the two updating events on the neighboring hypersurfaces at times $t$ and ${t + \Delta t}$ are highlighted in green and blue, respectively, with the shortest path between them in the causal network highlighted in black. The length of the black path, which in this case consists of three causal edges, corresponds to the value of ${\Delta \tau = \alpha \Delta t}$, whilst the average geodesic distance between the green and blue events in the associated hypergraphs corresponds to the value of (the projection of) ${\beta^{i}}$.

Transformations between reference frames in general relativity (as parametrized in terms of transformations between lapse functions ${\alpha}$ and shift vectors ${\beta^{i}}$ for normal observers) are given by diffeomorphism transformations corresponding to arbitrary actions of the conformal group ${C \left( 1, n \right)}$ on a Lorentzian manifold, and hence causal invariance corresponds precisely to a statement of conformal invariance of the causal network (which one can see more explicitly by, for instance, defining all pairs of non-parallel causal edges emanating from a single updating event to be orthogonal, corresponding to a choice of natural units in which ${c = 1}$). Since both the Lorentz and Poincar\'e groups are subgroups of the conformal group:

\begin{equation}
O \left( 1, n \right) \subset C \left( 1, n \right), \qquad \text{ and } \qquad \left( \mathbb{R}^{1, n} \rtimes O \left( 1, n \right) \right) \subset C \left( 1, n \right),
\end{equation}
we can see that both Lorentz and Poincar\'e symmetry follow immediately from the requirement of causal invariance (since causal invariance guarantees that the ordering of non-spacelike-separated updating events is always preserved under arbitrary changes of foliation).

\begin{figure}[ht]
\centering
\includegraphics[width=0.495\textwidth]{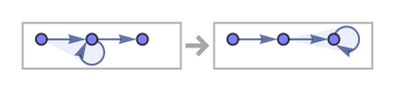}
\caption{A hypergraph transformation rule corresponding to the set substitution system ${\left\lbrace \left\lbrace x, y, y, \right\rbrace, \left\lbrace y, z \right\rbrace \right\rbrace \to \left\lbrace \left\lbrace x, y \right\rbrace, \left\lbrace y, z, z \right\rbrace \right\rbrace}$, which is known to produce a simple grid-like causal network. Example taken from \cite{wolfram2}.}
\label{fig:Figure24}
\end{figure}

If we wish to see a toy example of how this works more directly, we can consider the case of a simple grid-like causal network whose structure is analogous to that of 1+1-dimensional Minkowski space, such as the one produced by the rule shown in Figure \ref{fig:Figure24}. In this instance, we can use the standard combinatorial (geodesic) distance between points on each hypergraph, as well as the causal edge distance between non-spacelike-separated updating events, to label each updating event according to the following discrete coordinate scheme:

\begin{equation}
\mathbf{p} = \left( t, \mathbf{x} \right), \qquad \text{ where } t \in \mathbb{Z} \text{ and } \mathbf{x} = \left( x_1, \dots, x_n \right) \in \mathbb{Z}^{n},
\end{equation}
where, as usual:

\begin{equation}
\Delta \tau = \alpha \Delta t, \qquad x^{i} \left( t + \Delta t \right) = x^{i} \left( t \right) - \beta^{i} \left( t \right), \qquad \text{ etc.}
\end{equation}
This, in turn, allows us to define the following very na\"ive discrete analog of the Minkowski norm over the causal network:

\begin{definition}
The ``discrete Minkowski norm'' is given by:

\begin{equation}
\left\lVert \left( t, \mathbf{x} \right) \right\rVert = \left\lVert \mathbf{x} \right\rVert^2 - t^2 = \left( x_{1}^{2} + \cdots + x_{n}^{2} \right) - t^2.
\end{equation}
\end{definition}
By construction, this norm allows us to separate updating events in such a way that they are causally related (i.e. connected by a directed path in the causal network) if and only if they are timelike- or lightlike-separated, and spacelike-separated (i.e. connected by a family of paths in the hypergraph) otherwise:

\begin{definition}
Updating events ${\mathbf{p} = \left( t, \mathbf{x} \right)}$ are classified as either ``timelike'', ``lightlike'' or ``spacelike'' based upon their discrete Minkowski norm:

\begin{equation}
\mathbf{p} \sim \begin{cases}
\text{timelike}, \qquad &\text{ if } \left\lVert \left( t, \mathbf{x} \right) \right\rVert < 0,\\
\text{lightlike}, \qquad &\text{ if } \left\lVert \left( t, \mathbf{x} \right) \right\rVert = 0,\\
\text{spacelike}, \qquad &\text{ if } \left\lVert \left( t, \mathbf{x} \right) \right\rVert > 0.
\end{cases}
\end{equation}
\end{definition}

\begin{figure}[ht]
\centering
\includegraphics[width=0.325\textwidth]{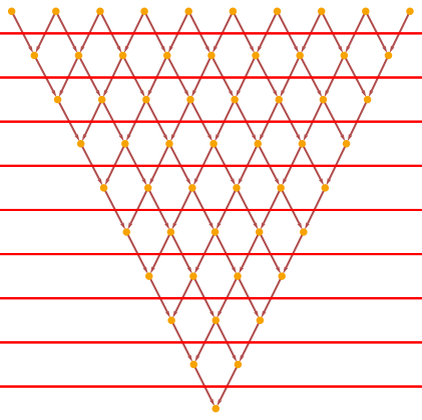}\hspace{0.1\textwidth}
\includegraphics[width=0.245\textwidth]{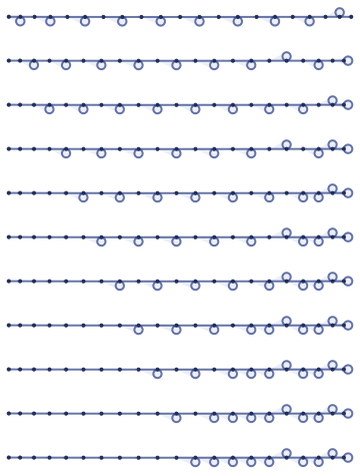}
\caption{The default choice of foliation of the causal network, along with a depiction of the standard updating order (as witnessed by an observer in a rest frame $F$), for the set substitution system ${\left\lbrace \left\lbrace x, y, y \right\rbrace, \left\lbrace y, z \right\rbrace \right\rbrace \to \left\lbrace \left\lbrace x, y \right\rbrace, \left\lbrace y, z, z \right\rbrace \right\rbrace}$. Example taken from \cite{wolfram2}.}
\label{fig:Figure25}
\end{figure}

\begin{figure}[ht]
\centering
\includegraphics[width=0.325\textwidth]{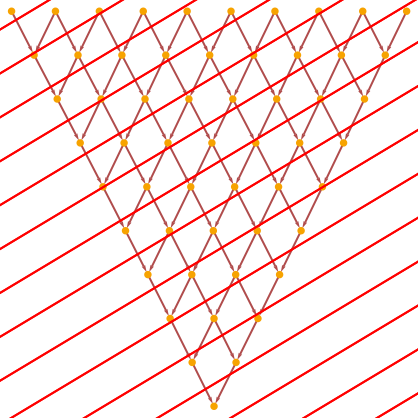}\hspace{0.1\textwidth}
\includegraphics[width=0.245\textwidth]{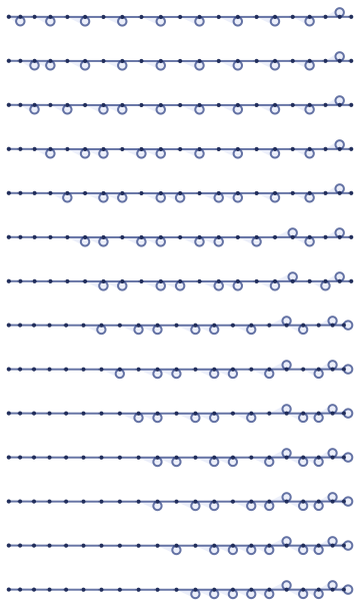}
\caption{A ``tipped'' foliation of the causal network, along with a depiction of the modified updating order (as witnessed by an observer in a boosted frame ${F^{\prime}}$), for the set substitution system ${\left\lbrace \left\lbrace x, y, y \right\rbrace, \left\lbrace y, z \right\rbrace \right\rbrace \to \left\lbrace \left\lbrace x, y \right\rbrace, \left\lbrace y, z, z \right\rbrace \right\rbrace}$. Example taken from \cite{wolfram2}.}
\label{fig:Figure26}
\end{figure}

\begin{definition}
Pairs of updating events ${\mathbf{p} = \left( t_1, \mathbf{x}_1 \right)}$, ${\mathbf{q} = \left( t_2, \mathbf{x}_2 \right)}$ can be classified as either ``timelike-separated'', ``lightlike-separated'' or ``spacelike-separated'' accordingly:

\begin{equation}
\left( \mathbf{p}, \mathbf{q} \right) \sim \begin{cases}
\text{timelike-separated}, \qquad &\text{ if } \left( \left( t_1, \mathbf{x}_1 \right) - \left( t_2, \mathbf{x}_2 \right) \right) \sim \text{ timelike},\\
\text{lightlike-separated}, \qquad &\text{ if } \left( \left( t_1, \mathbf{x}_1 \right) - \left( t_2, \mathbf{x}_2 \right) \right) \sim \text{ lightlike},\\
\text{spacelike-separated}, \qquad &\text{ if } \left( \left( t_1, \mathbf{x}_1 \right) - \left( t_2, \mathbf{x}_2 \right) \right) \sim \text{ spacelike}.
\end{cases}
\end{equation}
\end{definition}
Thus, if we now consider two idealized observers, embedded within inertial reference frames $F$ and ${F^{\prime}}$ (here corresponding to two different foliations of the causal network into flat spacelike hypersurfaces), then we can take the observer in frame ${F^{\prime}}$ to be moving with a constant discrete velocity ${\mathbf{v} \in \mathbb{Z}^n}$ relative to the observer in frame $F$:

\begin{equation}
\mathbf{v} = \left( \tanh \left( \rho \right) \right) \mathbf{u},
\end{equation}
for a vector ${\mathbf{u} \in \mathbb{Z}^{n}}$ representing the direction of motion, and a scalar:

\begin{equation}
\tanh \left( \rho \right) = \frac{\left\lVert \mathbf{v} \right\rVert}{\left\lVert \mathbf{u} \right\rVert} < 1,
\end{equation}
representing the magnitude (with appropriate normalization). If the $F$ and ${F^{\prime}}$ coordinates are given by ${\left( t, \mathbf{x} \right)}$ and ${\left( t^{\prime}, \mathbf{x}^{\prime} \right)}$, respectively, and if we assume that the two frames are synchronized at the point where they initially coincide:

\begin{equation}
t = t^{\prime} = 0,
\end{equation}
then we can construct the following discrete analog of the Lorentz transformation:

\begin{definition}
The ``discrete Lorentz transformation'' expresses the ${F^{\prime}}$ coordinate system in terms of the $F$ coordinate system as:

\begin{equation}
t^{\prime} = \left\lfloor \left( \cosh \left( \rho \right) \right) t - \left( \sinh \left( \rho \right) \right) \mathbf{x} \cdot \mathbf{u} \right\rfloor,
\end{equation}
where ${\cdot}$ denotes the standard (``dot'') inner product of vectors in ${\mathbb{Z}^{n}}$.
\end{definition}
Applying this discrete Lorentz transformation to the grid-like causal network shown in Figure \ref{fig:Figure25} with the default foliation (corresponding to the updating order as seen by an observer in the rest frame $F$) yields the ``tipped'' foliation choice shown in Figure \ref{fig:Figure26} (corresponding to the updating order as seen by an observer in the boosted frame ${F^{\prime}}$), illustrating the effects of time dilation on the resultant hypergraph evolution.

Thus, in the case of causal networks generated by causal invariant Wolfram model systems, we can prove that no equivariant measurable map exists that could be used to construct a preferred direction map on the causal network, for such a preferred direction map would imply a preferred foliation of the causal network (and hence a preferred updating order on the hypergraph), which is incompatible with the hypothesis that the causal network is invariant under different choices of foliation. However, one final concern that arises regarding the preservation of Lorentz symmetry in causal invariant Wolfram model systems relates to the known property of manifold-like causal sets produced by sprinkling in which the number of future \textit{links} emanating from any given element is almost surely infinite:

\begin{definition}
A ``link'' in a causal set ${\mathcal{C}}$ is a pair of elements ${e, e^{\prime} \in \mathcal{C}}$ with the property that:

\begin{equation}
\nexists e^{\prime \prime} \in \mathcal{C}, \qquad \text{ such that } e^{\prime \prime} \neq e, e^{\prime \prime} \neq e^{\prime} \text{ and } e \prec e^{\prime \prime} \prec e^{\prime}.
\end{equation}
\end{definition}
In other words, links correspond exactly to the directed edges in the corresponding Hasse diagram for the causal set, and are therefore directly analogous to directed edges in a causal network. We can see that elements of a causal set ${\mathcal{C}}$ that has been produced by a uniform  distribution of points on a Lorentzian manifold ${\left( \mathcal{M}, g \right)}$ will, on average, have an infinite number of future links, due to the non-compact nature of the hyperboloid of future-directed timelike unit vectors in Lorentzian manifolds. More precisely, for any given element ${e \in \mathcal{C}}$, the probability that a randomly chosen element ${x \in \mathcal{C}}$ such that ${x \prec e}$ will correspond to a link is equal to the probability that the Alexandrov interval ${\mathbf{A} \left[ e, x \right]}$ in the continuum approximation contains exactly no elements in the causal set ${\mathcal{C}}$, i.e:

\begin{equation}
P_v \left( 0 \right) = \exp \left( - \rho_c v \right),
\end{equation}
which is generally negligible, except at the order of the characteristic spacetime discretisation scale ${v \sim V_c}$. Most of the future links of element $e$ will hence lie inside the hyperboloid:

\begin{equation}
v = V_c \pm \sqrt{V_c},
\end{equation}
since, in the continuum $n$-dimensional Minkowski space ${\mathbb{R}^{1, n - 1}}$, the set of spacetime events lying within a proper time on the order of ${\left( v \right)^{\frac{1}{n}}}$ of an event $p$ lies entirely within the region between the future light cone of $p$ and the hyperboloid:

\begin{equation}
\left( \sum_{i = 1}^{n -  1} x_{i}^{2} \right) - t^2 \propto\left( v \right)^{\frac{2}{n}},
\end{equation}
where ${t > 0}$.

\begin{figure}[ht]
\centering
\includegraphics[width=0.395\textwidth]{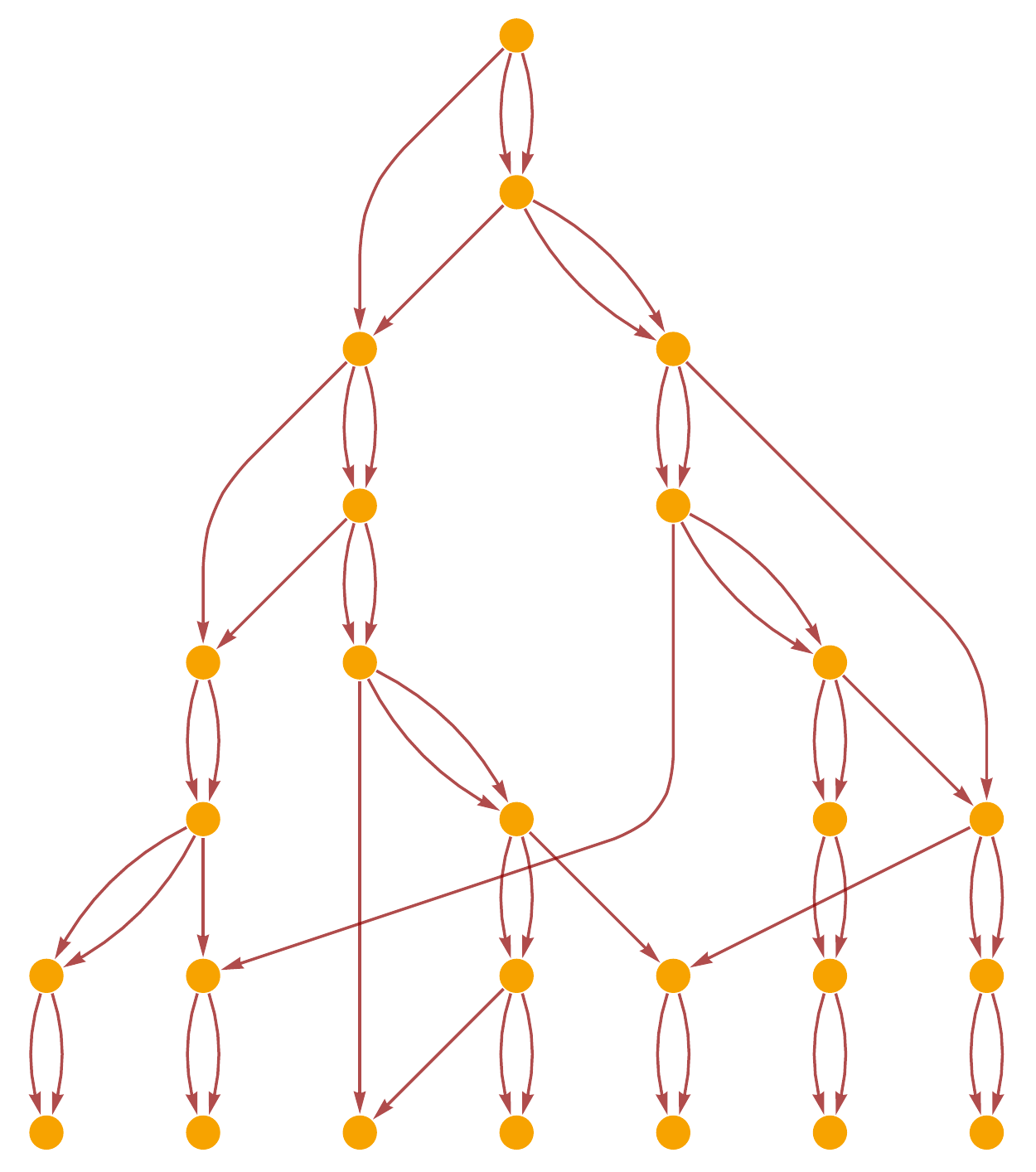}\hspace{0.1\textwidth}
\includegraphics[width=0.395\textwidth]{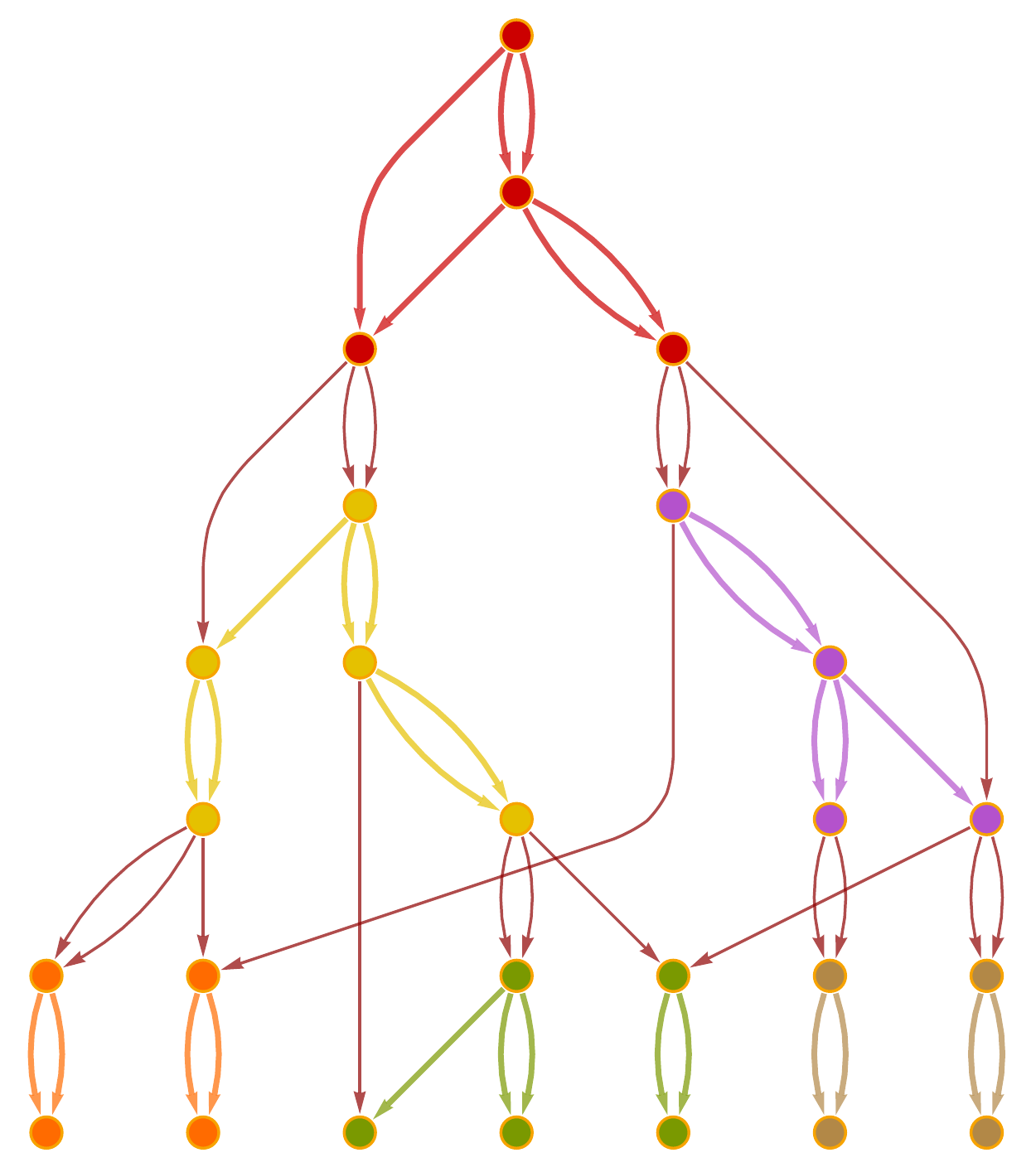}
\caption{On the left, the causal network after the first 8 evolution steps of the set substitution system ${\left\lbrace \left\lbrace x, y \right\rbrace, \left\lbrace z, y \right\rbrace \right\rbrace \to \left\lbrace \left\lbrace x, z \right\rbrace, \left\lbrace y, z \right\rbrace, \left\lbrace w, z \right\rbrace \right\rbrace}$. On the right, the same causal network with six equivalence classes defined over the updating events (highlighted in red, yellow, purple, orange, green and khaki).}
\label{fig:Figure27}
\end{figure}

\begin{figure}[ht]
\centering
\includegraphics[width=0.395\textwidth]{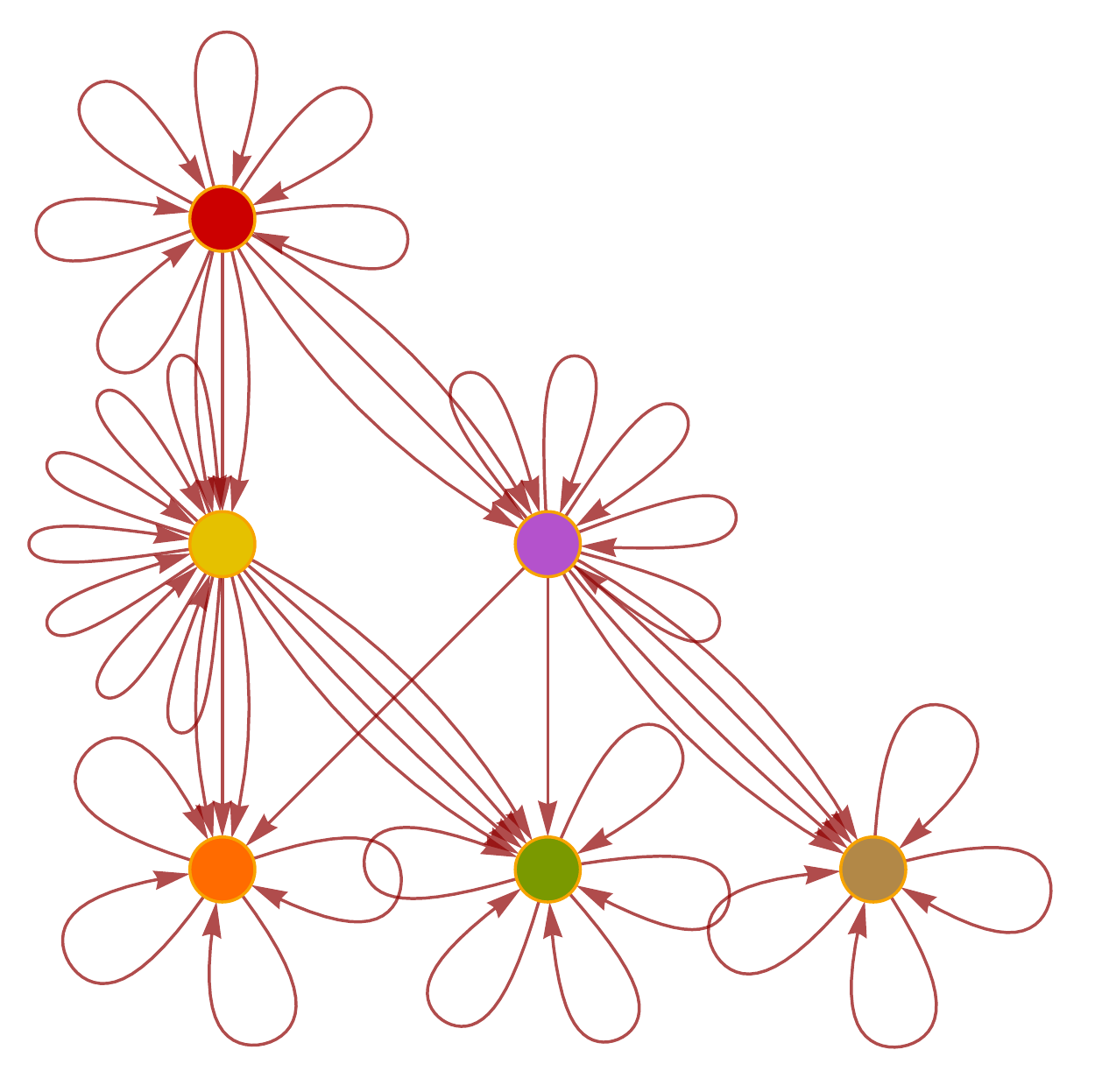}
\caption{The causal network after the first 8 evolution steps of the set substitution system ${\left\lbrace \left\lbrace x, y \right\rbrace, \left\lbrace z, y \right\rbrace \right\rbrace \to \left\lbrace \left\lbrace x, z \right\rbrace, \left\lbrace y, z \right\rbrace, \left\lbrace w, z \right\rbrace \right\rbrace}$, quotiented out by the six equivalence classes defined over the updating events (red, yellow, purple, orange, green and khaki) shown above.}
\label{fig:Figure28}
\end{figure}

Thus, it might at first appear that the average out-degree of any given updating event in a causal network would need to be infinite in order to maintain compatibility with Lorentz symmetry; however, such an argument would depend upon the (incorrect) assumption that the observer is able to discriminate between individual updating events (and hence to measure spacetime intervals on the order of an individual causal edge), which would be required in order for any putative violation of Lorentz symmetry to be macroscopically observable. Rather, as we have established above, the observer sees instead a progression from one spacelike hypersurface to the next, where the timelike distance between those hypersurfaces (i.e. the value of the lapse function ${\alpha}$) can be arbitrarily large, which is perfectly physically reasonable since elementary dimensional analysis\cite{wolfram2} suggests that the characteristic length scale of individual causal edges may be many orders of magnitude smaller than anything that can even \textit{in principle} be observed. It follows that one must instead construct an effective theory by essentially coarse-graining over the underlying (microscopic) causal network by using the specification of the foliation to define an equivalence relation over the set of updating events (hence dividing those events into finite equivalence classes), as shown in Figures \ref{fig:Figure27} and \ref{fig:Figure28}. By making the value of the lapse function ${\alpha}$ arbitrarily large, we have the effect of making these equivalence classes also arbitrarily large, such that the out-degree of vertices in the effective causal network may be unbounded, even if the out-degree of vertices in the underlying (microscopic) is bounded, as required.

\section{Dimension Estimation and Geodesics}
\label{sec:Section3}

One foundational question that arises both in the causal set program and in the Wolfram model approach to discrete spacetime is how best to determine the dimensionality $d$ of a continuum Lorentzian manifold ${\left( \mathcal{M}, g \right)}$ using only the discrete information provided by a causal set or causal network that is known to faithfully embed into ${\left( \mathcal{M}, g \right)}$. In the context of the causal set approach, Myrheim\cite{myrheim} demonstrated that the following quantity:

\begin{equation}
r = \frac{2 R}{n \left( n - 1 \right)},
\end{equation}
where $n$ denotes the number of elements in the finite causal set ${\mathcal{C}}$ and $R$ denotes the number of elements that are related by the causal partial order, i.e:

\begin{equation}
R = \left\lvert \left\lbrace \left( e_i, e_j \right) : e_i, e_j \in \mathcal{C} \text{ such that } e_i \prec e_j \right\rbrace \right\rvert,
\end{equation}
is a function purely of the dimensionality $d$ of the manifold ${\left( \mathcal{M}, g \right)}$ into which ${\mathcal{C}}$ faithfully embeds, where the maximum number of possible relations is ${n \choose 2}$. Thus, as pointed out by Meyer\cite{meyer}, we can therefore estimate the continuum dimensionality of the manifold (assuming in the first instance that the manifold is isometric to a $d$-dimensional flat/Minkowski spacetime ${\mathbb{R}^{1, d - 1}}$) by computing the expectation value of the random variable ${\hat{R}}$ associated with the number of relations $R$ over the ensemble ${\Omega}$ of causal sets representing the continuum Alexandrov interval:

\begin{equation}
\mathbf{A} \left[ p, q \right] \subset \mathcal{M},
\end{equation}
whose spacetime volume is much greater than the characteristic discreteness scale, ${v \gg \rho_{c}^{-1}}$. The probability that a randomly chosen pair of elements ${e_1, e_2 \in \mathbf{A} \left[ p, q \right]}$ are related by the causal partial order is a function of the probability of choosing ${e_1}$, which is proportional to the volume of the spacetime region ${\mathbf{A} \left[ p, q \right]}$, and the conditional probability of choosing an ${e_2}$ that lies in the future of ${e_1}$, which is proportional to the volume of the spacetime region ${J^{+} \left( e_1 \right) \cap J^{-} \left( p \right)}$.

\begin{figure}[ht]
\centering
\includegraphics[width=0.395\textwidth]{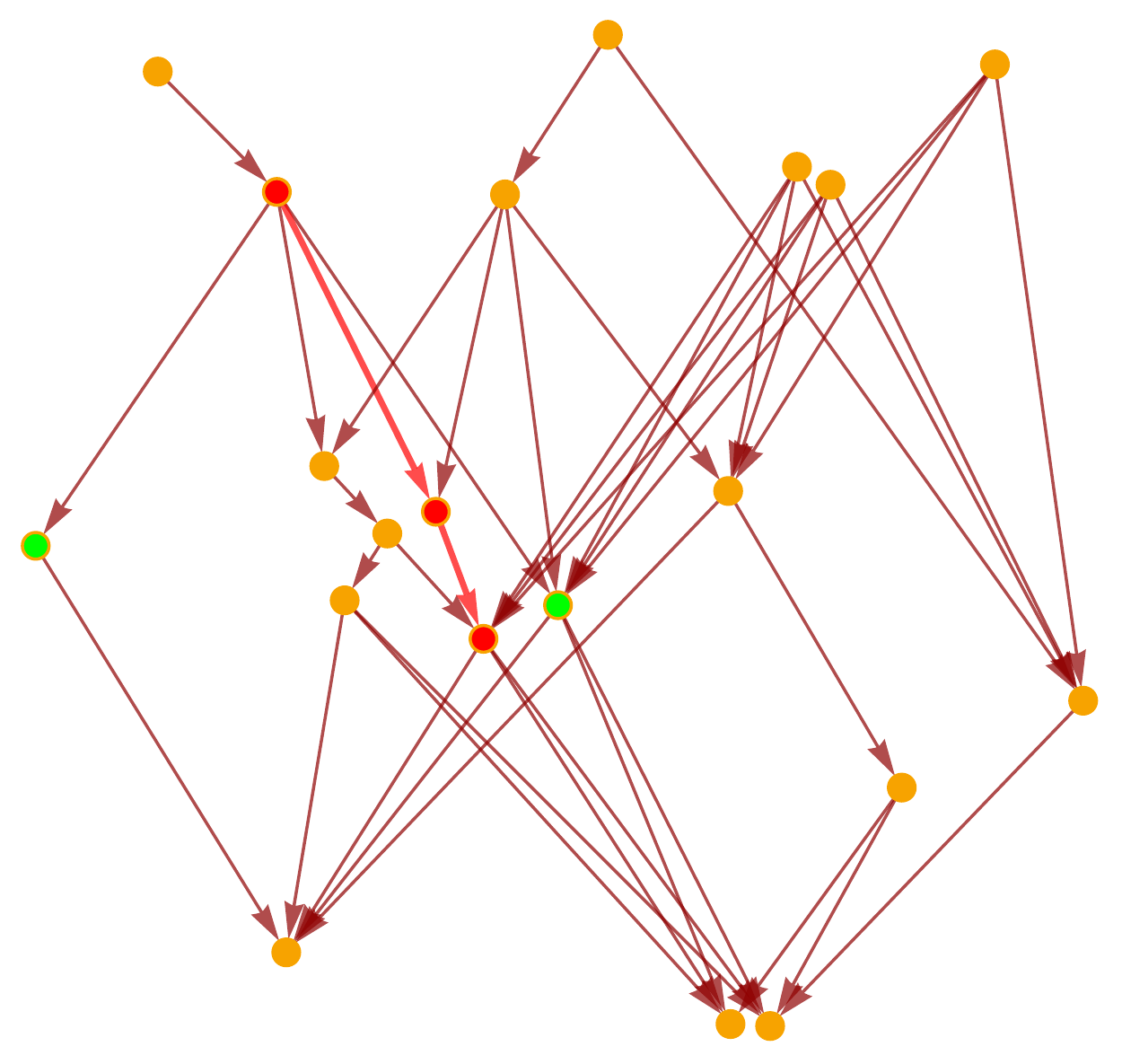}
\caption{The transitive reduction (i.e. the Hasse diagram) of the causal partial order graph for 20 uniformly sprinkled points in a rectangular region of ${1 + 1}$-dimensional flat (Minkowski) spacetime. Highlighted in red is a pair of randomly-chosen points which are related by the causal partial order (along with the corresponding path through the transitive reduction graph); conversely, highlighted in green is a pair of randomly-chosen points which are not related by the causal partial order.}
\label{fig:Figure29}
\end{figure}

\begin{figure}[ht]
\centering
\includegraphics[width=0.445\textwidth]{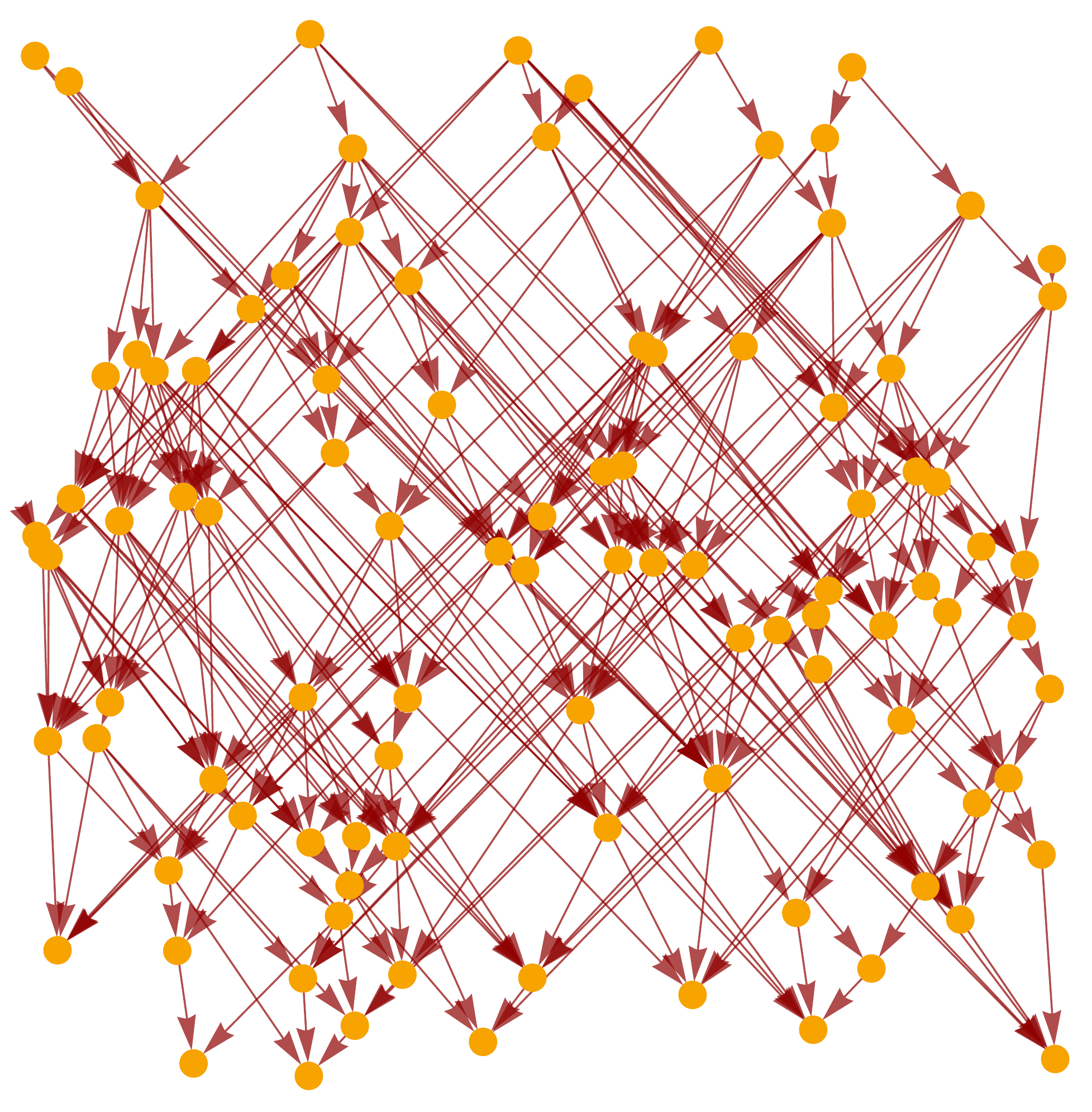}
\includegraphics[width=0.495\textwidth]{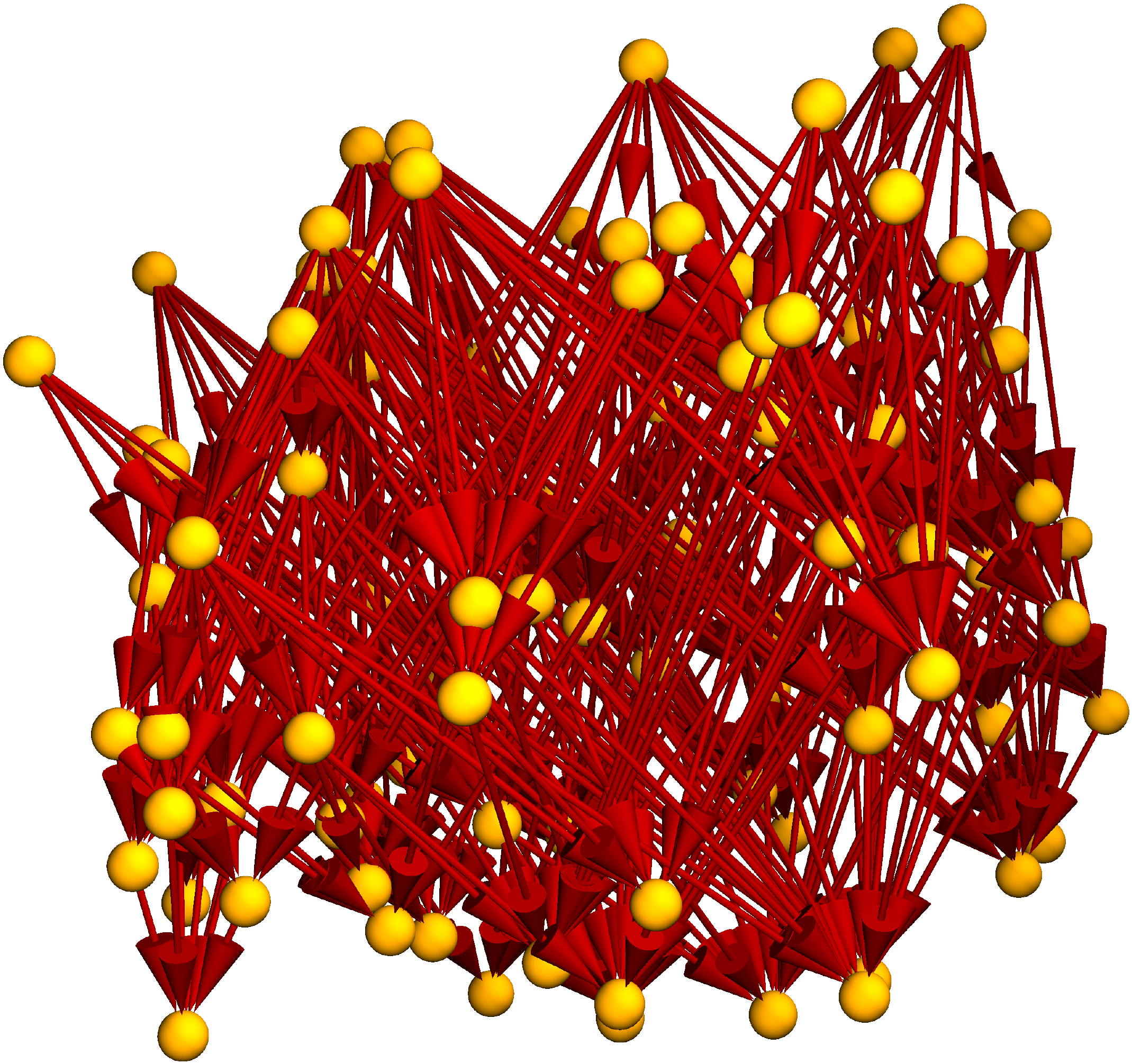}
\caption{The transitive reductions (i.e. the Hasse diagrams) of the causal partial order graphs for 100 uniformly sprinkled points in rectangular regions of ${1 + 1}$-dimensional and ${2 + 1}$-dimensional flat (Minkowski) spacetimes, yielding dimension estimates of ${2.07}$ and ${2.75}$, respectively, using the Myrheim-Meyer dimension estimation algorithm.}
\label{fig:Figure30}
\end{figure}

By Lorentz symmetry, we can select (without loss of generality):

\begin{equation}
p = \left( - \frac{T}{2}, 0, \dots, 0 \right), \qquad \text{ and } \qquad q = \left( \frac{T}{2}, 0, \dots, 0 \right),
\end{equation}
such that the total volume of the spacetime region ${\mathbf{A} \left[ p, q \right]}$ is given by:

\begin{equation}
v = \zeta_d T^d, \qquad \text{ where } \zeta_d = \frac{\mathcal{V}_{d - 2}}{2^{d - 1} d \left( d - 1 \right)},
\end{equation}
where ${\mathcal{V}_{d - 2}}$ is simply the volume of the unit ${\left( d - 2 \right)}$-sphere, yielding an expectation value for ${\hat{R}}$ of:

\begin{equation}
\left\langle \hat{R} \right\rangle = \rho_{c}^{2} \int_{\mathbf{A} \left[ p, q \right]} d x_1 \int_{J^{+} \left( x_1 \right) \cap J^{-} \left( q \right)} d x_2.
\end{equation}
Denoting the proper time between points ${x_1}$ and $q$ by ${T_1}$, we can rewrite this integral as:

\begin{equation}
\rho_{c}^{2} \int_{\mathbf{A} \left[ p, q \right]} d x_1 \int_{J^{+} \left( x_1 \right) \cap J^{-} \left( q \right)} d x_2 = \rho_{c}^{2} \zeta_d \int_{\mathbf{A} \left[ p, q \right]} d x_1 T_{1}^{d},
\end{equation}
which then evaluates to yield:

\begin{equation}
\left\langle \hat{R} \right\rangle = \rho_{c}^{2} v^2 \frac{\Gamma \left( d + 1 \right) \Gamma \left( \frac{d}{2} \right)}{4 \Gamma \left( \frac{3 d}{2} \right)}.
\end{equation}
Further exploiting the fact that:

\begin{equation}
\left\langle \hat{n} \right\rangle = \rho_c v,
\end{equation}
by the definition of the Poisson process, we can therefore estimate the dimensionality $d$ of the manifold ${\left( \mathcal{M}, g \right)}$ by simply solving the following equation for $d$:

\begin{equation}
\frac{\left\langle \hat{R} \right\rangle}{\left\langle \hat{n} \right\rangle^2} = \frac{\Gamma \left( d + 1 \right) \Gamma \left( \frac{d}{2} \right)}{4 \Gamma \left( \frac{3 d}{2} \right)} = f_0 \left( d \right).
\end{equation}
A basic illustration of how this sampling is performed is shown in Figure \ref{fig:Figure29}, and examples of the Myrheim-Meyer dimension estimation algorithm being applied to two causal sets consisting of 100 uniformly sprinkled points in rectangular regions of ${1 + 1}$-dimensional and ${2 + 1}$-dimensional flat (Minkowski) spacetimes, yielding dimension estimates of ${2.07}$ and ${2.75}$, respectively, are shown in Figure \ref{fig:Figure30}.

Indeed, Meyer's construction is sufficiently general that it actually yields an infinite parametric family of such dimension estimation algorithms, since although we have so far considered only \textit{chains} of length two, i.e. sequences of elements of the form ${e_1 \prec e_2}$, one can instead consider \textit{chains} of length $k$, i.e. sequences of the form ${e_1 \prec e_2 \prec \cdots \prec e_{k - 1} \prec e_{k}}$, for arbitrary $k$. Using ${C_k}$ to denote the number (or \textit{abundance}) of chains of length $k$ in the causal set ${\mathcal{C}}$:

\begin{equation}
C_k = \left\lvert \left\lbrace \left( e_i, e_{i + 1}, \dots, e_{i + k - 1} \right) : e_i, e_{i + 1}, \dots, e_{i + k - 1} \in \mathcal{C} \text{ such that } e_i \prec e_{i + 1} \prec \cdots \prec e_{i + k - 1} \right\rbrace \right\rvert,
\end{equation}
we can therefore compute the expectation value of the random variable ${\hat{C_k}}$ designating the abundance of chains of length $k$ in the ensemble ${\Omega}$ of causal sets representing the nested continuum Alexandrov intervals:

\begin{equation}
\mathcal{M} \supset \mathbf{A} \left[ p, q \right] \supset I \left( x_1, w \right) \supset I \left( x_2, q \right) \supset \cdots \supset I \left( x_k, q \right),
\end{equation}
by inductively evaluating the resulting $k$ nested integrals to yield:

\begin{equation}
\left\langle \hat{C_k} \right\rangle = \rho_{c}^{k} \chi_{k} v^k, \qquad \text{ where } \chi_{k} = \frac{1}{k} \left( \frac{\Gamma \left( d + 1 \right)}{2} \right)^{k - 1} \frac{\Gamma \left( \frac{d}{2} \right) \Gamma \left( d \right)}{\Gamma \left( \frac{k d}{2} \right) \Gamma \left( \frac{\left( k + 1 \right) d}{2} \right)},
\end{equation}
such that the ratio of ${\left\langle \hat{C_k} \right\rangle^{\frac{1}{k}}}$ to ${\left\langle \hat{C_k} \right\rangle^{\frac{1}{k^{\prime}}}}$ is purely a function of the manifold dimension $d$, for any choice of ${k, k^{\prime} \in \mathbb{N}}$. These constructions can, furthermore, be generalized to arbitrary (pseudo-) Riemannian manifolds ${\left( \mathcal{M}, g \right)}$ that are not isometric to $d$-dimensional Minkowski space ${\mathbb{R}^{1, d - 1}}$ as follows:

\begin{definition}
The neighborhood ${U_p}$ of a point $p$ in the (pseudo-) Riemannian manifold ${\left( \mathcal{M}, g \right)}$, equipped with an appropriate definition of geodesics, is known as a ``Riemann normal neighborhood''\cite{kobayashi} if we assign coordinates to each point $q$ in ${U_p}$ by following the geodesic emanating from point $p$, with tangent vector ${X_q}$, and connecting it to point $q$ within a unit distance of the affine parameter, thus defining an exponential map from the neighborhood ${V_0}$ of the origin of the tangent space ${T_p}$ to the neighborhood ${U_p}$ in ${\left( \mathcal{M}, g \right)}$, such that the coordinates of $q$ are the components of the tangent vector ${X_q}$. Such coordinates are known as ``Riemann normal coordinates''.
\end{definition}
As shown by Roy, Sinha and Surya\cite{roy}, the expectation value of the abundance of chains of length $k$ within a small causal diamond ${\mathbf{A} \left[ p, q \right]}$ in an arbitrary Lorentzian manifold ${\left( \mathcal{M}, g \right)}$ can thus be computed using the standard Riemann normal coordinate expansion of the metric in curved spacetime\cite{khetrapal}, yielding (up to the lowest-order correction):

\begin{equation}
\left\langle \hat{C_k} \right\rangle = \left\langle \hat{C_k} \right\rangle_{\eta} \left[ 1 + T^2 \alpha_k R \left( 0 \right) + T^2 \beta_k R_{0 0} \left( 0 \right) \right] + O \left( T^{k d + 3} \right),
\end{equation}
where:

\begin{equation}
\alpha_k = - \frac{d k}{12 \left( k d + 2 \right) \left( \left( k + 1 \right) d + 2 \right)}, \qquad \text{ and } \qquad \beta_k = \frac{d k}{12 \left( \left( k + 1 \right) d + 2 \right)},
\end{equation}
and where ${\left\langle \hat{C_k} \right\rangle}$ denotes the expectation value in flat spacetime, as given above. Therefore, when combined with the previous flat spacetime expansion, one obtains the following dimension estimator:

\begin{equation}
f_{0}^{2} \left( d \right) \left( - \frac{1}{3} \frac{\left( d + 2 \right)}{\left( 3d + 2 \right)} - \frac{\left( 4d + 2 \right)}{\left( 2d + 2 \right)} \left( \frac{\left\langle \hat{C_k} \right\rangle}{\chi_3} \right)^{\frac{4}{3}} \frac{1}{\left\langle \hat{C_k} \right\rangle^4} + \frac{1}{3} \frac{\left( 4d + 2 \right) \left( 5d + 2 \right)}{\left( 2d + 2 \right) \left( 3d + 2 \right)} \frac{\left\langle \hat{C_4} \right\rangle}{\chi_4} \frac{1}{\left\langle C_1 \right\rangle^4} \right) = - \frac{\left\langle C_2 \right\rangle^2}{\left\langle C_1 \right\rangle^4},
\end{equation}
where, as above:

\begin{equation}
f_0 \left( d \right) = \frac{\Gamma \left( d + 1 \right) \Gamma \left( \frac{d}{2} \right)}{4 \Gamma \left( \frac{3d}{2} \right)}.
\end{equation}
In order to justify this expansion, the causal diamond ${\mathbf{A} \left[ p, q \right]}$ must be \textit{small} in the sense that:

\begin{equation}
R \left( 0 \right) T^2 \ll 1, \qquad \text{ and } \qquad R_{0 0} \left( 0 \right) T^2 \ll 1,
\end{equation}
where ${R \left( 0 \right)}$ and ${R_{0 0} \left( 0 \right)}$ denote the Ricci scalar curvature and the time-time component of the Ricci curvature tensor at the center of the causal diamond, respectively, and $T$ denotes the proper time between events $p$ and $q$.

Conversely, if one already knows the limiting dimension $d$ of the causal set, then this expansion of the metric around a small causal diamond ${\mathbf{A} \left[ p, q \right]}$ in Riemann normal coordinates can be used to estimate the proper time distance $T$ between elements $p$ and $q$ as:

\begin{equation}
T^{3 d} = \frac{1}{2d^2 \rho_{c}^{3}} \left( J_1 - 2 J_2 + J_3 \right),
\end{equation}
where we define inductively:

\begin{equation}
J_k = \left( kd + 2 \right) \left( \left( k + 1 \right) d + 2 \right) \frac{1}{\zeta_{d}^{3}} \left( \frac{\left\langle \hat{C_k} \right\rangle}{\chi_k} \right)^{\frac{3}{k}},
\end{equation}
and where, as above, ${\hat{C_k}}$ denotes the expectation value for the abundance of chains of length $k$, and one has:

\begin{equation}
\zeta_d = \frac{\mathcal{V}_{d - 2}}{2^{d - 1} d \left( d - 1 \right)},
\end{equation}
for the volume of the unit ${\left( d - 2 \right)}$-sphere ${\mathcal{V}_{d - 2}}$, and:

\begin{equation}
\chi_k = \frac{1}{k} \left( \frac{\Gamma \left( d + 1 \right)}{2} \right)^{k - 1} \frac{\Gamma\left( \frac{d}{2} \right) \Gamma \left( d \right)}{\Gamma \left( \frac{kd}{2} \right) \Gamma \left( \frac{\left( k + 1 \right) d}{2} \right)}.
\end{equation}

\begin{figure}[ht]
\centering
\includegraphics[width=0.395\textwidth]{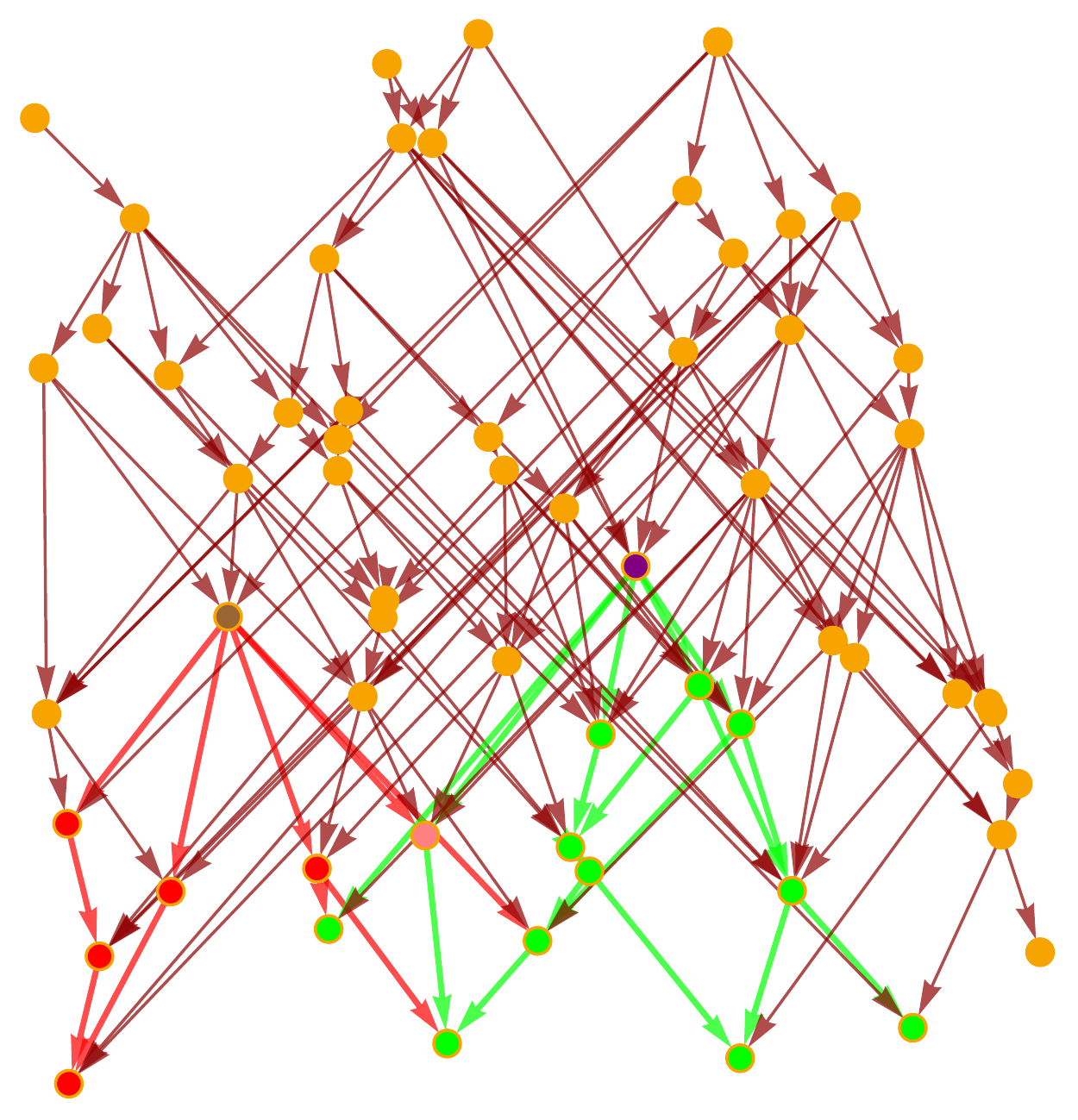}\hspace{0.1\textwidth}
\includegraphics[width=0.395\textwidth]{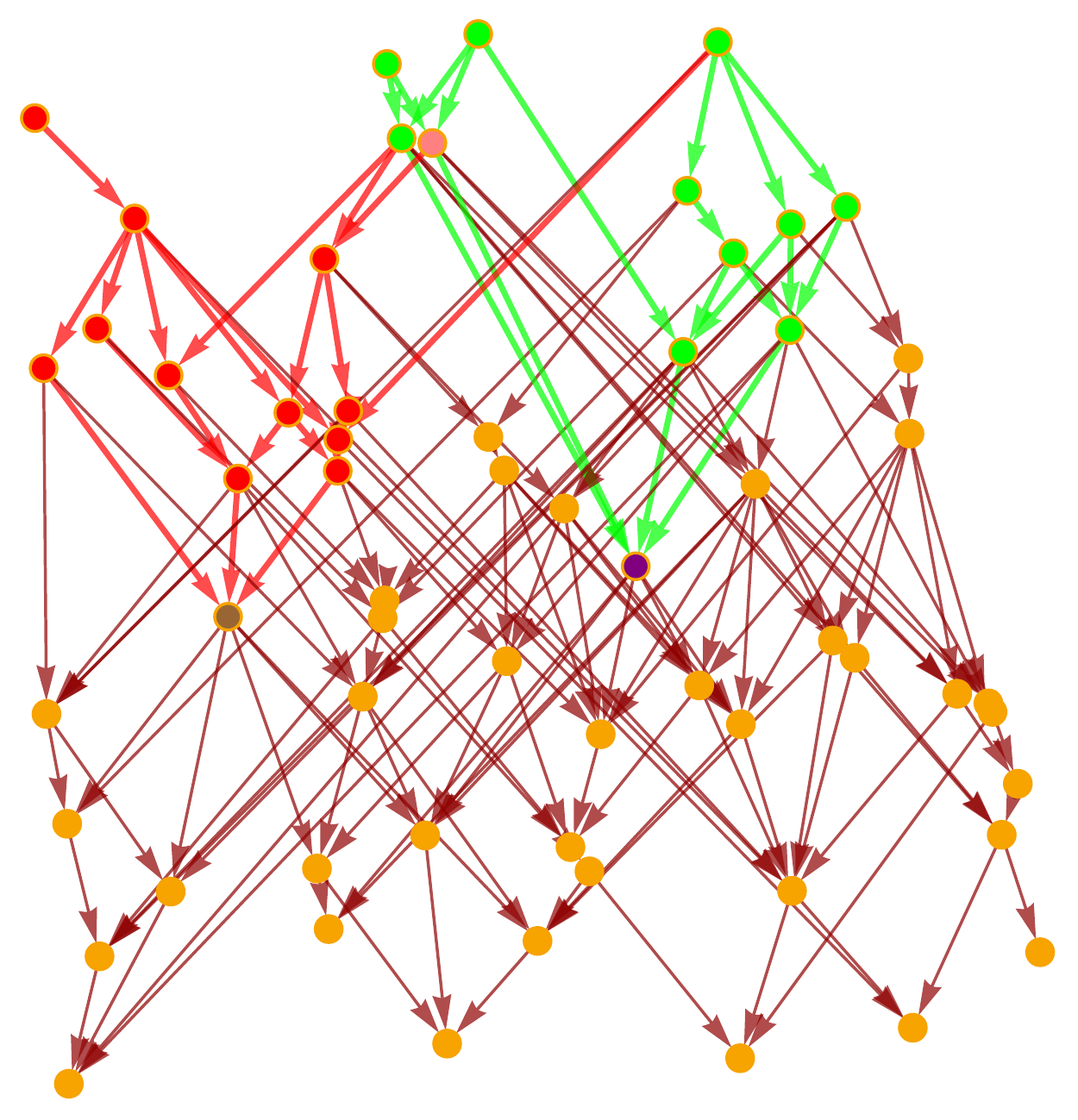}
\caption{The transitive reduction (i.e. the Hasse diagram) of the causal partial order graph for 60 uniformly sprinkled points in a rectangular region of ${1 + 1}$-dimensional flat (Minkowski) spacetime. Highlighted in brown and purple are two spacelike-separated events whose spatial distance we wish to estimate. In red we show the causal future/past of the brown event, and in green we show the causal future/past of the purple event. A common event in the causal future/past of both the brown and purple events, which happens to minimize the timelike distance over the set of all such events, is shown in pink.}
\label{fig:Figure31}
\end{figure}

\begin{figure}[ht]
\centering
\includegraphics[width=0.395\textwidth]{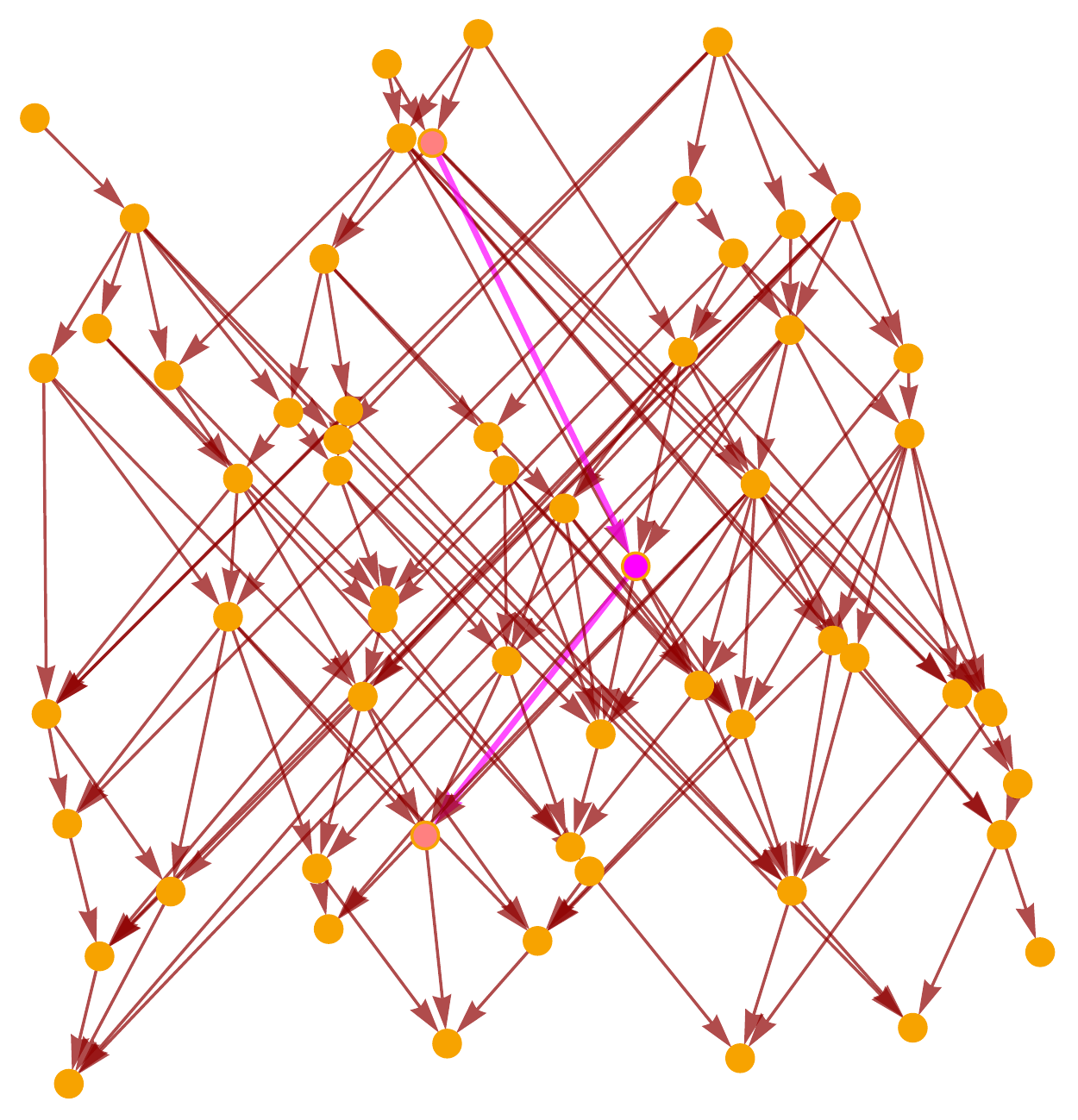}
\caption{A directed path showing the timelike distance between the two pink events (where this distance is minimal over the set of all events in the common future/past of the brown and purple events), indicating that the na\"ive estimate of the spacelike distance between the brown and purple events is equal to 2.}
\label{fig:Figure32}
\end{figure}

However, the spatial distance between two spacelike-separated events $p$ and $q$ in a causal set is far harder to estimate, since this information must somehow be inferred from the common causal future and past of events $p$ and $q$, i.e. from the sets:

\begin{equation}
J^{+} \left( p, q \right) = J^{+} \left( p \right) \cap J^{+} \left( q \right), \qquad \text{ and } J^{-} \left( p, q \right) = J^{-} \left( p \right) \cap J^{-} \left( q \right),
\end{equation}
respectively. One na\"ive choice of spatial \textit{predistance} function, as proposed by Brightwell and Gregory\cite{brightwell}, simply minimizes over the set of timelike distances ${\tau}$ between events in the common future and past of events $p$ and $q$ as follows:

\begin{equation}
d_s \left( p, q \right) = \min_{r \in J^{+} \left( p, q \right), s \in J^{-} \left( p, q \right)} \left[ \tau \left( r, s \right) \right].
\end{equation}
An example of the computation of this na\"ive spatial distance estimate for a causal set consisting of 60 uniformly sprinkled points in a rectangular region of ${1 + 1}$-dimensional flat (Minkowski) spacetime is shown in Figures \ref{fig:Figure31} and \ref{fig:Figure32}. However, although this is a somewhat reasonable approach to adopt in the 2-dimensional case of ${\mathbb{M}^2 = \mathbb{R}^{1, 1}}$, for ${\mathbb{M}^d = \mathbb{R}^{1, d - 1}}$ with ${d > 2}$ it suffers from a fundamental problem resulting from the fact that all pairs ${\left( r \in J^{+} \left( p, q \right), s \in J^{-} \left( p, q \right) \right)}$ which minimize the timelike distance ${\tau \left( p, q \right)}$ lie, once again, within the non-compact region between the light cone ${\tau = 0}$ and the hyperboloid:

\begin{equation}
\left( \sum_{i = 1}^{n - 1} x_{i}^{2} \right) - t^2 \propto \left( v \right)^{\frac{2}{n}},
\end{equation}
hence guaranteeing that, in the limit of a causal set of infinite cardinality, there will exist infinitely many such minimizing pairs. Therefore, there will almost surely exist a pair ${\left( r \in J^{+} \left( p, q \right), s \in J^{-} \left( p, q \right) \right)}$ whose timelike distance significantly underestimates the ``true'' spatial distance between events $p$ and $q$ (such that, in the limit of infinite cardinality, all spatial distances will equal 2, with probability 1).

Rideout and Wallden\cite{rideout} subsequently succeeded in circumventing this problem by computing an average of the timelike distance ${\tau}$ over such minimizing pairs, as opposed to performing a na\"ive minimization, in accordance with the following scheme:

\begin{definition}
A ``past $n$-link'' of a length-$n$ antichain ${x_1, x_2, \dots, x_n}$ in a causal set ${\mathcal{C}}$ is any element ${y \in \mathcal{C}}$ such that:

\begin{equation}
\forall 1 \leq m \leq n, \qquad y \prec x_m \text{ and } \nexists z \in \mathcal{C} \text{ such that } y \prec z \prec x_m.
\end{equation}
\end{definition}

\begin{definition}
A ``future $n$-link'' of a length $n$ antichain ${x_1, x_2, \dots x_n}$ in a causal set ${\mathcal{C}}$ is any element ${y \in \mathcal{C}}$ such that:

\begin{equation}
\forall 1 \leq m \leq n, \qquad x_m \prec y \text{ and } \nexists z \in \mathcal{C} \text{ such that } x_m \prec z \prec y.
\end{equation}
\end{definition}
Thus, the expectation value for the number of $n$-links ${\hat{N_n}}$ in a causal set ${\mathcal{C}}$ sprinkled over a $d$-dimensional spacetime is given by the straightforward integral:

\begin{equation}
\left\langle \hat{N_n} \right\rangle = \rho_c \int_{J^{+} \left( \left\lbrace x_i \right\rbrace \right)} \exp \left\lbrace - \rho_c \mathrm{Vol} \left( J^{-} \left( x \right) \cap \left[ \bigcup_{i = 1}^{n} J^{+} \left( x_i \right) \right] \right) \right\rbrace d x^d,
\end{equation}
assuming in the above a sprinkling density of ${\rho_c}$, a natural generalization of the common past/future of a pair of events to arbitrary sets as follows:

\begin{equation}
J^{\pm} \left( \left\lbrace x_i \right\rbrace \right) = \bigcap_{i = 1}^{n} J^{\pm} \left( x_i \right),
\end{equation}
and where ${\mathrm{Vol} \left( R \right)}$ designates the volume of a given continuum spacetime region $R$. This integral is derived by simply noting that the expectation value is a sum of independent random variables (namely the infinitesimal probability ${d x^d}$ of a sprinkled element existing at position $x$), each of which is multiplied by the probability that the given element is linked to each of the ${x_i}$ elements in the antichain, with the region of integration being given by the common future ${J^{\pm} \left( \left\lbrace x_i \right\rbrace \right)}$. Therefore, in particular, one can compute the 2-link distance between events $p$ and $q$ by averaging over the set of all minimizing pairs ${\left( r \in J^{+} \left( p, q \right), s \in J^{-} \left( p, q \right) \right)}$, in which one of either $r$ or $s$ forms a link with both $p$ and $q$ (i.e. such that either $r$ or $s$ is directly connected to both $p$ and $q$ by a single causal edge), and this predistance function has been shown by explicit numerical simulations to stabilize as a function of the sprinkling density ${\rho_c}$ (although it has the tendency to overestimate spatial distances consistently, when compared against the true continuum values).

Our rationale for referring to both the na\"ive spatial distance function of Brightwell and Gregory, and the 2-link distance function of Rideout and Wallden, as ``predistance'' functions is in order to reflect the fact that they fail to satisfy the axioms of a metric; more specifically, although the axioms of identity of indiscernibles, symmetry and non-negativity all hold trivially for purely set-theoretic reasons:

\begin{equation}
\forall p, q \in \mathcal{C}, \qquad d_s \left( p, q \right) = 0 \iff p = q, \qquad d_s \left( p, q \right) = d \left( q, p \right), \qquad d_s \left( p, q \right) \geq 0,
\end{equation}
it is easy to see that the triangle inequality:

\begin{equation}
\forall p, q, r \in \mathcal{C}, \qquad d_s \left( p, r \right) \leq d_s \left( p, q \right) + d_s \left( q, r \right),
\end{equation}
does not hold, as a consequence of the lack of sub-additivity of the associated continuous function for Lorentzian manifolds with non-vanishing extrinsic curvature. Eichhorn, Surya and Versteegen\cite{eichhorn} nevertheless succeeded in defining a one-parameter family of truly discrete spatial distance functions on inextendible antichains in causal sets, by noting that the volume ${\mathrm{Vol}_{past} \left( p \right)}$ of the past light cone of a point ${p \in J^{+} \left( \Sigma \right)}$, with a flat spacelike hypersurface ${\Sigma}$ corresponding to a constant-time slice in an inertial reference frame in flat (Minkowski) spacetime ${\mathbb{M}^d}$ (such that ${\Sigma \cong \mathbb{R}^{d -1}}$), namely:

\begin{equation}
\mathrm{Vol}_{past} \left( p \right) = \mathrm{Vol} \left( J^{-} \left( p \right) \cap J^{+} \left( \Sigma \right) \right),
\end{equation}
may be directly related to the diameter ${D_{base} \left( p \right)}$ of the base of the cone as it intersects ${\Sigma}$:

\begin{equation}
D_{base} \left( p \right) = D \left( J^{-} \left( p \right) \cap \Sigma \right),
\end{equation}
in the following rather straightforward way:

\begin{equation}
\mathrm{Vol}_{past} \left( p \right) = \zeta_d T^d = \zeta_d \left( \frac{D_{base} \left( p \right)}{2} \right).
\end{equation}
In the above, $T$ denotes the height of the cone, i.e. the proper time distance between event $p$ and ${\Sigma}$, ${\zeta_d}$ has its usual form:

\begin{equation}
\zeta_d = \frac{\pi^{\frac{\left( d - 1 \right)}{2}}}{d \Gamma \left( \frac{d + 1}{2} \right)},
\end{equation}
and this relationship holds because the boundary of the base of the intersection is an ${\left( n - 1 \right)}$-sphere ${\partial \left( J^{-} \left( p \right) \cap \Sigma \right) \cong S^{d - 2}}$ of diameter ${2 T}$. Therefore, any pair of antipodal points ${p, q \in \partial \left( J^{-} \left( p \right) \cap \Sigma \right)}$ will be separated by a spatial distance of ${d \left( p, q \right) = 2T}$, which can then be used to construct a general distance function.

\begin{figure}[ht]
\centering
\includegraphics[width=0.395\textwidth]{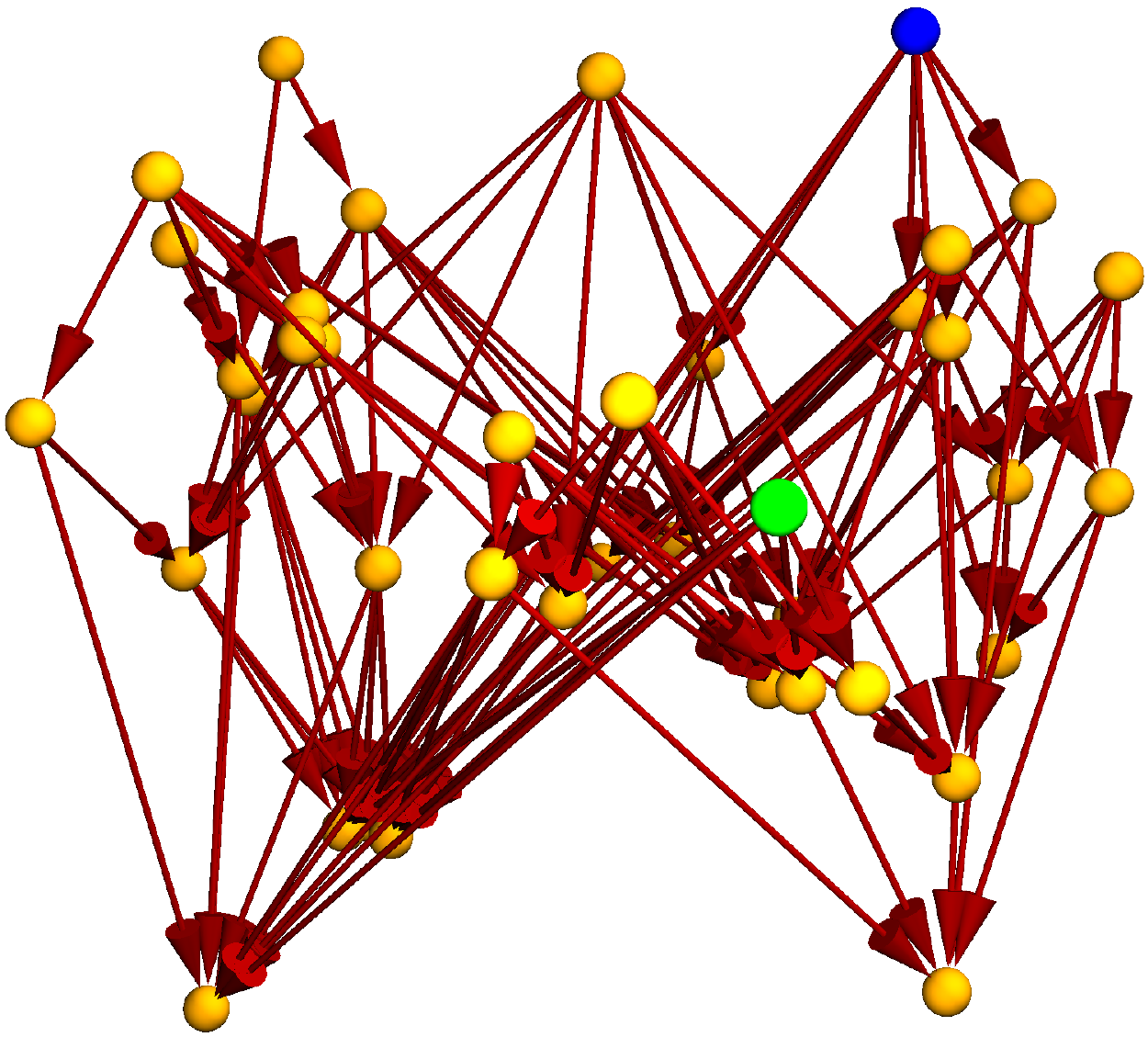}\hspace{0.1\textwidth}
\includegraphics[width=0.495\textwidth]{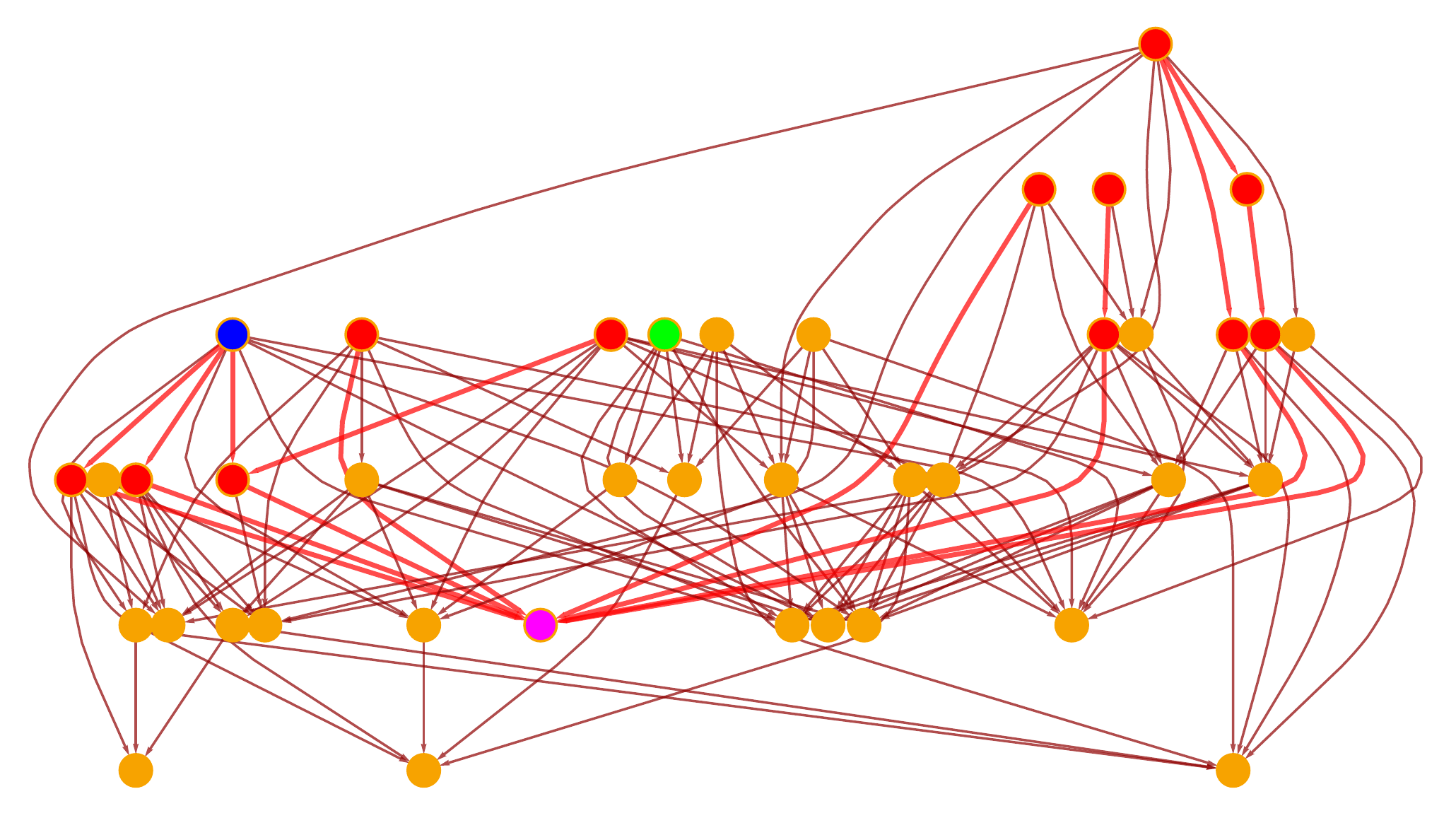}
\caption{The transitive reduction (i.e. the Hasse diagram) of the causal partial order graph for 40 uniformly sprinkled points in a rectangular region of ${2 + 1}$-dimensional flat (Minkowski) spacetime, with the left and right images showing the graph with and without vertex coordinate information, respectively. Two spacelike-separated points are highlighted in blue and green, and on the right a past light cone, emanating from a point highlighted in purple and containing the blue and green points on its boundary, is shown in red. The purple point is chosen so as to minimize the volume of the cone (in this case, it contains exactly 14 events), such that the blue and green vertices form antipodal points on its boundary, yielding an estimate of their spacelike separation distance of 4.75.}
\label{fig:Figure33}
\end{figure}

More explicitly, consider the Cartesian coordinate scheme ${\left( t, x_1, \dots, x_{d - 1} \right)}$, with a Cauchy hypersurface ${\left( \Sigma, \delta \right)}$ at time ${t = 0}$; without loss of generality, suppose that the events ${p, q \in \Sigma}$ are such that event $p$ lies at the origin ${\left( 0, 0, \dots, 0 \right)}$, and event $q$ lies on the positive ${x_1}$-axis ${\left( 0, x_q, 0, \dots, 0 \right)}$. The events in the spacelike hyperboloid:

\begin{equation}
\mathcal{H} \left( p, q \right) = \partial J^{+} \left( p \right) \cap \partial J^{+} \left( q \right),
\end{equation}
are all lightlike-separated from both $p$ and $q$, with the lightlike hypersurfaces:

\begin{equation}
\mathcal{H}_p = \partial J^{+} \left( p \right), \qquad \text{ and } \qquad \mathcal{H}_q = \partial J^{+} \left( q \right),
\end{equation}
given by:

\begin{equation}
\mathcal{H}_p = \left\lbrace r \left| \left( \sum_{i = 1}^{d - 1} x_{i}^{2} \left( r \right) \right) - t^2 \left( r \right) = 0 \right. \right\rbrace,
\end{equation}
and:

\begin{equation}
\mathcal{H}_q = \left\lbrace r \left| \left( x_1 \left( r \right) - x_q \right)^2 + \left( \sum_{i = 2}^{d - 1} x_{i}^{2} \left( r \right) \right) - t^2 \left( r \right) = 0 \right. \right\rbrace,
\end{equation}
respectively, such that the hyperboloid:

\begin{equation}
\mathcal{H} \left( p, q \right) = \left\lbrace r \left| x_1 \left( r \right) = \frac{x_q}{2} \text{ and } \left( \sum_{i = 2}^{d- 1} x_{i}^{2} \left( r \right) \right) - t^2 \left( r \right) = - \left( \frac{x_q}{2} \right)^2 \right. \right\rbrace,
\end{equation}
is simply the ${\left( d - 2 \right)}$-dimensional spacelike hyperboloid ${\mathcal{H}^{d - 2}}$. The volume ${\mathrm{Vol}_{past} \left( r \right)}$ increases monotonically with ${t \left( r \right)}$, with ${t \left( r \right)}$ taking its (unique) minimum value over all ${r \in \mathcal{H} \left( p, q \right)}$ when $r$ has the form ${\left( \frac{x_q}{2}, \frac{x_q}{2},0, \dots, 0 \right)}$, such that ${\mathrm{Vol}_{past} \left( r \right)}$ also takes its minimum value. Denoting this point by ${r_m}$, we therefore obtain a new predistance function:

\begin{equation}
\tilde{d} \left( p, q \right) = 2 \left( \frac{\mathrm{Vol}_{past} \left( r_m \right)}{\zeta_d} \right)^{\frac{1}{d}},
\end{equation}
but since:

\begin{equation}
2 t \left( r_m \right) = x_q,
\end{equation}
by definition, ${\tilde{d}}$ reduces exactly to the spatial distance function ${d_{\delta}}$ on ${\Sigma}$, and so (in particular) satisfies the triangle inequality, as required. An illustration of this estimation procedure, applied to a causal set consisting of 40 uniformly sprinkled points in a rectangular region of ${2 + 1}$-dimensional flat (Minkowski) spacetime, yielding a spacelike separation distance estimate of 4.75, is shown in Figure \ref{fig:Figure33}.

\begin{figure}[ht]
\centering
\includegraphics[width=0.395\textwidth]{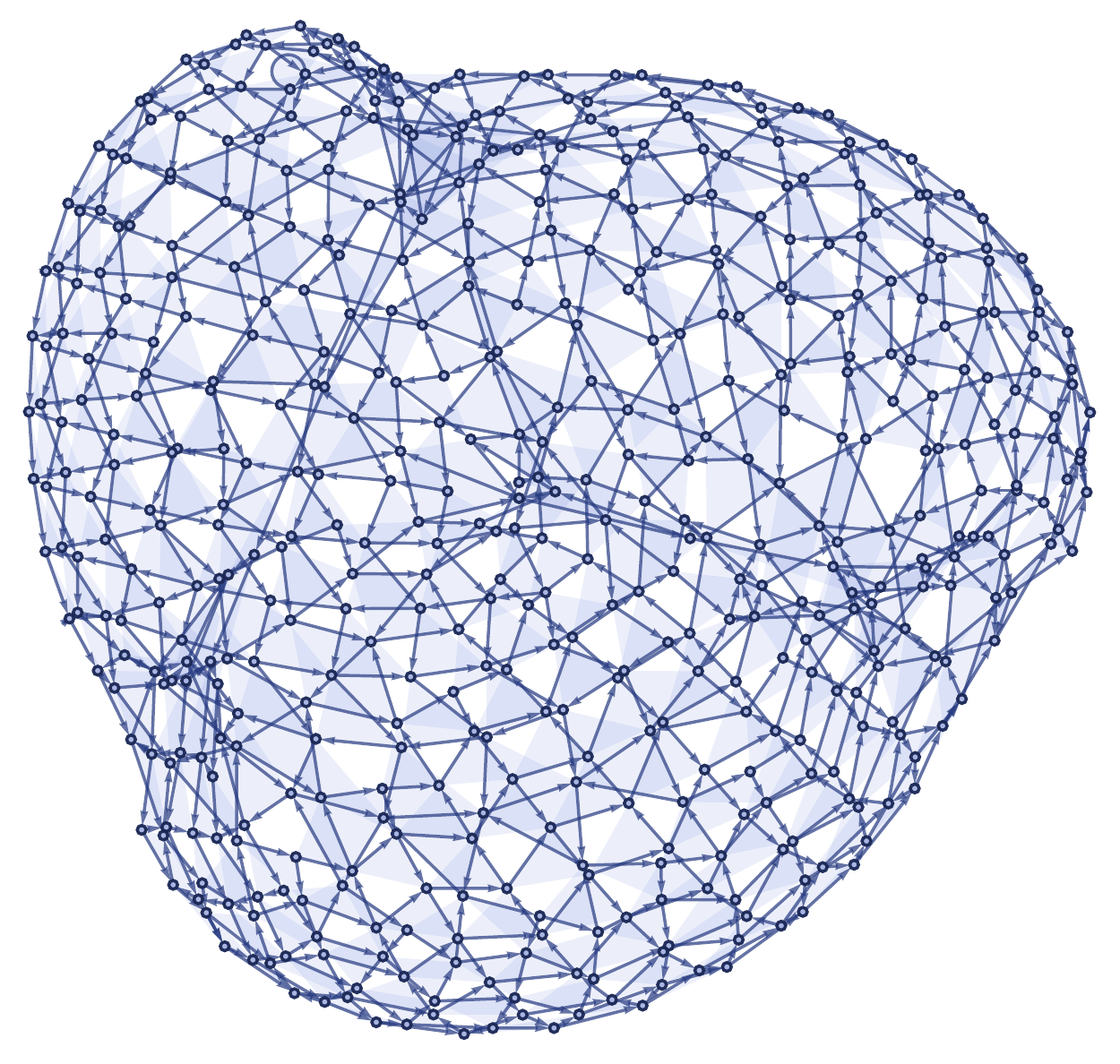}\hspace{0.1\textwidth}
\includegraphics[width=0.495\textwidth]{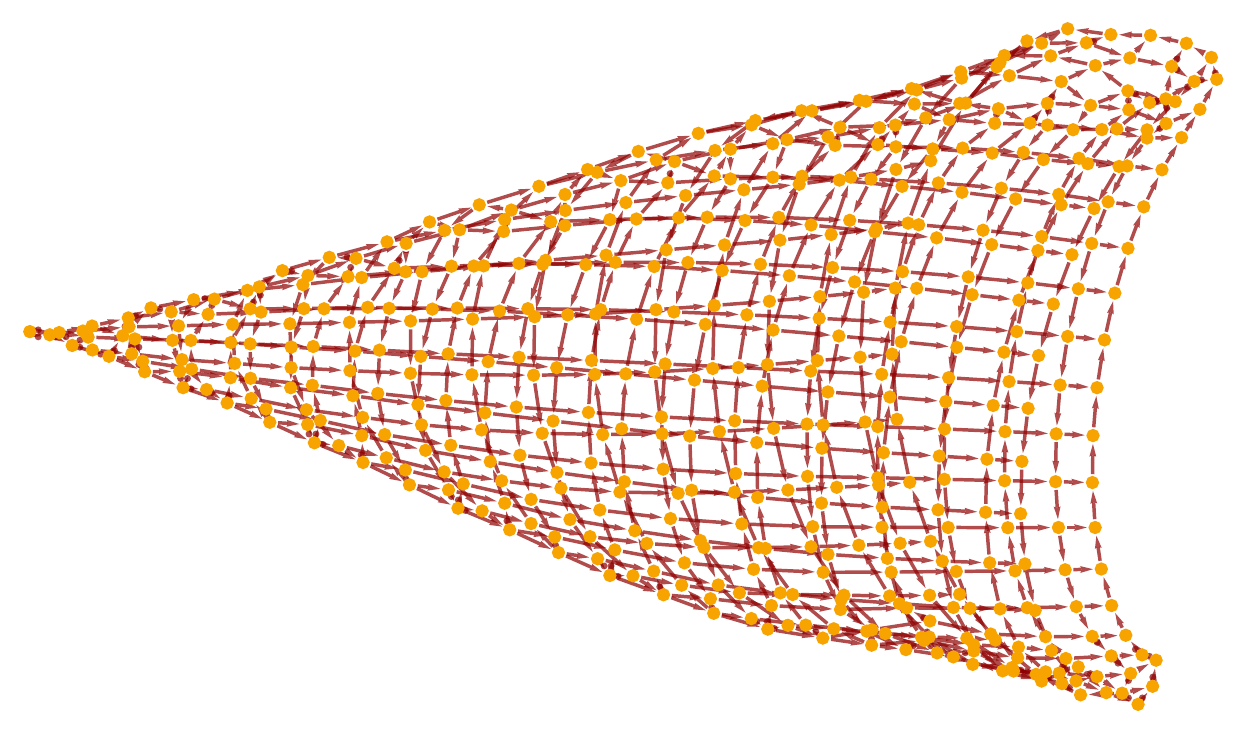}
\caption{The hypergraph and causal network after the first 500 evolution steps of the set substitution system ${\left\lbrace \left\lbrace x, y, y \right\rbrace, \left\lbrace x, z, u \right\rbrace \right\rbrace \to \left\lbrace \left\lbrace u, v, v \right\rbrace, \left\lbrace v, z, y \right\rbrace, \left\lbrace x, y, v \right\rbrace \right\rbrace}$, yielding approximations to an asymptotically-flat Riemannian manifold and an asymptotically-flat Lorentzian manifold, respectively. Example taken from \cite{wolfram2}.}
\label{fig:Figure34}
\end{figure}

\begin{figure}[ht]
\centering
\includegraphics[width=0.395\textwidth]{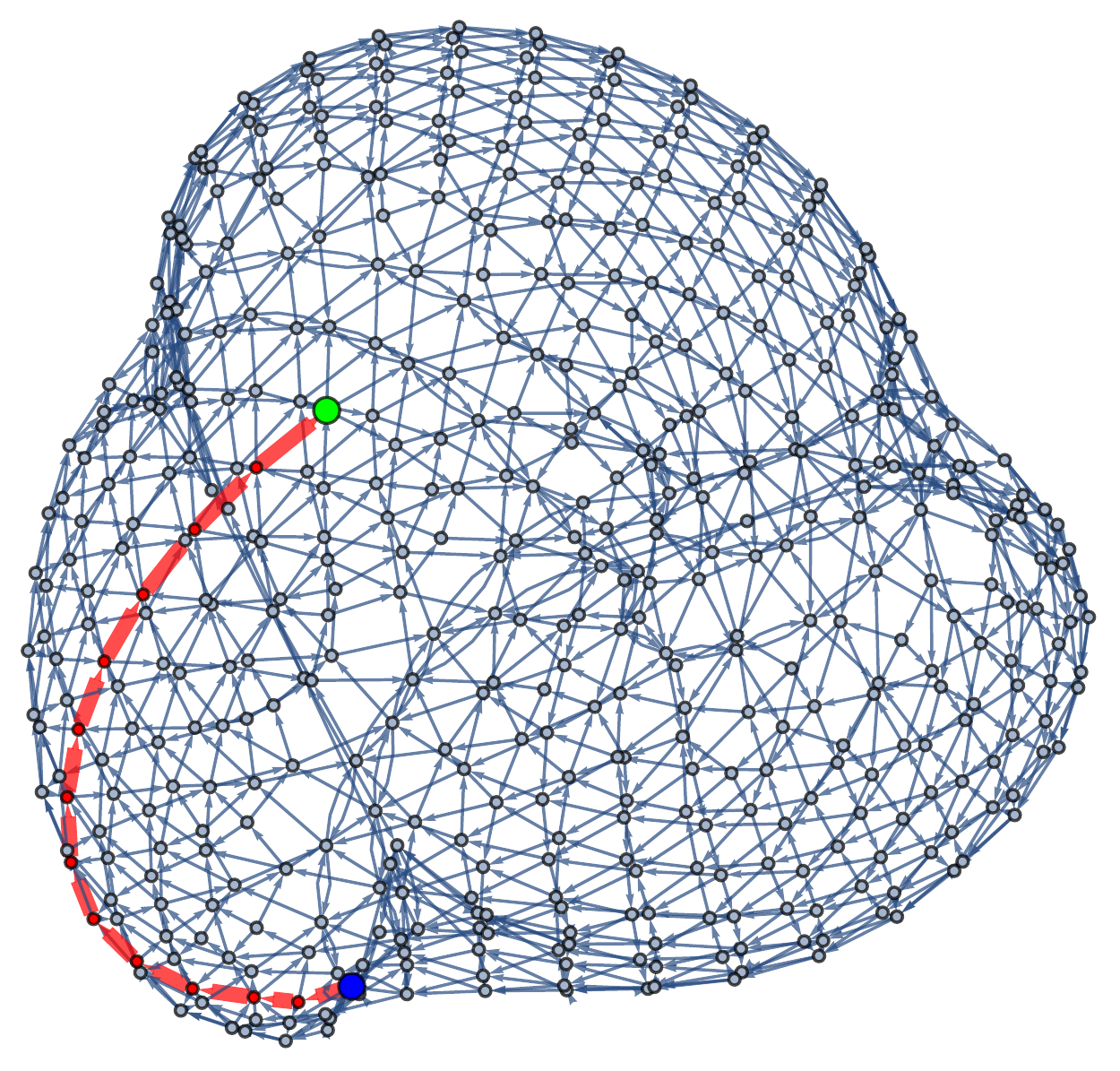}\hspace{0.1\textwidth}
\includegraphics[width=0.495\textwidth]{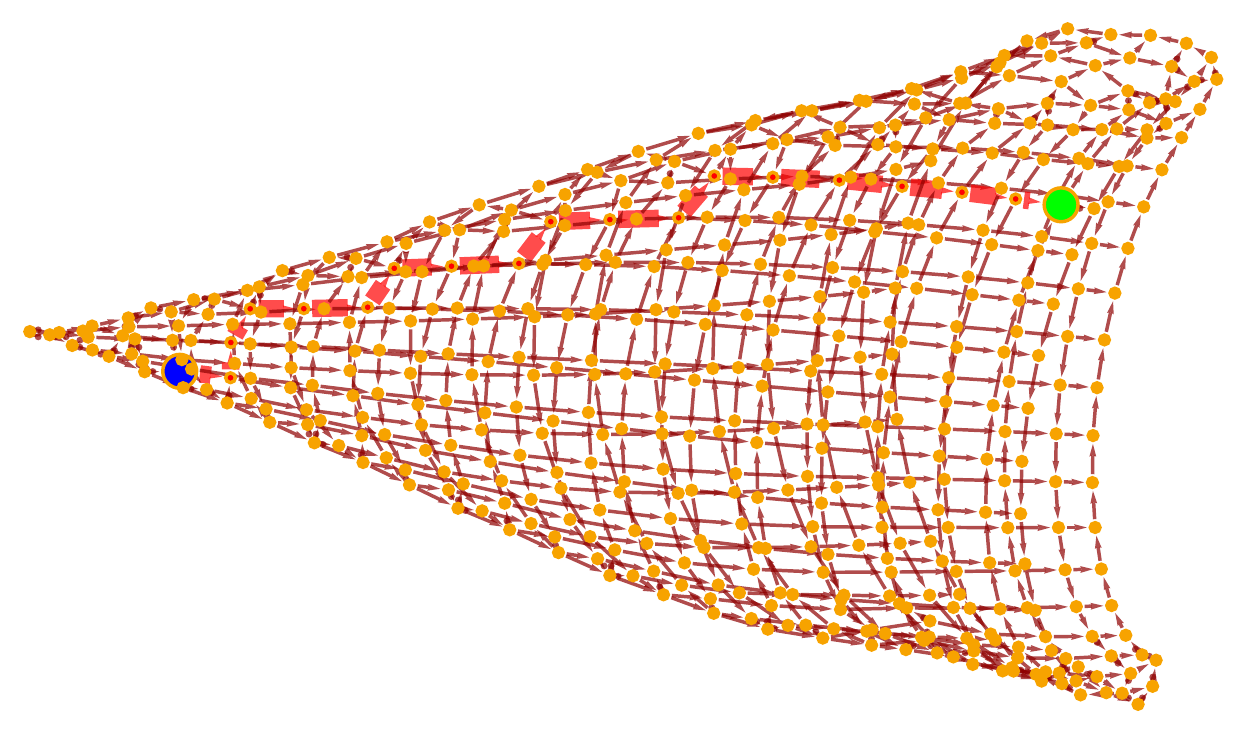}
\caption{Examples of geodesics on the hypergraph and causal network generated by the set substitution system ${\left\lbrace \left\lbrace x, y, y \right\rbrace, \left\lbrace x, z, u \right\rbrace \right\rbrace \to \left\lbrace \left\lbrace u, v, v \right\rbrace, \left\lbrace v, z, y \right\rbrace, \left\lbrace x, y, v \right\rbrace \right\rbrace}$, yielding approximations of the spacelike/timelike distances between the points highlighted in blue and green of 14 (spacelike) and 19 (timelike), respectively.}
\label{fig:Figure35}
\end{figure}

However, these mathematical and conceptual difficulties with defining spacelike distances simply do not arise in the Wolfram model case, since each hypergraph (corresponding to a particular spacelike hypersurface in the causal network) supplies us with an a priori spatial metric, as previously described, and as shown in Figure \ref{fig:Figure22}. Therefore, as discussed above, we can use the natural (undirected) geodesic distance ${\Delta l}$ between pairs of points in the hypergraph to infer a discrete spatial metric tensor ${\gamma^{i j}}$:

\begin{equation}
\Delta l^2 = \gamma_{i j} \Delta x^i \Delta x^j,
\end{equation}
and the (directed) geodesic distance ${\Delta t}$ between pairs of points in the causal network to infer timelike distances, yielding the overall discrete spacetime line element in ADM form:

\begin{equation}
\Delta s^2 = \left( - \alpha^2 + \beta^i \beta_i \right) \Delta t^2 + 2 \beta_i \Delta x^i \Delta t + \gamma_{i j} \Delta x^i \Delta x^j,
\end{equation}
for lapse function ${\alpha}$ and shift vector ${\beta^i}$. An example of a hypergraph/causal network pair which yield approximations to an asymptotically-flat Riemannian manifold and an asymptotically-flat Lorentzian manifold, respectively, is shown in Figure \ref{fig:Figure34}. An illustrative example of how spacelike and timelike distances can be approximated directly on these manifold-like structures using simple combinatorial geodesic distances is shown in Figure \ref{fig:Figure35}. Since all graphs and hypergraphs trivially satisfy the axioms of a metric space, this elementary construction suffices for our purposes.

\begin{figure}[ht]
\centering
\includegraphics[width=0.795\textwidth]{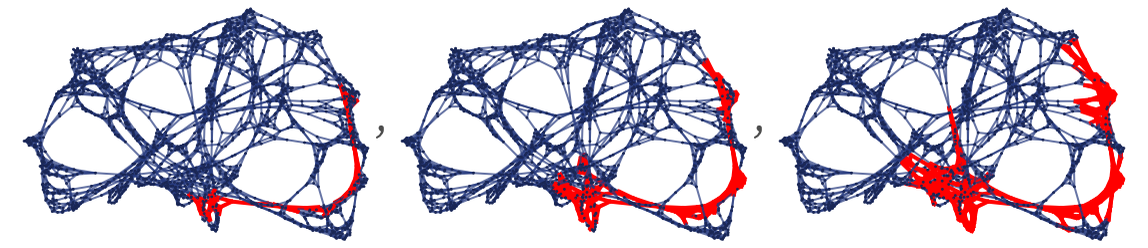}
\caption{Infinitesimal geodesic ``tubes'' of radius 1, 3 and 5, respectively, in the hypergraph generated by the set substitution system ${\left\lbrace \left\lbrace x, y \right\rbrace, \left\lbrace x, z \right\rbrace \right\rbrace \to \left\lbrace \left\lbrace x, z \right\rbrace, \left\lbrace x, w \right\rbrace, \left\lbrace y, w \right\rbrace, \left\lbrace z, w \right\rbrace \right\rbrace}$. Example taken from \cite{wolfram2}.}
\label{fig:Figure36}
\end{figure}

For the kinds of geometrical constructions that we shall consider in the subsequent sections, it is often helpful to consider not a \textit{single} geodesic (which is generally not uniquely defined), but rather a \textit{tube} of geodesics of finite radius. This is especially true in the Riemannian/hypergraph case, in which a single updating event will in general involve multiple hypergraph vertices, and so one must effectively compute averages in order to determine a unique spacelike distance between events. Note that, for a $d$-dimensional Riemannian manifold ${\left( \mathcal{M}, g \right)}$, the infinitesimal volume element ${d \mu_g}$ around a point $p$ is simply given by the square root of the determinant of the metric tensor:

\begin{equation}
d \mu_g \left( p \right) = \sqrt{\det \left( g \left( p \right) \right)},
\end{equation}
meaning that, for a nearby point ${p + \delta x}$, and assuming moreover that the manifold ${\mathcal{M}}$ is analytic, we can consider the following power series expansion in ${\delta x}$\cite{jost}\cite{gray}:

\begin{equation}
\sqrt{\det \left( g \left( p + \delta x \right) \right)} = \sqrt{\det \left( g \left( p \right) \right)} \left( 1 - \frac{1}{6} \sum_{i = 1}^{d} R_{i j} \left( p \right) \delta x^i \delta x^j + O \left( \delta x^3 \right) + \cdots \right).
\end{equation}
In the above, ${R_{i j}}$ denotes the standard Ricci curvature tensor, and ${\delta x^i}$, ${\delta x^j}$ are contravariant vectors denoting the orthogonal components of ${\delta x}$. By integrating over a ball ${B_{\epsilon} \left( p \right)}$ of infinitesimal radius ${\epsilon}$, centered at point $p$, one thus obtains the standard volume formula for a small geodesic ball:

\begin{equation}
\mathrm{Vol} \left( B_{\epsilon} \left( p \right) \right) = \int_{B_{\epsilon} \left( p \right)} \sqrt{\det \left( g \left( p + \delta x \right) \right)} d^d \left( \delta x \right),
\end{equation}
which, up to second-order in ${\epsilon}$, is given by:

\begin{equation}
\int_{B_{\epsilon} \left( p \right)} \sqrt{\det \left( g \left( p + \delta x \right) \right)} d^d \left( \delta x \right) = \frac{\pi^{\frac{d}{2}}}{\Gamma \left( \frac{d}{2} + 1 \right)} \epsilon^d \left( 1 - \frac{\epsilon^2}{6 \left( d + 2 \right)} \sum_{i = 1}^{d} R_{i}^{i} + O \left( \epsilon^4 \right) \right),
\end{equation}
with:

\begin{equation}
\sum_{i = 1}^{d} R_{i}^{i} = R,
\end{equation}
being the standard Ricci curvature scalar (i.e. the trace of the curvature tensor). We shall make extensive use of this formula in the next section. For the time being, however, consider instead integrating over a tube ${T_{\epsilon, \delta x} \left( p \right)}$ of infinitesimal radius ${\epsilon}$, starting at point p and extending an infinitesimal distance ${\delta}$ along a geodesic oriented in direction ${\delta x}$, in which case one finds the volume formula for a small geodesic tube:

\begin{equation}
\mathrm{Vol} \left( T_{\epsilon, \delta x} \left( p \right) \right) = \frac{\pi^{\frac{d - 1}{2}}}{\Gamma \left( \frac{d + 1}{2} \right)} \epsilon^{d - 1} \delta x \left( 1 - \left( \frac{d - 1}{d + 1} \right) \left( R - \sum_{i, j = 1}^{d} R_{i j} \hat{\delta x}^i \hat{\delta x}^j \right) \epsilon^2 + O \left( \epsilon^3 + \epsilon^2 \delta x \right) + \dots \right),
\end{equation}
with ${\hat{\delta x}^i, \hat{\delta x}^j}$ denoting the components of the unit vectors along the geodesic. An example of the construction of such infinitesimal geodesic ``tubes'' for the more general case of an arbitrary hypergraph is shown in Figure \ref{fig:Figure36}.

\begin{figure}[ht]
\centering
\includegraphics[width=0.795\textwidth]{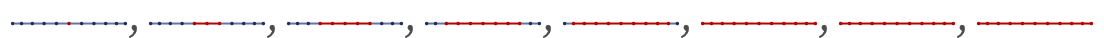}
\caption{The growth of a finite geodesic ball in a one-dimensional grid-like spatial hypergraph. Example taken from \cite{wolfram2}.}
\label{fig:Figure37}
\end{figure}

\begin{figure}[ht]
\centering
\includegraphics[width=0.795\textwidth]{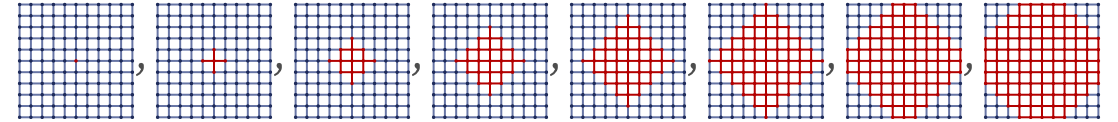}
\caption{The growth of a finite geodesic ball in a two-dimensional grid-like spatial hypergraph. Example taken from \cite{wolfram2}.}
\label{fig:Figure38}
\end{figure}

\begin{figure}[ht]
\centering
\includegraphics[width=0.795\textwidth]{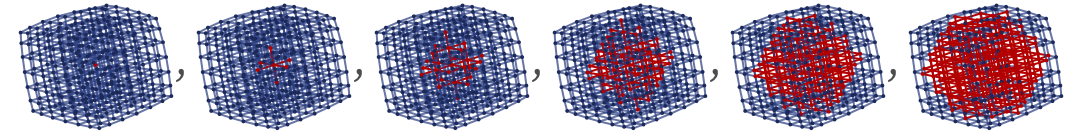}
\caption{The growth of a finite geodesic ball in a three-dimensional grid-like spatial hypergraph. Example taken from \cite{wolfram2}.}
\label{fig:Figure39}
\end{figure}

This analysis suggests that we ought to be able to probe certain features of the hypergraph geometry, namely limiting dimension and discrete curvature, by examining the growth rates of volumes of small geodesic balls and tubes. Here we shall concern ourselves primarily with dimension estimation; the forthcoming section will cover curvature estimation in much greater detail. For example, the growth of finite geodesic balls in grid-like hypergraphs corresponding to very coarse approximations to one-, two- and three-dimensional Euclidean space are shown in Figures \ref{fig:Figure37}, \ref{fig:Figure38} and \ref{fig:Figure39}, respectively, with growth rates given by:

\begin{equation}
\mathrm{Vol} \left( B_{\epsilon} \left( p \right) \right) = 2 \epsilon + 1,
\end{equation}

\begin{equation}
\mathrm{Vol} \left( B_{\epsilon} \left( p \right) \right) = 2 \epsilon^2 + 2 \epsilon + 1,
\end{equation}
and:

\begin{equation}
\mathrm{Vol} \left( B_{\epsilon} \left( p \right) \right) = \frac{4 \epsilon^3}{3} + 2 \epsilon^2 + \frac{8 \epsilon}{3} + 1,
\end{equation}
respectively, assuming infinite grids. More generally, a $d$-dimensional grid-like hypergraph will have a discrete volume element given by a terminating hypergeometric series, which we can write compactly in terms of the hypergeometric function ${{}_{2} F_{1} \left( a, b ; c ; z \right)}$ as follows\cite{wolfram}:

\begin{equation}
\mathrm{Vol} \left( B_{\epsilon} \left( p \right) \right) = {}_{2} F_{1} \left( -d, \epsilon + 1 ; -d + \epsilon + 1; -1 \right) {r \choose d},
\end{equation}
or more explicitly as:

\begin{equation}
{}_{2} F_{1} \left( -d, \epsilon + 1 ; -d + \epsilon + 1; -1 \right) {r \choose d} = \frac{2^d}{\Gamma \left( d + 1 \right)} \epsilon^d + \frac{2^{d - 1}}{\Gamma \left( d \right)} + \frac{2^{d - 2} \left( d + 1 \right)}{3 \Gamma \left( d - 1 \right)} \epsilon^{d - 2} + \cdots.
\end{equation}

\begin{figure}[ht]
\centering
\includegraphics[width=0.345\textwidth]{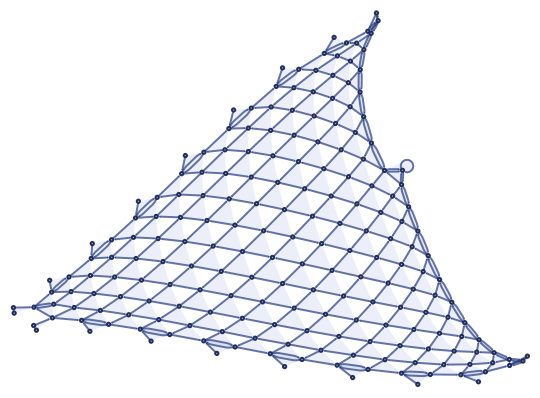}\hspace{0.1\textwidth}
\includegraphics[width=0.345\textwidth]{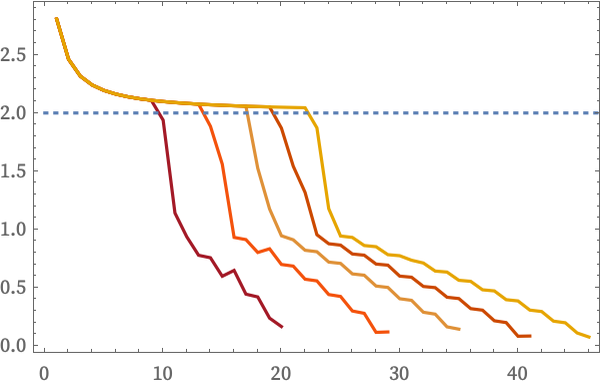}
\caption{Dimension estimates for an asymptotically-flat spatial hypergraph with a two-dimensional Riemannian manifold-like limiting structure, as generated by the set substitution system ${\left\lbrace \left\lbrace x, y, y \right\rbrace, \left\lbrace z, x, w \right\rbrace \right\rbrace \to \left\lbrace \left\lbrace y, v, y \right\rbrace, \left\lbrace y, z, v \right\rbrace, \left\lbrace w, v, v \right\rbrace \right\rbrace}$, indicating a limiting dimension of two. Example taken from \cite{wolfram2}.}
\label{fig:Figure40}
\end{figure}

\begin{figure}[ht]
\centering
\includegraphics[width=0.795\textwidth]{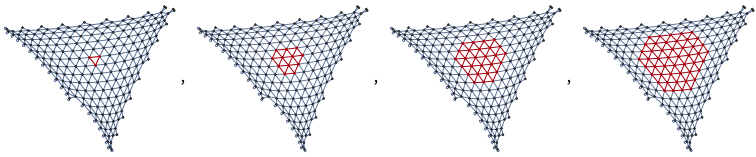}
\caption{The growth of a finite geodesic ball in an asymptotically-flat spatial hypergraph with a two-dimensional Riemannian manifold-like limiting structure, as generated by the set substitution system ${\left\lbrace \left\lbrace x, y, y \right\rbrace, \left\lbrace z, x, w \right\rbrace \right\rbrace \to \left\lbrace \left\lbrace y, v, y \right\rbrace, \left\lbrace y, z, v \right\rbrace, \left\lbrace w, v, v \right\rbrace \right\rbrace}$.}
\label{fig:Figure41}
\end{figure}

\begin{figure}[ht]
\centering
\includegraphics[width=0.395\textwidth]{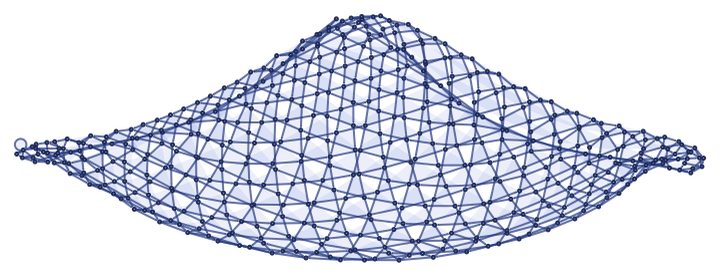}\hspace{0.1\textwidth}
\includegraphics[width=0.345\textwidth]{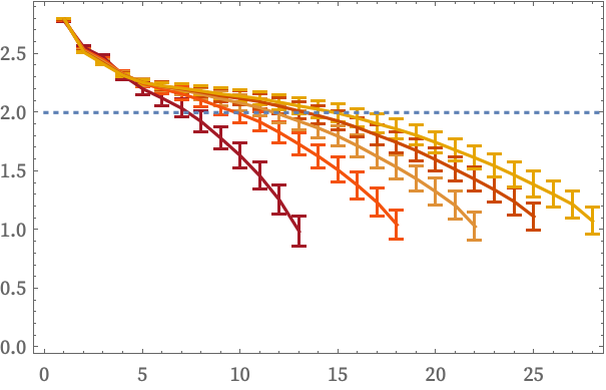}
\caption{Dimension estimates for a spatial hypergraph with a two-dimensional Riemannian manifold-like limiting structure, as generated by the set substitution system ${\left\lbrace \left\lbrace x, x, y \right\rbrace, \left\lbrace x, z, w \right\rbrace \right\rbrace \to \left\lbrace \left\lbrace w, w, v \right\rbrace, \left\lbrace v, w, y \right\rbrace, \left\lbrace z, y, v \right\rbrace \right\rbrace}$, exhibiting the effects of non-zero spatial curvature, and indicating a limiting dimension of two. Example taken from \cite{wolfram2}.}
\label{fig:Figure42}
\end{figure}

\begin{figure}[ht]
\centering
\includegraphics[width=0.795\textwidth]{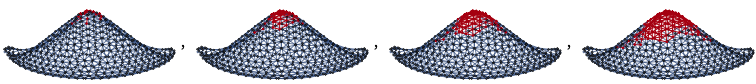}
\caption{The growth of a finite geodesic ball in a spatial hypergraph with a two-dimensional Riemannian manifold-like limiting structure, as generated by the set substitution system ${\left\lbrace \left\lbrace x, x, y \right\rbrace, \left\lbrace x, z, w \right\rbrace \right\rbrace \to \left\lbrace \left\lbrace w, w, v \right\rbrace, \left\lbrace v, w, y \right\rbrace, \left\lbrace z, y, v \right\rbrace \right\rbrace}$, exhibiting the effects of non-zero spatial curvature.}
\label{fig:Figure43}
\end{figure}

In general, the leading-order term in any such expansion (for a hypergraph with arbitrary topology) will be proportional to ${\epsilon^d}$; for instance, for a hypergraph whose metric perfectly approximates that of Euclidean space in the continuum limit, we would expect the volume element to scale exactly as:

\begin{equation}
\mathrm{Vol} \left( B_{\epsilon} \left( p \right) \right) = \frac{\pi^{\frac{d}{2}}}{\Gamma \left( \frac{d}{2} + 1 \right)} \epsilon^d.
\end{equation}
As such, we can construct a first-order estimator ${\Delta}$ of the hypergraph dimension $d$ by using a simple logarithmic difference approximation:

\begin{equation}
\Delta_{\epsilon} \left( p \right) = \frac{\log \left( \mathrm{Vol} \left( B_{\epsilon + 1} \left( p \right) \right) \right) - \log \left( \mathrm{Vol} \left( B_{\epsilon} \left( p \right) \right) \right)}{\log \left( \epsilon + 1 \right) - \log \left( \epsilon \right)},
\end{equation}
whose behavior as a function of ${\epsilon}$ is shown in Figures \ref{fig:Figure40} and \ref{fig:Figure42} for increasing numbers of evolution steps of two set substitution systems, namely ${\left\lbrace \left\lbrace x, y, y \right\rbrace, \left\lbrace z, x, w \right\rbrace \right\rbrace \to \left\lbrace \left\lbrace y, v, y \right\rbrace, \left\lbrace y, z, v \right\rbrace, \left\lbrace w, v, v \right\rbrace \right\rbrace}$ and ${\left\lbrace \left\lbrace x, x, y \right\rbrace, \left\lbrace x, z, w \right\rbrace \right\rbrace \to \left\lbrace \left\lbrace w, w, v \right\rbrace, \left\lbrace v, w, y \right\rbrace, \left\lbrace z, y, v \right\rbrace \right\rbrace}$, whose resulting hypergraphs are known to limit to two-dimensional Riemannian manifold-like structures (in the former case with asymptotic flatness, and in the latter case with non-zero spatial curvature), respectively. The limiting dimension of two is correctly estimated in both cases, with the estimates produced by considering the growth rates of small geodesic balls, as shown in Figures \ref{fig:Figure41} and \ref{fig:Figure43}.

\begin{figure}[ht]
\centering
\includegraphics[width=0.345\textwidth]{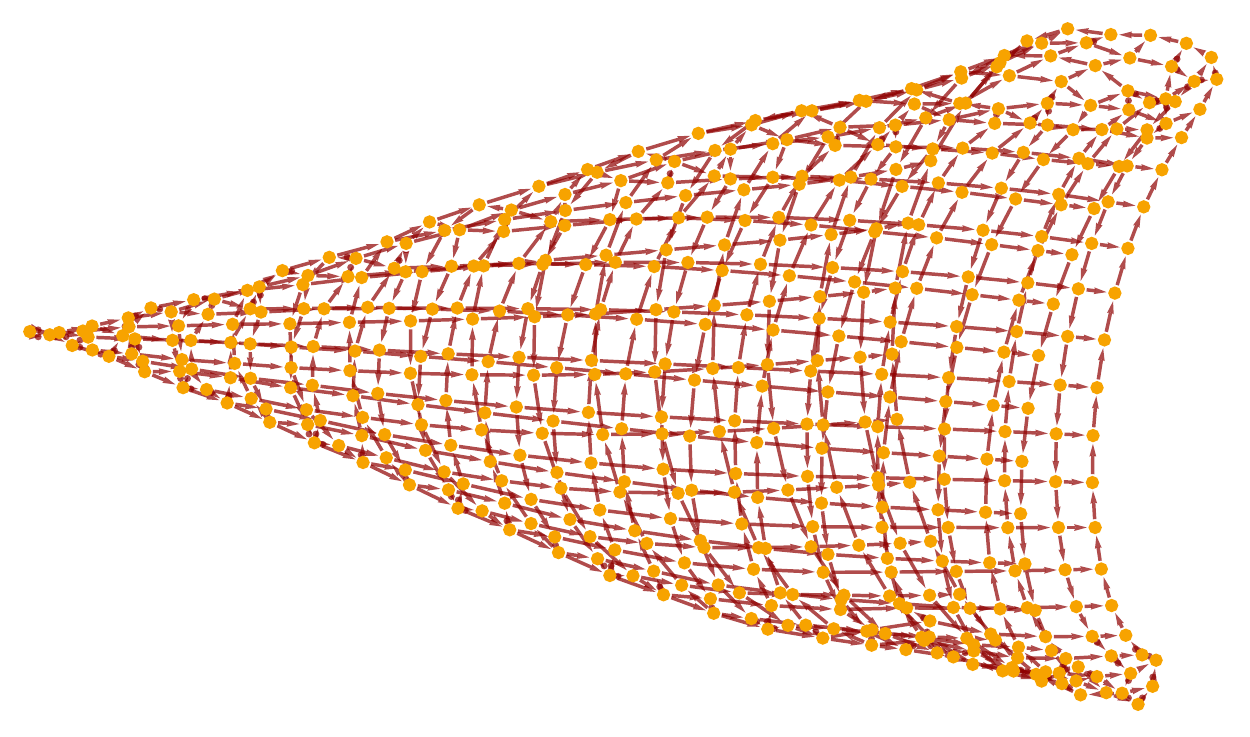}\hspace{0.1\textwidth}
\includegraphics[width=0.345\textwidth]{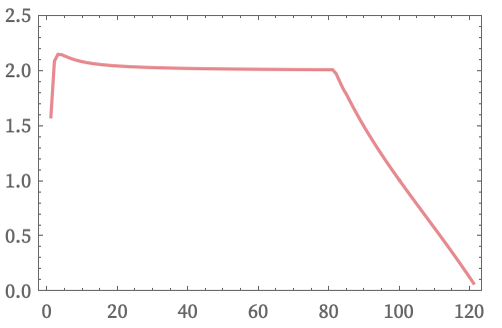}
\caption{Dimension estimates for an asymptotically-flat causal network with a two-dimensional Lorentzian manifold-like limiting structure, as generated by the set substitution system ${\left\lbrace \left\lbrace x, y, y \right\rbrace, \left\lbrace x, z, u \right\rbrace \right\rbrace \to \left\lbrace \left\lbrace u, v, v \right\rbrace, \left\lbrace v, z, y \right\rbrace, \left\lbrace x, y, v \right\rbrace \right\rbrace}$, indicating a limiting dimension of two. Example taken from \cite{wolfram2}.}
\label{fig:Figure44}
\end{figure}

\begin{figure}[ht]
\centering
\includegraphics[width=0.795\textwidth]{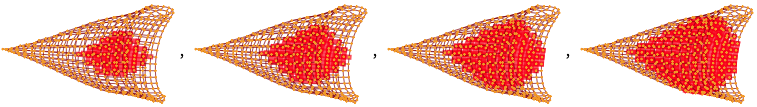}
\caption{The growth of a finite geodesic cone in an asymptotically-flat causal network with a two-dimensional Lorentzian manifold-like structure, as generated by the set substitution system ${\left\lbrace \left\lbrace x, y, y \right\rbrace, \left\lbrace x, z, u \right\rbrace \right\rbrace \to \left\lbrace \left\lbrace u, v, v \right\rbrace, \left\lbrace v, z, y \right\rbrace, \left\lbrace x, y, v \right\rbrace \right\rbrace}$.}
\label{fig:Figure45}
\end{figure}

\begin{figure}[ht]
\centering
\includegraphics[width=0.345\textwidth]{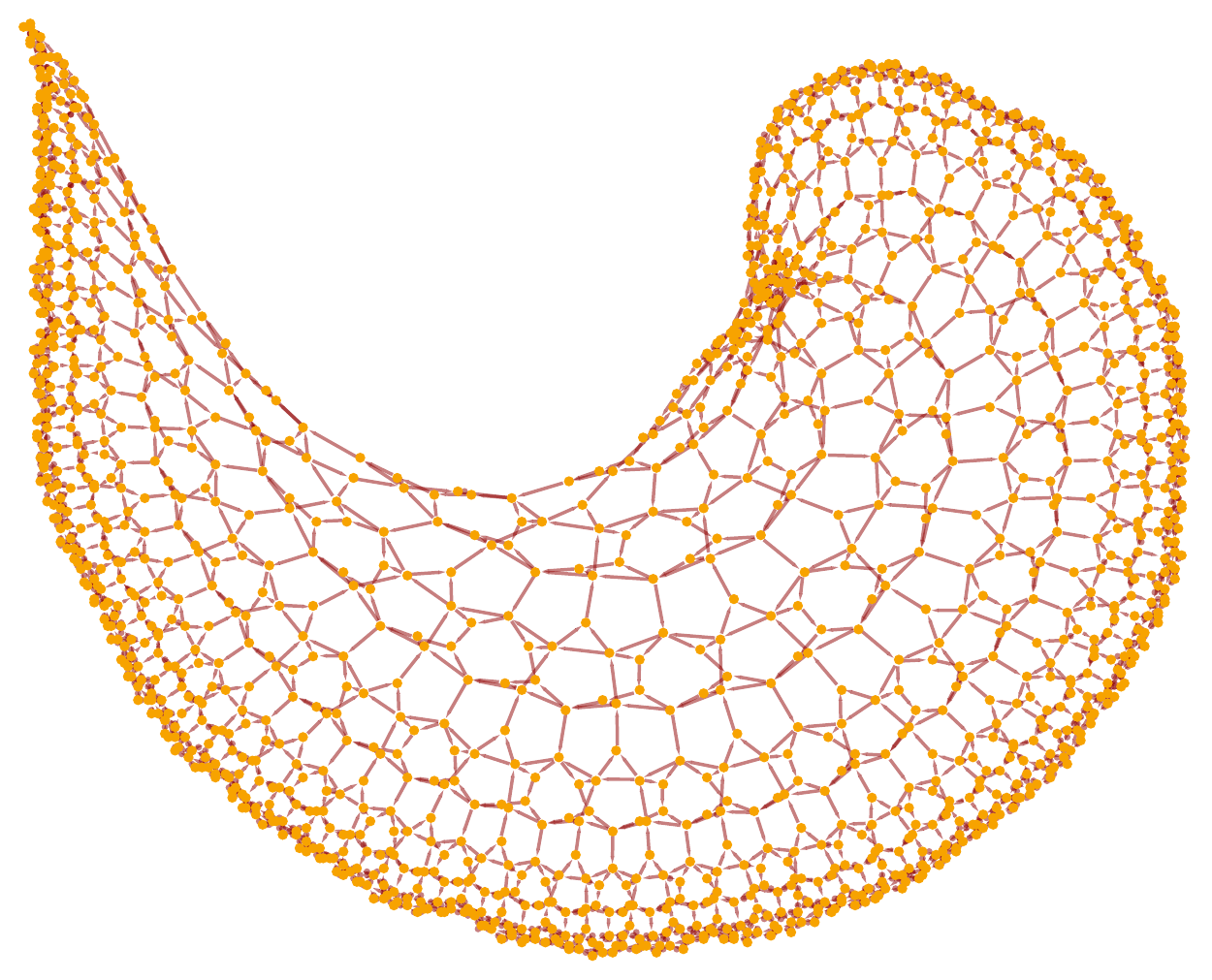}\hspace{0.1\textwidth}
\includegraphics[width=0.345\textwidth]{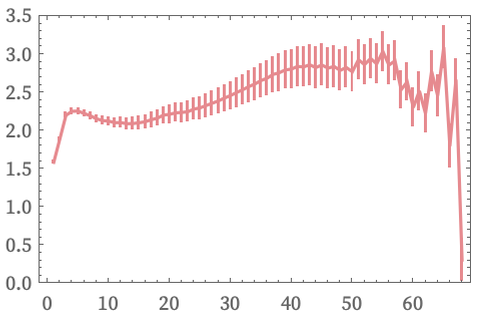}
\caption{Dimension estimates for a causal network with a two-dimensional Lorentzian manifold-like limiting structure, as generated by the set substitution system ${\left\lbrace \left\lbrace x, y, x \right\rbrace, \left\lbrace x, z, u \right\rbrace \right\rbrace \to \left\lbrace \left\lbrace u, v, u \right\rbrace, \left\lbrace v, u, z \right\rbrace, \left\lbrace x, y, v \right\rbrace \right\rbrace}$, exhibiting the effects of non-zero spacetime curvature, and indicating a limiting dimension of two. Example taken from \cite{wolfram2}.}
\label{fig:Figure46}
\end{figure}

\begin{figure}[ht]
\centering
\includegraphics[width=0.795\textwidth]{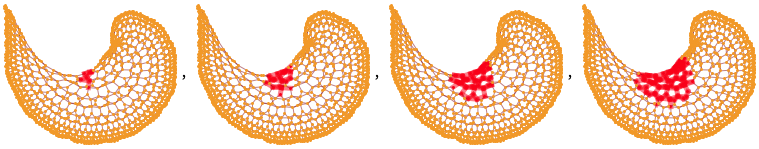}
\caption{The growth of a finite geodesic cone in a causal network with a two-dimensional Lorentzian manifold-like limiting structure, as generated by the set substitution system ${\left\lbrace \left\lbrace x, y, x \right\rbrace, \left\lbrace x, z, u \right\rbrace \right\rbrace \to \left\lbrace \left\lbrace u, v, u \right\rbrace, \left\lbrace v, u, z \right\rbrace, \left\lbrace x, y, v \right\rbrace \right\rbrace}$, exhibiting the effects of non-zero spacetime curvature.}
\label{fig:Figure47}
\end{figure}

\sloppy A very similar analysis may be performed in the (directed) causal network case, except that the discrete volume being measured is now that of an infinitesimal geodesic cone ${C_t}$ in spacetime as opposed to a geodesic ball ${B_{\epsilon}}$ in a spacelike hypersurface, and the form of the volume element is no longer Riemannian:

\begin{equation}
\mathrm{Vol} \left( B_{\epsilon} \left( p \right) \right) = a \epsilon^d \left[ 1 - \frac{R}{6 \left( d + 2 \right)} \epsilon^2 + O \left( \epsilon^4 \right) \right],
\end{equation}
with the spatial Ricci scalar $R$, but rather Lorentzian:

\begin{equation}
\mathrm{Vol} \left( C_{t} \left( p \right) \right) = a t^n \left[ 1 - \frac{1}{6} \sum_{i, j = 1}^{d} R_{i j} t^i t^j + O \left( \left\lVert \mathbf{t} \right\rVert^3 \right) \right],
\end{equation}
with spacetime Ricci tensor ${R_{i j}}$. Here, we are projecting the Ricci tensor in the timelike direction ${\mathbf{t}}$ corresponding to the orientation of the cone; the time vector can be written in the succinct form:

\begin{equation}
t^a = \alpha n^a + \beta^a,
\end{equation}
in terms of the ADM gauge variables ${\alpha}$ and ${\beta^i}$, and timelike normal vector ${\mathbf{n}}$. The behavior of the first-order logarithmic difference estimator ${\Delta_t \left( p \right)}$ as a function of time $t$ is shown in Figures \ref{fig:Figure44} and \ref{fig:Figure46} for two set substitution systems, namely ${\left\lbrace \left\lbrace x, y, y \right\rbrace, \left\lbrace x, z, u \right\rbrace \right\rbrace \to \left\lbrace \left\lbrace u, v, v \right\rbrace, \left\lbrace v, z, y \right\rbrace, \left\lbrace x, y, v \right\rbrace \right\rbrace}$ and ${\left\lbrace \left\lbrace x, y, x \right\rbrace, \left\lbrace x, z, u \right\rbrace \right\rbrace \to \left\lbrace \left\lbrace u, v, u \right\rbrace, \left\lbrace v, u ,z \right\rbrace, \left\lbrace x, y, v \right\rbrace \right\rbrace}$, whose resulting causal networks are known to limit to two-dimensional Lorentzian manifold-like structures (in the former case with asymptotic flatness, and in the latter case with non-zero spacetime curvature), respectively. As with the purely spatial examples in two dimensions, the limiting dimension of two is correctly estimated in each case, with estimates being produced by considering the growth rates of small geodesic cones, as shown in Figures \ref{fig:Figure45} and \ref{fig:Figure47}.

\begin{figure}[ht]
\centering
\includegraphics[width=0.395\textwidth]{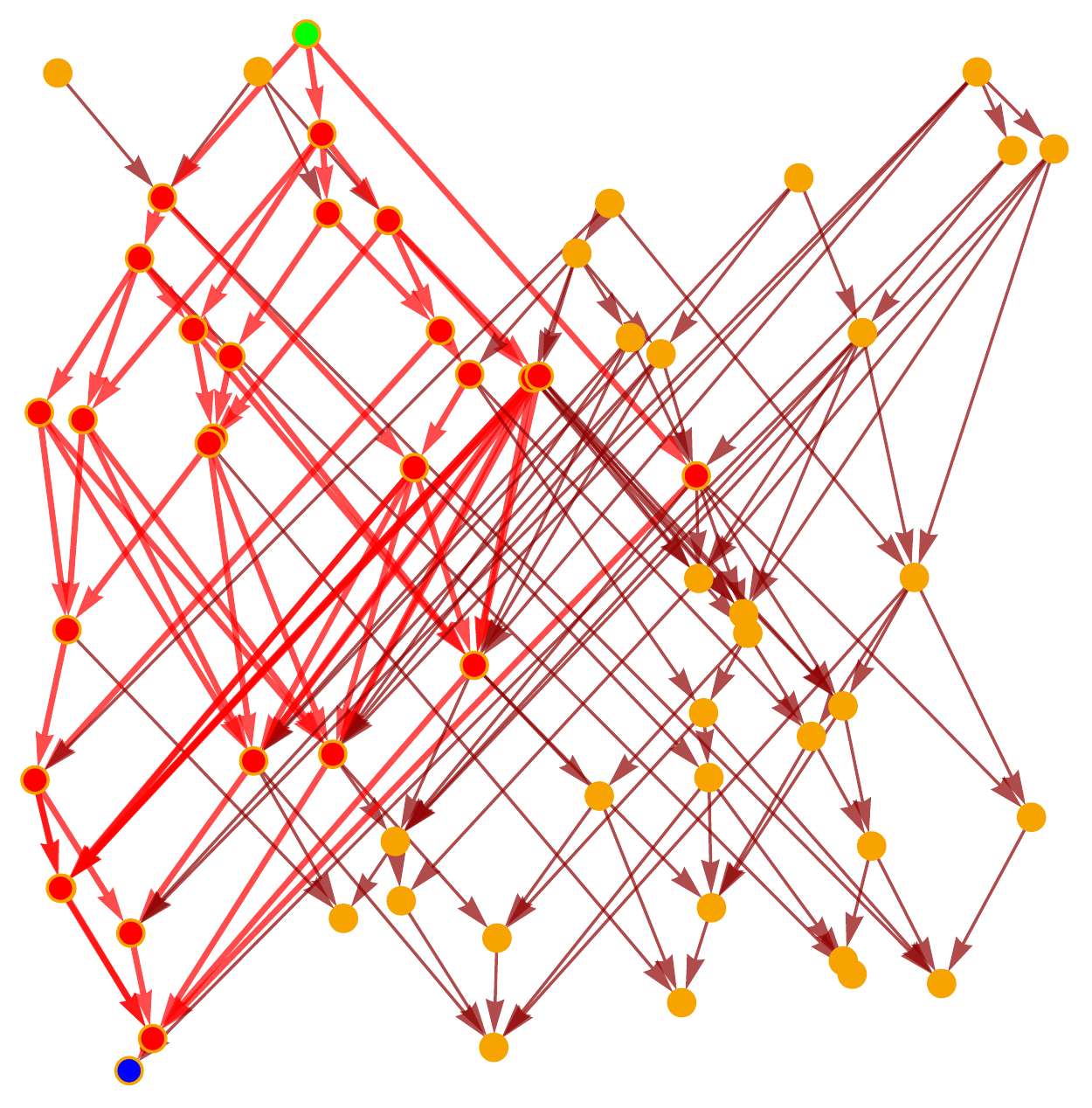}\hspace{0.1\textwidth}
\includegraphics[width=0.395\textwidth]{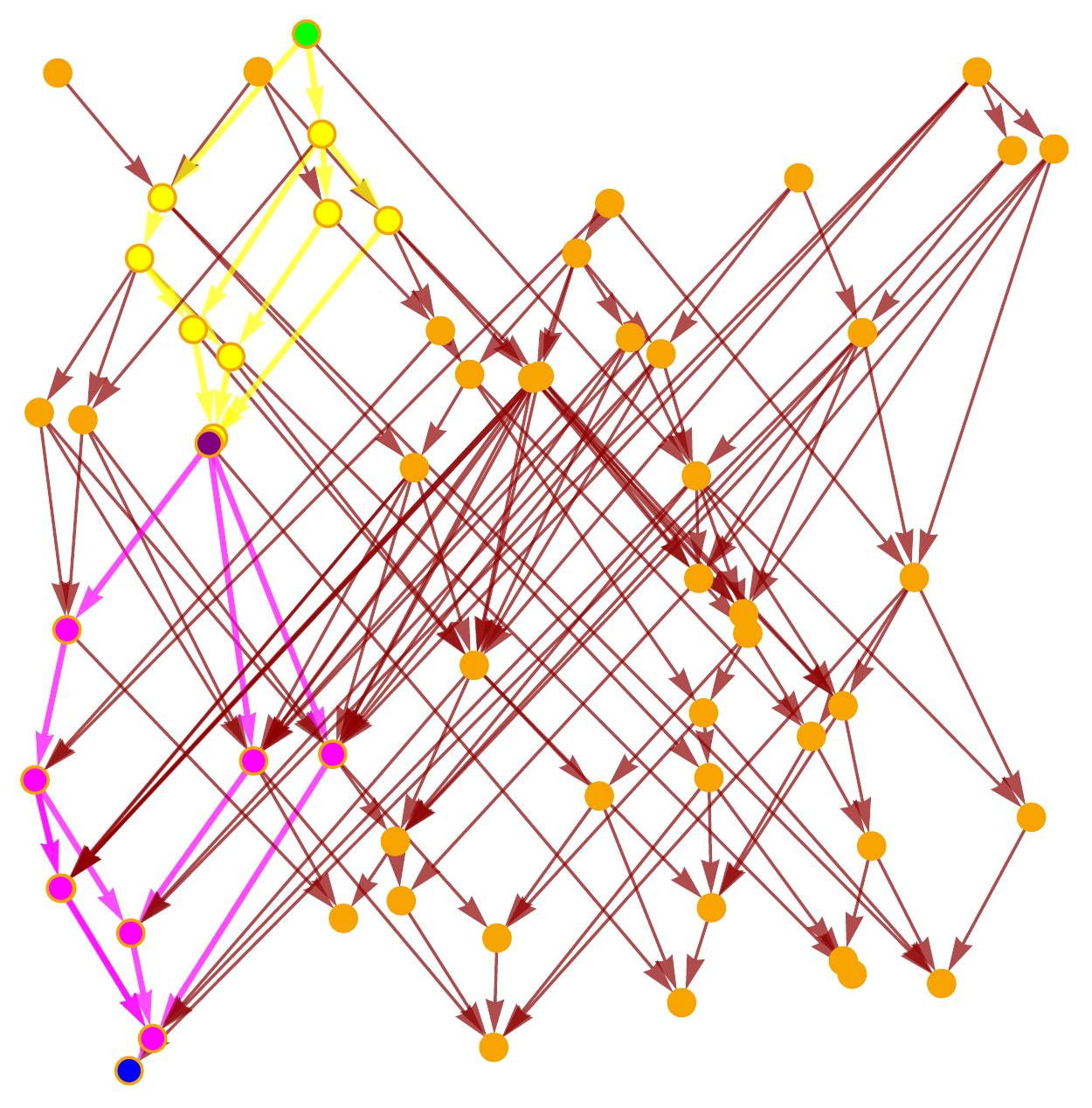}
\caption{The transitive reduction (i.e. the Hasse diagram) of the causal partial order for 60 uniformly sprinkled points in a rectangular region of ${1 + 1}$-dimensional flat (Minkowski) spacetime. Highlighted in green and blue are two timelike-separated events that constitute the endpoints of a discrete spacetime order interval, highlighted in red (on the left). Highlighted in purple (on the right) is a midpoint of the interval that divides it into two subintervals, highlighted in yellow and magenta. The yellow and magenta subintervals both have a cardinality of 10, as compared to the overall red interval with a cardinality of 28, yielding a dimension estimate of ${\log_2 \left( \frac{28}{10} \right) \approx 1.49}$.}
\label{fig:Figure48}
\end{figure}

It is worth noting that the Myrheim-Meyer dimension estimators significantly underestimate the limiting dimension of the causal networks shown in Figures \ref{fig:Figure44} to \ref{fig:Figure47}, as compared to the geodesic cone estimator, and converges much more gradually (and this effect is observed in a variety of both randomly-generated and algorithmically-generated causal sets). Indeed, the geodesic cone estimation technique is much closer in spirit to the \textit{midpoint scaling} dimension estimation technique proposed by Bombelli\cite{bombelli7}\cite{reid}, in which one exploits the fact that the volume of the discrete spacetime order interval ${\mathbf{I} \left[ p, q \right]}$ in $d$-dimensional flat (Minkowski) spacetime is proportional to $T$, the proper time between events $p$ and $q$, as follows:

\begin{equation}
\mathrm{Vol} \left( \mathbf{I} \left[ p, q \right] \right) = \frac{\pi^{\frac{\left( d - 1 \right)}{2}}}{2^{d - 2} d \left( d - 1 \right) \Gamma \left[ \frac{\left( d - 1 \right)}{2} \right]} T^d.
\end{equation}
Since any discrete interval ${\mathbf{I} \left[ p, q \right]}$ of cardinality $N$ can be divided into a pair of sub-intervals ${\mathbf{I}_1 \left[ p, r \right]}$ and ${\mathbf{I} \left[ r, q \right]}$ of cardinalities ${N_1}$ and ${N_2}$, respectively, then by denoting the smaller of the two cardinalities ${N_1}$ and ${N_2}$ by ${N_{small}}$, we can see that the causal set element $r$ corresponds to the midpoint of ${\mathbf{I} \left[ p, q \right]}$ if and only if ${N_{small}}$ is as large as possible. This procedure has the effect of scaling all spacetime intervals by a factor of ${\frac{1}{2}}$, such that:

\begin{equation}
\frac{T}{T_{small}} = 2, \qquad \implies \frac{\mathrm{Vol} \left( \mathbf{I} \left[ p, q \right] \right)}{\mathrm{Vol} \left( \mathbf{I} \left[ p, r \right] \right)} = 2^d,
\end{equation}
in the continuous manifold, and therefore we can conclude that:

\begin{equation}
\frac{N}{N_{small}} \approx 2^d,
\end{equation}
in the causal set approximation. Thus, we can compute a first-order logarithmic estimator ${\Delta}$ of the spacetime dimension $d$ by evaluating:

\begin{equation}
\Delta = \log_2 \left( \frac{N}{N_{small}} \right),
\end{equation}
which is essentially computing a logarithmic difference estimate of the spacetime dimension exactly as above, but using the growth rate of the causal diamond ${\mathbf{I} \left( p, q \right)}$ corresponding to a spacetime interval, rather than a geodesic cone ${C_t \left( p \right)}$. An example of the computation of the midpoint scaling dimension estimator for a causal set consisting of 60 uniformly sprinkled points in a rectangular region of ${1 + 1}$-dimensional flat (Minkowski) spacetime is shown in Figure \ref{fig:Figure48}, yielding a dimension estimate of ${1.49}$. We find from explicit numerical experiments (again, performed over both randomly-generated and algorithmically-generated causal sets) that the midpoint scaling dimension exhibits approximately the same scaling relation as the geodesic cone dimension for a randomly sprinkled causal set, just as one would reasonably expect, but often provides an underestimate as compared to the geodesic cone dimension estimator for an algorithmically-generated causal set. The reasons for this discrepancy are not currently clear, and are to be investigated in future work.

\section{Discrete Ricci Curvature, the Benincasa-Dowker Action and the Einstein Field Equations}
\label{sec:Section4}

Henceforth, Einstein summation convention is adopted. Despite the intrinsically non-local nature of causal sets (as a consequence of the infinite out-degree of each vertex in the associated causal network), it is nevertheless possible to construct discrete differential operators which provably approximate their local continuum analogs in appropriate cases. A good example is the discrete d'Alembertian operator $B$ on free (real) scalar fields ${\phi : \mathcal{C} \to \mathbb{R}}$ that was first proposed by Sorkin and Henson\cite{sorkin3}\cite{henson} for causal sets ${\mathcal{C}}$ produced by Poisson sprinklings in 2-dimensional flat (Minkowski) spacetime ${\mathbb{M}^2 = \mathbb{R}^{1, 1}}$:

\begin{equation}
B^{\left( 2 \right)} \phi \left( e \right) = \frac{1}{l^2} \left[ - 2 \phi \left( x \right) + 4 \left( \sum_{e^{\prime} \in L_0 \left( e \right)} \phi \left( e^{\prime} \right) - 2 \sum_{e^{\prime} \in L_1 \left( e \right)} \phi \left( e^{\prime} \right) + \sum_{e^{\prime} \in L_2 \left( e \right)} \phi \left( e^{\prime} \right) \right) \right],
\end{equation}
and was subsequently extended by Benincasa and Dowker\cite{benincasa} to the 4-dimensional flat spacetime case ${\mathbb{M}^4 = \mathbb{R}^{1, 3}}$:

\begin{equation}
B^{\left( 4 \right)} \phi \left( e \right) = \frac{1}{l^2} \left[ - \frac{4}{\sqrt{6}} \phi \left( e \right) + \frac{4}{\sqrt{6}} \left( \sum_{e^{\prime} \in L_0 \left( e \right)} \phi \left( e^{\prime} \right) - 9 \sum_{e^{\prime} \in L_1 \left( e \right)} \phi \left( e^{\prime} \right) + 16 \sum_{e^{\prime} \in L_2 \left( e \right)} \phi \left( e^{\prime} \right) - 8 \sum_{e^{\prime} \in L_3 \left( e \right)} \phi \left( e^{\prime} \right) \right) \right],
\end{equation}
where, in the above, $l$ is a characteristic length scale that is analogous to a lattice spacing and the sums are evaluated over past ``layers'' ${L_k \left( e \right)}$ (i.e. sets of $k$-nearest neighbors in the past of event ${e \in \mathcal{C}}$):

\begin{equation}
L_k \left( e \right) = \left\lbrace e^{\prime} \prec e : n \left[ e, e^{\prime} \right] = k \right\rbrace,
\end{equation}
where ${n \left[ p, q \right]}$ is a measure of the cardinality of the discrete spacetime order interval ${\mathbf{I} \left[ p, q \right]}$:

\begin{equation}
n \left[ p, q \right] = \left\lvert \mathbf{I} \left[ p, q \right] \right\rvert - 2, \qquad \text{ where } \mathbf{I} \left[ p, q \right] = \left\lbrace r \in \mathcal{C} : q \prec r \prec p \right\rbrace.
\end{equation}
The non-local nature of the operator can be inferred from its dependence on the number of $k$-nearest neighbors (in the ${d = 4}$ case, for ${k = 0, 1, 2, 3}$), which is potentially unbounded. As noted by Dowker and Glaser\cite{dowker}, this construction of the discrete d'Alembertian may be extended to ${\mathbb{M}^d = \mathbb{R}^{1, d- 1}}$ (for arbitrary values of $d$) by use of the following ansatz:

\begin{equation}
B^{\left( d \right)} \phi \left( e \right) = \frac{1}{l^2} \left( \alpha_d \phi \left( e \right) + \beta_d \sum_{i = 0}^{n_d - 1} C_{i}^{\left( d \right)} \sum_{y \in L_i} \phi \left( e \right) \right),
\end{equation}
in which ${n_d}$ denotes the number of layers to sum over, ${\alpha_d}$, ${\beta_d}$ and ${C_{i}^{\left( d \right)}}$ (for ${i = 0, \dots, n_d - 1}$) are undetermined dimension-dependent constants, and the initial coefficient ${C_{0}^{\left( d \right)}}$ is always fixed to be equal to 1.

When considering a $d$-dimensional Lorentzian manifold ${\left( \mathcal{M}, g \right)}$ equipped with a real scalar field ${\phi : \mathcal{M} \to \mathbb{R}}$ of compact support, the discrete d'Alembertian operator ${B^{\left( d \right)}}$ will yield a random variable ${\hat{B}^{\left( d \right)} \phi \left( x \right)}$ for every point ${x \in \mathcal{M}}$ that is added to a (random) causal set ${\mathcal{C}}$ sprinkled into ${\mathcal{M}}$ with density ${\rho = l^{-d}}$. As such, ${\phi}$ is restricted to a real discrete scalar field on the random causal set ${\mathcal{C}}$, and its expectation value ${\left\langle \hat{B}^{\left( d \right)} \phi \left( x \right) \right\rangle}$ can be computed via the Poisson distribution (for general dimension $d$) as:

\begin{multline}
\left\langle \hat{B}^{\left( d \right)} \phi \left( x \right) \right\rangle = \alpha_d l^{-2} \phi \left( x \right)\\
+ \beta_d l^{- \left( d + 2 \right)} \int_{J^{-} \left( x \right)} \sqrt{-g \left( y \right)} \phi \left( y \right) \sum_{i = 0}^{n_d - 1} C_{i}^{\left( d \right)} \frac{\left( \mathrm{Vol} \left( \mathbf{I} \left[ y, x \right] \right) l^{-d} \right)^{i - 2}}{\Gamma \left( i - 1 \right)} \exp \left( - \mathrm{Vol} \left( \mathbf{I} \left[ y, x \right] \right) l^{-d} \right) d^d y.
\end{multline}
In the above integral, the ${l^{-d} \sqrt{- g \left( y \right)}}$ contribution denotes the probability that a point ${x \in \mathcal{C}}$ is sprinkled inside the infinitesimal volume element at point ${y \in \mathcal{M}}$, and the factor ${\frac{\left( \mathrm{Vol} \left( \mathbf{I} \left[ y, x \right] \right) l^{-d} \right)^{i - 2}}{\Gamma \left( i - 1 \right)} \exp \left( - \mathrm{Vol} \left( \mathbf{I} \left[ y, x \right] \right) l^{-d} \right)}$ designates the probability that the element $x$ is sprinkled in the $i$-th ``layer'' (i.e. the probability that exactly $i$ elements are sprinkled in the interval ${\mathbf{I} \left[ y, x \right]}$). By introducing the following family of integrals, parametrized by dimension $d$:

\begin{equation}
I_{d} \left( l \right) = \int_{J^{-} \left( x \right)} \sqrt{-g \left( y \right)} \exp \left( - \mathrm{Vol} \left( \mathbf{I} \left[ y, x \right] \right) l^{-d} \right) \phi \left( y \right) d^d y,
\end{equation}
one obtains, in the ${d = 2}$ and ${d = 4}$ cases:

\begin{equation}
\left\langle \hat{B}^{\left( 2 \right)} \phi \left( x \right) \right\rangle = \alpha_2 l^{-2} \phi \left( x \right) + \beta_2 l^{-4} \mathcal{O}_2 I_2 \left( l \right),
\end{equation}
and:

\begin{equation}
\left\langle \hat{B}^{\left( 4 \right)} \phi \left( x \right) \right\rangle = \alpha_4 l^{-2} \phi \left( x \right) + \beta_4 l^{-6} \mathcal{O}_4 I_4 \left( l \right),
\end{equation}
respectively, where we have introduced the following pair of differential operators:

\begin{equation}
\mathcal{O}_2 = \frac{1}{8} \left( H + 2 \right) \left( H + 4 \right), \qquad \text{ and } \qquad \mathcal{O}_4 = \frac{1}{48} \left( H + 2 \right) \left( H + 4 \right) \left( H + 6 \right),
\end{equation}
with:

\begin{equation}
H = -l \frac{\partial}{\partial l}.
\end{equation}
Thus, for the particular case in which the causal set is sprinkled into 4-dimensional flat (Minkowski) spacetime ${\mathbb{M}^4}$, the expectation value becomes:

\begin{multline}
\frac{1}{\sqrt{\rho_c}} \left\langle \hat{B}^{\left( 4 \right)} \phi \left( x \right) \right\rangle = \frac{4 \sqrt{\rho_c}}{\sqrt{6}} \left[ - \phi \left( x \right) + \rho_c \int_{y \in J^{-} \left( x \right)} \phi \left( y \right) \exp \left( - \rho_c \mathrm{Vol} \left( \mathbf{I} \left[ y, x \right] \right) \right) \left( 1 - 9 \rho_c \mathrm{Vol} \left( \mathbf{I} \left[ y, x \right] \right) \right. \right. \\
\left. \left. + 8 \left( \rho_c \mathrm{Vol} \left( \mathbf{I} \left[ y, x \right] \right) \right)^2 - \frac{4}{3} \left( \rho_c \mathrm{Vol} \left( \mathbf{I} \left[ y, x \right] \right) \right)^3 \right) d^4 y \right].
\end{multline}

\begin{figure}[ht]
\centering
\includegraphics[width=0.395\textwidth]{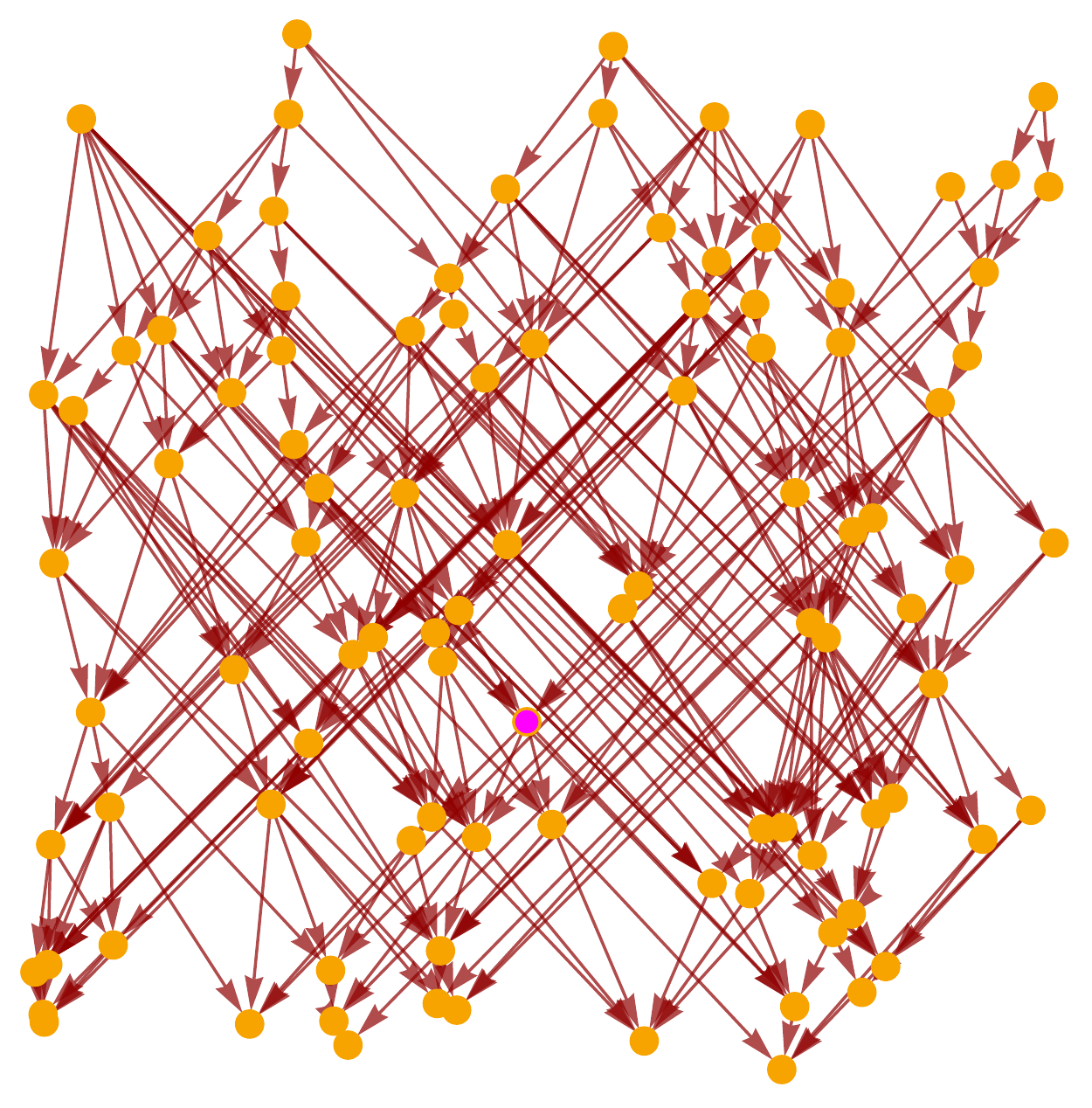}\hspace{0.1\textwidth}
\includegraphics[width=0.395\textwidth]{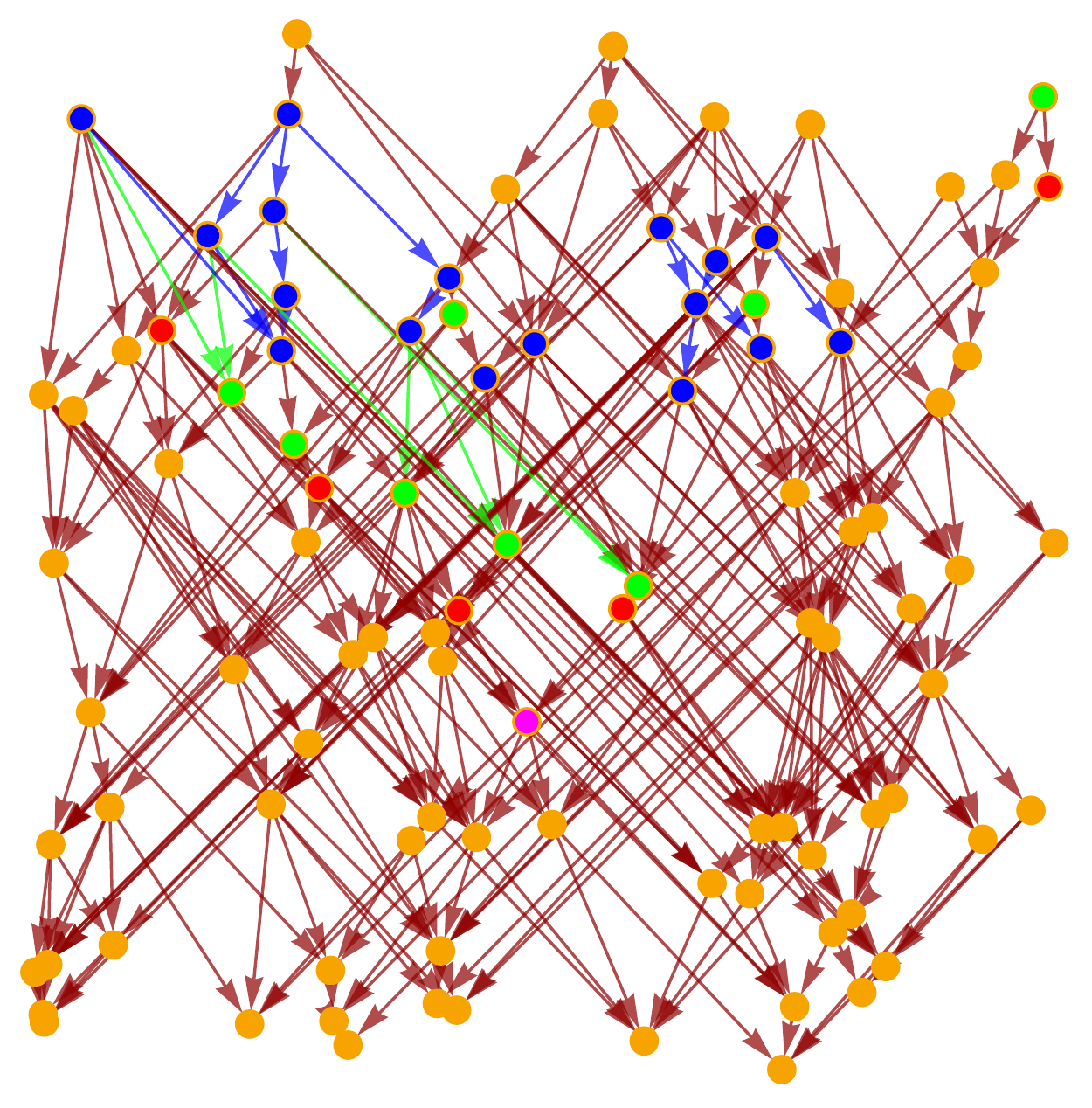}
\caption{The transitive reduction (i.e. the Hasse diagram) of the causal partial order graph for 100 uniformly sprinkled points in a rectangular region of ${1 + 1}$-dimensional flat (Minkowski) spacetime. Highlighted in magenta is a point for which the discrete d'Alembertian is being computed. On the right, highlighted in red, green and blue, are the ``layers'' ${L_0}$, ${L_1}$ and ${L_2}$ of $0$-nearest, $1$-nearest and $2$-nearest neighbors, respectively, whose values of the scalar field ${\phi}$ are then summed over.}
\label{fig:Figure49}
\end{figure}

Since the operator ${\mathcal{O}_2}$ will annihilate both ${l^2}$ and ${l^4}$, we can conclude that any contributions to the expectation value ${\left\langle \hat{B}^{\left( 4 \right)} \phi \left( x \right) \right\rangle}$ that would not otherwise converge to zero in the limit ${l \to 0}$ are eliminated. More specifically, we can evaluate the integral by first realizing that the integration region is compact, since the function ${\phi}$ is of compact support by hypothesis. If we now choose a reference frame (and hence a foliation of spacetime) denoted ${\mathcal{F}_{\phi}}$ in which ${\phi \left( y \right)}$ is made to vary slowly in the vicinity of the immediate past of event $x$ (with respect to the coordinate time in ${\mathcal{F}_{\phi}}$), then we can perform a series expansion in powers of small ${\left\lvert y - x \right\rvert}$ of ${\phi \left( y \right)}$ in the vicinity of ${\phi \left( x \right)}$, with respect to the frame ${\mathcal{F}_{\phi}}$. The non-compact region ${J^{-} \left( x \right)}$ can be subdivided into non-overlapping sub-regions ${\mathcal{W}_1}$, ${\mathcal{W}_2}$ and ${\mathcal{W}_3}$:

\begin{equation}
J^{-} \left( x \right) = \mathcal{W}_1 \cup \mathcal{W}_2 \cup \mathcal{W}_3, \qquad \text{ such that } \left( \mathcal{W}_1 \cap \mathcal{W}_2 \right) = \left( \mathcal{W}_1 \cap \mathcal{W}_3 \right) = \left( \mathcal{W}_2 \cap \mathcal{W}_3 \right) = \emptyset,
\end{equation}
where ${\mathcal{W}_1}$ is a neighborhood of event $x$, ${\mathcal{W}_2}$ is a neighborhood of the region boundary ${\partial J^{-} \left( x \right)}$ (bounded away from the origin, i.e. the region ``down the light cone''), and ${\mathcal{W}_3}$ is bounded away from the boundary ${\partial J^{-} \left( x \right)}$ (i.e. the ``deep chronological past''). We can see that the integrand is negligible for all:

\begin{equation}
\rho_c \mathrm{Vol} \left( \mathbf{I} \left[ y, x \right] \right) > \alpha^4,
\end{equation}
where ${\alpha}$ is a parameter chosen such that:

\begin{equation}
e^{- \alpha^4} \ll 1,
\end{equation}
and that the integrals over the regions ${\mathcal{W}_3}$ and ${\mathcal{W}_2}$ (further down the light cone) are bounded above by integrals that go to zero faster than any power of ${\rho_{c}^{-1}}$ and ${\rho_{c}^{- \frac{3}{2}}}$, respectively. This is due to the fact that the majority of the contribution to ${\left\langle \hat{B}^{\left( d \right)} \phi \left( x \right) \right\rangle}$ will come from the integration range that lies between the past light cone of event $x$ and the hyperboloid:

\begin{equation}
\rho_c \mathrm{Vol} \left( \mathbf{I} \left[ y, x \right] \right) = \alpha^4,
\end{equation}
comprising a neighborhood of event $x$ of size ${\alpha l}$ (i.e. region ${\mathcal{W}_1}$) and regions further down the light cone (i.e. region ${\mathcal{W}_2}$) and in the deep chronological past (i.e. region ${\mathcal{W}_3}$). The integral over the non-local region ${\mathcal{W}_2 \cup \mathcal{W}_3}$ can be parametrized by ${\rho_c \mathrm{Vol} \left( \mathbf{I} \left[ y, x \right] \right)}$ and the coordinates ${\eta^a}$ on the hyperboloid ${\mathcal{H}}$, yielding:

\begin{multline}
\int_{y \in \left( \mathcal{W}_2 \cap \mathcal{W}_3 \right)} \phi \left( y \right) \exp \left( - \rho_c \mathrm{Vol} \left( \mathbf{I} \left[ y, x \right] \right) \right) \left( 1 - 9 \rho_c \mathrm{Vol} \left( \mathbf{I} \left[ y, x \right] \right) + 8 \left( \rho_c \mathrm{Vol} \left( \mathbf{I} \left[ y, x \right] \right) \right)^2 \right.\\
\left. - \frac{4}{3} \left( \rho_c \mathrm{Vol} \left( \mathbf{I} \left[ y, x \right] \right) \right)^3 \right) d^4 y \propto \int_{\mathcal{H}} \int_{0}^{\alpha^4} \exp \left( - \rho_c \mathrm{Vol} \left( \mathbf{I} \left[ y, x \right] \right) \right) \left( 1 - 9 \rho_c \mathrm{Vol} \left( \mathbf{I} \left[ y, x \right] \right) \right.\\
\left. + 8 \left( \rho_c \mathrm{Vol} \left( \mathbf{I} \left[ y, x \right] \right) \right)^2 - \frac{4}{3} \left( \rho_c \mathrm{Vol} \left( \mathbf{I} \left[ y, x \right] \right) \right)^3 \right) \phi \left( \rho_c \mathrm{Vol} \left( \mathbf{I} \left[ y, x \right] \right), \eta^a \right) d \left( \rho_c \mathrm{Vol} \left( \mathbf{I} \left[ y, x \right] \right) \right) d^3 \eta.
\end{multline}
Given that, by assumption, the field ${\phi}$ is almost constant over the spacetime distance scale ${\alpha l}$, it follows that the integral with respect to ${\rho_C \mathrm{Vol} \left( \mathbf{I} \left[ y, x \right] \right)}$ is approximately zero, so all contributions from the non-local region ${\mathcal{W}_2 \cup \mathcal{W}_3}$ are suppressed, yielding:

\begin{equation}
\lim_{\rho_c \to \infty} \left[ \frac{1}{\sqrt{\rho_c}} \left\langle \hat{B}^{\left( 4 \right)} \phi \left( x \right) \right\rangle \right] = \Box \phi \left( x \right),
\end{equation}
as required. Thus, we have remarkably been able to recover an ``effectively local'' kinematics, despite the explicitly non-local nature of the discrete d'Alembertian operator ${B^{\left( 4 \right)}}$, since ${\hat{B}^{\left( 4 \right)} \phi \left( x \right)}$is determined only by a small compact neighborhood of $x$, as opposed to the entirety of ${J^{-} \left( x \right)}$ (since the computation is dominated by the restrictions of ``layers'' ${L_k}$ to the region ${\mathbf{I} \left[ p, q \right] \cap J^{-} \left( x \right)}$, at least in frame ${\mathcal{F}_{\phi}}$). An example of the computation of the discrete d'Alembertian on a causal set consisting of 100 uniformly sprinkled points in a rectangular region of ${1 + 1}$-dimensional flat (Minkowski) spacetime is shown in Figure \ref{fig:Figure49}.

This entire analysis may be generalized to arbitrary numbers of dimensions $d$ by making an appropriate choice of differential operator ${\mathcal{O}_d}$ that fixes the constants ${C_{i}^{\left( d \right)}}$ in accordance with the following scheme:

\begin{equation}
\mathcal{O}_d \exp \left( - l^{-d} \mathrm{Vol} \left( \mathbf{I} \left[ y, x \right] \right) \right) = \sum_{i = 0}^{n_d - 1} C_{i}^{\left( d \right)} \frac{\left( l^{-d} \mathrm{Vol} \left( \mathbf{I} \left[ y, x \right] \right) \right)^{i}}{\Gamma \left( i - 1 \right)} \exp \left( -l^{-d} \mathrm{Vol} \left( \mathbf{I} \left[ y, x \right] \right) \right),
\end{equation}
and simply solving for the remaining constants ${\alpha_d}$ and ${\beta_d}$. Dowker and Glaser\cite{dowker} proposed the following family of operators:

\begin{equation}
\mathcal{O}_d = \begin{cases}
\frac{\left( H + 2 \right) \left( H + 4 \right) \cdots \left( H + 2n + 2 \right)}{2^{n + 1} \Gamma \left( n + 2 \right)}, \qquad & \text{ if } d = 2n \text{ even},\\
\mathcal{O}_{2 n}, \qquad & \text{ if } d = 2n + 1 \text{ odd},
\end{cases}
\end{equation}
which generalize the previously described cases:

\begin{equation}
\mathcal{O}_2 = \frac{1}{8} \left( H + 2 \right) \left( H + 4 \right), \qquad \text{ and } \qquad \mathcal{O}_4 = \frac{1}{48} \left( H + 2 \right) \left( H + 4 \right) \left( H + 6 \right),
\end{equation}
in a fairly natural way. To evaluate the integral ${I_d \left( l \right)}$ in $d$-dimensional flat (Minkowski) spacetime ${\mathbb{M}^d}$ for general ${d > 2}$, we can choose the following radial null coordinate system:

\begin{equation}
v = \frac{1}{\sqrt{2}} \left( t + r \right), \qquad \text{ and } \qquad u = \frac{1}{\sqrt{2}} \left( t - r \right),
\end{equation}
with the parameter $r$ denoting the radius of the ${\left( d - 1 \right)}$-dimensional spherical coordinate system centered on $x$ (chosen here as the origin of the coordinate frame), such that the integral becomes simply:

\begin{equation}
I_d \left( l \right) = \int_{- \infty}^{0} \int_{u}^{0} r^{d - 2} \int \phi \left( y \right) \exp \left( - l^{-d} \mathrm{Vol} \left( \mathbf{I} \left[ y, x \right] \right) \right) d \Omega_{d - 2} \cdot d v \cdot d u,
\end{equation}
with ${d \Omega_{d - 2}}$ denoting integration over the ${\left( d - 2 \right)}$-sphere.

\begin{figure}[ht]
\centering
\includegraphics[width=0.495\textwidth]{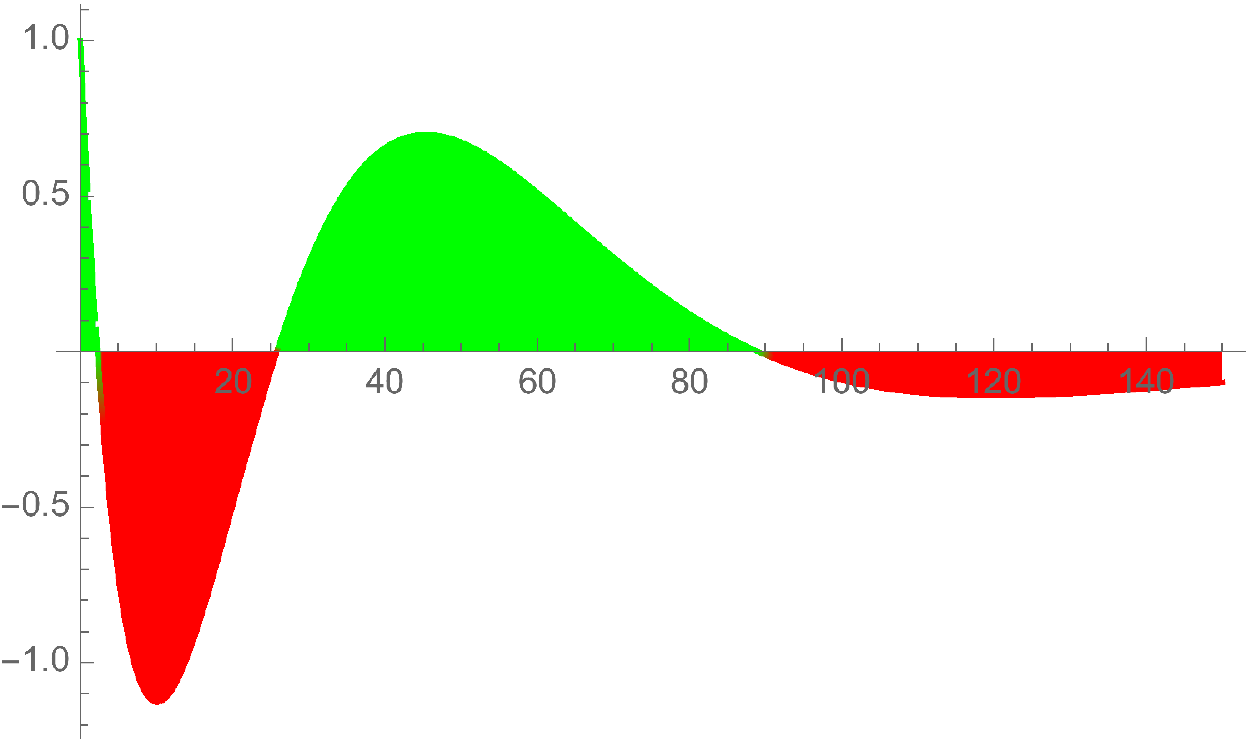}
\caption{A plot of the ``smearing'' function ${f \left( n, 0. 05 \right)}$ for ${0 \leq n \leq 150}$ in a 4-dimensional spacetime, showing the four ``smeared out'' layers corresponding to the four regions of alternating sign (with the two positive regions shown in green and the two negative regions shown in red).}
\label{fig:Figure50}
\end{figure}

We can generalize the above calculation of ${\left\langle \hat{B}^{\left( d \right)} \phi \left( x \right) \right\rangle}$ to any Riemann normal neighborhood in an arbitrary curved spacetime ${\left( \mathcal{M}, g \right)}$, at least in dimensions ${d = 2}$ and ${d = 4}$, as follows. First, we denote the volumes of causal intervals ${\mathbf{I} \left[ x, y \right]}$ in two and four dimensions by ${\mathrm{Vol}_2 \left( \mathbf{I} \left[ x, y \right] \right)}$ and ${\mathrm{Vol}_4 \left( \mathbf{I} \left[ x, y \right] \right)}$, respectively. Following the techniques of Sorkin\cite{sorkin3}, we introduce an intermediate length scale ${l_k \geq l}$ so as to dampen the microscopic fluctuations in ${B^{\left( d \right)} \phi \left( x \right)}$, with, moreover:

\begin{equation}
\xi_2 = \mathrm{Vol} \left( \mathbf{I} \left[ x, y \right] \right) l_{k}^{-2}, \qquad \text{ and } \qquad \xi_4 = \mathrm{Vol} \left( \mathbf{I} \left[ x, y \right] \right) l_{k}^{-4},
\end{equation}
and a new ``smoothed-out'' discrete operator ${B_k}$ whose expectation value is given by the usual integral, but with the length scale $l$ replaced with ${l_k}$:

\begin{multline}
\left\langle \hat{B}_{k}^{\left( d \right)} \phi \left( x \right) \right\rangle = \alpha_d l_{k}^{-2} \phi \left( x \right)\\
+ \beta_d l_{k}^{- \left( d + 2 \right)} \int_{J^{-} \left( x \right)} \sqrt{- g\left( y \right)} \phi \left( y \right) \sum_{i = 0}^{n_d - 1} C_{i}^{\left( d \right)} \frac{\left( \mathrm{Vol} \left( \mathbf{I} \left[ y, x \right] \right) l_{k}^{-d} \right)^{i - 2}}{\Gamma \left( i - 1 \right)} \exp \left( - \mathrm{Vol} \left( \mathbf{I} \left[ y, x \right] \right) l_{k}^{-d} \right) d^d y.
\end{multline}
For instance, in the familiar ${d = 4}$ case, if we set a \textit{non-locality parameter} ${\epsilon = \left( \frac{l}{l_k} \right)^4}$, then we can work backwards to deduce that the discrete operator itself is given by:

\begin{equation}
\left\langle \hat{B}_{k}^{\left( 4 \right)} \phi \left( x \right) \right\rangle = \frac{4}{\sqrt{6} l_{k}^{2}} \left[ - \phi \left( x \right) + \epsilon \sum_{y \prec x} f \left( n \left( x, y \right), \epsilon \right) \phi \left( y \right) \right],
\end{equation}
where:

\begin{equation}
f \left( n, \epsilon \right) = \left( 1 - \epsilon \right)^{n} \left[ 1 - \frac{9 \epsilon n}{1 - \epsilon} + \frac{8 \epsilon^2 \Gamma \left( n + 1 \right)}{\Gamma \left( n - 1 \right) \left( 1 - \epsilon \right)^2} - \frac{4 \epsilon^3 \Gamma \left( n + 1 \right)}{3 \Gamma \left( n - 2 \right) \left( 1 - \epsilon \right)^3} \right],
\end{equation}
and ${n \left( x, y \right)}$ simply denotes the cardinality of the discrete spacetime order interval:

\begin{equation}
n \left( x, y \right) = \left\lvert \mathbf{I} \left( x, y \right) \right\rvert,
\end{equation}
as usual, such that the smoothed operator ${B_k}$ reduces to the original operator ${B}$ whenever ${\epsilon = 1}$ (since the smoothed operator ${B_k}$ is essentially ``smearing out'' the contributions by sampling values of ${\phi}$ over four ``layers'' of elements with alternating sign, each with a characteristic depth of ${l_k}$, as shown in Figure \ref{fig:Figure50}). The ```smearing'' function $f$ in four spacetime dimensions can be generalized to a function ${f_d}$ in an arbitrary number of spacetime dimensions $d$:

\begin{equation}
f_d \left( n, \epsilon \right) = \left( 1 - \epsilon \right)^n \sum_{i = 0}^{n_d} C_{i}^{\left( d \right)} {n \choose i - 1} \left( \frac{\epsilon}{1 - \epsilon} \right)^{i - 1}.
\end{equation}

From here, one can infer that, in the presence of non-zero spacetime curvature, one has:

\begin{equation}
\left\langle \hat{B}_{k}^{\left( 2 \right)} \phi \left( x \right) \right\rangle = \frac{2}{l_{k}^{2}} \left[ - \phi \left( x \right) + \frac{2}{l_{k}^{2}} \int_{y \in J^{-} \left( x \right)} \sqrt{-g} \exp \left( - \xi_2 \right) \left( 1 - 2 \xi_2 + \frac{1}{2} x_{2}^{2} \right) \phi \left( y \right) d^2 y \right],
\end{equation}
and:

\begin{equation}
\left\langle \hat{B}_{k}^{\left( 4 \right)} \phi \left( x \right) \right\rangle = \frac{4}{\sqrt{6} l_{k}^{2}} \left[ - \phi \left( x \right) + \frac{1}{l_{k}^{2}} \int_{y \in J^{-} \left( x \right)} \sqrt{-g} \exp \left( - \xi_4 \right) \left( 1 - 9 \xi_4 + 8 \xi_{4}^{2} - \frac{4}{3} \xi_{4}^{3} \right) \phi \left( y \right) d^4 y \right],
\end{equation}
in the ${d = 2}$ and ${d = 4}$ cases, respectively, with ${\xi_2}$ and ${\xi_4}$ defined as above. To evaluate these expressions using Riemann normal coordinates, we divide the region of integration into the non-overlapping sub-regions ${\mathcal{W}_1}$, ${\mathcal{W}_2}$ and ${\mathcal{W}_3}$ (just as in the flat spacetime case), and we therefore see that the contribution to the integral from the near region ${\mathcal{W}_1}$ (or, more precisely, the Riemann normal neighborhood near event $x$) will be given in both cases by:

\begin{equation}
\lim_{\rho_c \to \infty} \left[ \frac{1}{\sqrt{\rho_c}} \left. \left\langle \hat{B}_{k}^{\left( d \right)} \phi \left( x \right) \right\rangle \right\rvert_{\mathcal{W}_1} \right] = \Box \phi \left( x \right) - \frac{1}{2} R \left( x \right) \phi \left( x \right),
\end{equation}
for ${d = 2}$ or ${d = 4}$, in the limit of an infinite sprinkling density, where ${R \left( x \right)}$ denotes the spacetime Ricci scalar curvature at event $x$. Clearly, the approximation is strongest whenever the radius of curvature ${r \gg l_k}$, and the scalar field ${\phi}$ varies slowly over scales on the order of the intermediate length ${l_k}$. Much as in the cases described above, the contribution to the integral from the deep chronological past ${\mathcal{W}_3}$ (i.e. the region away from the neighborhood of the boundary ${\partial J^{-} \left( x \right)}$) is bounded above by an integral that goes to zero faster than any power of ${\rho_{c}^{-1}}$.

However, the question of whether the contributions from the ``down the light cone'' region ${\mathcal{W}_2}$ (i.e. the region in the neighborhood of the boundary of ${\partial J^{-} \left( x \right)}$) vanish in the limit ${\rho_c \to \infty}$ of infinite sprinkling density is somewhat non-trivial to answer. It was proved by Belenchia, Benincasa and Dowker\cite{belenchia} that, in the limit of an approximately flat region of 4-dimensional spacetime, this cancellation does indeed occur, using the technique of \textit{null Fermi normal coordinates}:

\begin{definition}
The ``null Fermi normal coordinates'' of a null (lightlike) geodesic ${\gamma}$ in an arbitrary $d$-dimensional Lorentzian manifold ${\left( \mathcal{M}, g \right)}$ are produced in accordance with the following geometrical construction. Begin by introducing (via parallel transport) a pseudo-orthonormal frame ${E_{\mu}^{A}}$ along the geodesic ${\gamma}$:

\begin{equation}
\left. d s^2 \right\rvert_{\gamma} = 2 E^{+} E^{-} +\delta_{a b} E^a E^b,
\end{equation}
in which we define ${E_{+}^{\mu} = \dot{\gamma}^{\mu}}$, i.e. the derivative of ${\gamma^{\mu}}$ with respect to the affine parameter. If we now consider the family ${\beta \left( s \right)}$ of geodesics emanating from ${\gamma}$:

\begin{equation}
\beta \left( s \right) = \left( x^{\mu} \left( s \right) \right), \qquad \text{ with } \beta \left( 0 \right) = x_0 \in \gamma,
\end{equation}
for which the following relation holds:

\begin{equation}
g_{\mu \nu} \left( x_0 \right) x^{\mu^{\prime}} \left( 0 \right) E_{-}^{\nu} \left( x_0 \right) = x^{\mu^{\prime}} \left( 0 \right) E_{\mu}^{+} \left( x_0 \right) = 0,
\end{equation}
then the null Fermi coordinates:

\begin{equation}
\left( x^A \right) = \left( x^{+}, x^{-}, x^{a} \right),
\end{equation}
for a point ${x = \beta \left( s \right)}$ along such a geodesic are given by:

\begin{equation}
\left( x^A \right) = \left( x^{+}, x^{\bar{a}} = s E_{\mu}^{\bar{a}} \left( x_0 \right) x^{\mu^{\prime}} \left( 0 \right) \right),
\end{equation}
where we have defined:

\begin{equation}
\gamma \left( x^{+} \right) = x_0, \qquad \text{ and } \qquad \bar{a} = \left( -, a \right).
\end{equation}
\end{definition}
Note that the Fermi coordinates ${\left( x^A \right)}$ can therefore be straightforwardly related to the original geodesic coordinates ${\left( x^{\mu} \right)}$ using:

\begin{equation}
\left. \left( \frac{\partial x^A}{\partial x^{\mu}} \right) \right\rvert_{\gamma} = E_{\mu}^{A}, \qquad \text{ and } \qquad \left. \left( \frac{\partial x^{\mu}}{\partial x^A} \right) \right\rvert_{\gamma} = E_{A}^{\mu},
\end{equation}
due to the following pair of relationships between the derivatives:

\begin{equation}
E_{\mu}^{\bar{a}} \left( x_ 0 \right) x^{\mu^{\prime}} \left( 0 \right) = \left. \left( \frac{\partial x^{\bar{a}}}{\partial s} \right) \right\rvert_{s = 0} = \left. \left( \frac{\partial x^{\bar{a}}}{\partial x^{\mu}} \right) \right\rvert_{s = 0} x^{\mu^{\prime}} \left( 0 \right),
\end{equation}
and:

\begin{equation}
\left. \left( \frac{\partial x^{\mu}}{\partial x^{+}} \right) \right\rvert_{\gamma} = \dot{\gamma}^{\mu} = E_{+}^{\mu}.
\end{equation}
Now, if we define a light cone region ${LC}$ by the intersection of the boundary ${\partial J^{-} \left( x \right)}$ and the support of ${\phi}$:

\begin{equation}
LC = \partial J^{-} \left( x \right) \cap \mathrm{Supp} \left( \phi \right),
\end{equation}
then, assuming that every point in the region ${LC}$ lies along some unique past-directed null geodesic emanating from $x$ (which is a strong condition, since generally ${LC}$ will contain caustics, will complicate the calculation), every null geodesic generator ${\gamma \left( \theta, \varphi \right)}$ of ${LC}$ is labeled by the pair of polar coordinates ${\left( \theta, \varphi \right)}$, with an associated tangent vector ${T \left( \theta, \varphi \right)}$ at point $x$, whose components in Riemann normal coordinates are given by:

\begin{equation}
T^{\mu} \left( \theta, \varphi \right) = \frac{1}{\sqrt{2}} \left( -1, \sin \left( \theta \right) \cos \left( \varphi \right), \sin \left( \theta \right) \sin \left( \varphi \right), \cos \left( \theta \right) \right).
\end{equation}
All past-directed null tangent vectors then appear as antipodal pairs ${\left( T \left( \theta, \varphi \right), T \left( \pi - \theta, \varphi + \pi \right) \right)}$, such that:

\begin{equation}
T^{\mu} \left( \theta, \varphi \right) T_{\mu} \left( \pi - \theta, \varphi + \pi \right) = -1,
\end{equation}
and this overall construction defines the ``null Gaussian normal coordinates'' ${\left\lbrace V, U, \theta, \varphi \right\rbrace}$ in a neighborhood of the boundary ${\partial J^{-} \left( x \right)}$, denoted ${N_{LC}}$, that contains the light cone region ${LC}$ and is bounded away from point $x$. Here, coordinates ${\theta}$ and ${\varphi}$ label the null geodesic generators ${\gamma \left( \theta, \varphi \right)}$ of the boundary ${\partial J^{-} \left( x \right)}$, while coordinates $V$ and $U$ designate the affine parameters along each ${\gamma \left( \theta, \varphi \right)}$ and along certain past-directed, incoming null geodesics from every point in the light cone region ${LC}$, respectively. The past-directed null geodesics associated with the $U$ coordinate have the property that the tangent vector at a point $p$ to the incoming geodesic ${\gamma \left( \theta, \varphi \right)}$ is itself the vector ${T \left( \pi - \theta, \varphi + \pi \right)}$ at point $x$, having been mapped via parallel transport to $p$ along the geodesic ${\gamma \left( \theta, \varphi \right)}$. Then, for the region ${\mathcal{W}_2}$ down the light cone, any point ${y \in N_{LC}}$ with null Gaussian normal coordinates ${\left\lbrace V, U, \theta, \varphi \right\rbrace}$ will be such that ${U^{-2} V \left( y \right)}$ has a finite limit as ${U \to 0}$:

\begin{equation}
\lim_{U \to 0} \left[ U^{-2} V \left( y \right) \right] = f_0 \left( V, \theta, \varphi \right),
\end{equation}
and, moreover, if the discrete spacetime order interval ${\mathbf{I} \left[ y, x \right]}$ is contained within a tubular neighborhood of some null geodesic ${\gamma \left( \theta, \varphi \right)}$ on which null Fermi normal coordinates can be defined, then we have:

\begin{equation}
V \left( y \right) = U^2 f_0 \left( V, \theta, \varphi \right) + U^3 G \left( V, U, \theta, \varphi \right),
\end{equation}
for some continuous function $G$. This completes the proof. However, the problem of how to extend this argument to arbitrary spacetimes without approximate flatness still remains open.

Nevertheless, under the assumption that:

\begin{equation}
\lim_{\rho_c \to \infty} \left[ \frac{1}{\sqrt{\rho_c}} \left. \left\langle \hat{B}^{\left( d \right)} \phi \left( x \right) \right\rangle \right\rvert_{\mathcal{W}_2} \right] = 0,
\end{equation}
for our particular choice of spacetime, and setting ${\phi \left( x \right) = 1}$ for all events $x$ (which, strictly speaking, violates the requirement that ${\phi}$ should be of compact support, so more correctly we should say that ${\phi}$ is equal to 1 only in a neighborhood of region ${\mathcal{W}_1}$, with regions ${\mathcal{W}_2}$ and ${\mathcal{W}_3}$ contributing negligibly by hypothesis), we obtain:

\begin{equation}
\lim_{\rho_c \to \infty} \left[ \frac{1}{\sqrt{\rho_c}} \left\langle \hat{B}^{\left( d \right)} \phi \left( x \right) \right\rangle \right] = - \frac{1}{2} R \left( x \right).
\end{equation}
In (for instance) the 4-dimensional flat spacetime case ${\mathbb{M}^4}$, we can therefore write the following discrete action by summing the discrete Ricci scalar over the entire causal set ${\mathcal{C}}$:

\begin{equation}
S^{\left( 4 \right)} \left( \mathcal{C} \right) = \sum_{e \in \mathcal{C}} R \left( e \right),
\end{equation}
and since we have the following explicit form of of the discrete d'Alembertian ${B^{\left( 4 \right)}}$ from Benincasa and Dowker\cite{benincasa}:

\begin{equation}
B^{\left( 4 \right)} \phi \left( e \right) = \frac{1}{l^2} \left[ - \frac{4}{\sqrt{6}} \phi \left( e \right) + \frac{4}{\sqrt{6}} \left( \sum_{e^{\prime} \in L_0 \left( e \right)} \phi \left( e^{\prime} \right) - 9 \sum_{e^{\prime} \in L_1 \left( e \right)} \phi \left( e^{\prime} \right) + 16 \sum_{e^{\prime} \in L_2 \left( e \right)} \phi \left( e^{\prime} \right) - 8 \sum_{e^{\prime} \in L_3 \left( e \right)} \phi \left( e^{\prime} \right) \right) \right],
\end{equation}
we can expand out the Ricci scalar $R$ at element ${e \in \mathcal{C}}$ in terms of ${N_k \left( e \right)}$, i.e. the number of $k$-nearest neighbors in the past of event $e$:

\begin{equation}
N_k \left( e \right) = \left\lvert L_k \left( e \right) \right\rvert = \left\lvert \left\lbrace e^{\prime} \prec e : n \left[ e, e^{\prime} \right] = k \right\rbrace \right\rvert,
\end{equation}
as follows:

\begin{equation}
S^{\left( 4 \right)} = \frac{4}{\sqrt{6}} \left[ 1 - N_0 \left( e \right) + 9 N_1 \left( e \right) - 16 N_2 \left( e \right) + 8 N_3 \left( e \right) \right].
\end{equation}
This yields the following explicit form of the discrete \textit{Benincasa-Dowker action}:

\begin{equation}
S^{\left( 4 \right)} \left( \mathcal{C} \right) = \frac{4}{\sqrt{6}} \left[ n - N_0 \left( \mathcal{C} \right) + 9 N_1 \left( \mathcal{C} \right) - 16 N_2 \left( \mathcal{C} \right) + 8 N_3 \left( \mathcal{C} \right) \right],
\end{equation}
in which ${N_k \left( \mathcal{C} \right)}$ denotes the number of discrete spacetime order intervals containing $k$ elements in the causal set ${\mathcal{C}}$, and $n$ denotes the cardinality of the causal set ${n = \left\lvert \mathcal{C} \right\rvert}$. As argued by Sorkin\cite{sorkin3}, Benincasa and Dowker\cite{benincasa}, and shown numerically by Cunningham\cite{cunningham} for certain restricted cases, under the assumption that the contribution to the integral from the ${\mathcal{W}_2}$ region down the light cone vanishes, the discrete action ${S^{\left( 4 \right)}}$ limits (assuming zero surface terms) to the continuum Einstein-Hilbert action for the Lorentzian manifold ${\left( \mathcal{M}, g \right)}$ in the limit of infinite sprinkling density:

\begin{equation}
\lim_{\rho_c \to \infty} \left[ \hbar \frac{l^2}{l_{p}^{2}} \left\langle \hat{S}^{\left( 4 \right)} \left( \mathcal{C} \right) \right\rangle \right] = S_{EH} \left( g \right),
\end{equation}
where ${l_{p}}$ denotes the Planck length, ${\hbar}$ denotes the reduced Planck constant, and we are assuming that the intermediate length scale ${l_k}$ is chosen such that:

\begin{equation}
l_k \gg \left( l^2 L \right)^{\frac{1}{3}},
\end{equation}
where length $L$ is on the order of the Hubble scale. In other words, the assumption that the classical action is approximately local is only valid in the regime in which the Ricci curvature is approximately constant over length scales on the order of ${\left( l^2 L \right)^{\frac{1}{3}}}$.

On the other hand, the approach adopted thus far in the analysis of the limiting geometry of hypergraphs and causal networks from Wolfram model evolution has been much more direct\cite{gorard}, making use of the standard geometrical intuition for the Ricci scalar curvature $R$ (the simplest curvature invariant on Riemannian manifolds ${\left( \mathcal{M}, g \right)}$ of dimension $d$). As alluded to above, the Ricci curvature quantifies the discrepancy between the volume of an infinitesimal geodesic ball in ${\mathcal{M}}$ and the volume of a ball of the same radius in ordinary (flat) Euclidean space ${\mathbb{R}^d}$\cite{ricci}:

\begin{equation}
\frac{\mathrm{Vol} \left( B_{\epsilon} \left( p \right) \subset \mathcal{M} \right)}{\mathrm{Vol} \left( B_{\epsilon} \left( 0 \right) \subset \mathbb{R}^d \right)} = 1 - \frac{R}{6 \left( d + 2 \right)} \epsilon^2 + O \left( \epsilon^4 \right),
\end{equation}
in the limit as ${\epsilon \to 0}$, where ${B_{\epsilon} \left( p \right) \subset \mathcal{M}}$ designates a ball of radius ${\epsilon}$ in the (curved) Riemannian manifold ${\mathcal{M}}$, and ${B_{\epsilon} \left( 0 \right) \subset \mathbb{R}^d}$ designates a ball of radius ${\epsilon}$ in (flat) Euclidean space ${\mathbb{R}^d}$. Alternatively, one can reformulate this as a mathematically equivalent definition in terms of the discrepancy between the average distance $W$ between points on the ball ${B_{\epsilon} \left( p \right) \subset \mathcal{M}}$ and the corresponding points on the ball ${B_{\epsilon} \left( q \right) \subset \mathcal{M}}$ after parallel transport, as compared to the distance ${\delta}$ between the centers of the two balls:

\begin{equation}
W \left( B_{\epsilon} \left( p \right), B_{\epsilon} \left( q \right) \right) = \delta \left( 1 - \frac{\epsilon^2}{2 \left( d + 2 \right)} R + O \left( \epsilon^3 + \epsilon^2 \delta \right) \right),
\end{equation}
in the limit as ${\epsilon, \delta \to 0}$. In the above, ${\delta = d \left( p, q \right)}$ denotes the distance between the two centers $p$ and $q$. As described by Forman\cite{forman}, Ollivier\cite{ollivier}\cite{ollivier2}\cite{ollivier3}, Eidi and Jost\cite{eidi} and others, it is possible to extend this definition beyond Riemannian manifolds to arbitrary metric-measure spaces ${\left( X, d \right)}$ (including directed hypergraphs), in the latter cases by generalizing the notion of a Riemannian volume measure to a probability measure, and generalizing the average distance between two balls after parallel transport to the \textit{Wasserstein} (transportation) distance between the corresponding measures, as follows.

\begin{definition}
For a Polish metric space ${\left( X, d \right)}$, equipped with a Borel ${\sigma}$-algebra, a ``random walk'' on $X$, denoted $m$, corresponds to a family of probability measures:

\begin{equation}
m = \left\lbrace m_x : x \in X \right\rbrace,
\end{equation}
such that each ${m_x}$ satisfies the requirements of having a finite first moment, and the map ${x \to m_x}$ being measurable.
\end{definition}
If one now considers a set ${\Pi \left( m_x, m_y \right)}$ designating all possible \textit{transportations} of the measure ${m_x}$ to the measure ${m_y}$, then the (1-)Wasserstein distance computes the minimal cost (in terms of \textit{disassembling} the measure at $x$, \textit{transporting} it to $y$, and then \textit{reassembling} it at $y$) of this procedure:

\begin{definition}
The ``1-Wasserstein distance'', denoted ${W_1 \left( m_x, m_y \right)}$, between two probability measures ${m_x}$ and ${m_y}$ on the metric space $X$, designates the optimal transportation distance between those measures:

\begin{equation}
W_1 \left( m_x, m_y \right) = \inf_{\epsilon \in \Pi \left( m_x, m_y \right)} \left[ \int_{\left( x, y \right) \in X \times Y} d \left( x, y \right) d \epsilon \left( x, y \right) \right],
\end{equation}
where ${\Pi \left( m_x, m_y \right)}$ denotes the set of measures on the product space ${X \times Y}$ projecting onto ${m_x}$ and ${m_y}$ (i.e. the coupling between random walks projecting to ${m_x}$ and ${m_y}$).
\end{definition}
We can thus define the \textit{Ollivier-Ricci} scalar curvature directly in terms of the discrepancy between the Wasserstein distance between ${m_x}$ and ${m_y}$ and the ordinary metric distance between $x$ and $y$:

\begin{definition}
For a metric space ${\left( X, d \right)}$, equipped with a random walk $m$, the ``Ollivier-Ricci scalar curvature'' ${\kappa}$ in the direction ${\left( x, y \right)}$, assuming distinct points ${x, y \in X}$, is given by:

\begin{equation}
\kappa \left( x, y \right) = 1 - \frac{W_1 \left( m_x, m_y \right)}{d \left( x, y \right)}.
\end{equation}
\end{definition}
Clearly, in the special case where ${\left( X, d \right)}$ is a Riemannian manifold ${\left( \mathcal{M}, g \right)}$, such that the measure $m$ becomes the standard Riemannian volume measure ${\mathrm{Vol}}$, the Ollivier-Ricci scalar ${\kappa \left( p, q \right)}$ reduces to the ordinary Riemannian Ricci scalar ${R \left( p, q \right)}$, providing formal justification for our aforementioned intuition that ${m_p}$ and ${m_q}$ may be considered to be the \textit{volumes} of generalized \textit{balls} centered at points $p$ and $q$.

Let us now consider the first of our two important special cases, namely the case in which ${\left( X, d \right)}$ is a \textit{discrete} metric-measure space:

\begin{definition}
The ``discrete 1-Wasserstein distance'', denoted ${W_1 \left( m_x, m_y \right)}$, between two discrete probability measures ${m_x}$ and ${m_y}$, on the discrete metric space $X$, designates the discrete (otherwise known as multi-marginal) optimal transportation distance between those measures:

\begin{equation}
W_1 \left( m_x, m_y \right) = \inf_{\mu_{x, y} \in \Pi \left( m_x, m_y \right)} \left[ \sum_{\left( x^{\prime}, y^{\prime} \right) \in X \times X} d \left( x^{\prime}, y^{\prime} \right) \mu_{x, y} \left( x^{\prime}, y^{\prime} \right) \right].
\end{equation}
\end{definition}
Within this discrete set up, ${\Pi \left( m_x, m_y \right)}$ now denotes the set of all discrete probability measures ${\mu_{x, y}}$ satisfying the following pair of coupling conditions:

\begin{equation}
\sum_{y^{\prime} \in X} \mu_{x, y} \left( x^{\prime}, y^{\prime} \right) = m_x \left( x^{\prime} \right), \qquad \text{ and } \qquad \sum_{x^{\prime} \in X} \mu_{x, y} \left( x^{\prime}, y^{\prime} \right) = m_y \left( y^{\prime} \right).
\end{equation}

We can specialize further by considering the case of a directed hypergraph ${H \left( V, E \right)}$, in which each hyperedge ${e \in E}$ is assumed (for the sake of simplicity) to designate a directional relation between two sets of vertices, namely $A$ and $B$ (the \textit{tail set} and the \textit{head set}, respectively):

\begin{definition}
The ``discrete 1-Wasserstein distance'', denoted ${W_1 \left( \mu_{A^{in}}, \mu_{B^{out}} \right)}$, between two discrete probability measures ${\mu_{A^{in}}}$ and ${\mu_{B^{out}}}$ defined over a directed hypergraph ${H = \left( V, E \right)}$, designates the discrete (otherwise known as multi-marginal) optimal transportation distance between those measures:

\begin{equation}
W_1 \left( \mu_{A^{in}}, \mu_{B^{out}} \right) = \min_{u \in A^{in} \left( u \to A \right), v \in B^{out} \left( B \to v \right)} \left[\sum_{u \to A} \sum_{B \to v} d \left( u, v \right) \epsilon \left( u, v \right) \right],
\end{equation}
where ${d \left( u, v \right)}$ denotes the the minimum number of directed hyperedges that must be traversed when traveling from vertex ${u \in A^{in} \left( u \to A \right)}$ to vertex ${v \in B^{out} \left( B \to v \right)}$, and ${\epsilon \left( u, v \right)}$ denotes the coupling, i.e. the total ``mass'' being moved from vertex $u$ to vertex $v$.
\end{definition}
In the above, one is minimizing over the set of all couplings ${\epsilon}$ between measures ${\mu_{A^{in}}}$ and ${\mu_{B^{out}}}$ that satisfy the following pair of coupling constraints:

\begin{equation}
\sum_{u \to A} \epsilon \left( u, v \right) = \sum_{j = 1}^{m} \mu_{y_j} \left( v \right), \qquad \text{ and } \qquad \sum_{B \to v} \epsilon \left( u, v \right) = \sum_{i = 1}^{n} \mu_{x_i} \left( u \right),
\end{equation}
where, for a given vertex in the tail set ${x_i \in A}$, the number of incoming hyperedges to ${x_i}$, denoted ${d x_{i}^{in}}$, is the number of hyperedges that include ${x_i}$ as an element of their head set, and for a given vertex in the head set ${y_j \in B}$, the number of outgoing hyperedges from ${y_j}$, denoted ${d y_{j}^{out}}$, is the number of hyperedges that include ${y_j}$ as an element of their tail set. Henceforth, we shall assume (as previously) the simplest possible choice of combinatorial distance metric ${d \left( u, v \right)}$ and coupling constants ${\epsilon \left( u, v \right)}$, such that each hyperedge corresponds to a single unit of spatial distance, thus forcing the induced metric connection to be torsion-free (i.e. Levi-Civita). However, note that by considering more general choices of coupling (in particular, ones in which the hyperedge connecting vertices $u$ and $v$ does not necessarily have the same weight as the hyperedge connecting vertices $v$ and $u$), it may be possible to construct more general choices of connection, such as the metric-independent affine connections considered in the context of Riemann-Cartan geometry.

\begin{figure}[ht]
\centering
\includegraphics[width=0.395\textwidth]{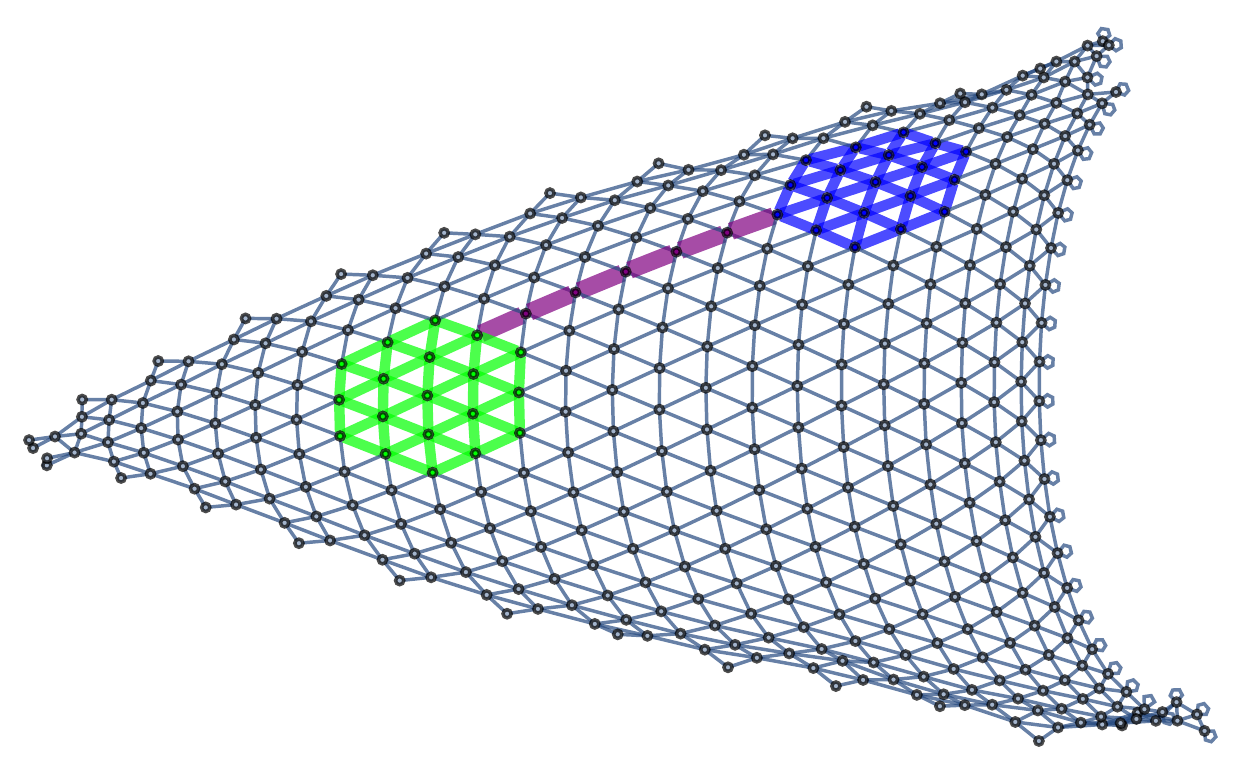}\hspace{0.1\textwidth}
\includegraphics[width=0.395\textwidth]{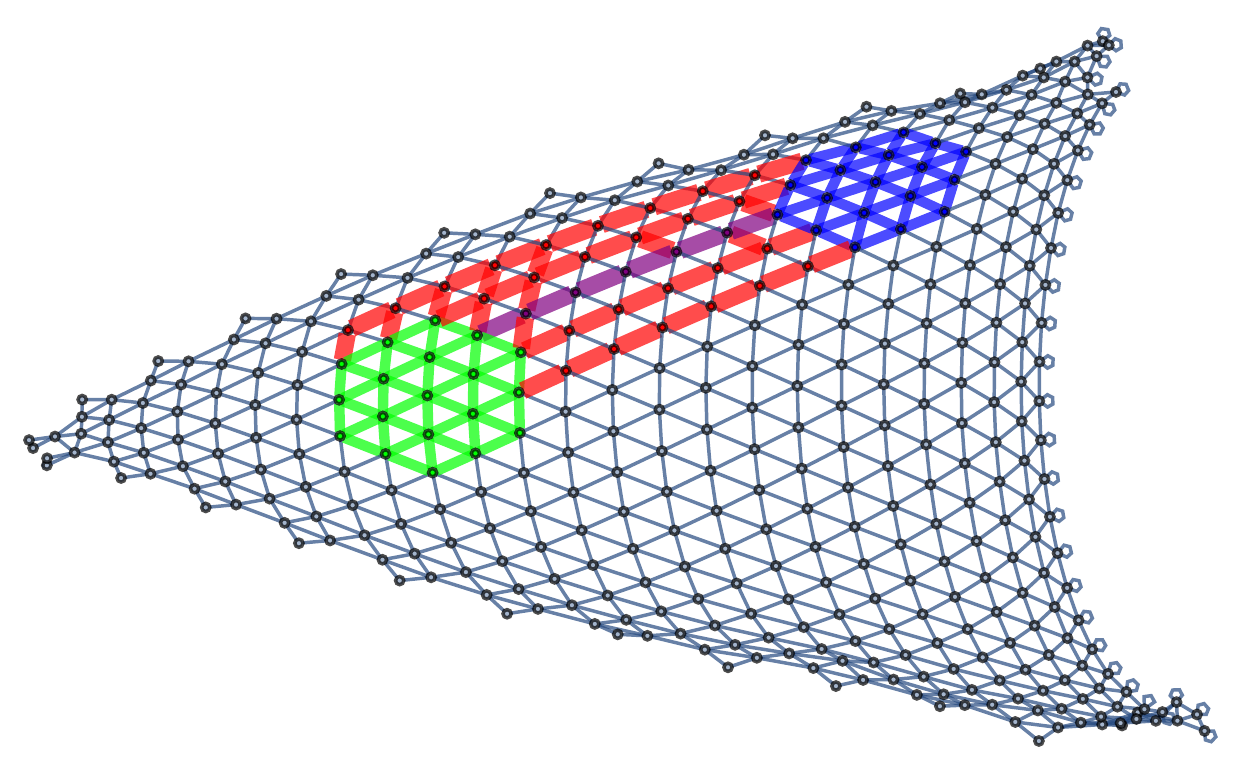}
\caption{On the left, a pair of nearby finite geodesic balls in an asymptotically-flat spatial hypergraph with a two-dimensional Riemannian manifold-like limiting structure, as generated by the set substitution system ${\left\lbrace \left\lbrace x, y, y \right\rbrace, \left\lbrace z, x, w \right\rbrace \right\rbrace \to \left\lbrace \left\lbrace y, v, y \right\rbrace, \left\lbrace y, z, v \right\rbrace, \left\lbrace w, v, v \right\rbrace \right\rbrace}$, with a purple path showing the distance between the centers. On the right, the family of all paths (red) between corresponding points on the surfaces of the two balls (after parallel transport). Since there is no net divergence or convergence of the red geodesics, it follows that the discrete Ollivier-Ricci scalar curvature ${\kappa}$ is zero along the purple path.}
\label{fig:Figure51}
\end{figure}

\begin{figure}[ht]
\centering
\includegraphics[width=0.395\textwidth]{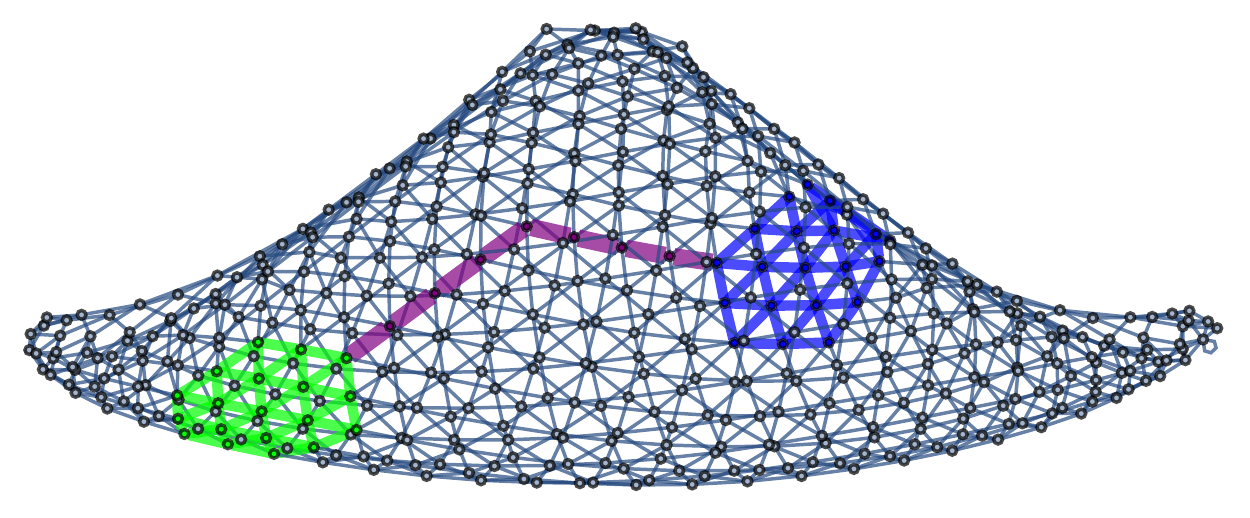}\hspace{0.1\textwidth}
\includegraphics[width=0.395\textwidth]{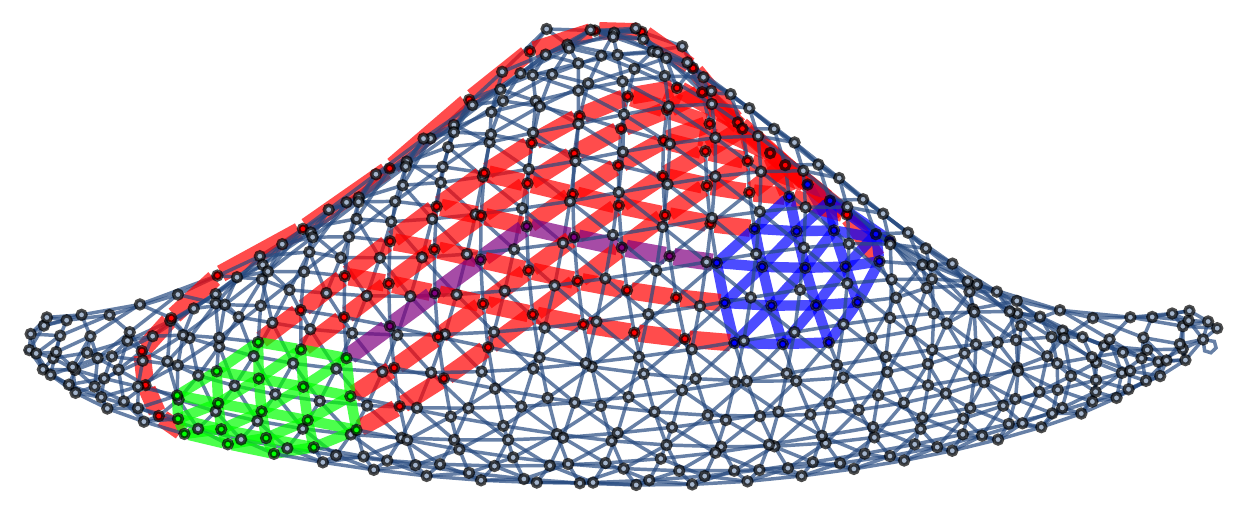}
\caption{On the left, a pair of nearby finite geodesic balls in a spatial hypergraph with a two-dimensional Riemannian manifold-like limiting structure with positive global curvature, as generated by the set substitution system ${\left\lbrace \left\lbrace x, x, y \right\rbrace, \left\lbrace x, z, w \right\rbrace \right\rbrace \to \left\lbrace \left\lbrace w, w, v \right\rbrace, \left\lbrace v, w, y \right\rbrace, \left\lbrace z, y, v \right\rbrace \right\rbrace}$, with a purple path showing the distance between the centers. On the right, the family of all paths (red) between corresponding points on the surfaces of the two balls (after parallel transport). Since there is a net divergence of the red geodesics, it follows that the discrete Ollivier-Ricci scalar curvature ${\kappa}$ is positive along the purple path.}
\label{fig:Figure52}
\end{figure}

\begin{figure}[ht]
\centering
\includegraphics[width=0.395\textwidth]{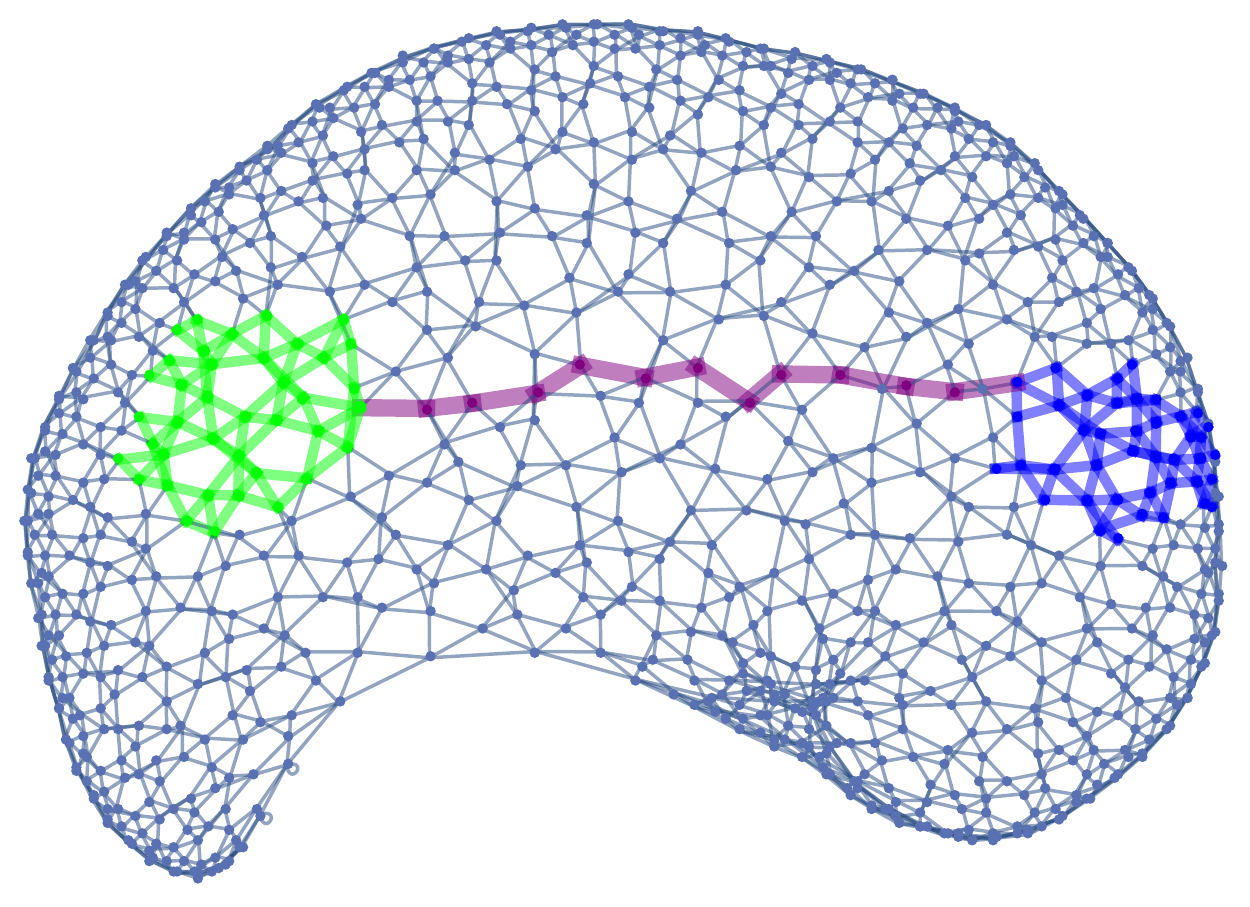}\hspace{0.1\textwidth}
\includegraphics[width=0.395\textwidth]{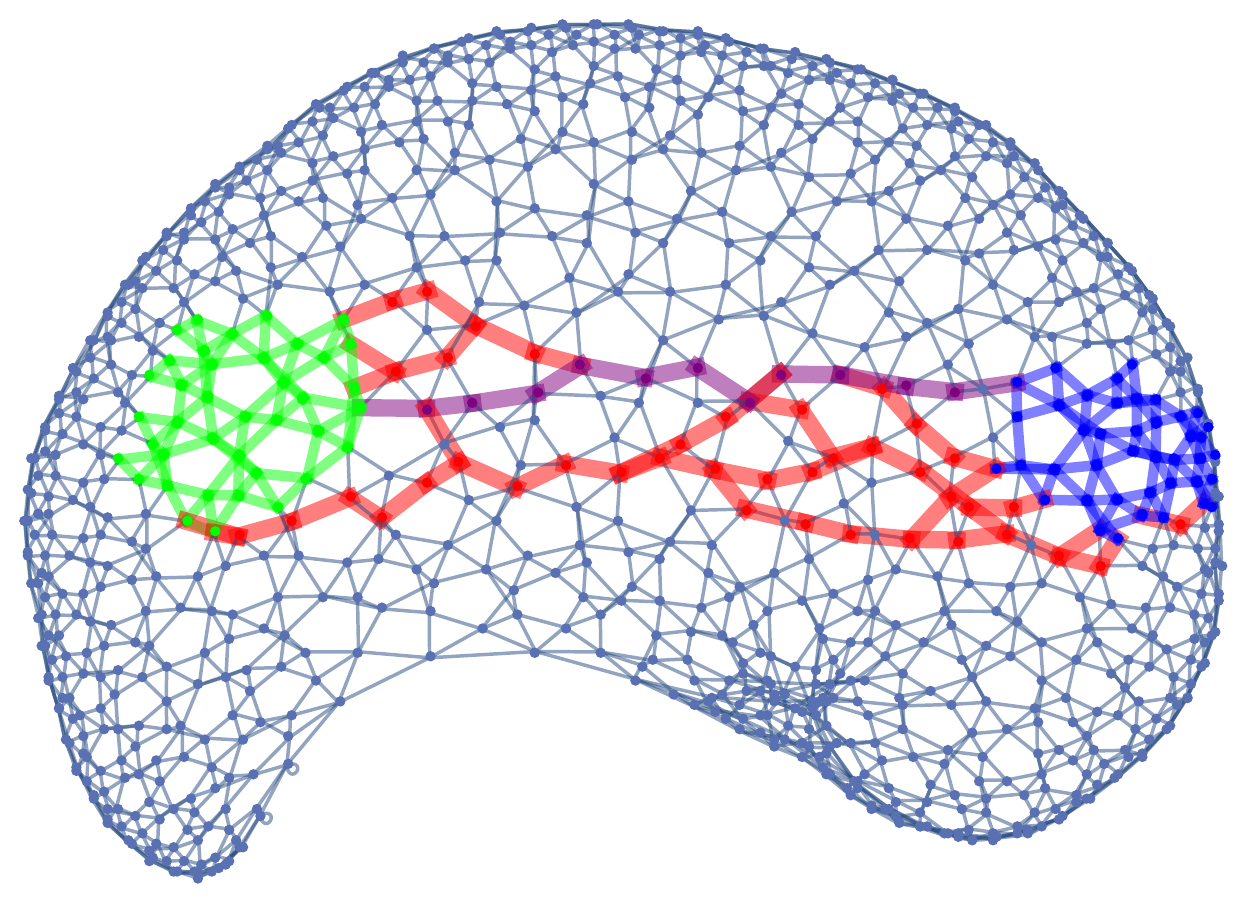}
\caption{On the left, a pair of nearby finite geodesic balls in a spatial hypergraph with a two-dimensional Riemann manifold-like limiting structure with negative global curvature, as generated by the set substitution system ${\left\lbrace \left\lbrace x, y, x \right\rbrace, \left\lbrace x, z, u \right\rbrace \right\rbrace \to \left\lbrace \left\lbrace u, v, u \right\rbrace, \left\lbrace v, u, z \right\rbrace, \left\lbrace x, y, v \right\rbrace \right\rbrace}$, with a purple path showing the distance between the centers. On the right, the family of all paths (red) between corresponding points on the surfaces of the two balls (after parallel transport). Since there is a net convergence of the red geodesics, it follows that the discrete Ollivier-Ricci scalar curvature ${\kappa}$ is negative along the purple path.}
\label{fig:Figure53}
\end{figure}

Explicitly, we therefore have the following definition of Ollivier-Ricci curvature in directed hypergraphs:

\begin{definition}
For a directed hypergraph ${H = \left( V, E \right)}$, the ``Ollivier-Ricci scalar curvature'' ${\kappa}$ of a single directed hyperedge ${e \in E}$, where:

\begin{equation}
A = \left\lbrace x_1, \dots, x_n \right\rbrace \to^{e} B = \left\lbrace y_1, \dots, y_m \right\rbrace,
\end{equation}
and where ${n, m \leq \left\lvert V \right\rvert}$, is:

\begin{equation}
\kappa \left( e \right) = 1 - W_1 \left( \mu_{A^{in}}, \mu_{B^{out}} \right),
\end{equation}
with probability measures ${\mu_{A^{in}}}$ and ${\mu_{B^{out}}}$ satisfying the coupling constraints:

\begin{equation}
\mu_{A^{in}} = \sum_{i = 1}^{n} \mu_{x_i}, \qquad \text{ and } \qquad \mu_{B^{out}} = \sum_{j = 1}^{m} \mu_{y_j}.
\end{equation}
\end{definition}
These probability measures can furthermore be written in a more explicit form as:

\begin{equation}
\forall 1 \leq i \leq n, z \in V, \qquad \mu_{x_i} \left( z \right) = \begin{cases}
0, \qquad & \text{ if } z = x_i \text{ and } d_{x_{i}^{in}} \neq 0,\\
\frac{1}{n}, \qquad & \text{ if } z = x_i \text{ and } d_{x_{i}^{in}} = 0,\\
\sum\limits_{e^{\prime} : z \to x_i} \frac{1}{n \times d_{x_{i}^{in}} \times \left\lvert \mathrm{tail} \left( e^{\prime} \right) \right\rvert}, \qquad & \text{ if } z \neq x_i \text{ and } \exists e^{\prime} : z \to x_i,\\
0, \qquad & \text{ if } z \neq x_i \text{ and } \nexists e^{\prime} : z \to x_i,
\end{cases}
\end{equation}
and:

\begin{equation}
\forall 1 \leq j \leq m, z \in V, \qquad \mu_{y_j} \left( z \right) = \begin{cases}
0, \qquad & \text{ if } z = y_j \text{ and } d_{y_{j}^{out}} \neq 0,\\
\frac{1}{m}, \qquad & \text{ if } z = y_j \text{ and } d_{y_{j}^{out}} = 0,\\
\sum\limits_{e^{\prime} : y_j \to z} \frac{1}{m \times d_{y_{j}^{out}} \times \left\lvert \mathrm{head} \left( e^{\prime} \right) \right\rvert}, \qquad & \text{ if } z \neq y_j \text{ and } \exists e^{\prime} : y_j \to z,\\
0, \qquad & \text{ if } z \neq y_j \text{ and } \nexists e^{\prime} : y_j \to z.
\end{cases}
\end{equation}
Examples of the computation of the discrete Ollivier-Ricci scalar curvature in hypergraphs which limit to asymptotically-flat, asymptotically positively curved and asymptotically negatively curved Riemannian manifold-like structures are shown in Figures \ref{fig:Figure51}, \ref{fig:Figure52} and \ref{fig:Figure53}, respectively.

However, in (continuum) Riemannian geometry, we know that the Ricci scalar curvature $R$ is itself merely the trace of the Ricci curvature tensor ${R_{a b}}$:

\begin{equation}
R = \mathrm{Tr} \left( R_{a b} \right) = g^{a b} R_{a b} = R_{a}^{a},
\end{equation}
assuming metric tensor ${g_{a b}}$. The standard geometrical intuition underlying the curvature tensor ${R \left( \boldsymbol\xi, \boldsymbol\xi \right)}$ at point $p$ is that it quantifies the discrepancy between the volume of a conical region in ${\mathcal{M}}$ oriented in the direction of the vector ${\boldsymbol\xi}$, consisting entirely of geodesic segments of length ${\epsilon}$ emanating from a single point ${p \in \mathcal{M}}$, and the volume of the corresponding region in ordinary (flat) Euclidean space ${\mathbb{R}^d}$. The Ricci curvature tensor is, in turn, merely a contraction of the full Riemann curvature tensor ${R_{a b c d}}$ between the first and third indices:

\begin{equation}
R_{a b} = R_{a c b}^{c},
\end{equation}
with all other contractions yielding either ${\pm R_{a b}}$ or zero, by symmetry. The intuition behind the full Riemann curvature tensor ${R_{b c d}^{a}}$ is that it quantifies the discrepancy between the direction of a vector ${\mathbf{v}}$ before and after parallel transport around a closed curve in the manifold ${\left( \mathcal{M}, g \right)}$:

\begin{equation}
\delta v^a = R_{b c d}^{a} d x^c d x^d v^b,
\end{equation}
in other words ${R_{b c d}^{a}}$ quantifies the degree to which (a local coordinate patch of) the manifold ${\left( \mathcal{M}, g \right)}$ fails to be \textit{holonomic}, and hence fails to preserve geometrical data as it is parallel-transported around a closed curve. Equivalently, one can think of ${R_{b c d}^{a}}$ in more algebraic terms as being the commutator of the covariant derivative operator, acting on an arbitrary vector:

\begin{equation}
\boldsymbol\nabla_c \boldsymbol\nabla_d v^a - \boldsymbol\nabla_d \boldsymbol\nabla_c v^a = R_{b c d}^{a} v^b.
\end{equation}

In order to be able to define notions such as parallel transport, holonomy, etc. in a rigorous fashion, it is first necessary to define a connection; as previously mentioned, the natural choice (both in the case of directed hypergraphs, and more broadly within general relativity) is the Levi-Civita/torsion-free/metric connection, whose components in a given basis are defined by the Christoffel symbols:

\begin{equation}
\Gamma_{\mu \nu}^{\rho} = \frac{1}{2} g^{\rho \sigma} \left( \partial_{\mu} g_{\sigma \nu} + \partial_{\nu} g_{\mu \sigma} - \partial_{\sigma} g_{\mu \nu} \right),
\end{equation}
i.e. by a simple linear combination of first derivatives of the metric tensor. Intuitively, the ${\Gamma_{\mu \nu}^{\rho}}$ are \textit{pseudo-tensors} that describe how the basis vectors change from point to point along ${\left( \mathcal{M}, g \right)}$. In the Levi-Civita connection, we can write out the Riemann tensor in the following explicit form as a linear combination of second-derivatives of the metric tensor:

\begin{equation}
R_{b c d}^{a} = \partial_c \Gamma_{b d}^{a} - \partial_d \Gamma_{b c}^{a} + \Gamma_{e c}^{a} \Gamma_{b d}^{e} - \Gamma_{e d}^{a} \Gamma_{b c}^{e}.
\end{equation}
In any locally flat (i.e. locally inertial) reference frame ${\mathcal{F}}$, the Christoffel symbols, although not their derivatives, all vanish, such that these components of the Riemann tensor can all be written in lower-index form as:

\begin{equation}
R_{a b c d} = \frac{1}{2} \left( \partial_b \partial_c g_{a d} - \partial_b \partial_d g_{a c} + \partial_a \partial_d g_{b c} - \partial_a \partial_c g_{b d} \right).
\end{equation}

In such a frame ${\mathcal{F}}$, if we now choose to use geodesic normal coordinates around a point ${p \in \mathcal{M}}$, then the metric tensor becomes approximately Euclidean:

\begin{equation}
g_{i j} = \delta_{i j} + O \left( \left\lVert \mathbf{x} \right\rVert^2 \right),
\end{equation}
with ${\delta_{i j}}$ denoting the standard Kronecker delta function, corresponding to the Euclidean metric (i.e. one finds that the geodesic distance is approximately the Euclidean distance). Next, we can perform a Taylor expansion of the metric tensor along a radial geodesic in normal coordinates to find:

\begin{equation}
g_{i j} = \delta_{i j} - \frac{1}{3} R_{i j k l} x^k x^l + O \left( \left\lVert \mathbf{x} \right\rVert^2 \right),
\end{equation}
indicating that the correction factor is proportional to a projection of the Riemann curvature tensor ${R_{i j k l}}$. In the above, we are Taylor expanding with respect to the tangent space of a given geodesic, ${\boldsymbol\gamma_0}$, in the space of all possible geodesics (described as a one-parameter family of geodesics ${\boldsymbol\gamma_{\tau}}$), i.e. we are performing the expansion in terms of Jacobi fields of the form:

\begin{equation}
\mathbf{J} \left( t \right) = \left. \left( \frac{\partial \boldsymbol\gamma_{\tau} \left( t \right)}{\partial \tau} \right) \right\rvert_{\tau = 0}.
\end{equation}
Finally, by expanding the square root of the determinant of the metric tensor, we obtain the previously described relationship between the metric volume element ${d \mu_g}$ of the manifold ${\left( \mathcal{M}, g \right)}$ and the volume element ${d \mu_{Euclidean}}$ of flat (Euclidean) space:

\begin{equation}
d \mu_g = \left[ 1 - \frac{1}{6} R_{j k} x^j x^k + O \left( \left\lVert \mathbf{x} \right\rVert^3 \right) \right] d \mu_{Euclidean},
\end{equation}
thus formally justifying the aforementioned relationship between the correction factor for the volume of an infinitesimal geodesic cone ${C_t}$ and a timelike projection of the Ricci curvature tensor:

\begin{equation}
\mathrm{Vol} \left( C_t \left( p \right) \right) = at^n \left[ 1 - \frac{1}{6} R_{i j} t^i t^j + O \left( \left\lVert \mathbf{t} \right\rVert^3 \right) \right],
\end{equation}
as required.

In order to generalize the Riemann curvature tensor, and hence the Ricci tensor, to arbitrary (potentially discrete) metric-measure spaces, we consider first the sectional curvature $K$ of a Riemannian manifold ${\left( \mathcal{M}, g \right)}$:

\begin{equation}
K \left( \mathbf{u}, \mathbf{v} \right) = \frac{\left\langle R \left( \mathbf{u}, \mathbf{v} \right) \mathbf{v}, \mathbf{u} \right\rangle}{\left\langle \mathbf{u}, \mathbf{u} \right\rangle \left\langle \mathbf{v}, \mathbf{v} \right\rangle - \left\langle \mathbf{u}, \mathbf{v} \right\rangle^2},
\end{equation}
where ${\mathbf{u}}$ and ${\mathbf{v}}$ are linearly independent tangent vectors at the point ${p \in \mathcal{M}}$, and $R$ denotes the full Riemann curvature tensor. In the particular case where ${\mathbf{u}}$ and ${\mathbf{v}}$ are orthonormal, this simply reduces to:

\begin{equation}
K \left( \mathbf{u}, \mathbf{v} \right) = \left\langle R \left( \mathbf{u}, \mathbf{v} \right) \mathbf{v}, \mathbf{u} \right\rangle.
\end{equation}
Note, in particular, that the components of the sectional curvature tensor ${K \left( \mathbf{u}, \mathbf{v} \right)}$ completely determine the components of the Riemann curvature tensor ${R \left( \mathbf{u}, \mathbf{v} \right)}$, and therefore so long as we can determine an appropriate means of generalizing the sectional curvature to arbitrary metric-measure spaces, our construction will be complete. Geometrically, ${K \left( \mathbf{u}, \mathbf{v} \right)}$ is quantifying the Gaussian curvature of a surface constructed by a particular family of geodesics emanating from point $p$, proceeding in the directions of a tangent plane ${\sigma_p}$ that is defined by the pair of tangent vectors ${\mathbf{u}}$ and ${\mathbf{v}}$.

\begin{figure}[ht]
\centering
\includegraphics[width=0.395\textwidth]{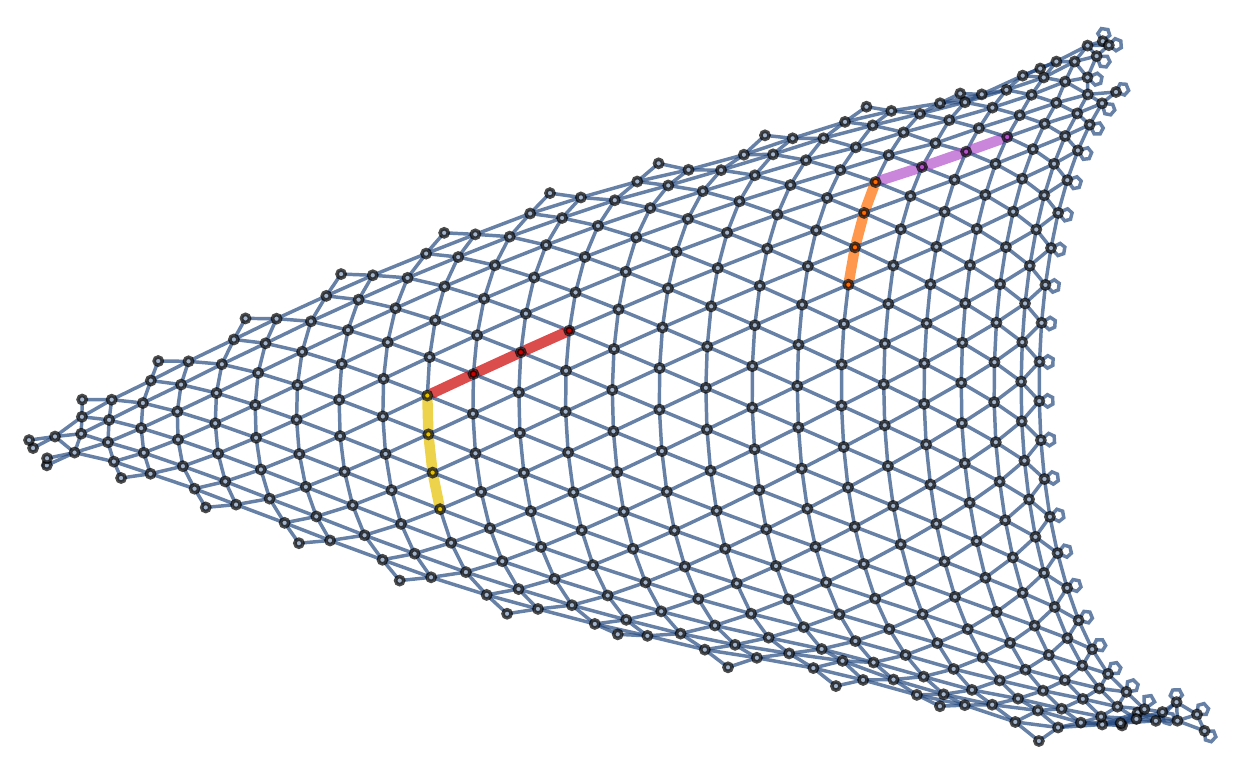}\hspace{0.1\textwidth}
\includegraphics[width=0.395\textwidth]{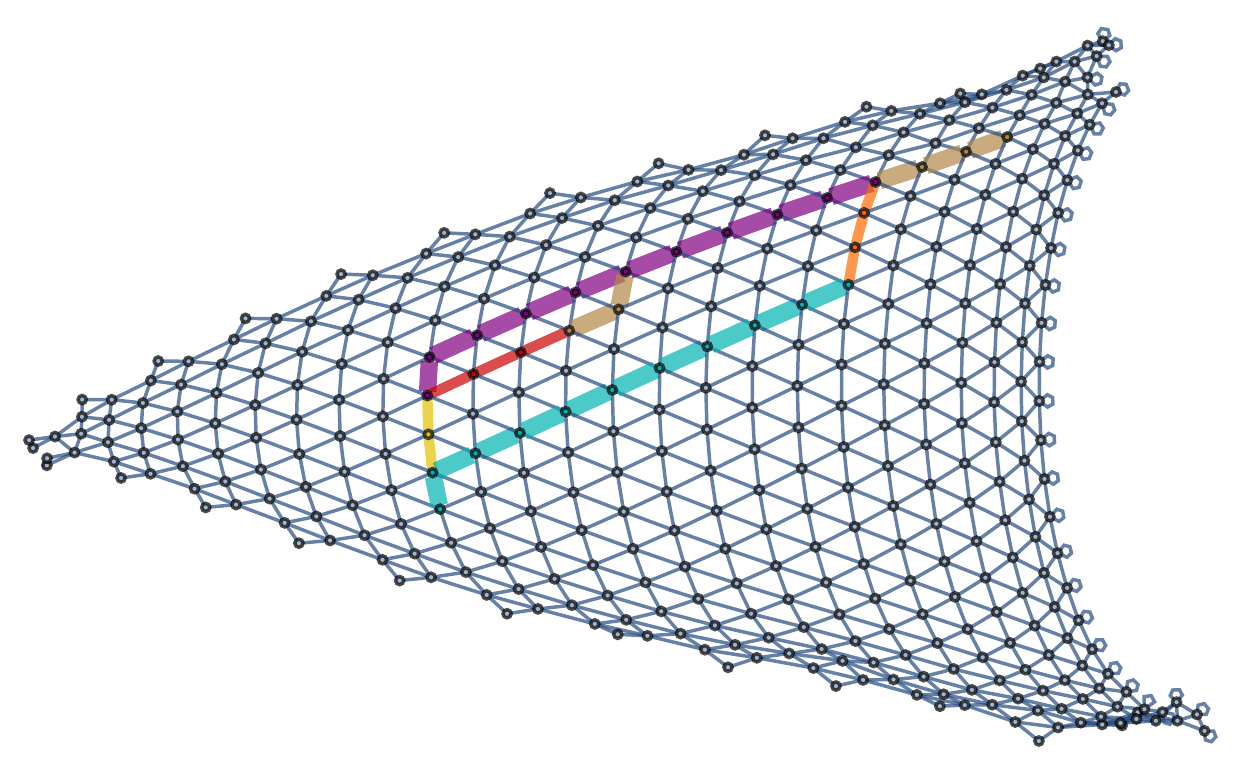}
\caption{On the left, a pair of nearby pairs of tangent vectors in an asymptotically-flat spatial hypergraph with a two-dimensional Riemannian manifold-like limiting structure, as generated by the set substitution system ${\left\lbrace \left\lbrace x, y, y \right\rbrace, \left\lbrace z, x, w \right\rbrace \right\rbrace \to \left\lbrace \left\lbrace y, v, y \right\rbrace, \left\lbrace y, z, v \right\rbrace, \left\lbrace w, v, v \right\rbrace \right\rbrace}$. On the right, a purple path shows the distance between the two origin points, with additional paths (light blue and brown) between the corresponding endpoints of the two pairs of geodesics. Since there is no net divergence or convergence of the light blue and brown geodesics, it follows that the projection of the discrete sectional curvature $K$ onto the yellow and red geodesics is equal to zero (along the purple path).}
\label{fig:Figure54}
\end{figure}

\begin{figure}[ht]
\centering
\includegraphics[width=0.395\textwidth]{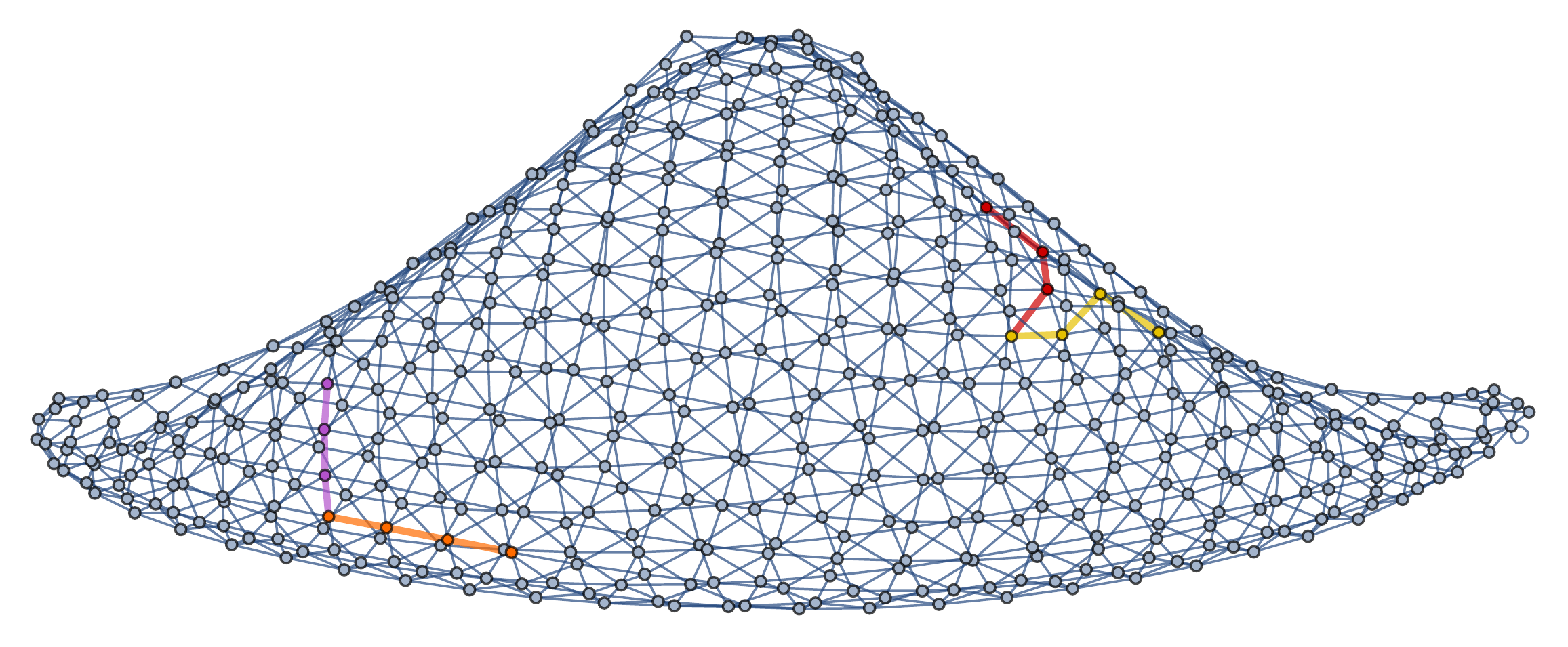}\hspace{0.1\textwidth}
\includegraphics[width=0.395\textwidth]{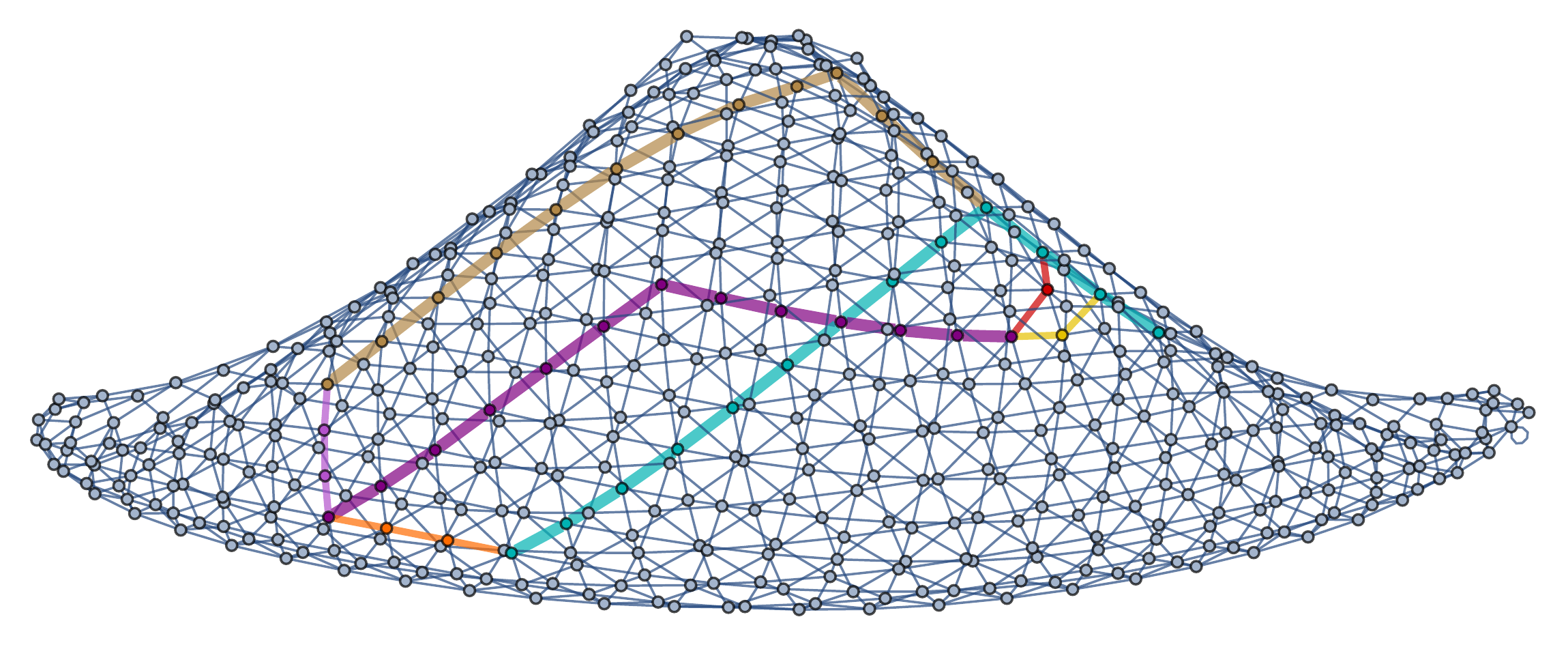}
\caption{On the left, a pair of nearby pairs of tangent vectors in a spatial hypergraph with a two-dimensional Riemannian manifold-like limiting structure with positive global curvature, as generated by the set substitution system ${\left\lbrace \left\lbrace x, x, y \right\rbrace, \left\lbrace x, z, w \right\rbrace \right\rbrace \to \left\lbrace \left\lbrace w, w, v \right\rbrace, \left\lbrace v, w, y \right\rbrace, \left\lbrace z, y, v \right\rbrace \right\rbrace}$. On the right, a purple path shows the distance between the two origin points, with additional paths (light blue and brown) between the corresponding endpoints of the two pairs of geodesics. Since there is a net divergence of the light blue and brown geodesics, it follows that the projection of the discrete sectional curvature $K$ onto the yellow and red geodesics is positive (along the purple path).}
\label{fig:Figure55}
\end{figure}

\begin{figure}[ht]
\centering
\includegraphics[width=0.395\textwidth]{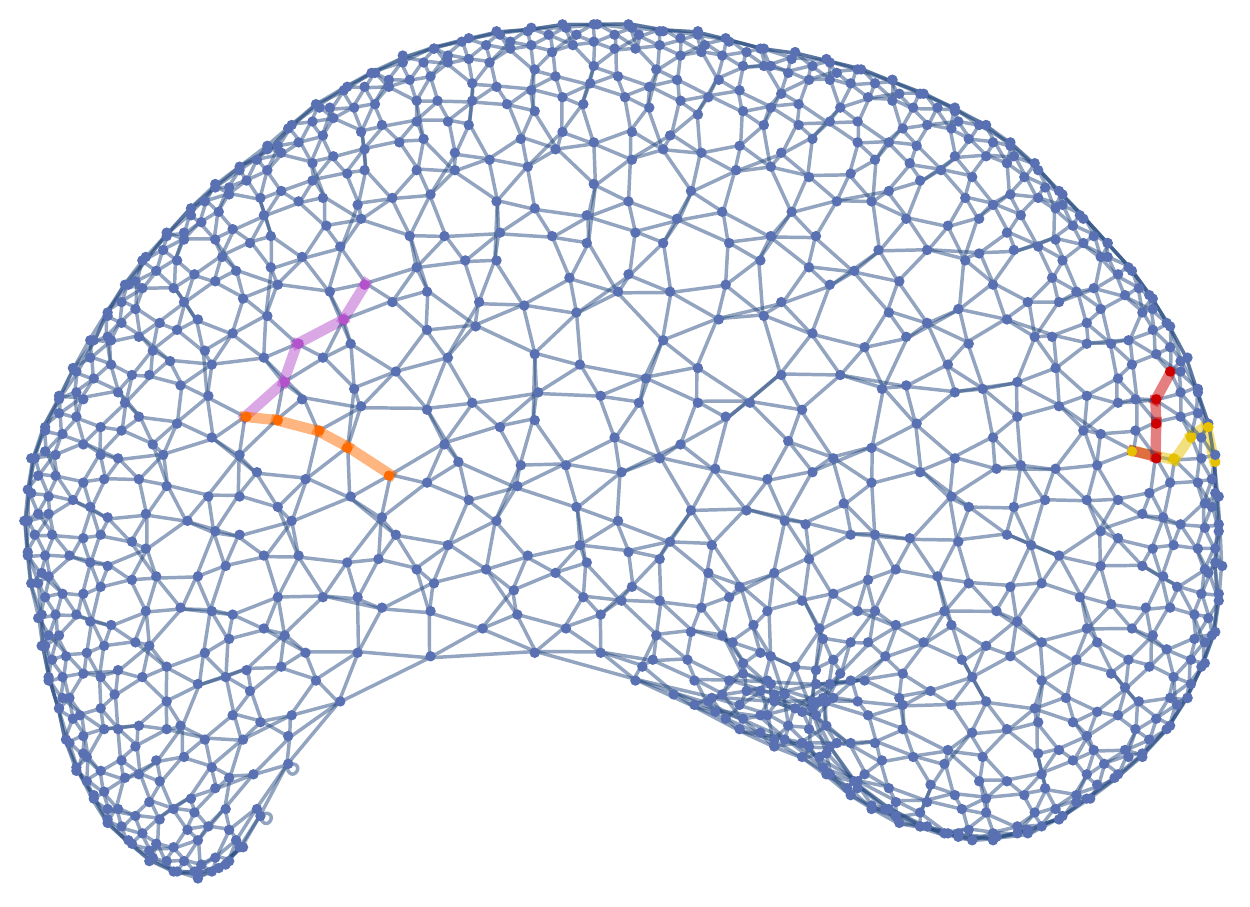}\hspace{0.1\textwidth}
\includegraphics[width=0.395\textwidth]{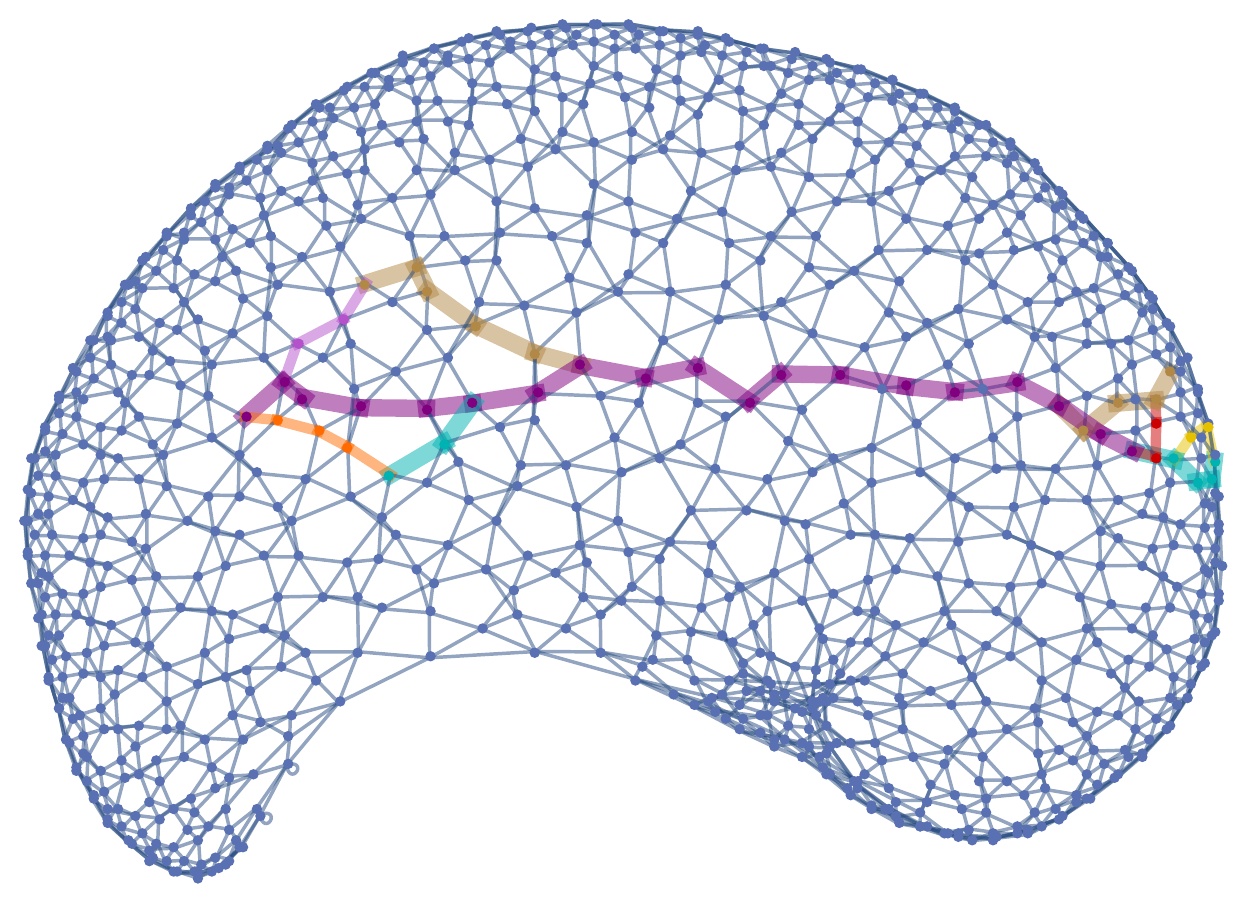}
\caption{On the left, a pair of nearby pairs of tangent vectors in a spatial hypergraph with a two-dimensional Riemannian manifold-like limiting structure with negative global curvature, as generated by the set substitution system ${\left\lbrace \left\lbrace x, y, x \right\rbrace, \left\lbrace x, z, u \right\rbrace \right\rbrace \to \left\lbrace \left\lbrace u, v, u \right\rbrace, \left\lbrace v, u, z \right\rbrace, \left\lbrace x, y, v \right\rbrace \right\rbrace}$. On the right, a purple path shows the distance between the two origin points, with additional paths (light blue and brown) between the corresponding endpoints of the two pairs of geodesics. Since there is a net convergence of the light blue and brown geodesics, it follows that the projection of the discrete sectional curvature $K$ onto the yellow and red geodesics is negative (along the purple path).}
\label{fig:Figure56}
\end{figure}

\begin{figure}[ht]
\centering
\includegraphics[width=0.395\textwidth]{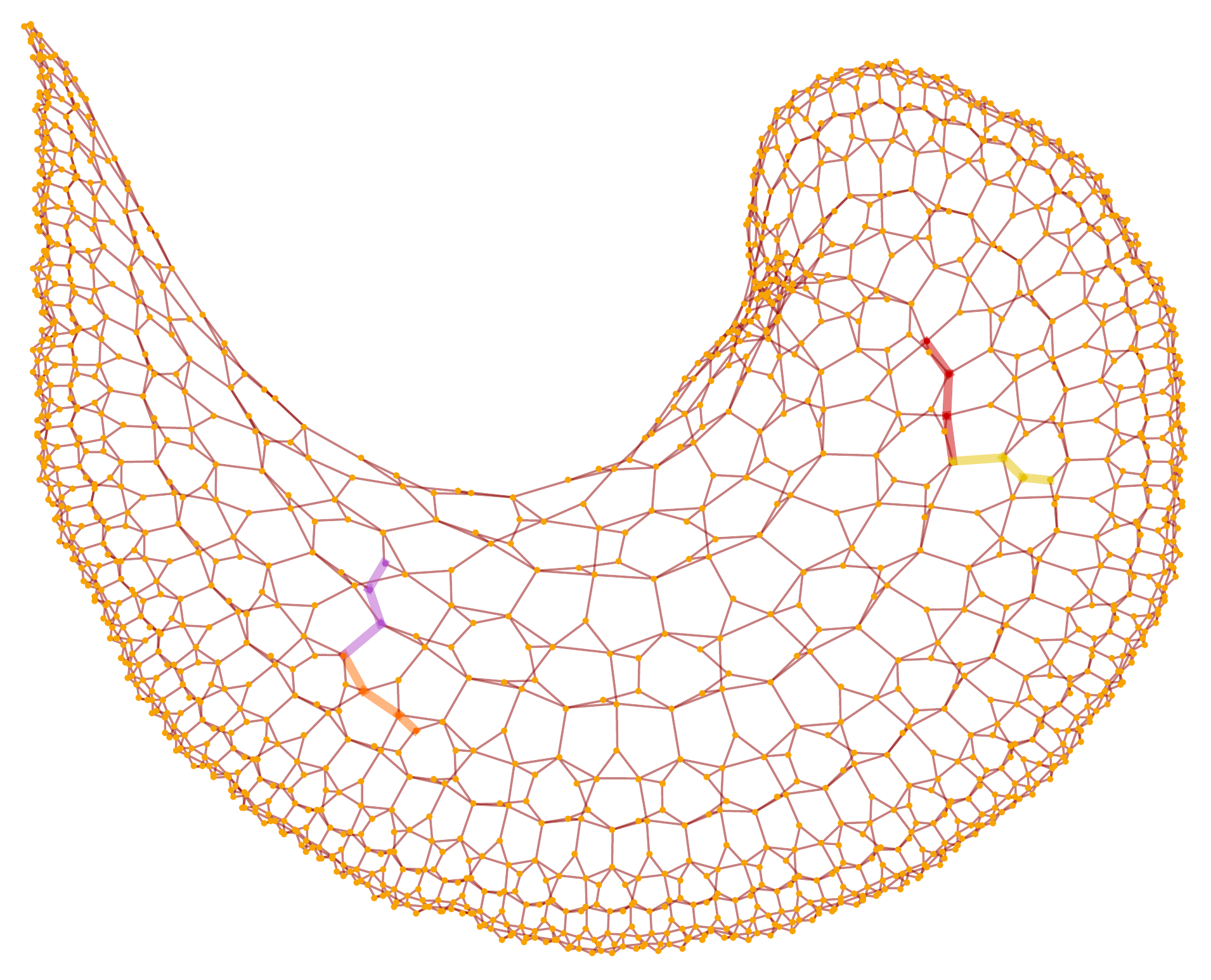}\hspace{0.1\textwidth}
\includegraphics[width=0.395\textwidth]{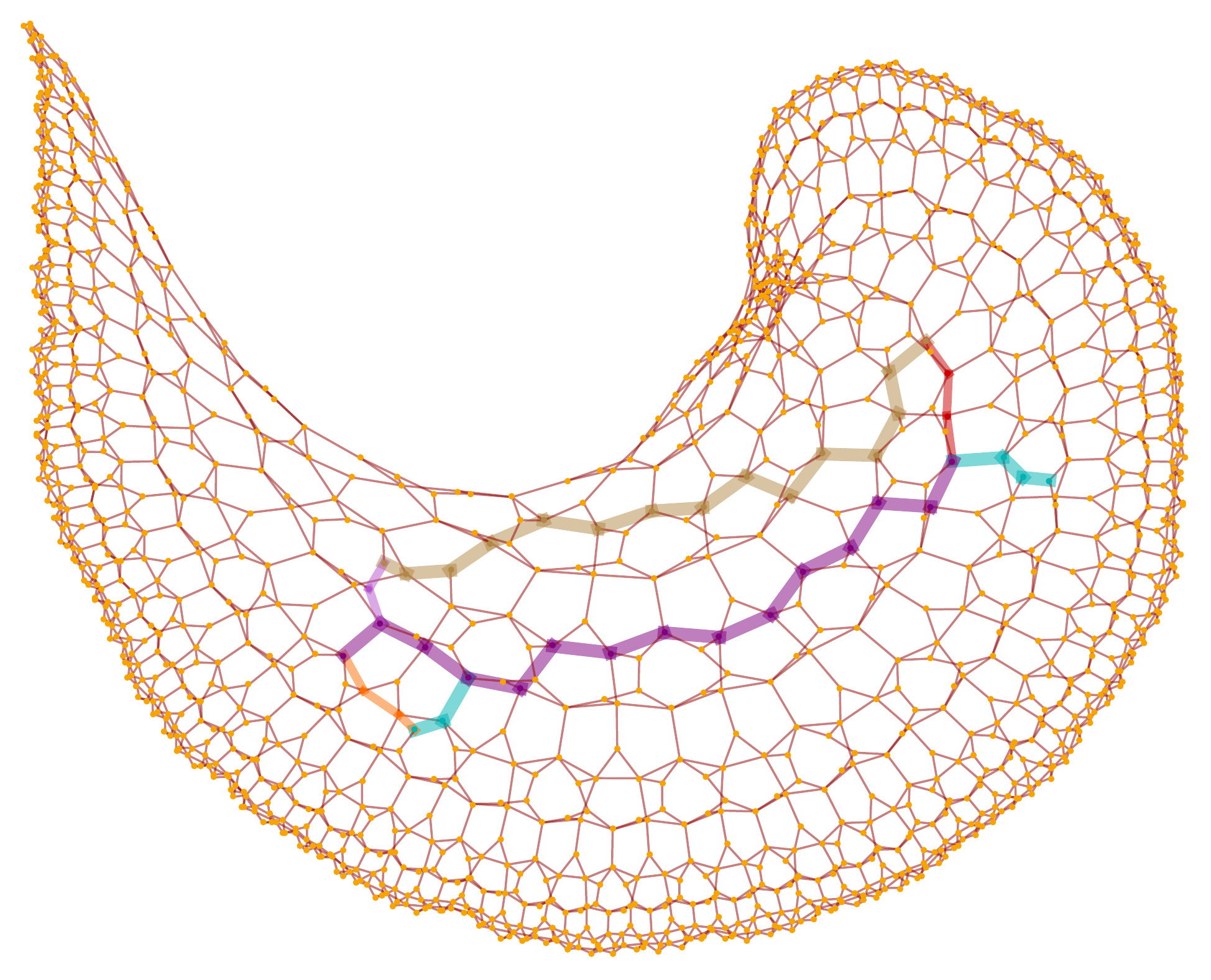}
\caption{On the left, a pair of nearby pairs of tangent vectors in a causal network with a two-dimensional Lorentzian manifold-like limiting structure with negative global curvature, as generated by the set substitution system ${\left\lbrace \left\lbrace x, y, x \right\rbrace, \left\lbrace x, z, u \right\rbrace \right\rbrace \to \left\lbrace \left\lbrace u, v, u \right\rbrace, \left\lbrace v, u, z \right\rbrace, \left\lbrace x, y, v \right\rbrace \right\rbrace}$. On the right, a purple path shows the distance between the two origin points, with additional paths (light blue and brown) between the corresponding endpoints of the two pairs of geodesics. Since there is a net convergence of the light blue and brown geodesics, it follows that the projection of the discrete sectional curvature $K$ onto the yellow and red geodesics is negative (along the purple path).}
\label{fig:Figure57}
\end{figure}

Consider now an arbitrary metric space ${\left( X, d \right)}$; the above geometrical definition of the sectional curvature may easily be extended to this more general case by considering the discrepancy between the distance between points $x$ and $y$ (where point $y$ is reached by traveling a distance ${\delta}$ away from $x$ along the tangent vector ${\mathbf{u}}$) and the average distance between the points ${\exp_x \left( \epsilon \mathbf{v}_x \right)}$ and ${\exp_y \left( \epsilon \mathbf{v}_y \right)}$ lying a distance of ${\epsilon}$ along geodesics in directions ${\mathbf{v}_x}$ and ${\mathbf{v}_y}$, respectively:

\begin{equation}
d \left( \exp_x \left( \mathbf{v}_x \right), \exp_y \left( \epsilon \mathbf{v}_y \right) \right) = \delta \left( 1 - \frac{\epsilon^2}{2} K \left( \mathbf{u}, \mathbf{v} \right) + O \left( \epsilon^3 + \epsilon^2 \delta \right) \right),
\end{equation}
in the limit as ${\epsilon, \delta \to 0}$. In the above, ${\exp_x \left( \mathbf{v} \right)}$ is used to denote the endpoint of the unit-speed geodesic ${\boldsymbol\gamma}$ whose origin point is ${x \in X}$ and whose initial direction is ${\mathbf{v}}$, and ${\delta = d \left( x, y \right)}$ denotes the usual (combinatorial) distance between points $x$ and $y$. Two discrete geodesics ${\gamma_1}$ and ${\gamma_2}$ emanating from a common point $p$ may be described as being \textit{tangent} if the shortest path from any point on ${\gamma_1}$ to any point on ${\gamma_2}$ (or vice versa) terminates at $p$. Examples of the computation of the discrete sectional curvature tensor in hypergraphs which limit to asymptotically-flat, asymptotically positively curved and asymptotically negatively curved Riemannian manifold-like structures are shown in Figures \ref{fig:Figure54}, \ref{fig:Figure55} and \ref{fig:Figure56}, respectively. As an important consistency condition, we can confirm that taking an average of ${K \left( \mathbf{u}, \mathbf{v} \right)}$ across all vectors ${\mathbf{v}}$ (which is equivalent to taking a trace) does indeed yield the previous definition of the Ollivier-Ricci scalar:

\begin{equation}
\delta \left( 1 - \frac{\epsilon^2}{2 \left( d + 2 \right)} R + O \left( \epsilon^3 + \epsilon^2 \delta \right) \right),
\end{equation}
since the sets of all tangent vectors of length ${\epsilon}$ at points ${x, y \in X}$ form finite geodesic balls ${S_x}$ and ${S_y}$, such that the average sectional curvature becomes exactly a measure of the discrepancy between the average distance between corresponding points on ${S_x}$ and ${S_y}$ (after parallel transport) and the distance between the centers $x$ and $y$, as required. The extension of the discrete sectional curvature tensor computation from the hypergraph (Riemannian) case to the directed causal graph (Lorentzian) case is shown for a causal network which limits to an asymptotically negatively curved Lorentzian manifold-like structure in Figure \ref{fig:Figure57}.

\begin{figure}[ht]
\centering
\includegraphics[width=0.595\textwidth]{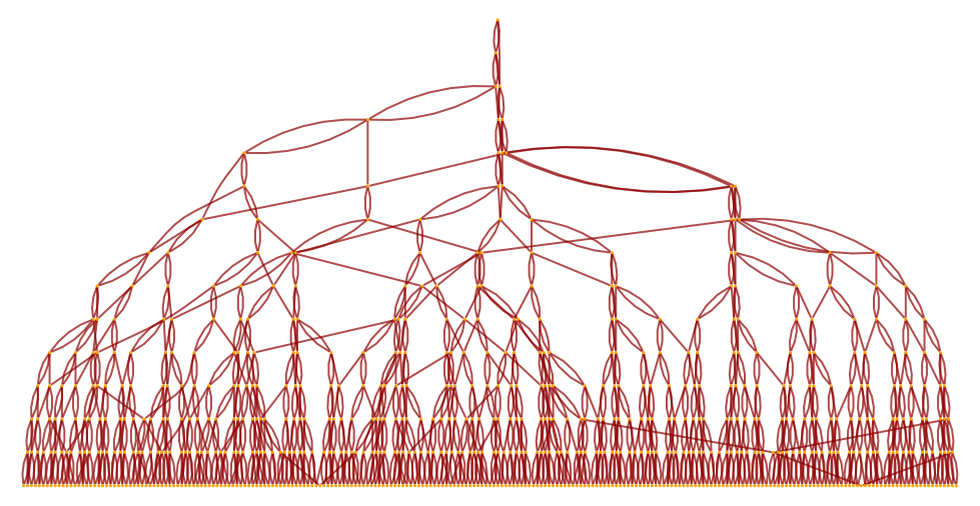}
\caption{An exponentially-growing causal network with no finite-dimensional limit (for which projections of the Ollivier-Ricci curvature tensor ${R_{a b}}$ grow without bound in all timelike directions), generated by the set substitution system ${\left\lbrace \left\lbrace x \right\rbrace, \left\lbrace x \right\rbrace \right\rbrace \to \left\lbrace \left\lbrace x \right\rbrace, \left\lbrace x \right\rbrace, \left\lbrace x \right\rbrace \right\rbrace}$.}
\label{fig:Figure58}
\end{figure}

In order for the causal network to converge to a Lorentzian manifold-like structure with finite dimensionality in the continuum limit, it cannot be the case that projections of the Ollivier-Ricci curvature tensor ${R_{a b}}$ are allowed to grow without bound (since, as we have shown above, each such projection contributes quadratically to the expansion of the Lorentzian metric volume element in a given timelike direction, so an unbounded increase in the curvature would be indistinguishable from a global increase in dimension in the continuum limit). An example of such a causal network exhibiting no finite-dimensional manifold-like limit is shown in Figure \ref{fig:Figure58}. In other words, we wish for the spacetime average $S$ of the Ollivier-Ricci curvature tensor (averaging over all timelike projection directions, and over all events in spacetime):

\begin{equation}
S \left[ g^{a b} \right] = \sum_{x} \sum_{j, k} d \mu_g R_{j k} x^j x^k,
\end{equation}
to remain bounded. However, as previously discussed, averaging out over all possible timelike projection directions is equivalent to taking the trace of the curvature tensor, such that the second sum simply reduces to the spacetime Ollivier-Ricci scalar curvature $R$ via index contraction:

\begin{equation}
S \left[ g^{a b} \right] = \sum_{x} d \mu_g R \left( x \right).
\end{equation}

Each addition of a causal edge to the causal network will correspond (in the continuum limit) to an infinitesimal metric perturbation\cite{gorard}, so this constraint is equivalent to the condition that the variation of $S$ (when considered as an action) should converge to zero in the limit ${\rho_c \to \infty}$:

\begin{equation}
\lim_{\rho_c \to \infty} \left[ \frac{\delta S \left[ g^{a b} \right]}{\delta g^{a b}} \right] = 0.
\end{equation}
Furthermore, if we make an additional \textit{weak ergodicity} assumption on the causal network dynamics, in which we suppose that, for any hypersurface of constant ${x^{\nu}}$ in the causal network, the net flux of causal edges through that hypersurface in the continuum limit converges to zero, this provides us with the analytic justification necessary to exchange this discrete sum for an integral in the limit ${\rho_c \to \infty}$, with the volume element thus replaced with the determinant of the standard (continuous) metric tensor:

\begin{equation}
\lim_{\rho_c \to \infty} \left[ S \left[ g^{a b} \right] \right] = \int_{\mathcal{M}} d^4 x \sqrt{-g} R.
\end{equation}
Thus, we see that the condition that the causal network converges to a finite-dimensional manifold-like structure becomes, in the continuum limit, equivalent to the statement of extremization of the classical (vacuum) Einstein-Hilbert action, with the standard choice of relativistic Lagrangian density:

\begin{equation}
\mathcal{L}_G = \sqrt{-g} R.
\end{equation}
The extremization can therefore be solved in the usual way - namely by taking a functional derivative with respect to the inverse metric tensor (assuming as before, zero boundary terms):

\begin{equation}
\frac{\delta S \left[ g^{a b} \right]}{\delta g^{a b}} = \sqrt{- g} \left( R_{a b} - \frac{1}{2} R g_{a b} \right),
\end{equation}
with minimization hence yielding the vacuum Einstein field equations:

\begin{equation}
R_{a b} - \frac{1}{2} R g_{a b} = 0.
\end{equation}

We consequently see that the Benincasa-Dowker action is recovered as a special case of this general construction of the discrete Einstein-Hilbert action, in which the couplings ${\epsilon}$ between measures ${\mu_{A^{in}}}$ and ${\mu_{B^{out}}}$:

\begin{equation}
\sum_{u \to A} \epsilon \left( u, v \right) = \sum_{j = 1}^{m} \mu_{y_j} \left( v \right), \qquad \text{ and } \qquad \sum_{B \to v} \epsilon \left( u, v \right) = \sum_{i = 1}^{n} \mu_{x_i} \left( u \right),
\end{equation}
are set to be equal to 1 everywhere. This effectively endows the causal network with a constant real scalar field ${\phi}$, such that the Ollivier-Ricci scalar curvature $R$ can be written up to third-order in terms of the (expectation value of the) discrete d'Alembertian ${\hat{B}}$ in the neighborhood region ${\mathcal{W}_1}$ as:

\begin{equation}
R \left( x \right) \phi \left( x \right) = 2 \left( \Box \phi \left( x \right) - \lim_{\rho_c \to \infty} \left[ \frac{1}{\sqrt{\rho_c}} \left. \left\langle \hat{B} \phi \left( x \right) \right\rangle \right\rvert_{\mathcal{W}_1} \right] \right),
\end{equation}
yielding the usual discrete action (assuming, for illustrative purposes, a ${3+1}$-dimensional  causal network ${\mathcal{C}}$):

\begin{equation}
S^{\left( 4 \right)} \left( \mathcal{C} \right) = \sum_{e \in \mathcal{C}} R \left( e \right) = \frac{4}{\sqrt{6}} \left[ n - N_0 + 9 N_1 - 16 N_2 + 8 N_3 \right],
\end{equation}
in the particular case where the weak ergodicity assumption stated above is replaced with a much stronger assumption of uniform (Poisson) distribution of causal set elements, as required.

\begin{figure}[ht]
\centering
\includegraphics[width=0.495\textwidth]{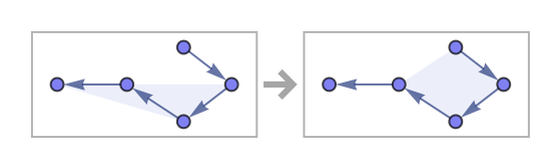}
\caption{A hypergraph transformation rule corresponding to the set substitution system ${\left\lbrace \left\lbrace x, y \right\rbrace, \left\lbrace y, z, u , v \right\rbrace \right\rbrace \to \left\lbrace \left\lbrace x, y, z, u \right\rbrace, \left\lbrace u, v \right\rbrace \right\rbrace}$. Example taken from \cite{wolfram2}.}
\label{fig:Figure59}
\end{figure}

\begin{figure}[ht]
\centering
\includegraphics[width=0.695\textwidth]{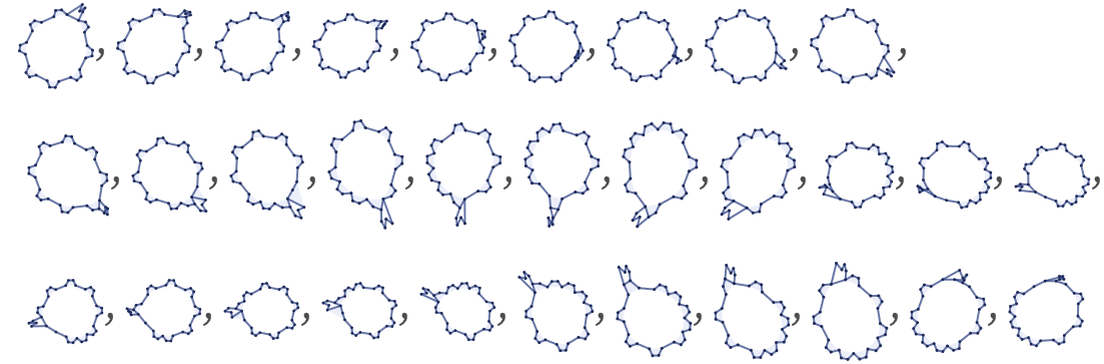}
\caption{The evolution history of the set substitution system ${\left\lbrace \left\lbrace x, y \right\rbrace, \left\lbrace y, z, u, v \right\rbrace \right\rbrace \to \left\lbrace \left\lbrace x, y, z, u \right\rbrace, \left\lbrace u, v \right\rbrace \right\rbrace}$, exhibiting the propagation of a localized topological obstruction. Example taken from \cite{wolfram2}.}
\label{fig:Figure60}
\end{figure}

\begin{figure}[ht]
\centering
\includegraphics[width=0.695\textwidth]{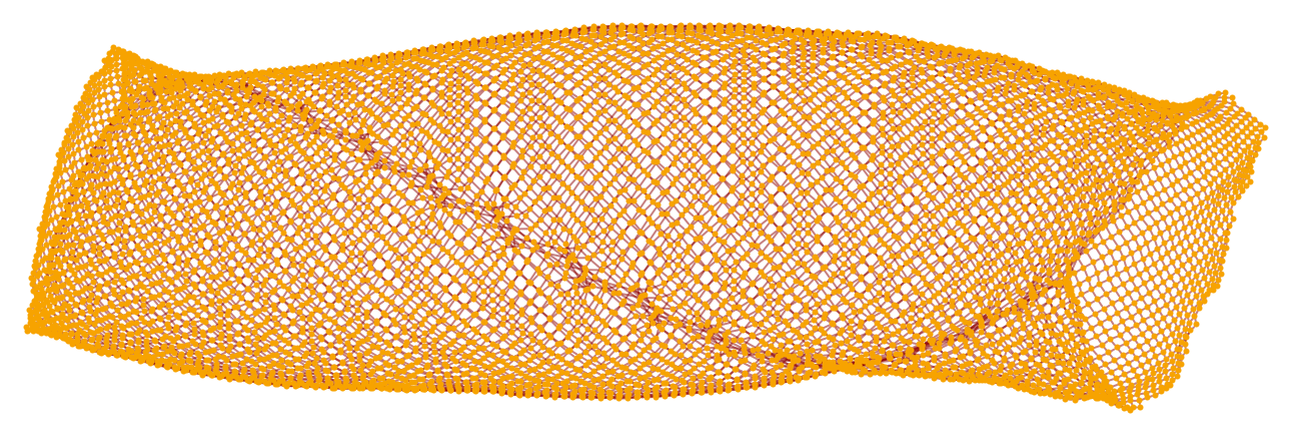}
\caption{The causal network for the set substitution system ${\left\lbrace \left\lbrace x, y \right\rbrace, \left\lbrace y, z, u, v \right\rbrace \right\rbrace \to \left\lbrace \left\lbrace x, y, z, u \right\rbrace, \left\lbrace u, v \right\rbrace \right\rbrace}$, exhibiting the propagation of a (spatially) localized topological obstruction. Example taken from \cite{wolfram2}.}
\label{fig:Figure61}
\end{figure}

Although it is somewhat beyond the scope of the present article, note that a derivation of the full form of the Einstein field equations is also possible\cite{wolfram2}\cite{gorard}, but involves an interpretation of baryonic matter contributions as persistent and localized topological obstructions in the hypergraph, as shown in Figures \ref{fig:Figure59} and \ref{fig:Figure60}, for the case of the set substitution rule ${\left\lbrace \left\lbrace x, y \right\rbrace, \left\lbrace y, z, u, v \right\rbrace \right\rbrace \to \left\lbrace \left\lbrace x, y, z, u \right\rbrace, \left\lbrace u, v \right\rbrace \right\rbrace}$. The propagation of such (spatially) localized topological obstructions can also be seen in the causal network, as shown in Figure \ref{fig:Figure61}. If we interpret the discrete upper-index energy-momentum tensor ${T^{\mu \nu}}$ as quantifying the flux of causal edges (i.e. the analog of the relativistic 4-momentum ${P^{\mu}}$) through a hypersurface of constant ${x^{\nu}}$ in the causal network, such that energy corresponds to the flux of causal edges through spacelike hypersurfaces, momentum corresponds to the flux of causal edges through timelike hypersurfaces, etc., the presence of topological obstructions thus has the effect of adding in a matter field term to the previous form of the relativistic Lagrangian density:

\begin{equation}
\mathcal{L}_G + C_M \mathcal{L}_M,
\end{equation}
for some arbitrary constant ${C_M}$, which (for the sake of convention) we shall assume to be equal to ${16 \pi}$ for scalar field matter.

Therefore, by assuming that these topological obstructions obey an equation of state that forces ${T_{a b}}$ to take the following form in terms of the matter Lagrangian density:

\begin{equation}
T_{a b} = - \frac{C_M}{8 \pi \sqrt{-g}} \frac{\partial \mathcal{L}_M \left[ g^{a b} \right]}{\partial g^{a b}},
\end{equation}
or, in terms of the effective matter action:

\begin{equation}
T_{a b} = - \frac{C_M}{8 \pi \sqrt{-g}} \frac{\delta S_M \left[ g^{a b} \right]}{\delta g^{a b}},
\end{equation}
the minimization of the combined action hence yields the full (non-vacuum) Einstein field equations:

\begin{equation}
R_{a b} - \frac{1}{2} R g_{a b} = 8 \pi T_{a b}.
\end{equation}
Note that our choice to distinguish between the density of causal edges involved in maintaining the background hypergraph (i.e. those involved in the maintenance of space itself) and those involved in the propagation of our ``baryonic matter'' contributions from localized topological obstructions is somewhat arbitrary; in other words, this derivation defines the Einstein field equations only up to an integration constant. The spacetime average $S$ of the Ollivier-Ricci curvature tensor could just as equally have been given by:

\begin{equation}
S \left[ g^{a b} \right] = \sum_x d \mu_g \left( R \left( x \right) + 2 \Lambda \right),
\end{equation}
for some arbitrary additive constant ${2 \Lambda}$, yielding, in the continuum limit, the updated (vacuum) relativistic Lagrangian density:

\begin{equation}
\mathcal{L}_G = \sqrt{-g} \left( R + 2 \Lambda \right),
\end{equation}
from which the full Einstein field equations (including a potentially non-zero cosmological constant term) can correspondingly be derived:

\begin{equation}
R_{a b} - \frac{1}{2} R g_{a b} + \Lambda g_{ a b} = 8 \pi T_{a b}.
\end{equation}
Therefore, we see that the trade-off between the energy-momentum terms $T$ and the cosmological constant terms ${\Lambda}$ reflects the trade-off between the causal edges involved in the propagation of topological obstructions, and those involved in the maintenance of space. The correspondence between this extended derivation and the treatment of scalar matter fields of the form ${\phi : \mathcal{C} \to \mathbb{R}}$ in causal set theory (approximating continuous scalar fields of the form ${\phi : \mathcal{M} \to \mathbb{R}}$), whose discrete Klein-Gordon operator ${B_{k}}$, as defined above, can be used to construct a discrete Green's function ${B_{k}^{-1}}$, etc., remains a topic for future investigation.

Note that a highly related calculation for the discrete Ricci scalar has previously been considered in the context of causal set theory (for instance by Sverdlov and Bombelli\cite{sverdlov}), using the volume of the small causal diamond ${\mathbf{A} \left[ p, q \right]}$ in Riemann normal coordinates ${x^{\mu}}$, for which the metric ${g_{\mu \nu}}$ has, as discussed above, the elegant form:

\begin{equation}
g_{i j} = \delta_{i j} - \frac{1}{3} R_{i j k l}  x^k x^l + O \left( \left\lVert \mathbf{x} \right\rVert^2 \right).
\end{equation}
Thus, as described by Myrheim\cite{myrheim} and Gibbons and Solodukhin\cite{gibbons}, one obtains the following general formula for the volume ${\mathrm{Vol} \left( \mathbf{A} \left[ p, q \right] \right)}$:

\begin{equation}
\mathrm{Vol} \left( \mathbf{A} \left[ p, q \right] \right) = \frac{\pi}{24} T^4 \left( 1 + \alpha T^2 R + \beta R_{i j} T^j T^k + \cdots \right),
\end{equation}
where ${T^{\mu}}$ denotes the components of the timelike vector lying tangent to the geodesic ${\mathbf{\gamma}}$ at the origin, with magnitude:

\begin{equation}
g_{j k} T^j T^k = - T^2,
\end{equation}
such that $T$ denotes the proper time between events $p$ and $q$ (as before), yielding, for our specific purposes:

\begin{equation}
\mathrm{Vol} \left( \mathbf{A} \left[ p, q \right] \right) = \mathrm{Vol}_0 \left[ p, q \right] \left( 1 - \frac{d}{24 \left( d + 1 \right) \left( d + 2 \right)} R T^2 + \frac{d}{24 \left( d + 1 \right)} R_{0 0} T^2 \right),
\end{equation}
where ${\mathrm{Vol}_0 \left[ p, q \right]}$ denotes the volume of the corresponding interval in flat spacetime. Since the volume ${\mathrm{Vol} \left( \mathbf{A} \left[ p, q \right] \right)}$ can be approximated by the volume of the discrete interval ${\mathrm{Vol} \left( \mathbf{I} \left[ p, q \right] \right)}$, and the timelike distance $T$ can be approximated by the length of the longest chain between elements $p$ and $q$, one thus obtains an approximation for the Ricci scalar $R$.

Note that in both the Benincasa-Dowker and the general Wolfram model cases considered above, the boundary terms in the extremization of the discrete action have been neglected. These terms can indeed be computed explicitly in the case of causal sets sprinkled into flat spacetimes\cite{buck}, for instance by using the Gibbons-Hawking-York (GHY)\cite{gibbons2}\cite{york}\cite{jubb}\cite{lehner} null boundary term for a closed and compact spacelike hypersurface ${\Sigma \subset \mathcal{M}}$:

\begin{equation}
S_{GHY} \left[ \Sigma, M^{\pm} \right] = \mp \frac{1}{l_{p}^{d - 2}} \int_{\Sigma} d^{d - 1} x \sqrt{h} K,
\end{equation}
where ${M^{\pm}}$ are defined as the causal future and past sets of the hypersurface ${\Sigma}$:

\begin{equation}
M^{\pm} = J^{\pm} \left( \Sigma \right), \qquad \text{ such that } M^{+} \cap M^{-} = \Sigma,
\end{equation}
$h$ is the induced metric on the hypersurface ${\Sigma}$, $K$ is the trace of the extrinsic curvature tensor on ${\Sigma}$:

\begin{equation}
K_{\mu \nu} = h_{\mu}^{\rho} h_{\nu}^{\sigma} \boldsymbol\nabla_{\rho} n_{\sigma},
\end{equation}
and ${l_p}$ is the rationalized Planck length (with ${\hbar = 1}$):

\begin{equation}
l_p = \left( 8 \pi G \right)^{\frac{1}{d - 2}}.
\end{equation}
However, a systematic comparison of treatments of the boundary terms in the causal set and Wolfram model cases (and the possibility of their extension to the general curved spacetime case) remains beyond the scope of the present article.

\section{Classical/Quantum Sequential Growth and Causal Multiway Systems}
\label{sec:Section5}

\begin{figure}[ht]
\centering
\includegraphics[width=0.495\textwidth]{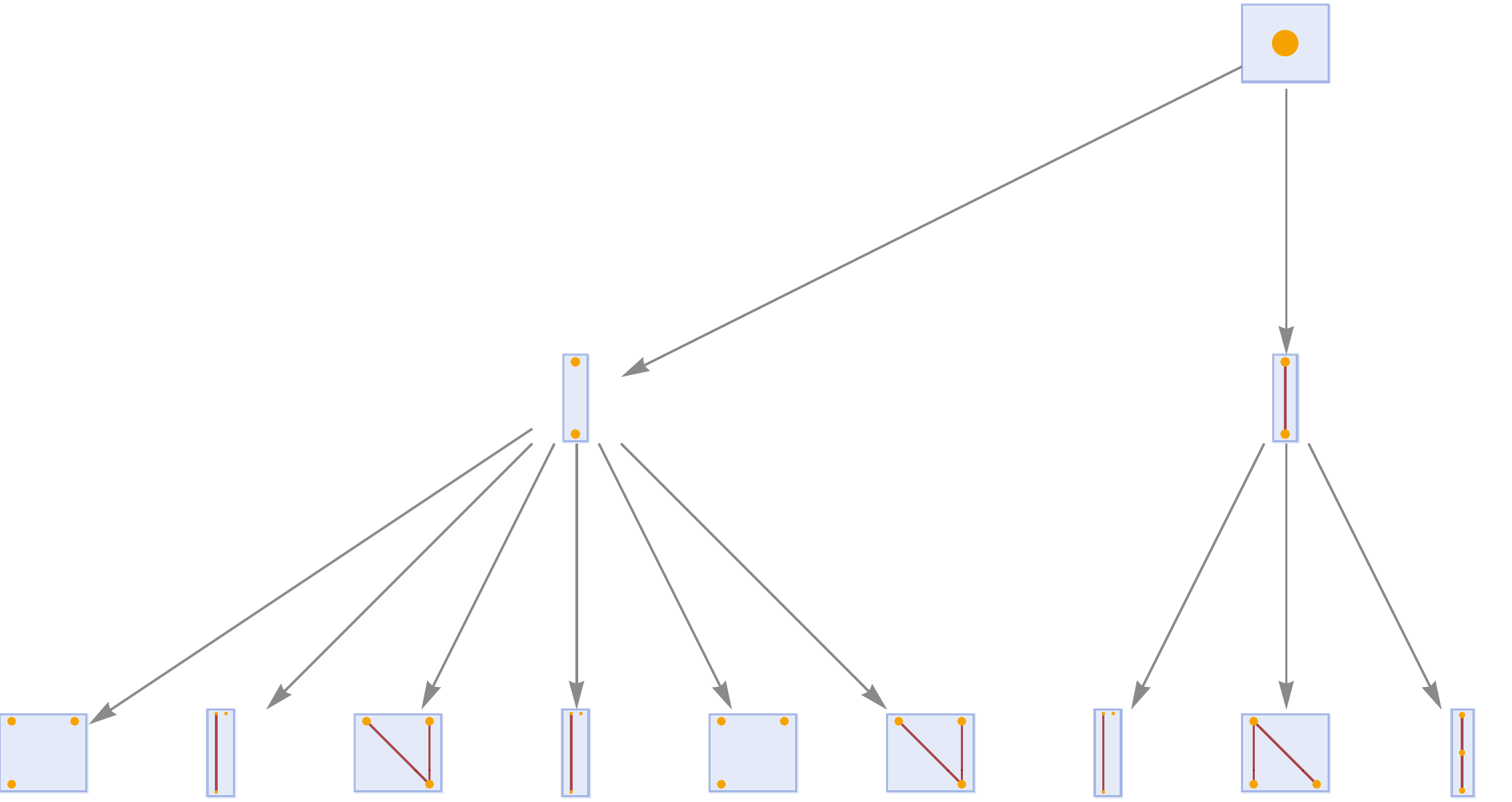}
\includegraphics[width=0.495\textwidth]{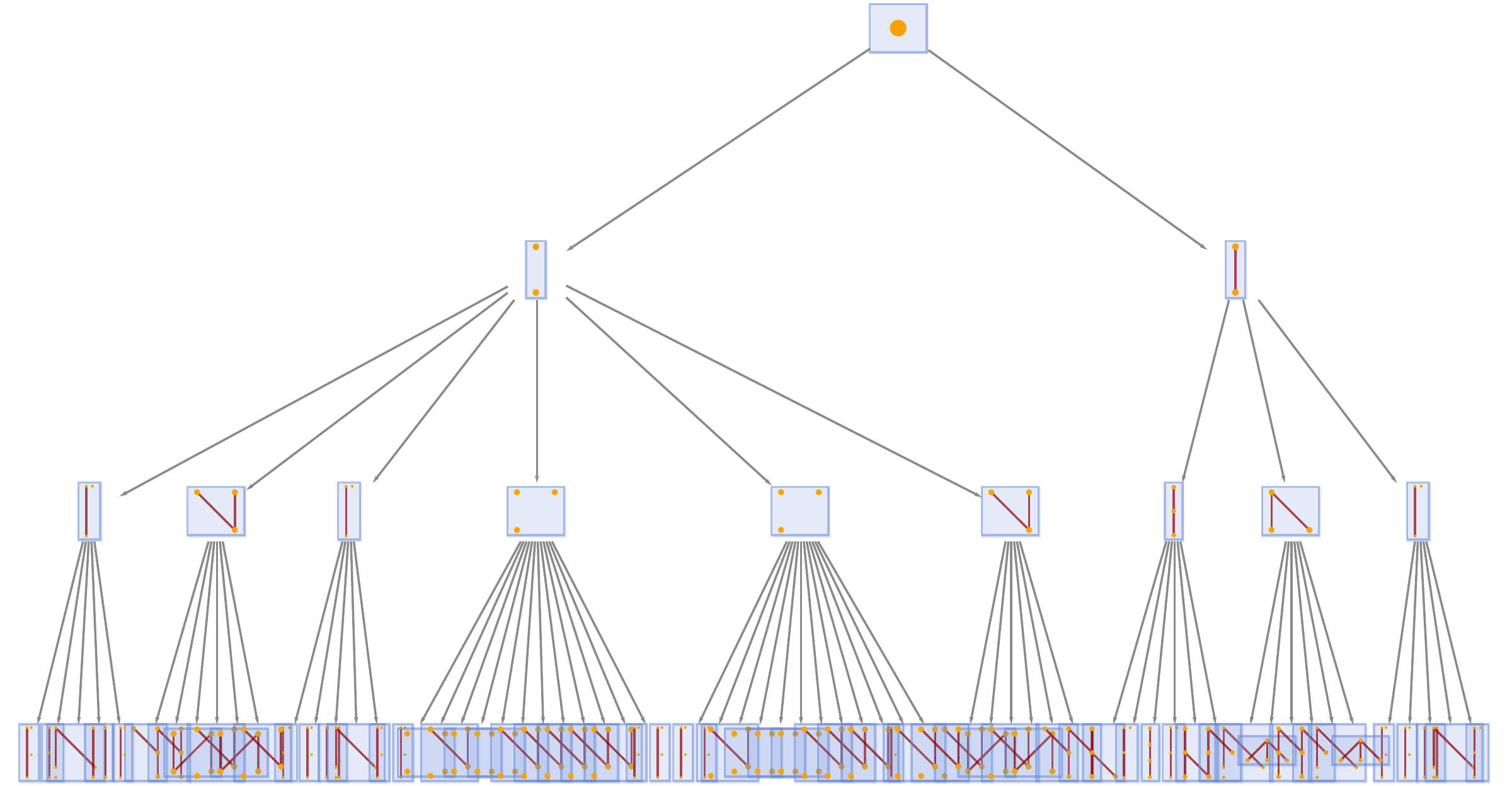}
\caption{Classical sequential growth trees produced after two and three steps of classical sequential growth evolution, respectively, represented here by multiway evolution graphs whose state vertices are causal networks. The resulting multiway system defines a trivial partial order on causal sets, i.e. a \textit{postcau}.}
\label{fig:Figure62}
\end{figure}

The primary focus of this article thus far has been on the relationship between the kinematic/geometrical aspects of causal sets and the Wolfram model, but it is worth devoting at least a little attention to the correspondence between their dynamics (beyond merely the discrete action principles considered in the previous section). Rideout and Sorkin\cite{rideout2} proposed a \textit{classical sequential growth} (CSG) dynamics in 2000, as part of a prior attempt to define an (in this case, stochastic) algorithmic dynamics for causal sets. In the context of classical sequential growth dynamics, if one uses the notation ${\mathcal{C}_n}$ to denote a causal set of cardinality $n$ (i.e. ${\left\lvert \mathcal{C}_n \right\rvert = n}$), then new causal set elements are ``accreted'' via an update rule of the form ${\mathcal{C}_n \to \mathcal{C}_{n + 1}}$, in such a way that the newly added element in ${\mathcal{C}_{n + 1}}$ always obeys the causal partial order relation, i.e. it is permitted to lie in the causal future of elements in ${\mathcal{C}_n}$ (or otherwise to be spacelike-separated from them), but can never lie in the causal past of elements in ${\mathcal{C}_n}$. Thus, the applications of the update rule ${\mathcal{C}_n \to \mathcal{C}_{n + 1}}$ define a new partial order of causal sets, or \textit{postcau}, ordered by the ``external'' time implicit in the rule application; in other words, one has a multiway evolution graph (otherwise known as a classical sequential growth tree), in which each state vertex is a single causal set, as shown in Figure \ref{fig:Figure62}.

\begin{figure}[ht]
\centering
\includegraphics[width=0.495\textwidth]{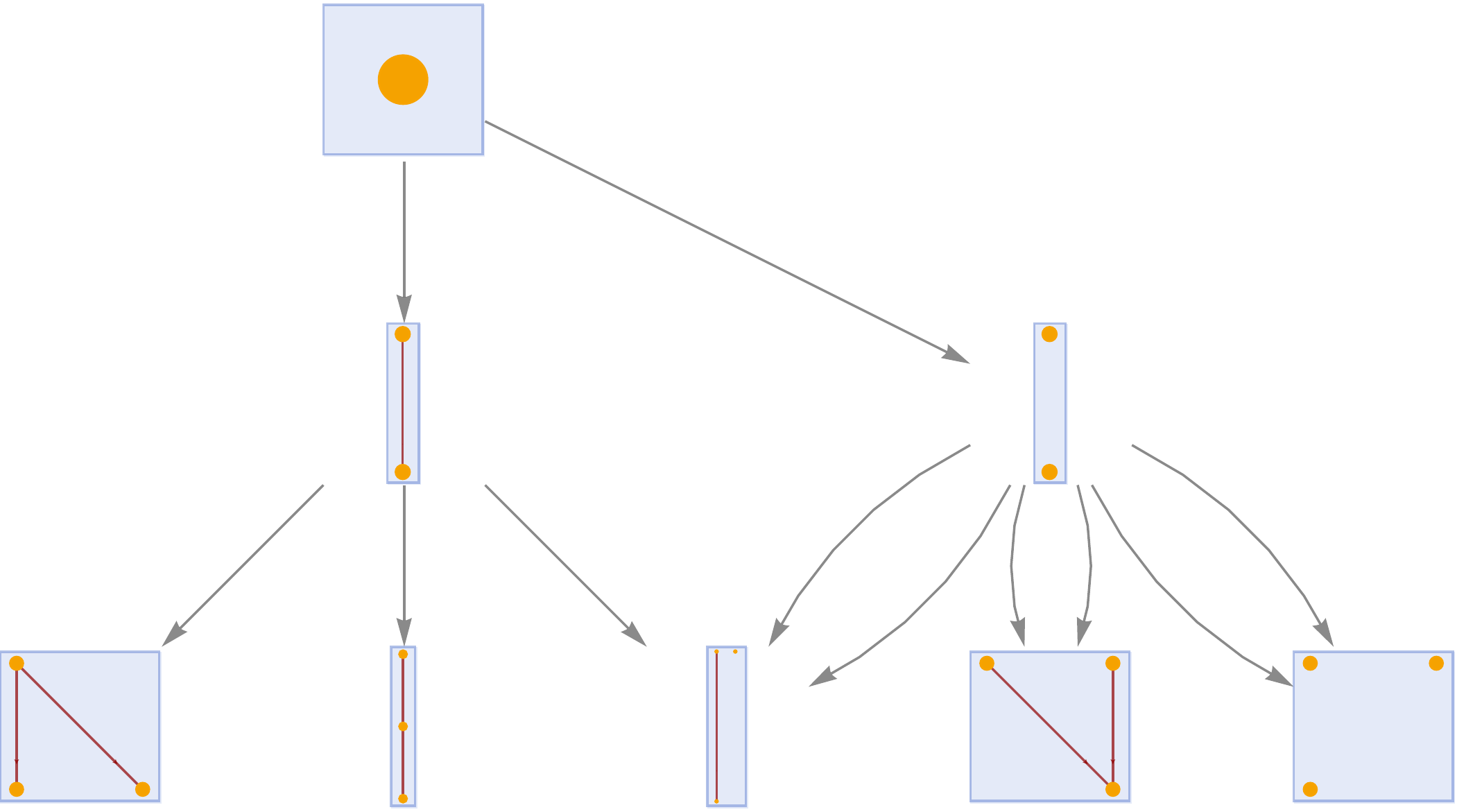}
\includegraphics[width=0.495\textwidth]{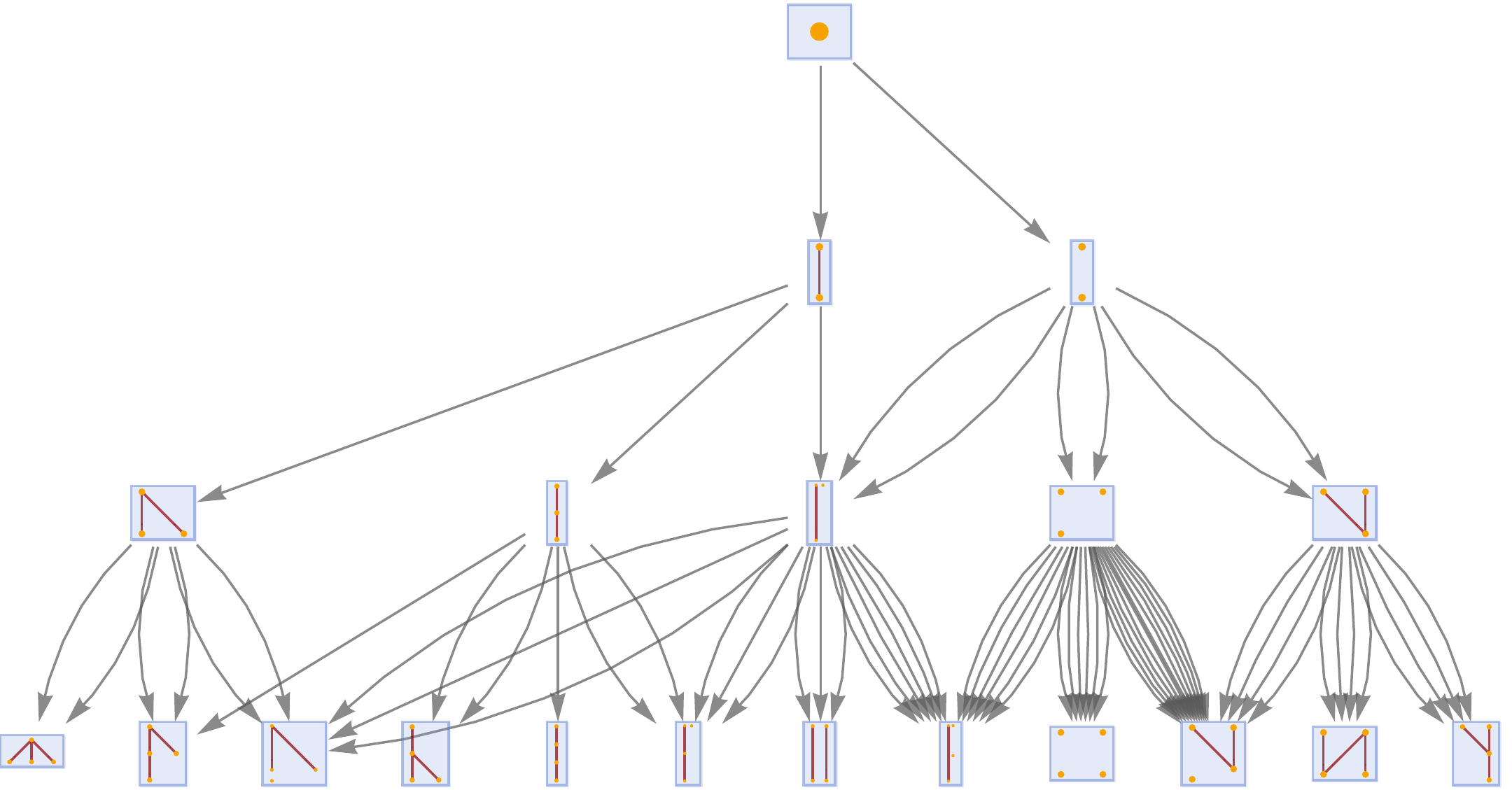}
\caption{Non-trivial postcaus produced after two and three steps of classical sequential growth evolution, respectively, represented here by multiway evolution graphs whose state vertices are causal networks, and in which states have been merged based on the criterion of causal network isomorphism (i.e. causal set relabeling). The multiplicities of edges here designate the number of mergings/relabelings that have occurred.}
\label{fig:Figure63}
\end{figure}

In the continuum limit, this classical sequential growth tree is then taken  to generate a continuous sample space ${\Omega}$ of countable causal sets that are \textit{past finite}. However, by introducing this ``external'' partial order relation on causal sets, we have effectively made an arbitrary gauge choice by introducing a certain natural labeling on the causal set elements, which for each causal set ${\mathcal{C} \in \Omega}$ in the sample space, we can represent as an injective function $L$ from the causal set to the natural numbers:

\begin{equation}
L : \mathcal{C} \to \mathbb{N},
\end{equation}
which is itself compatible with the partial order relation ${\prec}$ on ${\mathcal{C}}$:

\begin{equation}
\forall e, e^{\prime} \in \mathcal{C}, \qquad e \prec e^{\prime} \implies L \left( e \right) < L \left( e^{\prime} \right).
\end{equation}
In other words, by introducing a notion of coordinate time, we have also introduced a canonical foliation of each causal network (given by the level surfaces of this labeling function $L$), which now apparently violates the discrete analog of general covariance. Hence, in order for the stochastic dynamics defined over a classical sequential growth tree to be discretely covariant, it must be the case that the transition probability between causal sets ${\mathcal{C}_n}$ and ${\mathcal{C}_{n + 1}}$ must be the same as the transition probability between causal sets ${\mathcal{C}_n}$ and ${\mathcal{C}_{n + 1}^{\prime}}$, where ${\mathcal{C}_{n + 1}^{\prime}}$ is simply a different labeling of the set ${\mathcal{C}_{n + 1}}$, consequently ensuring that the gauge choice (i.e. labeling) does not yield any observable consequences. We can therefore achieve discrete covariance by merging state vertices in the multiway system in accordance with isomorphism of the underlying causal networks, at which point one obtains a non-trivial postcau, as shown in Figure \ref{fig:Figure63}.

\begin{figure}[ht]
\centering
\includegraphics[width=0.495\textwidth]{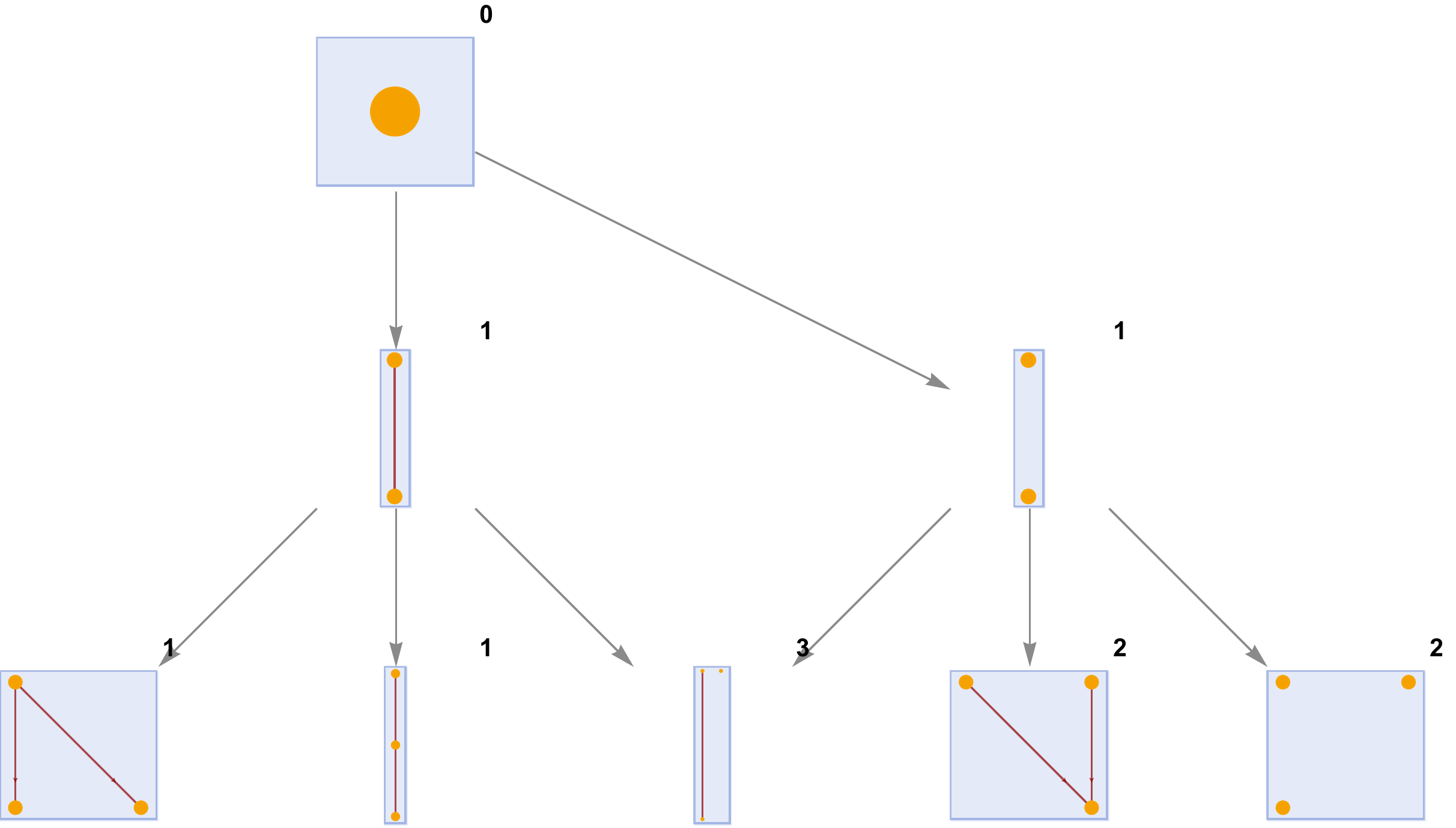}
\includegraphics[width=0.495\textwidth]{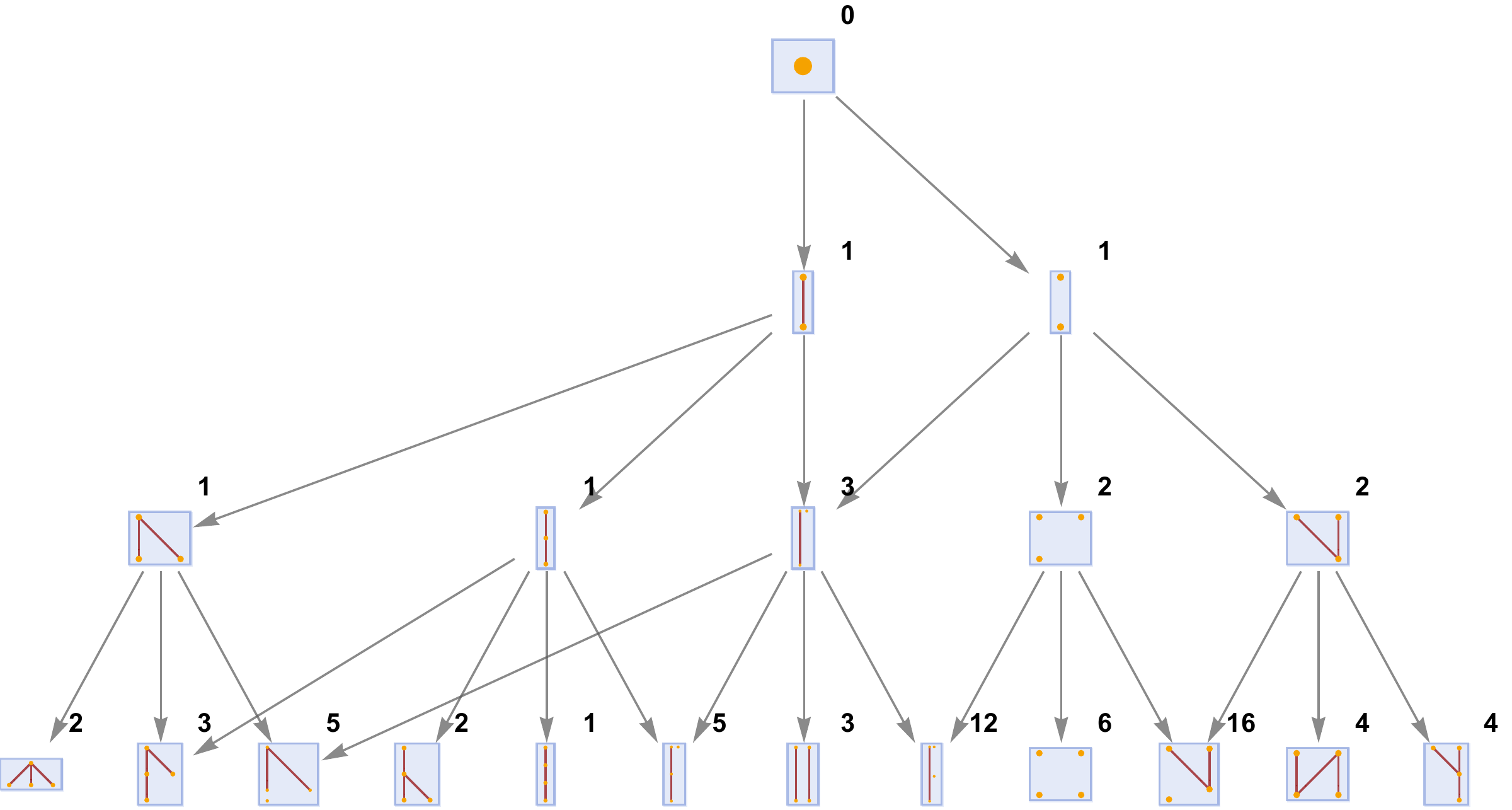}
\caption{Non-trivial postcaus produced after two and three steps of classical sequential growth evolution, respectively, represented here by multiway evolution graphs whose state vertices are causal networks, and in which states have been merged based on the criterion of causal network isomorphism (i.e. causal set relabeling). The weights of vertices here designate the relative transition probabilities from the initial state vertex.}
\label{fig:Figure64}
\end{figure}

Clearly, in order for the stochastic evolution to be Markovian, the set of all transition probabilities arising from a single causal set must be equal to 1, but the question still remains as to whether causal sets that have been merged on the basis of relabeling (i.e. causal network isomorphism) should be treated as being distinct with respect to this summation condition. Here, we follow the convention of Rideout and Sorkin\cite{rideout3} by assuming that they should, thus yielding the relative transition probabilities (rendered as vertex weights) shown in Figure \ref{fig:Figure64}. However, the final consistency condition that must be satisfied by a classical sequential growth dynamics is that of ``Bell causality'', which states loosely that, when adding a new causal set element to ${\mathcal{C}_n}$ to obtain ${\mathcal{C}_{n + 1}}$, the transition probability should be independent of those causal set elements in ${\mathcal{C}_n}$ that do not lie in the causal past of the element being added (so-called as a consequence of its similarity to the internal causality condition that must be satisfied in the standard derivation of Bell's theorem). However, in order to see precisely how this condition may be translated most naturally into the formalism of Wolfram model multiway systems, it is first necessary to see how quantum dynamics may be reproduced in the context of the Wolfram model itself.

\begin{figure}[ht]
\centering
\includegraphics[width=0.895\textwidth]{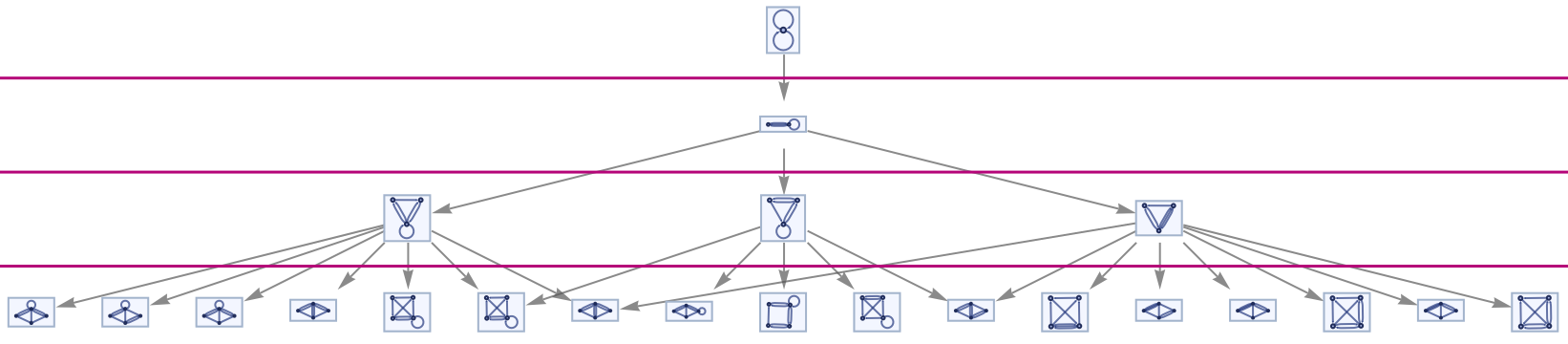}
\caption{The default choice of foliation of the multiway evolution graph for the set substitution system ${\left\lbrace \left\lbrace x, y \right\rbrace, \left\lbrace y, z \right\rbrace \right\rbrace \to \left\lbrace \left\lbrace w, y \right\rbrace, \left\lbrace y, w \right\rbrace, \left\lbrace x, w \right\rbrace \right\rbrace}$. Example taken from \cite{wolfram2}.}
\label{fig:Figure65}
\end{figure}

\begin{figure}[ht]
\centering
\includegraphics[width=0.395\textwidth]{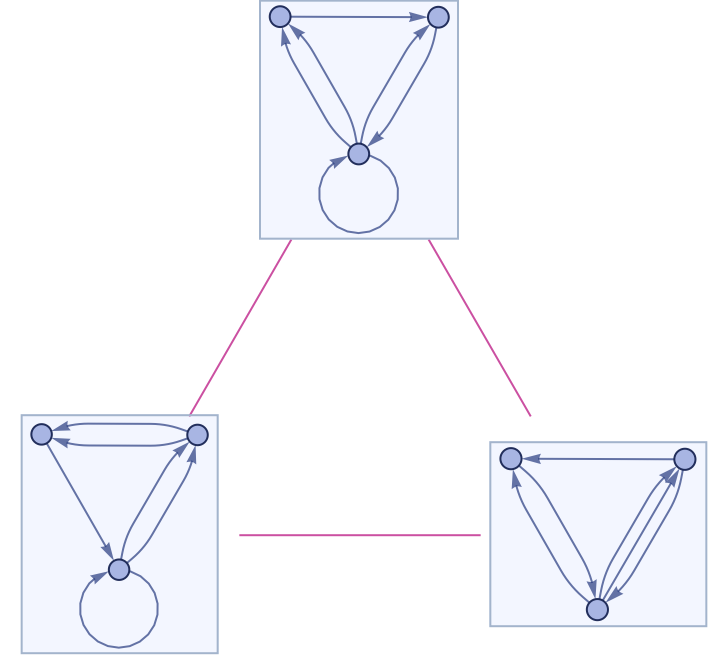}\hspace{0.1\textwidth}
\includegraphics[width=0.495\textwidth]{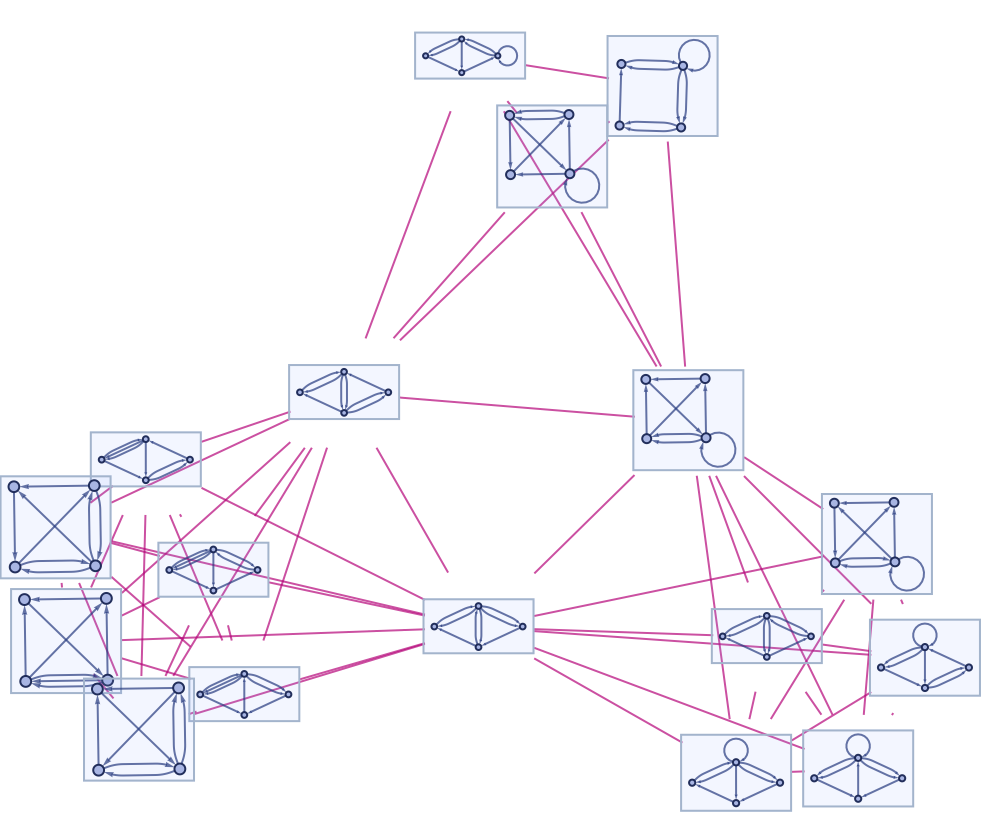}
\caption{The corresponding branchial graphs (representing branchlike hypersurfaces), as witnessed by an ``observer'' embedded within the default foliation of the multiway evolution graph for the set substitution system ${\left\lbrace \left\lbrace x, y \right\rbrace, \left\lbrace y, z \right\rbrace \right\rbrace \to \left\lbrace \left\lbrace w, y \right\rbrace, \left\lbrace y, w \right\rbrace, \left\lbrace x, w \right\rbrace \right\rbrace}$. Example taken from \cite{wolfram2}.}
\label{fig:Figure66}
\end{figure}

In much the same way as relativistic observers may be interpreted as foliating a causal network into discrete spacelike hypersurfaces, we now consider a quantum mechanical analog in which ``observers'' foliate a multiway evolution graph into what we shall term ``branchlike hypersurfaces'' (which we shall represent as ``branchial graphs'' - analogous to the discrete Cauchy surfaces obtained by foliating a causal network), as described in \cite{gorard2} and \cite{gorard4}:

\begin{definition}
An ``observer'' in a multiway system is an ordered sequence of non-intersecting ``branchlike hypersurfaces'' ${\Sigma_t}$ that covers the full multiway evolution graph, with the ordering determined by a choice of universal time function:

\begin{equation}
t : \mathcal{M} \to \mathbb{Z}, \qquad \text{ such that } \Delta t \neq 0 \text{ everywhere},
\end{equation}
such that each branchlike hypersurface is a level set of this function, satisfying the following pair of conditions:

\begin{equation}
\forall t_1, t_2 \in \mathbb{Z}, \qquad \Sigma_{t_1} = \left\lbrace p \in \mathcal{M} : t \left( p \right) = t_1 \right\rbrace, \text{ and } \Sigma_{t_1} \cap \Sigma_{t_2} = \emptyset \iff t_1 \neq t_2,
\end{equation}
with ${\mathcal{M}}$ denoting the set of vertices in the multiway evolution graph.
\end{definition}
The branchial graphs are, in turn, equipped with a natural discrete metric given by ancestry distance between multiway states, such that vertices $A$ and $B$ (corresponding to two states with the same value of the universal time function) are connected by a single undirected edge in the branchial graph if and only if they share an immediate common ancestor $C$ in the multiway evolution graph. An example of the default foliation of the multiway evolution graph for a simple Wolfram model system is shown in Figure \ref{fig:Figure65}, followed by the corresponding sequence of branchial graphs (representing branchlike hypersurfaces) as witnessed by an ``observer'' embedded within that same foliation in Figure \ref{fig:Figure66}.

\begin{figure}[ht]
\centering
\includegraphics[width=0.495\textwidth]{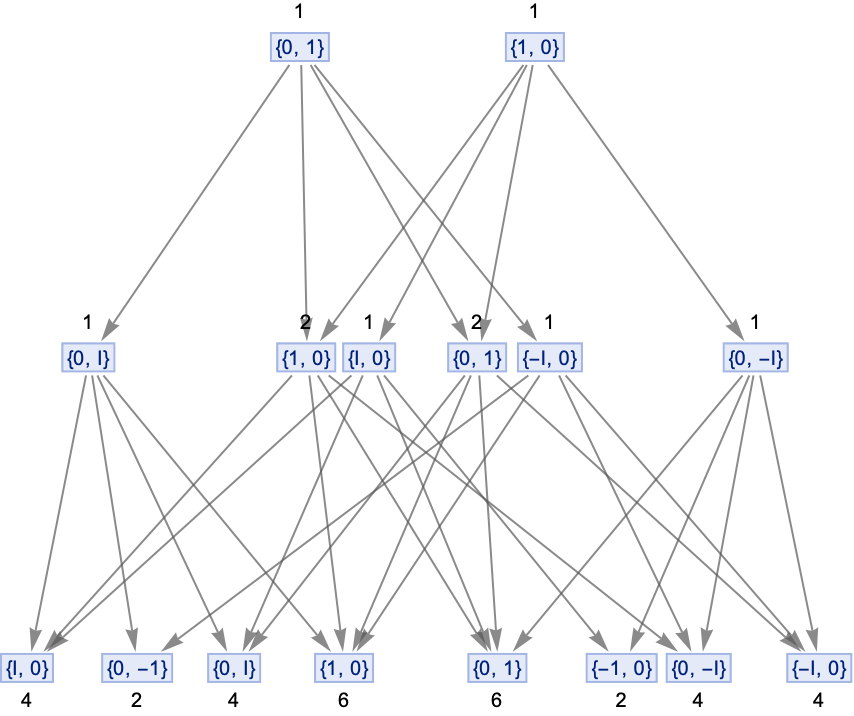}
\caption{The multiway evolution graph for a toy quantum system based upon repeated application of a root-NOT quantum gate to an initial superposition state of the form ${\frac{1}{\sqrt{2}} \left( \ket{0} + \ket{1} \right)}$, with vertex weights given by the number of distinct evolution paths leading to a particular state.}
\label{fig:Figure67}
\end{figure}

\begin{figure}[ht]
\centering
\includegraphics[width=0.395\textwidth]{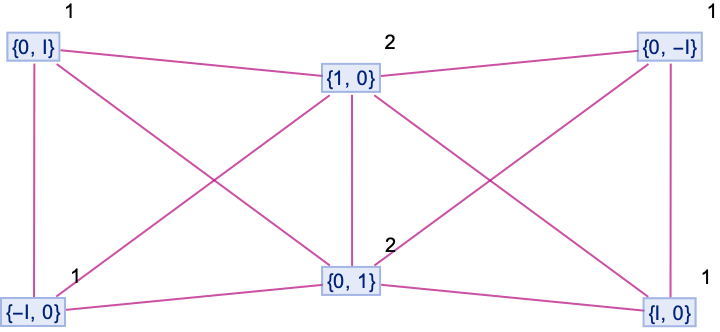}\hspace{0.1\textwidth}
\includegraphics[width=0.395\textwidth]{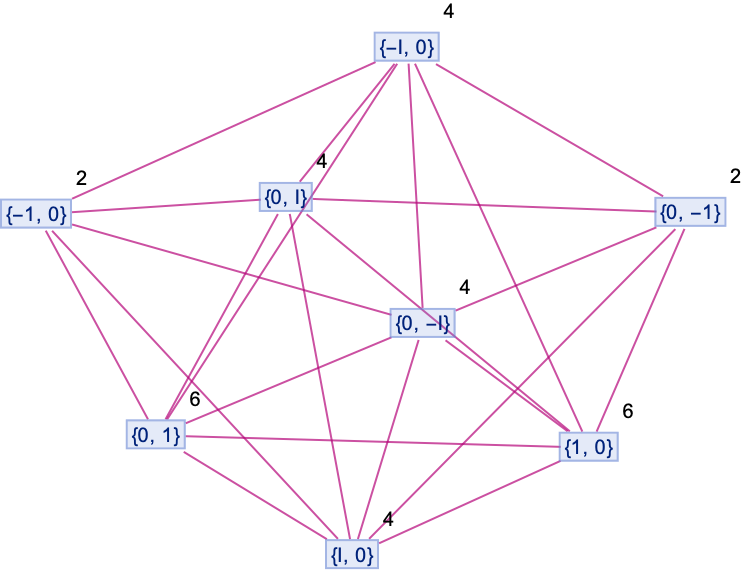}
\caption{The corresponding sequence of branchial graphs (representing branchlike hypersurfaces), as witnessed by an ``observer'' embedded within the default foliation of the multiway evolution graph for the toy root-NOT quantum system, starting from an initial superposition state of the form ${\frac{1}{\sqrt{2}} \left( \ket{0} + \ket{1} \right)}$, with vertex weights given by the number of distinct evolution paths leading to a particular state.}
\label{fig:Figure68}
\end{figure}

The conjectural relationship between branchial graphs and quantum evolution lies in a formal correspondence between state vertices in a multiway system and pure states of a quantum system, such that each branchial graph is a representation of an instantaneous superposition of certain eigenstates at a given moment in time, with the evolution from one branchlike hypersurface to the next corresponding to the unitary evolution of some generalized Hartle-Hawking wave function (describing the overall state of the universe). One can easily make this conjecture precise in certain toy cases, for instance by considering a multiway evolution graph representing state transitions between qubits obtaining by repeated application of a root-NOT quantum gate:

\begin{equation}
\sqrt{NOT} = \frac{1}{2} \begin{bmatrix}
1 + i & 1 - i\\
1 - i & 1 + i
\end{bmatrix},
\end{equation}
to an initial (superposition) state of the form ${\frac{1}{\sqrt{2}} \left( \ket{0} + \ket{1} \right)}$, as shown in Figures \ref{fig:Figure67} and \ref{fig:Figure68}.

In this toy example, it is trivial to prove that the multiway representation of the quantum evolution is faithful, in the sense that, when one sums over the collection of state vertices on each branchlike hypersurface (with states vectors being multiplied by path weights of the respective vertices), the multiway formulation of the evolution is equivalent to that obtained by ordinary matrix multiplication. However, in \cite{gorard4} the authors demonstrated that this correspondence could be proved much more generally, by formulating (reversible) Wolfram model rules of the form ${L \to R}$ in the context of adhesive categories ${\mathbf{C}}$ as spans of monomorphisms of the form:

\begin{equation}
\rho = \left( l : K \to L, r : K \to R \right),
\end{equation}
such that each updating event is equivalent to a statement of existence for the following pair of pushout diagrams\cite{ehrig}\cite{habel}:

\begin{equation}
\begin{tikzcd}
L \arrow[d, "m"] & K \arrow[l, "l"] \arrow[d, "n"] \arrow [r, "r"] & R \arrow[d, "\rho"]\\
G & D \arrow[l, "g"] \arrow[r, "h"] & H
\end{tikzcd}.
\end{equation}
More specifically, it was shown that by a straightforward modification of the techniques developed by Dixon and Kissinger\cite{dixon}, one can prove that the categories of both multiway evolution graphs and branchial graphs are equipped with a natural symmetric monoidal structure\cite{baez} given by parallel compositions of multiway rules that generalizes the ordinary tensor product, i.e. one obtains a bifunctor ${\otimes}$ of the form:

\begin{equation}
\otimes : \mathbf{C} \times \mathbf{C} \to \mathbf{C},
\end{equation}
equipped with a natural ``associator'' isomorphism ${\alpha}$ with components:

\begin{equation}
\alpha_{A, B, C} : A \otimes \left( B \otimes C \right) \cong \left( A \otimes B \right) \otimes C,
\end{equation}
a natural ``symmetry'' isomorphism ${\sigma}$ with components:

\begin{equation}
\sigma_{A, B} : A \otimes B \cong B \otimes A,
\end{equation}
and a pair of left and right ``unitor'' isomorphisms ${\lambda}$ and ${\rho}$ with components:

\begin{equation}
\lambda_{A} : I \otimes A \cong A, \qquad \text{ and } \qquad \rho_A : A \otimes I \cong A,
\end{equation}
for a designated ``identity object'' $I$. Although in the interests of concision we do not attempt to reproduce the argument here, it suffices to say that these properties are essentially inherited from the associativity, commutativity and identity properties of the disjoint union operation over set substitution rules. Moreover, one can show by a similar technique that the category is also equipped with a natural dagger structure given by reversal of arrows in the multiway evolution graph that generalizes the ordinary Hermitian adjoint operation, i.e. one obtains an involutive functor ${\dag}$ of the form:

\begin{equation}
\dag : \mathbf{C}^{op} \to \mathbf{C},
\end{equation}
such that the adjoint of the identity is always the identity:

\begin{equation}
id_{A} = id_{A}^{\dag} : A \to A,
\end{equation}
the adjoint is always involutive:

\begin{equation}
f^{\dag \dag} = f : A \to B,
\end{equation}
and the adjoint reverses the order of composition of morphisms ${f : A \to B}$ and ${g : B \to C}$:

\begin{equation}
\left( g \circ f \right)^{\dag} = f^{\dag} \circ g^{\dag} : C \to A.
\end{equation}
This dagger structure is trivially compatible with the symmetric monoidal structure (although the question of whether the associated category is compact closed still remains unresolved).

\begin{figure}[ht]
\centering
\includegraphics[width=0.695\textwidth]{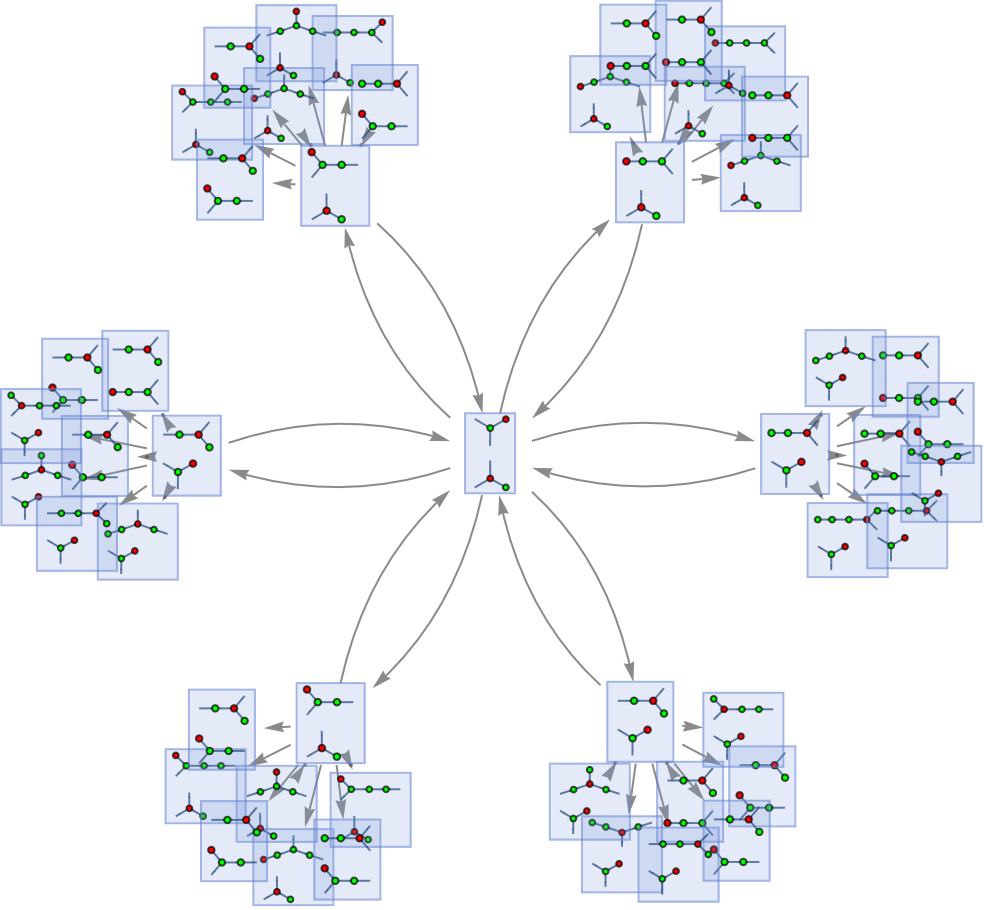}
\caption{The multiway states graph (i.e. the variant of the multiway evolution graph in which cycles are permitted) corresponding to the first two steps in the evolution of a multiway operator system based on the diagrammatic rewriting rules of the ZX-calculus, starting from a four-spider initial ZX-diagram (obtained as a tensor product of two two-spider ZX-diagrams), using the input arity-2 variant of the Z-spider identity rule.}
\label{fig:Figure69}
\end{figure}

The significance of this argument for the present discussion lies in the fact that the category of finite-dimensional Hilbert spaces, denoted ${\mathbf{FdHilb}}$, is itself a dagger-symmetric monoidal category\cite{selinger}, with the dagger structure corresponding to the ordinary Hermitian adjoint operation and the monoidal structure corresponding to the ordinary tensor product of Hilbert spaces. This formally justifies our interpretation of each branchial graph as being a subspace of a finite-dimensional Hilbert space, and leads, moreover, to a rigorous derivation of the limiting metric on branchlike hypersurfaces. Specifically, just as the points on discrete spacelike hypersurfaces can be interpreted as points on a continuous Riemannian manifold (with the combinatorial distance metric converging to a smooth Riemannian metric), points on branchlike hypersurfaces can be shown to correspond to points in a continuous projective Hilbert space, with the combinatorial distance on branchial graphs converging to a smooth \textit{Fubini-Study} metric on ${\mathbb{CP}^n}$\cite{gorard2}\cite{gorard4}:

\begin{definition}
The ``Fubini-Study'' metric on a complex projective Hilbert space ${\mathbb{CP}^n}$, when written in terms of the homogeneous coordinate system:

\begin{equation}
\mathbf{Z} = \left[ Z_0, \dots, Z_n \right],
\end{equation}
corresponding to the standard coordinate notation for projective varieties in algebraic geometry, may be defined abstractly in terms of the line element:

\begin{equation}
ds^2 = \frac{\left\lvert \mathbf{Z} \right\rvert^2 \left\lvert d \mathbf{Z} \right\rvert^2 - \left( \mathbf{\bar{Z}} \cdot d \mathbf{\bar{Z}} \right) \left( \mathbf{Z} \cdot d \mathbf{\hat{Z}} \right)}{\left\lvert \mathbf{Z} \right\rvert^4},
\end{equation}
or given in terms of explicit indices as:

\begin{equation}
ds^2 = \frac{Z_{\alpha} \hat{Z}^{\alpha} d Z_{\beta} \hat{Z}^{\beta} - \hat{Z}^{\alpha} Z_{\beta} d Z_{\alpha} \hat{Z}^{\beta}}{\left( Z_{\alpha} \hat{Z}^{\alpha} \right)^2}.
\end{equation}
\end{definition}
This, in turn, formally justifies an interpretation of the discrete distance metric on branchial graphs as a measure of entanglement distance between pure states, since if each point on a branchlike hypersurface corresponds to a pure state of the form:

\begin{equation}
\ket{\psi} = \sum_{k = 0}^{n} Z_k \ket{e_k} = \left[ Z_0 : Z_1 : \cdots : Z_n \right],
\end{equation}
with an orthonormal basis set ${\left\lbrace \ket{e_k} \right\rbrace}$ in some projective Hilbert space, then the general Fubini-Study metric can be written as an infinitesimal line element:

\begin{equation}
ds^2 = \frac{\braket{\delta \psi}{\delta \psi}}{\braket{\psi}{\psi}} - \frac{\braket{\delta \psi}{\psi} \braket{\psi}{\delta \psi}}{\braket{\psi}{\psi}^2},
\end{equation}
which makes manifest its correspondence with the quantum Bures metric on density matrix operators.

This formal justification of the interpretation of path weights of state vertices in the multiway evolution graph as quantum amplitudes has particular significance for our present discussion, since it allows us to modify the covariant probability space ${\left( \Omega, \Sigma, \mu_c \right)}$ associated to the multiway evolution graph of causal sets produced by classical sequential growth dynamics, and replace the classical probability measure ${\mu_c}$ with a \textit{quantum} probability measure ${\mu : \Sigma \to \mathbb{R}^{+}}$ based on path weights of state vertices (a similar construction of a quantum measure from discrete algorithmic dynamics was previously explored by the authors in \cite{shah}). Due to the merging of state vertices based on causal network isomorphism, the function ${\mu}$ is not straightforwardly additive:

\begin{equation}
\mu \left( \alpha \cup \beta \right) \neq \mu \left( \alpha \right) + \mu \left( \beta \right),
\end{equation}
for generic ${\alpha, \beta \in \Sigma}$ such that ${\mu \left( \alpha \cap \beta \right) = \emptyset}$; rather, one must subtract off the measure associated with states which are removed by the isomorphism testing, yielding what (following Sorkin's terminology\cite{sorkin4}\cite{sorkin5}) we shall refer to as a \textit{quantum additivity rule}:

\begin{equation}
\mu \left( \alpha \cup \beta \cup \gamma \right) = \mu \left( \alpha \cup \beta \right) + \mu \left( \alpha \cup \gamma \right) + \mu \left( \beta \cup \gamma \right) - \mu \left( \alpha \right) - \mu \left( \beta \right) - \mu \left( \gamma \right),
\end{equation}
for generic ${\alpha, \beta, \gamma \in \Sigma}$ such that ${\mu \left( \alpha \cap \beta \right) = \mu \left( \alpha \cap \gamma \right) = \mu \left( \beta \cap \gamma \right) = \emptyset}$. More formally, this measure ${\mu : \Sigma \to \mathbb{R}^{+}}$ can be derived from  a decoherence functional ${D : \Sigma \times \Sigma \to \mathbb{C}}$\cite{salgado}\cite{sorkin6}:

\begin{equation}
\forall \alpha \in \Sigma, \qquad \mu \left( \alpha \right) = D \left( \alpha, \alpha \right),
\end{equation}
with the (partial) additivity of path weights following causal network isomorphism yielding the following pair of (countable) biadditivity conditions:

\begin{equation}
\forall \alpha \in \Sigma, \qquad D \left( \alpha, \bigcup_{i} \beta_i \right) = \sum_{i} D \left( \alpha, \beta_i \right),
\end{equation}
and:

\begin{equation}
\forall \beta \in \Sigma, \qquad D \left( \bigcup_{i} \alpha_i, \beta \right) = \sum_i D \left( \alpha_i, \beta \right).
\end{equation}
Furthermore, by normalizing the sum of the path weights across each branchlike hypersurface (as one does implicitly in the case of the toy root-NOT quantum evolution shown above), one obtains normalization of the decoherence functional of the sample space:

\begin{equation}
D \left( \Omega, \Omega \right) = 1,
\end{equation}
and by noting that path weights can only ever add (i.e. there is no subtractive operation on path weights in the multiway evolution graph), it follows that, for all finite collections ${\left\lbrace \alpha_i \right\rbrace}$, the associated matrix:

\begin{equation}
M_{i j} = D \left( \alpha_i, \alpha_j \right),
\end{equation}
is positive semi-definite. Finally, the existence of the involutive dagger structure on the category of multiway evolution graphs guarantees the Hermiticity of the functional:

\begin{equation}
D \left( \alpha, \beta \right) = D^{\dag} \left( \beta, \alpha \right),
\end{equation}
thus demonstrating that the resultant dynamics corresponds precisely to a \textit{quantum sequential growth model}, of exactly the kind described by Sorkin. This leads us to the welcome realization that classical sequential growth dynamics on causal sets may be reproduced when a particular branch of the multiway system is followed in accordance with a stochastic process (with classical transition probabilities given by state path weights), whilst quantum sequential growth dynamics may be recovered when one instead traverses a superposition of \textit{all} possible branches of the multiway system (with quantum amplitudes given by state path weights).

\begin{figure}[ht]
\centering
\includegraphics[width=0.395\textwidth]{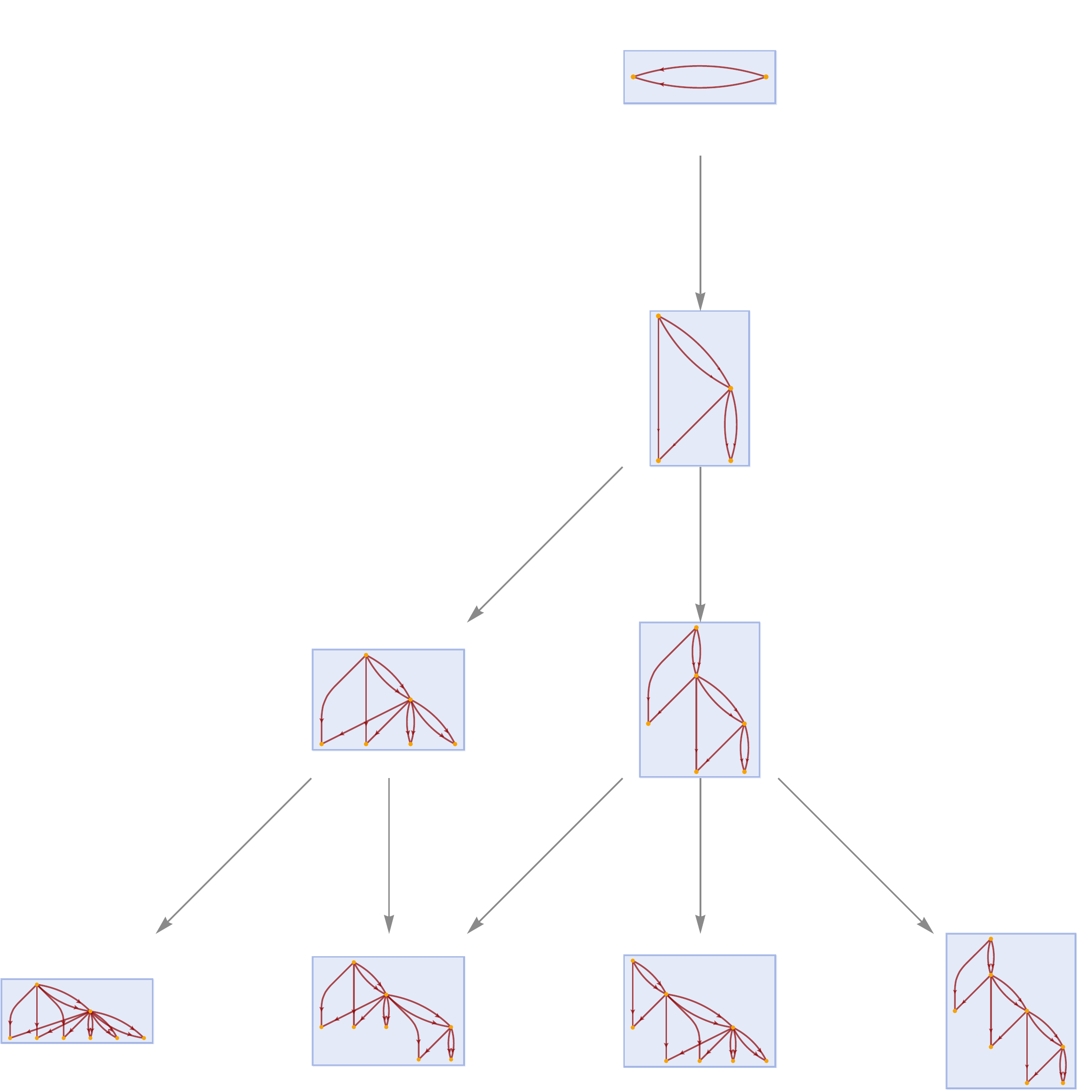}\hspace{0.1\textwidth}
\includegraphics[width=0.495\textwidth]{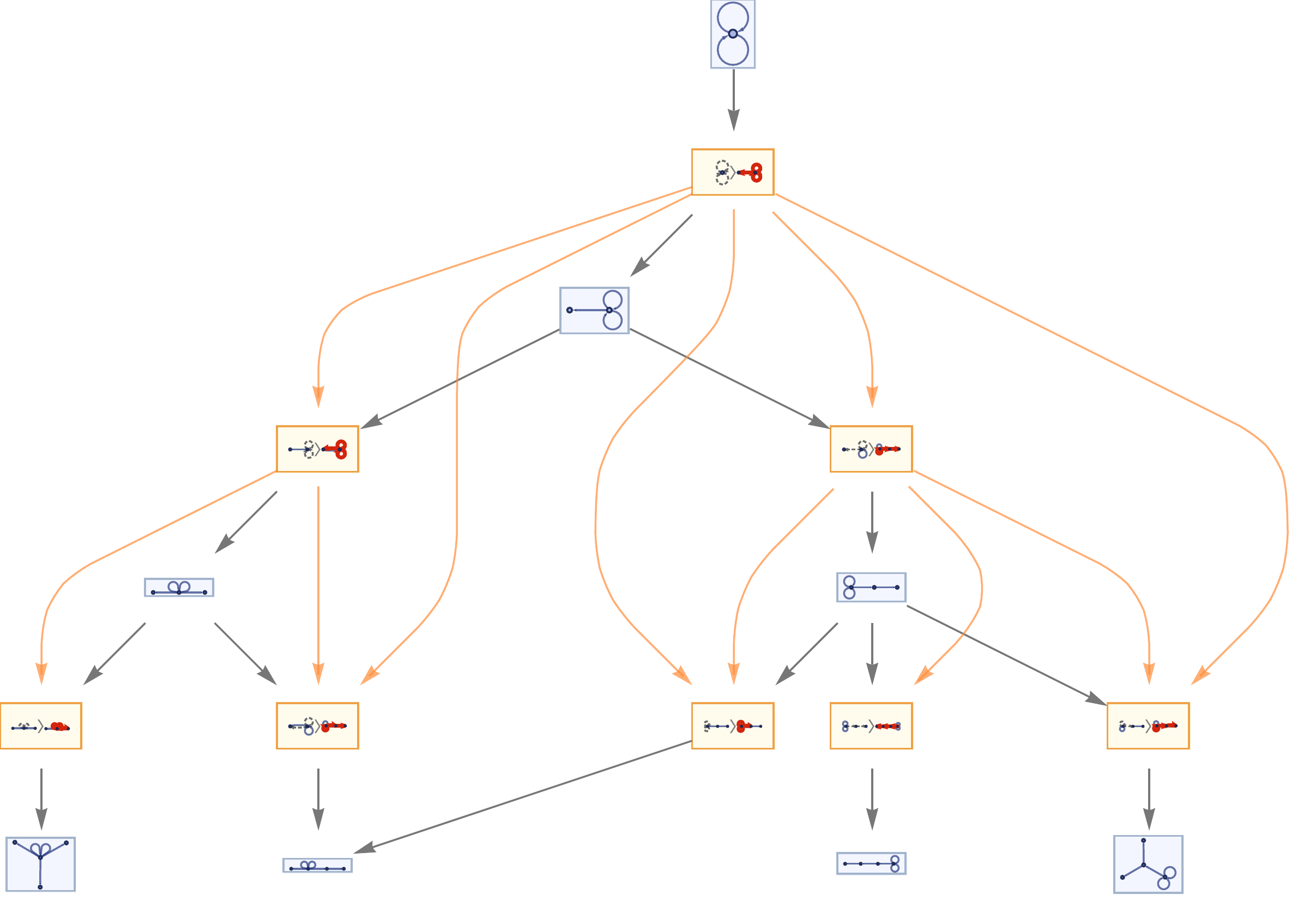}
\caption{On the left, the causal multiway system generated by the evolution of the set substitution system ${\left\lbrace \left\lbrace x, y \right\rbrace, \left\lbrace y, z \right\rbrace \right\rbrace \to \left\lbrace \left\lbrace x, y \right\rbrace, \left\lbrace y, z \right\rbrace, \left\lbrace z, w \right\rbrace \right\rbrace}$. On the right, the corresponding multiway evolution causal graph (with evolution edges shown in gray and causal edges shown in orange) for the same system.}
\label{fig:Figure70}
\end{figure}

However, there still remains a key piece of the argument missing if one wishes to make this correspondence precise: namely, the multiway analog of Bell causality. We begin by noting that the kinds of multiway evolution graph shown above (in which each state vertex is actually a representation of a causal set/causal network) are rather general, and can be generated by many processes other than classical sequential growth dynamics. In particular, since each application of a rewrite rule in a Wolfram model evolution has the effect of adding a single vertex to the corresponding causal network, one can also construct such \textit{causal multiway systems} via pure Wolfram model evolution, in which a superposition of all possible causal histories of the evolution is represented, as shown in Figure \ref{fig:Figure70}.

\begin{figure}[ht]
\centering
\includegraphics[width=0.395\textwidth]{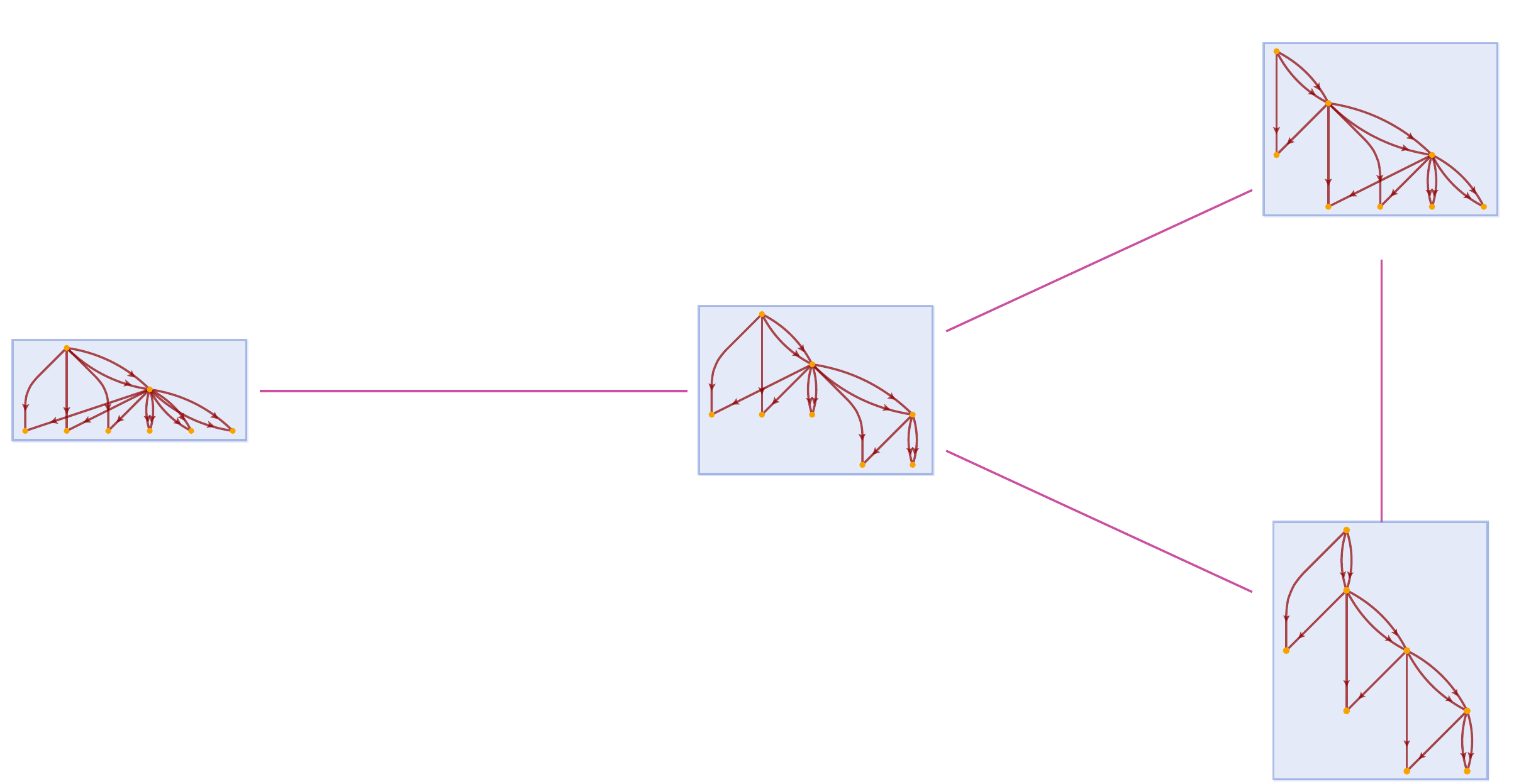}\hspace{0.1\textwidth}
\includegraphics[width=0.495\textwidth]{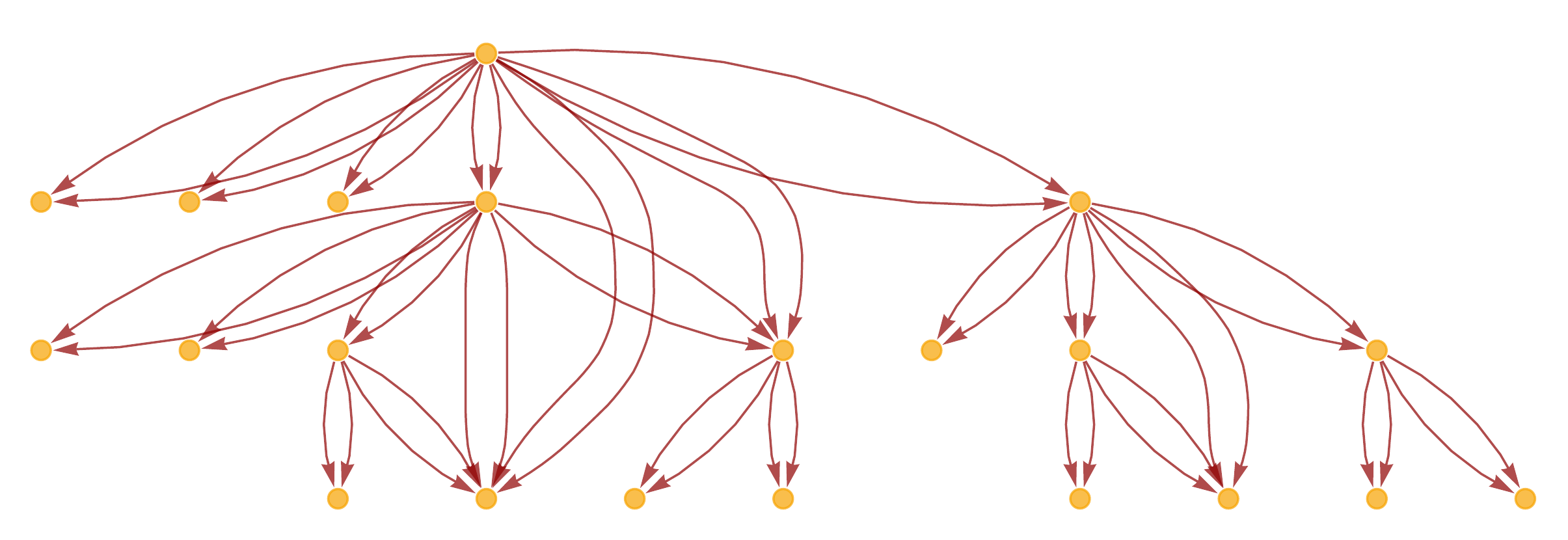}
\caption{On the left, the branchial graph witnessed after three evolution steps within the default foliation of the causal multiway system generated by the evolution of the set substitution system ${\left\lbrace \left\lbrace x, y \right\rbrace, \left\lbrace y, z \right\rbrace \right\rbrace \to \left\lbrace \left\lbrace x, y \right\rbrace, \left\lbrace y, z \right\rbrace \left\lbrace z, w \right\rbrace \right\rbrace}$. On the right, the corresponding multiway causal graph for the same system (produced by ``gluing'' the states in the branchial graph together).}
\label{fig:Figure71}
\end{figure}

Such \textit{causal multiway systems} have a great variety of very pleasant mathematical properties. For instance, the subtle distinction between global confluence of the underlying rewriting system and causal invariance of the multiway evolution is rendered trivial; since the merging criterion for state vertices is exactly causal network isomorphism, the multiway evolution is causal invariant if and only if the associated abstract rewriting system on the causal network is globally confluent. Moreover, the branchial graphs generated by causal multiway systems provide a very neat ``factorization'' of the multiway causal graph (i.e. the directed acyclic graph showing causal relationships between all events in the multiway system, including those on different branches of evolution history) into several ``singleway'' causal graphs (i.e. causal networks associated with single paths through the multiway system), as shown in Figure \ref{fig:Figure71}. Here, the undirected branchial edges correspond to the gluing maps between distinct singleway causal graphs, associated with common causal ancestry in the multiway system.

Within the context of this construction, the condition of Bell causality becomes simply a locality constraint, but rather than enforcing locality in physical space (as defined through spacelike hypersurfaces), it is enforcing locality in \textit{branchial space} (as defined through branchlike hypersurfaces). Specifically, we can formalize the condition of Bell causality as the statement that the classical transition probability (or, in the quantum case, the quantum transition measure) associated with the transition ${\mathcal{C}_n \to \mathcal{C}_{n + 1}}$, denoted ${\alpha_n}$, should depend only upon a certain ``precursor set'', denoted ${p_n \subset \mathcal{C}_n}$, and should be entirely independent of a ``spectator set'', denoted ${s_n \subset \mathcal{C}_n}$. Here, the ``precursor set'' is defined as being the set of all elements in the causal past of the element ${e_n}$ (i.e. the element being added in the transition ${\mathcal{C}_n \to \mathcal{C}_{n + 1}}$), and the ``spectator set'' is defined as being all elements in ${\mathcal{C}_n}$ that are not precursor events. For any non-empty spectator set ${s_n}$ satisfying ${\left\lvert s_n \right\rvert < n}$, the associated precursor sets:

\begin{equation}
\mathcal{C}_m = \mathcal{C}_n \setminus s_n \qquad \text{ and } \qquad \mathcal{C}_{m + 1} = \mathcal{C}_{n + 1} \setminus s_n,
\end{equation}
satisfying ${m + \left\lvert s_n \right\rvert = n}$, are related by a transition probability ${\alpha_m}$ for ${\mathcal{C}_m \to \mathcal{C}_{m + 1}}$ that is proportional to ${\alpha_n}$, with the Bell causality constraint thus being given by:

\begin{equation}
\frac{\alpha_n \left( \mathcal{C}_n \to \mathcal{C}_{n + 1} \right)}{\alpha_{n}^{\prime} \left( \mathcal{C}_n \to \mathcal{C}_{n + 1}^{\prime} \right)} = \frac{\alpha_m \left( \mathcal{C}_m \to \mathcal{C}_{m + 1} \right)}{\alpha_{m}^{\prime} \left( \mathcal{C}_m \to \mathcal{C}_{m + 1}^{\prime} \right)},
\end{equation}
with ${\mathcal{C}_n \to \mathcal{C}_{n + 1}^{\prime}}$ being an alternative transition from ${\mathcal{C}_n}$, and with the sets ${p_{n}^{\prime}}$, ${s_{n}^{\prime}}$ and:

\begin{equation}
\mathcal{C}_{m + 1}^{\prime} = \mathcal{C}_{n + 1}^{\prime} \setminus s_{n}^{\prime},
\end{equation}
and transition probability/measure ${\alpha_{n}^{\prime}}$ defined in the analogous way.

From here, we can see immediately that the precursor sets correspond precisely to those sets of events which share common causal ancestry in the causal multiway system, and are therefore directly connected by branchial edges. Thus, we can reformulate the condition of Bell causality as the condition that path weights of vertices (and hence, both classical probabilities and quantum measures) in the causal multiway system must depend solely upon those vertices that are immediately local in the associated branchial graph. A more thorough exploration of the properties of causal multiway systems and their implications for both classical and quantum causal set dynamics is currently planned as a future publication.

\section{Concluding Remarks}

It is hoped that the present article will constitute an initial step towards a much deeper investigation of the formal correspondence between the discrete spacetime formalisms presented by the causal set and Wolfram model programs. In particular, although some preliminary analytical and numerical work is presented here in an attempt to related various aspects of the two models (such as dimension estimation techniques, spatial metric estimation techniques, approaches to deriving discrete gravitational actions, approaches to deriving quantum dynamics, etc.), there remain many details still to be clarified. For instance, it would be extremely interesting to conduct far more systematic numerical stability and convergence comparisons of the Myrheim-Meyer, midpoint scaling, geodesic ball, geodesic cone, etc., dimension estimation algorithms; the Brightwell-Gregory, Rideout-Wallden, Eichhorn-Surya-Versteegen, hypergraph geodesic, etc. spacelike distance estimation algorithms; the Ollivier-Ricci, Forman-Ricci and discrete D'Alembertian approaches to discrete Ricci curvature estimation; the discrete Einstein-Hilbert (Wolfram model) and Benincasa-Dowker (causal set) actions; etc. (all across a suitably representative sample of both randomly- and algorithmically-generated causal sets).

On the more mathematical side, one might also conduct rigorous comparisons between the Green's functions on free scalar fields defined over causal sets and the description of matter fields in terms of localized topological obstructions in Wolfram model hypergraphs (a topic which has been entirely omitted here). A complete treatment of the GHY boundary terms from both the Benincasa-Dowker action and the full discrete Einsten-Hilbert action in causal set and Wolfram model systems, respectively, is also potentially worth pursuing (and also omitted here). Finally, a full investigation of the formalism of causal multiway systems, their implications for the causal structure of ordinary Wolfram model multiway systems, and the extent of their relationship to classical and quantum growth dynamics in causal sets is also planned as part of future work in this area.

\section*{Acknowledgments}

The author is deeply indebted to Stephen Wolfram for his encouragement and numerous suggestions, as well as to Tommaso Bolognesi and Fay Dowker for some useful conversations during the early stages of the present work.

\end{document}